%% file: polame.tex
\documentclass[a4paper,fleqn,usenatbib]{mnras}

\usepackage[T1]{fontenc}
\usepackage{ae,aecompl}

\usepackage{graphicx}	
\usepackage{amsmath}	
\usepackage{amssymb}	

\usepackage{rotating}
\usepackage{adjustbox}
\usepackage{lscape}
\usepackage{tabularx}
\usepackage{xcolor}
\usepackage{soul} 

\usepackage{newtxtext,newtxmath} 

\title[QUIJOTE-MFI wide-survey Galactic AME sources]
{QUIJOTE Scientific Results -- VII. Galactic AME sources in the QUIJOTE-MFI Northern Hemisphere Wide-Survey}
\input{polame_authors_and_institutes.tex}

\date{Accepted XXX. Received YYY; in original form ZZZ}

\pubyear{2023}

\begin{document}
\label{firstpage}
\pagerange{\pageref{firstpage}--\pageref{lastpage}}
\maketitle

\begin{abstract}
  The QUIJOTE-MFI Northern Hemisphere Wide-Survey has provided
  maps of the sky above declinations $-30^\circ$ at 11, 13, 17 and 19$\,$GHz. 
  These data are combined 
  with ancillary data to produce Spectral Energy Distributions 
  in intensity in the frequency range 0.4--3\,000$\,$GHz on a 
  sample of 52 candidate compact sources harbouring anomalous microwave emission (AME).
 We apply a component separation analysis at 1$^\circ$ scale on the full sample from which we identify 44 sources with high AME significance. We explore correlations between different fitted parameters on this last sample.
QUIJOTE-MFI data contribute to notably improve the characterisation of the AME spectrum, and its separation from the other components. In particular, ignoring the 10--20\,GHz data produces on average an underestimation of the AME amplitude, and an overestimation of the free-free component. We find an average AME peak frequency of 23.6 $\pm$ 3.6$\,$GHz, about 4$\,$GHz lower than the value reported in previous studies.
  The strongest correlation is found
  between the peak flux density of the thermal dust and of the 
  AME component. A mild correlation is found between 
  the AME emissivity  ($A_{\rm AME}/\tau_{250}$) and the interstellar radiation field.
  On the other hand no correlation is found between the 
  AME emissivity and the free-free radiation Emission Measure. 
  Our statistical results suggest that the interstellar 
  radiation field could still be the main driver of the 
  intensity of the AME as regards spinning dust 
  excitation mechanisms. On the other hand, it is not clear 
  whether spinning dust would be most likely associated with 
  cold phases of the interstellar medium rather than with hot 
  phases dominated by free-free radiation.
\end{abstract}








\section{Introduction}

A detailed knowledge of the sky emission properties in the frequency
range $\sim$1--3\,000$\,$GHz, from low-frequency (LF) bands at which the Galactic synchrotron 
emission generally dominates, to high frequency (HF) bands
at which the Galactic dust emission dominates, is crucial for a state-of-the art
characterization of the Cosmic Microwave Background (CMB) radiation both in
intensity and in polarization \citep[e.g.][]{planck2018, litebird2022}.
Understanding the properties of the Galactic foregrounds is essential in order to measure a possibly intrinsic polarization
signature in the CMB emission that could give insights about inflation 
scenarios.
This task is considered to be a serious challenge by both the community of
astronomers in quest of a B-mode detection  \citep[see][]{CLASS2015,polarbear2017,simons2019,actpol2020,cmbs4_2020,planck2018,bicep2021,lspe2021,groundbird2021,litebird2022,qubic2022} and the community of
astronomers interested to understand the spatial and spectral
 variations of the Galactic emission \citep[see][]{cbass2018,spass2019,mfiwidesurvey}.

In addition to synchrotron emission and thermal dust emission, the Galactic sky
also emits thermal \textit{bremmstrahlung} or free-free radiation, a radiation produced
by deceleration of electrons, and supposedly unpolarized \citep[][]{rybicki1979, trujillo-bueno2002}.  
Another type of radiation is the so-called Anomalous Microwave
Emission (AME) that was discovered about twenty five years ago
\citep[see][]{leitch1997,kogut1997,deoli98}. 
The AME is a diffuse component showing a
spectral bump detected over
almost the full sky in the frequency range 10--60$\,$GHz and peaking in flux density
around a central frequency of $\sim$ 30$\,$GHz. In this frequency range,
the synchrotron and free-free emission can dominate over the AME
emission while the thermal dust emission is expected to be negligible. 
The carriers and physical mechanisms producing AME
are not conclusively known yet, however theoretical emission mechanisms have been proposed
based on phenomenological interpretations of
correlations found between the AME radiation and other Galactic
template components. A review of these aspects and of the
proposed models in the literature is given by \cite{dickinson18}.
The main current paradigm is that electric dipole
emission from very small fast rotating spinning dust grains out of
thermal equilibrium could be the origin of this emission
\citep[see][]{draine98,ali2009,hoang10,ysard10}.
Recent advances on the development of another model, initially
proposed by \cite{jones2009} and exploring the
possibility that AME can be produced instead by thermal amorphous
dust are discussed by \cite{nashimoto2020a}
and \cite{nashimoto2020b}. The majority of these models predict very low levels of polarisation for the AME, this being supported by observational data \citep{lopez-caraballo2011,dickinson2011,rubino12,W44}.

Given its twofold role as a CMB contaminant and as a source of information about the physics of the ISM, it is important to make progress on the study of the observational properties of AME, and confronting them with theoretical models.
Galactic candidate AME sources were intensively discussed in
\cite{pir15} (herefater PIRXV). In that work the analysis of a sample of 98 compact candidate AME
sources distributed over the full sky provides significant detection ($>
5\sigma$) of AME for 42 sources, which reduces to safe detection of
AME for 27 sources once the potential contribution of thick free-free
emission from ultra compact \sc{H\,ii} \rm{} regions has been integrated to the analysis. In this work, we complete and revisit the sample of sources
observable from the Northern hemisphere. For this we use the QUIJOTE-MFI
wide-survey maps \citep{mfiwidesurvey}, which are crucial to pin down the AME spectrum at low frequencies, thence allowing a more reliable separation between the AME and free-free amplitudes \citep[e.g.,][]{poidevin18} than previous works, which systematically have overestimated the free-free emission and underestimated the AME amplitude.
Some of the sections in the present article closely follow those in PIRXV. In such cases we tried to use
similar section names so that the reader can easily refer to the
information provided by PIRXV and, as much as possible, we tried to avoid
redundancy with their explanations. All the calculations made for
our analysis are independent of those done by PIRXV.

The structure of the article is as follows: the data used for the analysis are presented in
Section~\ref{data}. The sample selection and fitting procedure used
for the Spectral Energy Distribution (SED) analysis are detailed in 
Section~\ref{sample_sel_and_fit}. Consistency checks obtained from the
comparison of our method with that used by PIRXV are also presented in that Section. 
The significance of the AME detection obtained from our analysis, 
potential contamination by UC\sc{H\,ii} \rm{} regions and robustness
and validation of our method are discussed in
Section~\ref{ame_regions}. Statistics on the parameters characterizing
the sample of regions that passed the validation tests 
are investigated in Section~\ref{statistics}.
A discussion is given in Section~\ref{discussion}.
Our results and conclusions are summarized in Section~\ref{summary}.
Additional plots showing low Spearman rank correlation 
coefficients (SRCCs) between some of 
the parameters obtained from the modelling of the SEDs, 
and mentioned in some of the above
sections, are presented in Appendix~\ref{append_additional_plots}.
All the parameters estimates obtained from the modelling 
of the SEDs, and additional information, obtained on the 
full sample, are tabulated in Appendix~\ref{append_seds_params}.
All the plots of the SEDs and the multicomponents models
are shown in Appendix~\ref{append_seds}. Finally, a summary 
of the SRCCs obtained between
all the pairs of parameters used to model the SEDs are given in
Appendix~\ref{append_tables}.

\section{Data}\label{data}

The maps used in this analysis are listed in Table~\ref{tab:surveydata}.
Details about the maps are given in the following subsections.


\begin{table*}
\begin{center}
\begin{tabular}{ccccccc}
\hline\hline
\noalign{\smallskip}
Frequency & Wavelength & Telescope$/$   & Angular Resolution  & Original & Calibration & References\\
\noalign{\smallskip}
[GHz]& [mm] & survey & [$^{\prime}$] & Units & Uncertaintiy [$\%$] & \\
\noalign{\smallskip}
\hline
\noalign{\smallskip}
  0.408 & 735.42& JB$/$Eff$/$Parkes & $\approx$ 60 &   [K$_{\rm  RJ}$]    & 10 &\citet{haslam82}\\
&&&&&& \citet{remazeilles15}  \\
  0.820 & 365.91& Dwingeloo               & 72      &    [K$_{\rm RJ}$]  & 10 &  \citet{berkhuijsen72}\\
  1.420 & 211.30& Stockert/Villa-Elisa & 36 &  [K$_{\rm RJ}$] & 10 &\citet{reich82}\\ 
&&&&&& \citet{reich86}\\
&&&&&& \citet{reich01}   \\
11.1 & 28.19 & QUIJOTE & 55.4 &[mK$_{\rm CMB}$] & 5 & \citet{mfiwidesurvey} \\
12.9 & 23.85 & QUIJOTE & 55.8 &[mK$_{\rm CMB}$] & 5 & \citet{mfiwidesurvey} \\
16.8 & 18.24 & QUIJOTE & 38.9 & [mK$_{\rm CMB}$]& 5 & \citet{mfiwidesurvey} \\
18.8 & 16.32 & QUIJOTE & 40.3 &[mK$_{\rm CMB}$] & 5 & \citet{mfiwidesurvey} \\
  22.8 & 13.16& WMAP 9-yr & $\approx$ 49 &  [mK$_{\rm CMB}$]& 3 &\citet{bennett13}     \\
  28.4 & 10.53& $Planck$ LFI & $32.29$ &   [K$_{\rm CMB}$]& 3 &\citet{cpp2015-1}    \\
  33.0 & 9.09& WMAP 9-yr & $\approx 40$ &     [mK$_{\rm CMB}$]& 3 &\citet{bennett13}    \\ 
  40.6 & 7.37& WMAP 9-yr & $\approx 31$ &    [mK$_{\rm CMB}$]& 3 &\citet{bennett13}   \\
  44.1 & 6.80& $Planck$ LFI &$27$ &   [K$_{\rm  CMB}$]& 3 &\citet{cpp2015-1}   \\
  60.8 & 4.94& WMAP 9-yr & $\approx 21$ &   [mK$_{\rm CMB}$]& 3 &\citet{bennett13}    \\
  70.4& 4.27& $Planck$ LFI & $13.21$ &  [K$_{\rm CMB}$]& 3 &\citet{cpp2015-1}    \\
  93.5 & 3.21& WMAP 9-yr & $\approx 13$&     [mK$_{\rm CMB}$]& 3 &\citet{bennett13}  \\
  100 & 3.00& $Planck$ HFI &$9.68$ & [K$_{\rm CMB}$]& 3 &\citet{cpp2015-1}   \\
  143& 2.10& $Planck$ HFI &$7.30$ & [K$_{\rm CMB}$]& 3 &\citet{cpp2015-1}   \\
  217 & 1.38& $Planck$ HFI & $5.02$& [K$_{\rm CMB}$]& 3 &\citet{cpp2015-1}   \\
  353 & 0.85& $Planck$ HFI & $4.94$ & [K$_{\rm CMB}$]& 3 &\citet{cpp2015-1}      \\
  545 & 0.55& $Planck$ HFI & $4.83$ &  [MJy/sr]& 6.1 &\citet{cpp2015-1}     \\
  857& 0.35& $Planck$ HFI & $4.64$&  [MJy/sr]& 6.4 &\citet{cpp2015-1}     \\
  1249 & 0.24& COBE-DIRBE & $\approx 40$ &  [MJy/sr]& 11.9 &\citet{hauser98}      \\
  2141 & 0.14& COBE-DIRBE & $\approx 40$ &  [MJy/sr]& 11.9 &\citet{hauser98}      \\
  2998 & 0.10& COBE-DIRBE & $\approx 40$  &  [MJy/sr]& 11.9&\citet{hauser98}       \\
\noalign{\smallskip}
\hline\hline
\end{tabular}
\end{center}
\normalsize
\caption{List of surveys and maps used in our analysis.
}
\label{tab:surveydata}
\end{table*}

\subsection{QUIJOTE Data}

The data used at frequencies 11, 13, 17 and 19 GHz come from the first release of the QUIJOTE wide survey maps \citep[][]{mfiwidesurvey}. These maps were obtained from $9\,200$\,h of data collected over 6 years of observations from 2012 to 2018 with the Multi-Frequency Instrument (MFI) on the first QUIJOTE telescope, from the Teide Observatory in Tenerife, Canary Islands, Spain at an altitude of $2\,400$ meters above sea level, at 28.3$^\circ$ N and 16.5$^\circ$ W. These observations were performed at constant elevations and with the telescope continuously spinning around the azimuth axis (the so-called ``nominal mode’’) to obtain daily maps of the full northern sky. After combination of all these data we obtained maps covering $\sim 70\%$ of the sky and with sensitivities in total intensity between 60 and 200~$\mu$K/deg, depending on the horn and frequency and sensitivites, down to $\sim 35\mu$K/deg, in polarisation. Full details on these maps, and multiple characterisation and validation tests, are given in \citet{mfiwidesurvey}, while the general MFI data processing pipeline will be described in \citet{mfipipeline}.

The MFI consists of 4 horns, two of them (horns 1 and 3) covering a 10-14~GHz band with two outputs channels centred at 11 and 13~GHz, and two other ones (horns 2 and 4) covering the 16-20~GHz band with two output channels at 17 and 19~GHz \citep[][]{mfipipeline}. Due to a malfunctioning of horn 1 in polarization during some periods, all the scientific QUIJOTE papers associated with this release make use of horn 3 only at 11 and 13~GHz. Although this paper uses intensity data only, we follow the same criterion and use only horn 3, which is much better characterised\footnote{Note that the analysis in intensity presented in this paper benefits from a sufficiently large signal-to-noise ratio and therefore a good characterisation of systematics is more relevant.}. At 17 and 19 GHz we combine data from horns 2 and 4 through a weighted mean, using predefined constant weights\footnote{Instead of doing a pixel-by-pixel combination at the map level, we extract flux densities independently and combine the derived flux densities.} \citep[][]{mfiwidesurvey}. Finally, it must be noted that, due to the use of the same low-noise amplifiers, the noises from the lower and upper frequency bands of each horn are significantly correlated \citep[see Section 4.3.3 in][]{mfiwidesurvey}. In principle this correlation should be accounted for in any scientific analysis that uses spectral information. However, we have checked that neglecting them introduces a small effect on the results presented in this paper. AME parameters are the most affected, and we have checked that accounting for this correlation introduces differences in these parameters that are typically below the 3\% level. Therefore, for the sake of simplicity we decided to use the four frequency points (nominal frequencies 11.1, 12.9, 16.8 and 18.8~GHz) in the analysis as independent data points. We assume a 5\% overall calibration uncertainty of the QUIJOTE MFI data, which is added in quadrature to the statistical error bar. There is compelling evidence that this $5\,\%$ value, which is driven by uncertainties in the calibration models, is sufficiently conservative \citep{mfipipeline,mfiwidesurvey}. 

\subsection{Ancillary Data}

\subsubsection{Low frequency ancillary data}\label{sec:low_ancillary_data}

At low frequencies we use a destriped version \citep[][]{platania2003} of the all-sky 408$\,$MHz map of \citet{haslam82}, 
the Dwingeloo survey map at 0.820$\,$GHz of \citet{berkhuijsen72}, 
and the 1.420$\,$GHz map of \citet{reich82}.
Since our study is focused on compact candidate AME sources 
we prefer to use the all-sky 408$\,$MHz
destriped map of \citet{haslam82}. The \citet{platania2003} version 
of this map is used for consistency with previous
QUIJOTE papers, but we have checked that the results 
are consistent with those obtained using the map 
provided by \citet{remazeilles15}. The \cite{jonas1998} map at 2.326$\,$GHz, which was used in PIRXV, measures $I+Q$. Therefore it would lead to residuals in polarised regions, and we prefer not to use it.

Some of the considered sources are not well sampled or not
included in the footprint of some of the ancillary maps.
Therefore, for a given source a map is used only if all 
pixels 
within a circular region of $3^{\circ}$ radius are covered. 
We noted that, for a subset of compact sources, the 
map at 1.420$\,$GHz shows a misscentering of the emission by more than half a degree with respect to other low-frequency maps. 
For that reason we prefer not to use that map in the analysis 
of G$059.42-00.21$, G$061.47+00.11$ and G$099.60+03.70$. 
The 1.420$\,$GHz map is calibrated to the full beam, and therefore 
we apply the full-beam to main-beam recalibration factor of 1.55 
for compact sources derived by \citet{reich88}. 
Overall, we assume a 10$\,\%$ uncertainty in the radio data at 
low frequency, which encompasses intrinsic calibration uncertainties as well as issues related with beam uncertainties and recalibration factors.

\subsubsection{WMAP Maps}

At frequencies of 23, 33, 41, 61, and 94$\,$GHz, we use the intensity
maps from the 9-year data release of the \textit{WMAP} satellite \citep{bennett13}.
All the maps were retrieved from the  LAMBDA 
database.\footnote{Legacy Archive for
  Microwave Background Data Analysis, {\tt  http://lambda.gsfc.nasa.gov/}.}
For all the maps we assume a 3$\%$ overall calibration uncertainty. 
The uncertainty in \textit{WMAP}'s amplitude calibration is much better, however here we use 3$\%$ to account for other systematic effects like uncertainties in the beams or bandpasses (which in turn lead to uncertainties in the colour corrections) that will have a direct effect on our derived flux densities.

\subsubsection{Planck Maps}

Below 100 GHz intensity 
maps are available at frequencies
28, 44, and 70$\,$GHz. They were obtained 
with the Low-Frequency Instrument
(LFI) on board of the \textit{Planck} satellite \citep{cpp2015-1}.  
We use the second public release version of the intensity maps as provided by the 
\textit{Planck} Legacy Archive (PLA\footnote{\textit{Planck} Legacy Archive (PLA) {\tt
    http://pla.esac.esa.int/pla/}.}).
Above 100 GHz we use the second data release version
of the intensity maps obtained with the 
High-Frequency Instrument (HFI) on board 
the \textit{Planck} satellite \citep{cpp2015-1} at 
frequencies centred at
100, 143, 217, 353, 545, and 857$\,$GHz. We have checked that using the third data release (PR3) leads to differences in the derived flux densities typically below $0.3\,\%$ for most of the frequencies and therefore have no impact in the final results presented in this paper.
The Type 1 CO maps \citep{cpp13} 
were used to correct the 100, 217, and 353~GHz intensity maps for 
contamination introduced by the CO rotational transition lines
(1-0), (2-1) and (3-2), respectively. 
We assume an overall calibration uncertainty of 3 $\%$ for the
LFI data, and also for the HFI data at frequencies lower than 
or equal to 353$\,$GHz,
a value of 6.1$\%$ at 545$\,$GHz, and
a value of 6.4$\%$ at 857$\,$GHz \citep{Planck2015-viii}.

\subsubsection{High frequency ancillary data}\label{sec:high_ancillary_data}

In the FIR range we use the Zodi-Subtracted Mission Average
(ZSMA) \textit{COBE}-DIRBE maps \citep{hauser98} at 240~$\mu$m (1249$\,$GHz), 
140$\,\mu$m (2141$\,$GHz), and 100$\,\mu$m (2997$\,$GHz).
We assume an 11.9$\%$ overall calibration uncertainty in the data at
these frequencies\footnote{11.9$\%$ is the calibration uncertainty for the 240~$\mu$m according to \cite{hauser98}, and we consider the same value for all bands.}.

\section{Sample selection and SED fitting}\label{sample_sel_and_fit}

In the following section we describe the process followed to build the sample of the candidate compact Galactic AME sources. Details about aperture photometry used to build the SEDs are given in Section~\ref{aperture_photometry}. The modelling used to analyse the SED of each candidate AME source is detailed in Section~\ref{model_fitting}. Finally, a consistency test is investigated and a comparison of our analysis, including the QUIJOTE maps, with the analysis obtained by \citet{pir15} on the sample of sources common to both studies is given in Section~\ref{comparisons_with_pirxv}.

\input{polametex_table_sources_2.txt}

\begin{figure*}
\begin{center}
\vspace*{2mm}
\centering
\includegraphics[width=180mm,angle=0]{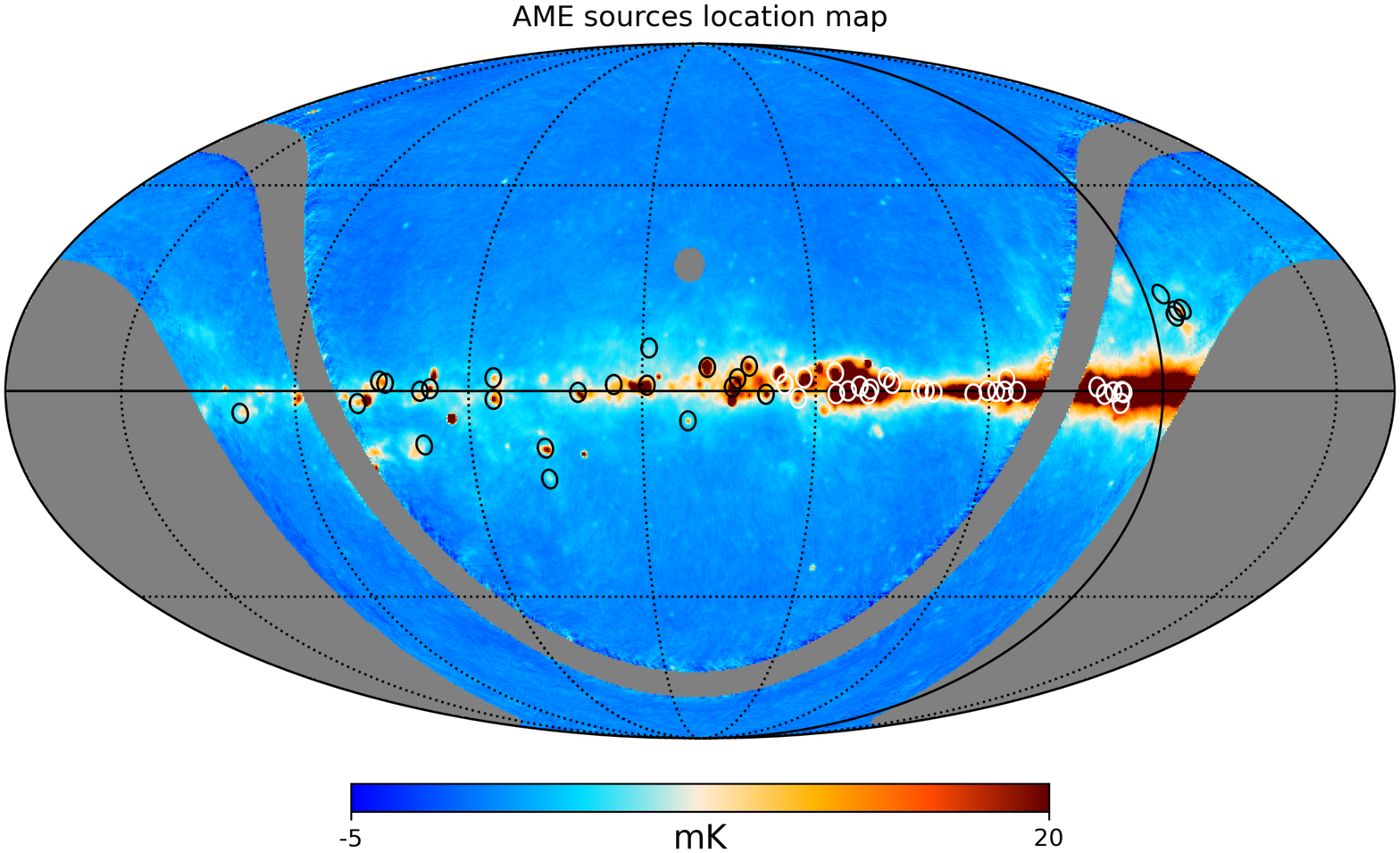}
\vspace*{-0.5cm}
\caption{AME sources location in the Galaxy displayed on top of the QUIJOTE-MFI 11\,GHz wide survey map at 1 degree resolution. Coordinates are listed in Table~\ref{tab:listofclouds}. The map is centred at position $(l,b)= (120^\circ, 0^\circ)$.}
\label{fig:cloudslocation}
\end{center}
\end{figure*}

\subsection{AME sources sample}

To build the sample of candidate AME sources, 
we use the list of sources selected and discussed 
in PIRXV as a reference. In their work, this list 
was obtained by using 3 different methods. One method was 
to identify sources already known from the literature and 
add them to a sample.
Another method was to produce a 1$^{\circ}$-smoothed 
map of residuals at 28.4$\,$GHz,
by subtracting off synchrotron, free-free, thermal dust,
and CMB components. A 5$^{\circ}$-smoothed version of this map was also created and subtracted from the 1$^{\circ}$-map in order to minimise diffuse emission. Bright and relatively compact sources were then identified in that map.
In a third method, an initial sample was built by using the
SExtractor \citep{bertin1996} software to detect bright sources 
in the 70$\,$GHz \textit{Planck} CMB-subtracted map. 
This sample was cross-correlated with 28.4$\,$GHz and 
100$\,$GHz catalogs obtained using the same technique. 
The output catalog was filtered to remove sources associated with
radio galaxies, including a small number of known 
bright supernova remnants and planetary
nebulae. Visual inspection was conducted on preliminary SEDs
obtained from the 1$^{\circ}$-smoothed maps in order to filter 
out the regions that were
not showing a peak at 30$\,$GHz on scales $\lesssim 2 ^{\circ}$ 
and to define the final sample of 98 candidate AME 
sources analysed and discussed in PIRXV.  

\begin{figure*}
\begin{center}
\vspace*{2mm}
\centering
\includegraphics[width=155mm,angle=0]{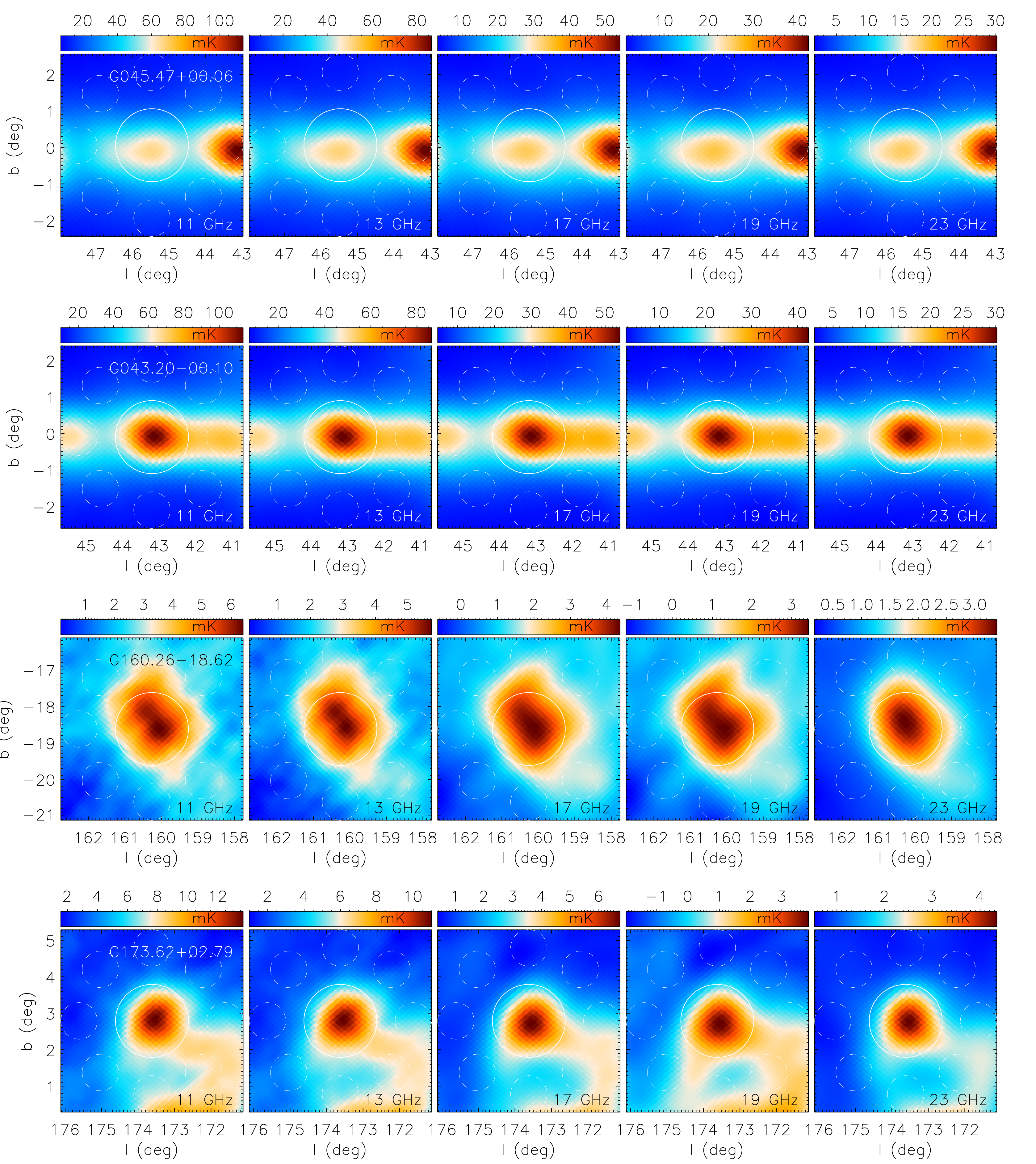}
 \vspace{0.02cm}
\caption{Subsample of $1^{\circ}$ smoothed intensity maps in Galactic coordinates. Fom left to right: QUIJOTE-MFI intensity maps at 11$\,$GHz (horn 3), 13$\,$GHz (horn 3), 17$\,$GHz (horn
  2 and 4), 19$\,$GHz (horn 2 and 4),
  and \textit{WMAP} intensity maps at 23$\,$GHz. From top to bottom
  the sources shown are the well-known Galactic 
  supernova remnant NRAO601 (G045.47+00.06),
 star forming region W49 (G043.20-00.10),
  Perseus molecular cloud (G160.26-18.62) and the cluster S235 (G173.62+02.79). 
In each plot the central circle shows the aperture used to obtain the density flux estimates. The eight dashed circles show the positions of the apertures used to calculate the uncertainties on these fluxes as explained in section~\ref{aperture_photometry}. 
}
\label{fig:quijote_int_maps9}
\end{center}
\end{figure*}

Of these 98 sources, 42 are well observed at all QUIJOTE 
frequencies of the MFI wide survey and are 
therefore included in our sample.
Additional sources that are not included in the sample analysed 
by PIRXV have been identified from
catalogs and lists of molecular clouds regions available
in the literature. This was done with the SCUPOL catalog 
that compiles thermal dust polarimetry information on small 
scales ($\approx 14{\arcsec}$) provided by \cite{matthews2009}, 
with the list of molecular clouds toward which
Zeeman measurements provide magnetic field line-of-sight (LOS)
estimates obtained by \cite{crutcher1999}, and with 
the molecular cloud catalog of \cite{lee2016}. 
In this way 10 additional candidate AME sources have been
identified. The maps of these sources that are
not already included in PIRXV's catalog were
inspected by eye at all available frequencies 
between 0.4$\,$GHz and 3000$\,$GHz and preliminary SEDs 
were built in order to look for the 
presence of a bump in the frequency range 10 -- 60$\,$GHz. 
The location of the final sample of candidate AME regions selected 
for our analysis is shown superimposed on the QUIJOTE 11$\,$GHz
Galactic full sky map in Figure~\ref{fig:cloudslocation}.
Their names, coordinates and additional information 
are displayed in Table~\ref{tab:listofclouds}. The final sample
contains a total of 52 sources. QUIJOTE-MFI intensity 
maps at 11, 13, 17 and 19 $\,$GHz and WMAP 22.7$\,$GHz 
intensity maps are displayed in 
Figure~\ref{fig:quijote_int_maps9} for a sample of
sources. Each source clearly shows similar intensity distribution patterns across the different frequency survey.

\subsection{Aperture photometry} \label{aperture_photometry}

In this work we conduct a component separation
analysis of the various components in intensity contributing to the
total emission of each source based on a SED analysis.
In intensity this method consists in calculating the total 
emission of a given source at each frequency. Once a SED has been
calculated one can use modelling to assess the fraction of the
total intensity emission associated with the different 
components (synchrotron, free--free, thermal dust, and AME) 
at all frequencies. SED modelling analysis has
been widely used in the literature \citep[e.g.,][]{watson2005,
  per20, lopez-caraballo2011,pir15,perseus,W44, poidevin18}.

The maps of pixel size $N_{\mathrm{side}}=512$ in the 
HEALPix\footnote{ \url{https://sourceforge.net/projects/healpix/}} 
pixelization scheme \citep[see][]{Gorski05} are first smoothed 
to $1^{\circ}$. To calculate the total emission at each 
frequency, the maps in CMB thermodynamic units (K$_{\rm CMB}$) 
are first converted to Rayleigh-Jeans (RJ) units
(K$_{\rm RJ}$) at the central frequency, then all 
the maps are converted to units of 
Jy pixel$^{-1}$ using $S = 2k_{b} T_{\rm RJ}\Omega\nu^{2}/c^{2}$, 
where $k_{b}$ is the Boltzmann constant, $T_{\rm RJ}$, is 
the Rayleigh-Jeans temperature, $\Omega$ is the solid angle 
of the pixel, $\nu$ is the frequency and $c$ is the 
speed of light. 
The pixels are then summed in the aperture covering the region
of interest to obtain an integrated flux density. An estimate 
of the background is subtracted using a median estimator of 
pixels lying in the region defined as the background region.

In Section~\ref{comparisons_with_pirxv} we provide some comparisons
with the results obtained by PIRXV. To do so, we use the 
same apertures and annulus used in that paper, i.e.
$r_{\rm APERTURE}=60^{\arcmin}, 
r_{\rm  ANNULUS(IN)}=80^{\arcmin}$ and
$r_{\rm  ANNULUS(OUT)}=100^{\arcmin}$. 
This method, also used in previous works, relies on the 
pixel-to-pixel scatter in the background annulus to obtain an 
estimate of the uncertainty in the flux density estimate. This 
technique is straightforward in the case of uncorrelated noise. 
However, in our case there is pixel-to-pixel correlated noise, 
due to instrumental 1/f noise and to beam-averaged sky background 
fluctuations, whose correlation function is not easy to be reliably 
characterised. We instead apply aperture photometry at the central
position of each source in the standard manner, and then the
calculations are repeated eight times such that we perform flux-density integrations on eight independent disks of radius $r_{\rm APERTURE}=30^{\arcmin}$ with central coordinates distributed along a circle with radius 2$^\circ$ around the source (as shown in
Figure~\ref{fig:quijote_int_maps9}). The final uncertainty is 
obtained from the scatter of these eight flux-density estimates.
This procedure is used for all sources except for the 
California region for which the background structure is 
complex and was producing bad fits such that $\nu_{\rm AME}$= 60.0 $\pm$0.0$\,$GHz, i.e. the prior upper limit. For that 
region we therefore use the same aperture and background annulus 
as in PIRXV
and we expect our uncertainties on the fluxes of this region
to be slightly underestimated.

\subsection{Model fitting} \label{model_fitting}

For each source the flux density $S$ from the aperture photometry is
fitted by a simple model consisting of the free-free, synchrotron (if appropriate), thermal dust, AME and CMB components:
 \begin{equation}
S_{\rm total} = S_{\rm ff} + S_{\rm sync} + S_{\rm td} + S_{\rm AME}
+S_{\rm CMB}. 
\end{equation}

The free--free spectrum shape is fixed and the free-free flux density,
$S_{\rm ff}$, is calculated from the brightness 
temperature, $T_{\rm ff}$, using the expression:   
\begin{equation}
S_{\rm ff} =\frac{2 k_{b} T_{\rm ff} \Omega \nu^{2}}{c^{2}},
\end{equation}
where $\Omega$ is the solid angle of our $60^{\arcmin}$ aperture. 
The brightness temperature is calculated with
the expression: 
\begin{equation}
T_{\rm ff} =T_{\rm e} (1-e^{-\tau_{\rm ff}}),
\end{equation}
where, following \cite {draine11} the optical depth, $\tau_{\rm ff}$, is given by 
\begin{equation}
\tau_{\rm ff} =5.468 \times 10^{-2} T_{\rm e}^{-1.5} \nu_{9}^{-2} \text{EM} g_{\rm ff},
\label{eq:tau_ff}
\end{equation}
where $T_{\rm e}$ is the electron temperature in 
Kelvin, $\nu_9$ is the frequency in GHz units,
EM is the Emission Measure in units of pc cm$^{-6}$, 
and $g_{\rm ff}$ is the
Gaunt factor, which is approximated as:
\begin{equation}
g_{\rm ff} =\rm ln ( \exp [ 5.960 - \frac{\sqrt3}{\pi} \rm ln ({Z_{i}
  \nu_{9}} T_{\rm e,4}^{-3/2}) ]+ e),
\end{equation} 
where the charge is assumed to be 
$Z_{i}=1$ (i.e., hydrogen plasma) and $T_{\rm e,4}$ 
is in units of $10^{4}$\,K.
Our best estimate for the electron temperature is the median
value of the \texttt{Commander} template within the aperture used on each source \citep[][]{Planck2015results_x}. These values lie in the range 5\,458--7\,194\,K. 
The only remaining free parameter associated with the
free--free component is the free--free amplitude, which can
be parameterized by the effective EM.

Equation~\ref{eq:tau_ff} tells that the turnover frequency that marks the transition between the optically-thick and optically-thin regimes ($\tau_{\rm ff}\approx 1$) depends on the emission measure (as EM$^{1/2}$) and on the electron temperature. In order to properly trace the degeneration between the free-free amplitude and the turnover frequency, instead of working with integrated quantities we would have to reconstruct EM along individual lines of sight inside each region and then integrate. Given the non-linear dependency of the flux density on EM, the two procedures are not equivalent, and this typically results in our fitted spectra having smaller turnover frequencies. For this reason, in cases where the data clearly shows the turnover frequency to be above 0.408 GHz (see e.g. G$015.06-00.69$ in Figure~\ref{fig:sed_int1}), in order to avoid the free-free (AME) amplitude to be biased low (high) we do not use in the fit the points with frequencies below 1.42$\,$GHz (depicted in these cases by a blue asterisk in Figure~\ref{fig:sed_int1}).

The synchrotron component is fitted by a single power law given by:
\begin{equation}
S_{\rm sync} = S_{\rm synch, 1GHz} \cdot \left(\frac{\nu}{\rm GHz}\right)^{\alpha_{\rm synch, int}}, 
\end{equation}
where the two parameters that are fitted for are the spectral index, $\alpha_{\rm synch, int}$, and the amplitude at 1$\,$GHz, $S_{\rm synch, 1GHz}$. 
This synchrotron component is included in the 
fits only for a few sources (G$010.19-00.32$, G$012.80-00.19$, 
G$037.79-00.11$, G$040.52+02.53$, G$041.03-00.07$ and
G$045.47+00.06$) as indicated in 
Table~\ref{tab:seds_fit_parameters1}. This choice was based on the slope of the low-frequency flux densities. The first three and the last of these sources are SNRs, as listed in Table~\ref{tab:listofclouds}. There is yet another source classified as SNR in our sample, G$118.09+04.96$. However the low-frequency data do not show any hint of synchrotron emission in this source, and actually the addition of this component to the fit has no impact on the fitted AME spectrum.

The CMB is modelled using the differential of a blackbody 
at $T_{\rm CMB}$ = 2.7255 K \citep[][]{fixsen2009}:
\begin{equation}
S_{\rm CMB} = \eta \frac{2 k_{b} \Omega \nu^{2}}{c^{2}} \Delta T_{\rm CMB}, 
\end{equation}
where 
$\eta = x^{2} \cdot $exp$(x)/($exp$(x)-1)^{2}$ and  $x=h\nu/(k_{b}T_{\rm CMB})$ is the conversion between thermodynamic and RJ brightness temperature, and
$\Delta T_{\rm CMB}$ 
is the CMB fluctuation temperature in thermodynamic units. 

 Spinning dust models have many free parameters, which are extremely difficult to constrain jointly. As a result, using a phenomelogical model, which traces well the data and the typical spinning dust models, is common practice in the field. In this work the AME component is fitted by the 
 phenomenological model consisting of an
 empirical log-normal approximation, first proposed by
 \cite{stevenson2014}. 
The log-normal model is described by the following equation:
\begin{equation} \label{eq_parabola}
S_{\rm AME} = A_{\rm AME}  \cdot \rm exp \left( -\frac{1}{2} \cdot \left[
    \frac{\rm ln(\nu / \nu_{\rm AME})}{W_{\rm AME}} \right ]^2, \right)
\end{equation} 
where the three free parameters are the width of the
parabola $W_{\rm AME}$, 
the peak frequency $\nu_{\rm AME}$, and the amplitude of the
parabola at the peak frequency $A_{\rm AME}$. 
Some previous works \citep[e.g.,][]{W44} have used a different phenomelogical model proposed by \citet{bonaldi07}. However we note that in this model the AME peak frequency and the AME width are not independent parameters. Hence, we prefer to use the \citet{stevenson2014} model, which does not have this coupling.

The thermal dust emission is modelled by a single-component
modified blackbody relation of the form, 
\begin{equation} \label{eq_thermal_dust}
S_{\rm td} = \tau_{250} (\nu/1200 \rm GHz)^{\beta_{\rm dust}}\it B_{\nu}(T_{\rm dust}), 
\end{equation} 
where $\tau_{250}$ is the
averaged dust optical depth at 250$\,\mu$m, $\beta_{\rm dust}$ 
is the averaged thermal dust emissivity, and $B_{\nu}$ is 
the Planck's law of the black-body radiation at the temperature, $T_{\rm dust}$,
which is the averaged dust temperature.


The fit procedure includes priors on some of the parameters and
consists of a minimization process using non-linear least-squares
fitting in Interactive Data Language (IDL) with MPFIT \citep[][]{markwardt2009}. The errors on the fitted parameters in this method are computed from the input data covariance, and neither the goodness of the fit nor parameter degeneracies are taken into account. It must then be noted that parameter errors are sometimes underestimated. This is the case for instance when it is hard to separate the free-free and the spinning dust components. In those cases the errors on EM and $A_{\rm AME}$ will tend to be underestimated. A more reliable error estimate would require full sampling of the probability distribution and will be considered in future similar studies. 
Such a method should help to refine our results but would not change our main conclusions.

Flat priors are used on the following list of parameters: $T_{\rm dust}$, $\beta_{\rm dust}$, $\Delta T_{\rm CMB}$, $A_{\rm AME}$, $\nu_{\rm AME}$ and $W_{\rm AME}$. Dust temperatures, $T_{\rm dust}$, are allowed in the temperature range 10--35$\,$K while dust index emissivities, $\beta_{\rm dust}$, are allowed in the range 1.2--2.5. Both priors are representative of average dust physical conditions in the diffuse interstellar medium (ISM) and molecular clouds. The CMB fluctuation temperatures, $\Delta T_{\rm CMB}$, are allowed to vary in the temperature range $\pm$125$\,$K. This range of values is representative of the CMB fluctuation temperatures one can expect when operating aperture photometry including a background subtraction. The AME amplitude, $W_{\rm AME}$, is allowed to vary in the range 0--300$\,$Jy. The AME frequency, $\nu_{\rm AME}$, is allowed to vary in the frequency range 10--60$\,$GHz, and for the width parameter $W_{\rm AME}$, we use a prior 0.2--1.0. While spinning dust models computed for representative ISM environments \citep{draine98,ali2009} typically have maximum widths corresponding to $W_{\rm AME}\approx 0.7$ we prefer not to be so strongly model constrained and allow for slightly wider AME spectra. More details on the effect of the priors used to model the AME are discussed in Section~\ref{robustness} and Table~\ref{tab:seds_fit_parameters_dr21}, in Section~\ref{ame_peak_frequency_hist}, and in Section~\ref{ame_width_hist}.

Colour corrections for QUIJOTE, \textit{WMAP}, \textit{Planck} and DIRBE, which depend on the fitted spectral models, have been applied using an iterative procedure that involves calls to a specifically developed software package. This code, which will be described in more detail in \citet{mfipipeline}, uses as input the fitted spectral model in each iteration, which is convolved with the experiment bandpass. Colour corrections are typically $\lesssim 2\%$ for QUIJOTE, WMAP and \textit{Planck-LFI}, and $\lesssim 10\%$ for \textit{Planck-HFI} and DIRBE, which have considerably larger bandwidths. Colour corrections for low-frequency surveys, which have much narrower bandpasses, are not necessary.

\begin{figure*}
\begin{center}
\vspace*{2mm}
\hspace*{-0.8cm}
\centering
\includegraphics[width=200mm,angle=0]{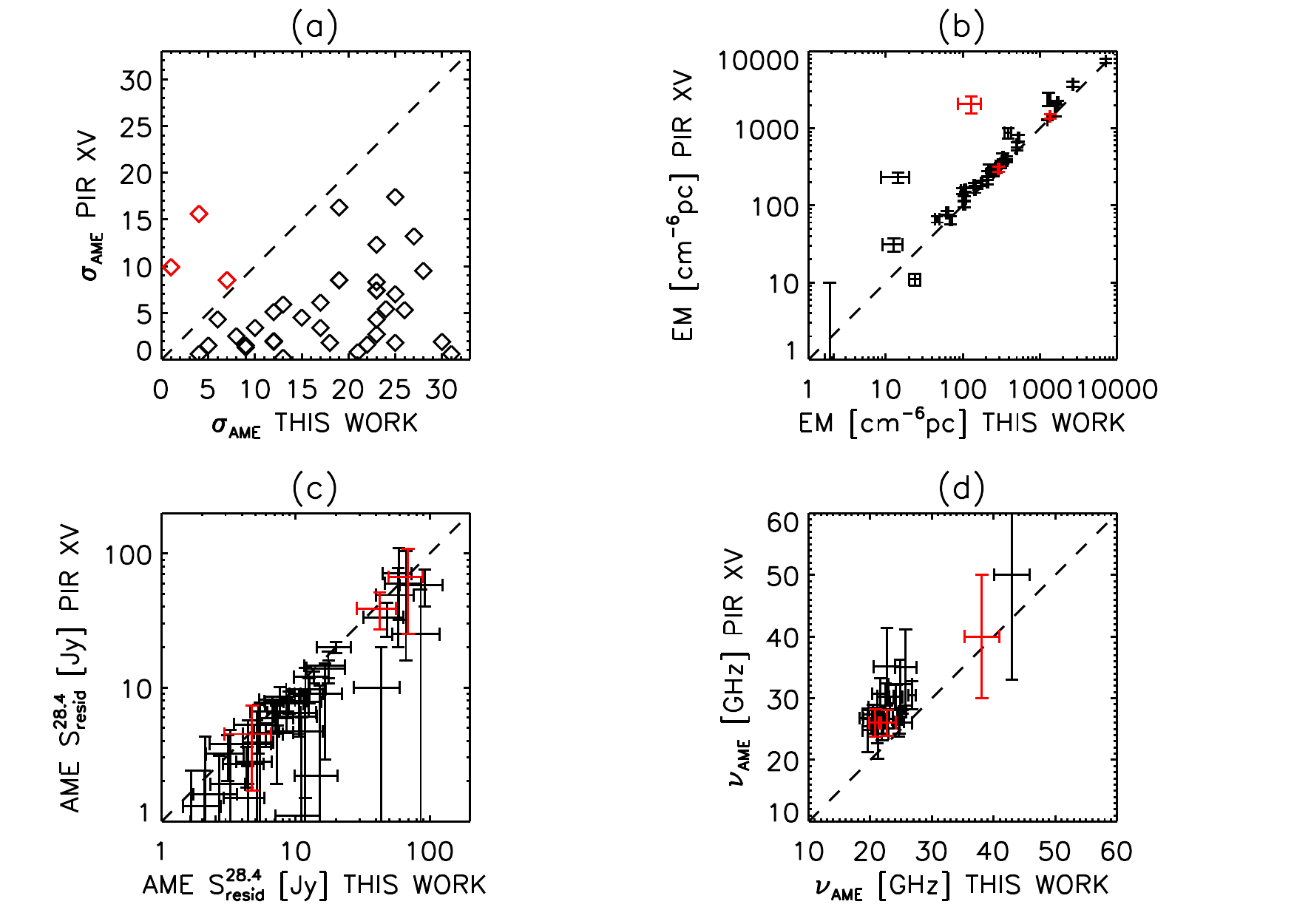}
\vspace*{-0.5cm}
\caption{Comparison between the results obtained with our analysis and in PIRXV for the AME significance $\sigma_{\rm AME}$ defined as the ratio of the flux density of AME at the frequency peak position divided by the uncertainty on this estimate 
(a), the emission measure EM (b), the residual AME flux density at 28.4$\,$GHz (c) and the AME peak frequency (d). Our analysis includes the QUIJOTE-MFI data. The 
  data shown in red correspond to sources for which the
  significance of the AME detection is higher in PIRXV than in our analysis.}
\label{fig:comp_pirxv}
\end{center}
\end{figure*}

\begin{figure*}
\begin{center}
\vspace*{2mm}
\centering
\hspace*{0.cm}
\includegraphics[width=80mm,angle=0]{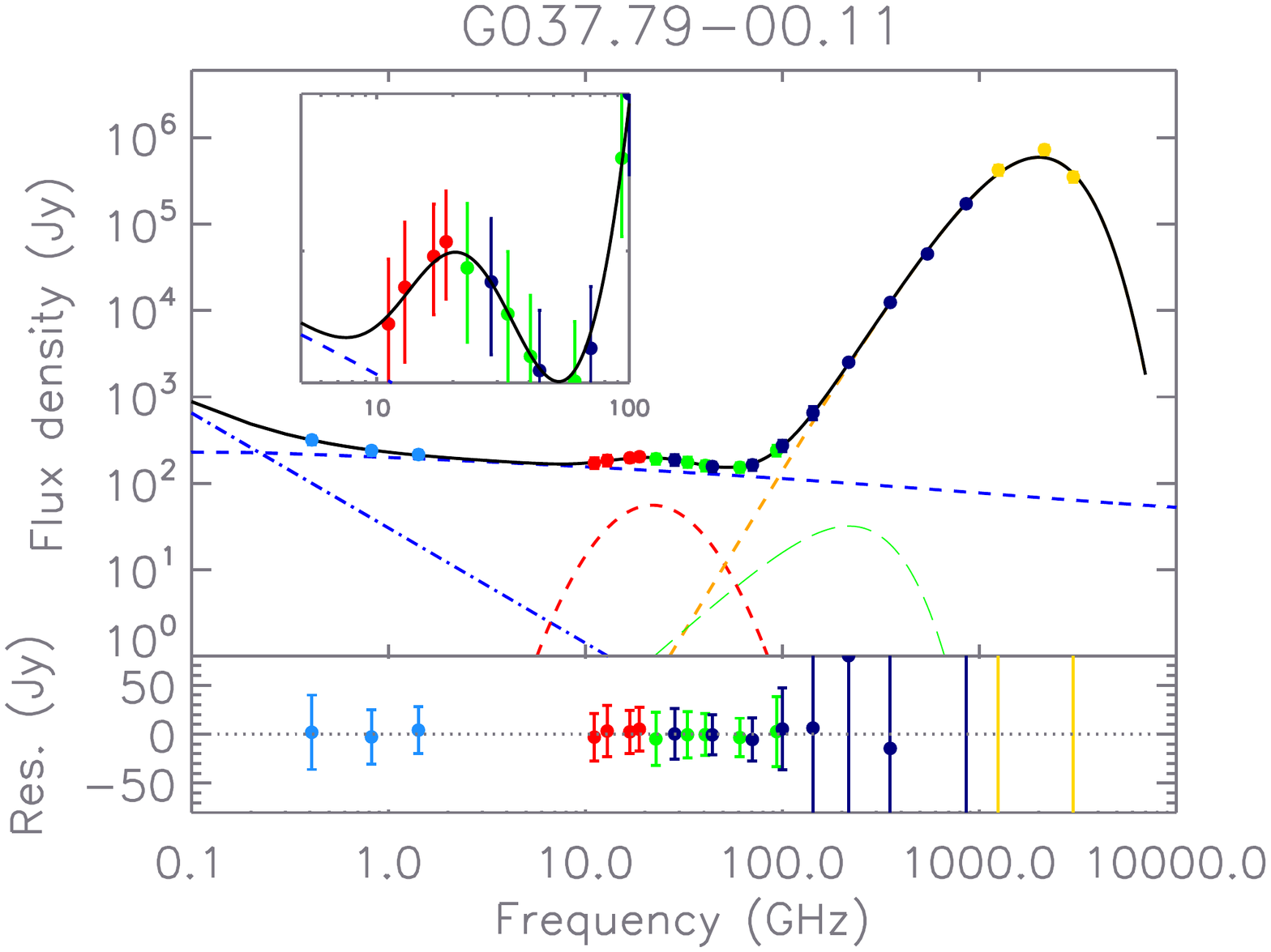}
\hspace*{1.cm}
\vspace*{-3.5cm}
\includegraphics[width=80mm,angle=0]{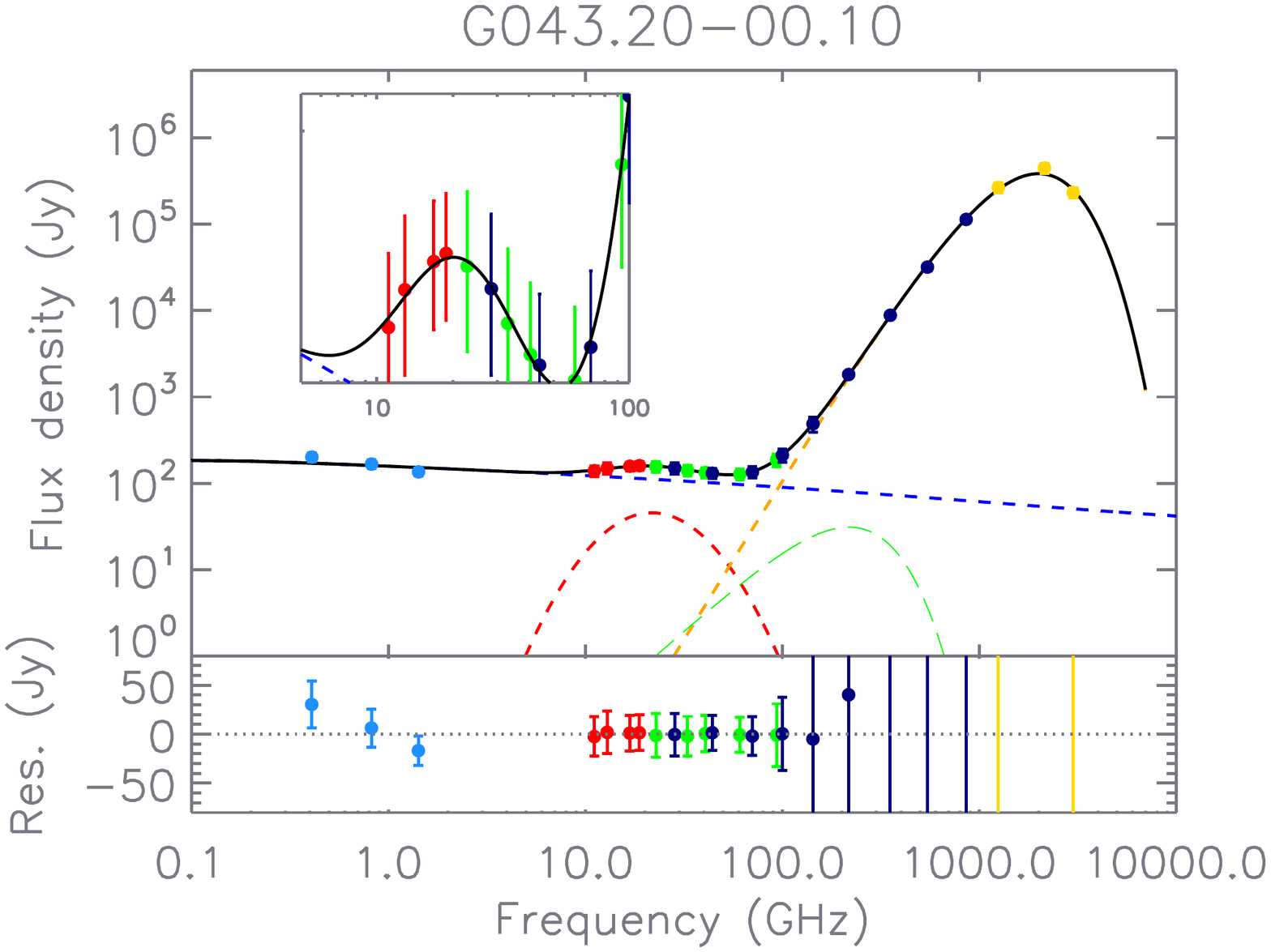}
\hspace*{0.cm}
\includegraphics[width=80mm,angle=0]{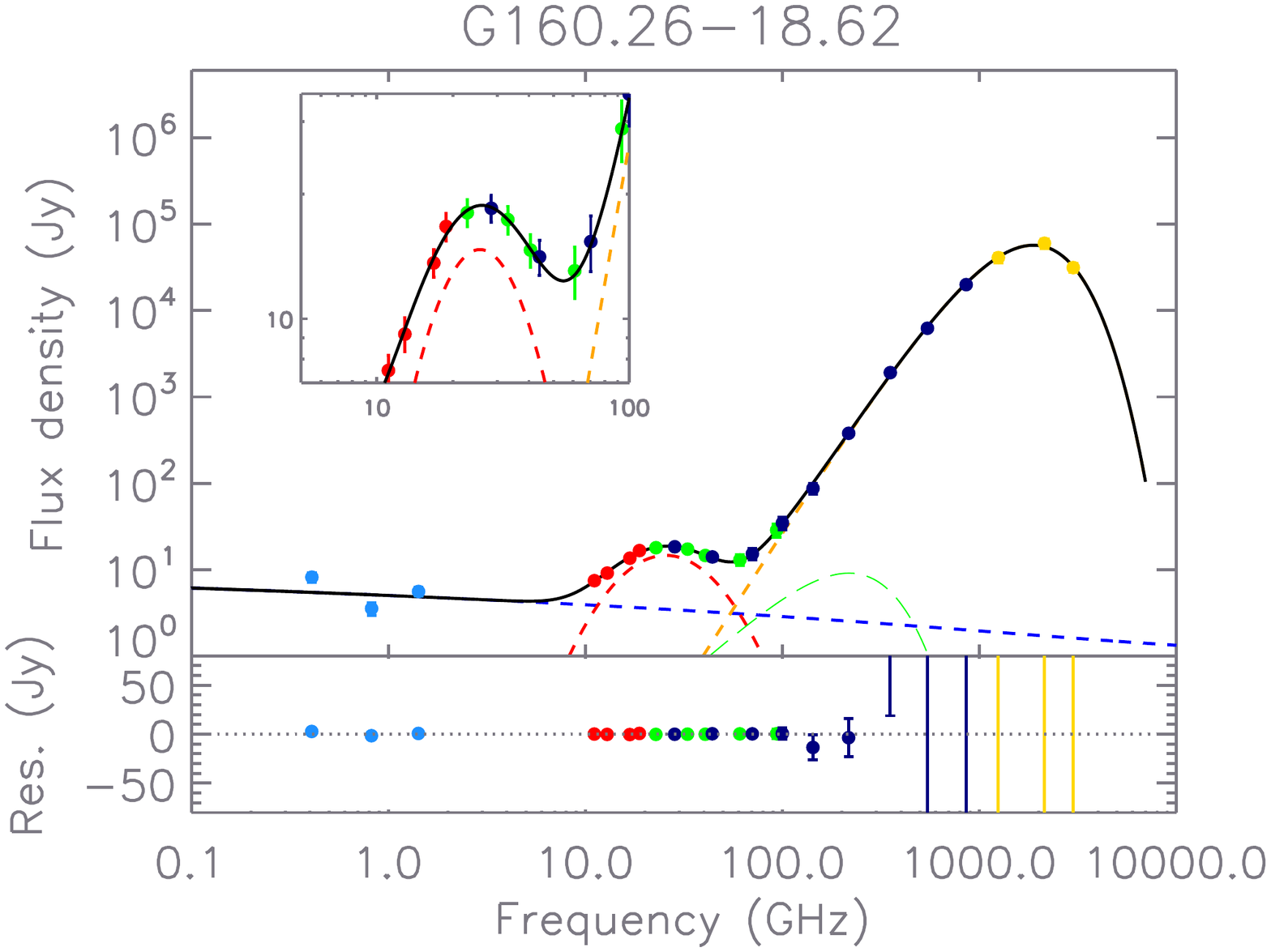}
\hspace*{1.cm}
\includegraphics[width=80mm,angle=0]{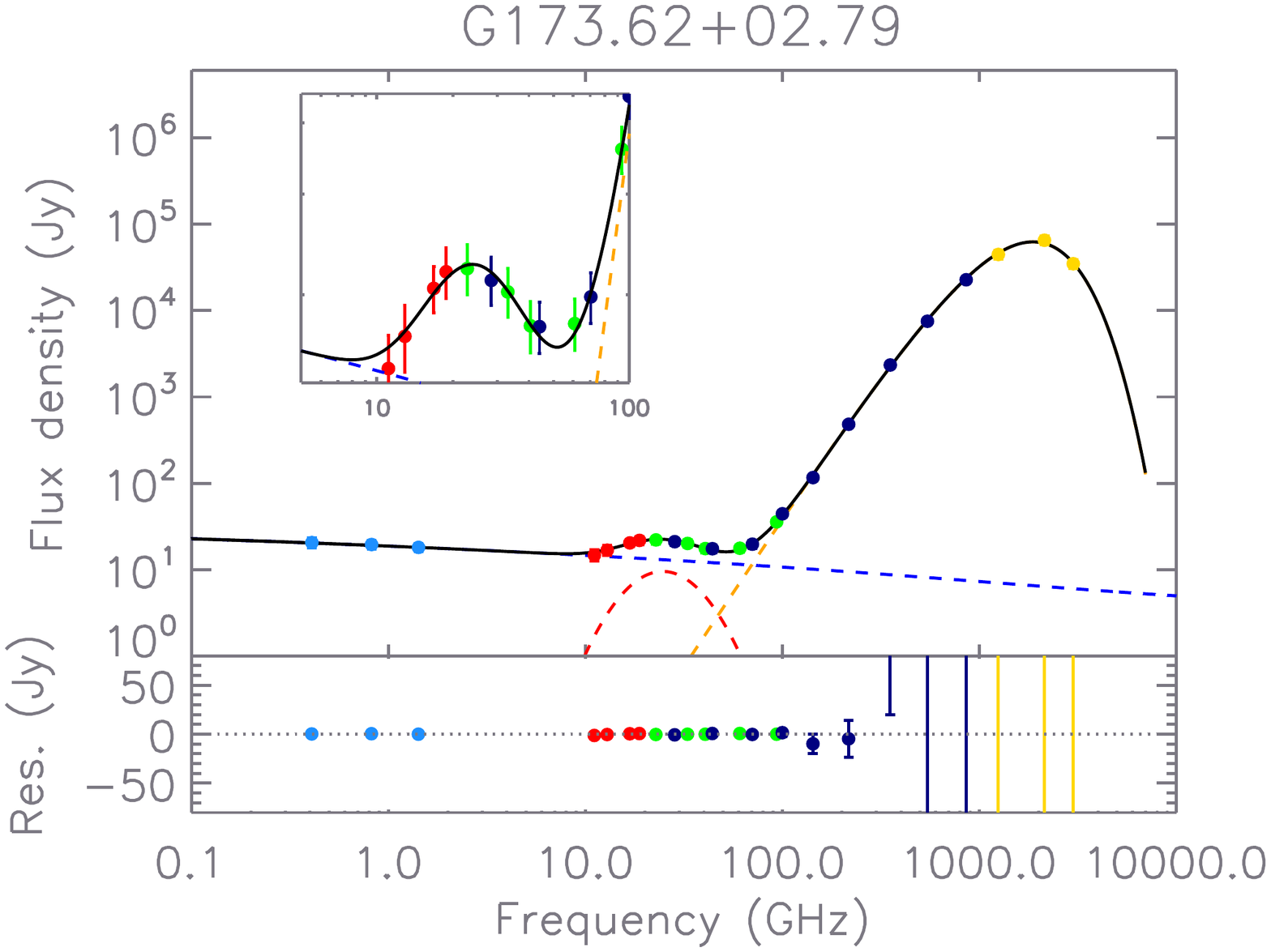}
\vspace*{-3.5cm}
\caption{\small SED of the sample of regions shown in
  Figure~\ref{fig:quijote_int_maps9}.
  The QUIJOTE intensity flux densities
are shown with red points, and the \textit{WMAP}, \textit{Planck}, and
DIRBE intensity flux densities are shown with green, blue, and
yellow< points, respectively. The low frequency points are 
shown in pale blue. The result of the multicomponent fit is 
illustrated by the continuous black curve. The fit to the AME component
is shown with the dashed red line. The fit to the free--free component is shown with the dashed blue line. 
The fit to the thermal dust component is shown with
the dashed yellow line. The fit to the CMB component is shown with
the dashed green line. A zoom on the AME bump is shown in the subpanel. Residuals to the fits are shown in the bottom plots.}
\label{fig:sed_int_subsample}
\end{center}
\end{figure*}

\subsection{Comparison with AME sources previously characterized in Planck Intermediate Results XV}\label{comparisons_with_pirxv}

Before making an analysis of the full sample of 52 candidate AME sources
displayed in Table~\ref{tab:listofclouds} we first compare the results
obtained with a multicomponent analysis of the SEDs calculated on 
the sample of 42 sources already studied by PIRXV. 
The AME model used by PIRXV assumes a spinning dust 
model corresponding to the warm ionized medium (WIM) with a peak at 
28.1\,GHz to give the generic shape for which only 
the amplitude of the peak and the peak frequency were fitted for.
This horizontal shift in frequency is artificial, as the WIM model, with the parameters that have been used do produce that model, predicts a precise value for $\nu_{\rm peak}$.
On the contrary, as explained before, the AME model
used in our analysis is a phenomenological model with 
three parameters including one
parameter to fit for the width of the bump of the AME. 

To build the SEDs in the same way as PIRXV, 
as mentioned before, we use an aperture of radius
$60\arcmin$ and an annulus of internal and external radii of sizes
$80\arcmin$ and $100\arcmin$, respectively. 
For this comparison, we then use the parameters 
obtained by PIRXV on the CMB and
thermal dust components as fixed input parameters and then we fit our
model of AME, free-free and synchrotron (in the cases where
the synchrotron was considered in the fits by PIRXV, i.e. on
sources G$010.19-00.32$, G$012.80-00.19$, G$037.79-00.11$,
G$045.47+00.06$ and G$118.09+04.96$).
From these fits we calculate the 
AME significance ($\sigma_{\rm AME}$) as the ratio 
of the flux density of AME at the frequency peak position 
divided by the uncertainty on this estimate.
The results are displayed in Figure~\ref{fig:comp_pirxv} (a).
Three points show a higher AME significance in PIRXV than in 
our analysis (data shown with red 
colour in the plots).
Overall, however, our analysis shows that for most of the sources
the AME amplitude, and its significance are higher once the 
QUIJOTE data
are included (data shown with black colour in the plot). 
This trend can be explained by the level of free-free detection to
be generally higher in the PIRXV analysis 
than in our component
separation analysis as shown in Figure~\ref{fig:comp_pirxv} (b). 
This point is also confirmed by the higher level of 
AME obtained with our analysis compared 
to the level of AME detected
by PIRXV at a frequency of 28.4$\,$GHz as displayed in 
Figure~\ref{fig:comp_pirxv} (c). In this plot 
AME $S^{28.4}_{\rm resid}$ is the AME flux obtained 
from the modelling at 28.4$\,$GHz. This general trend is
consistent with the results obtained by \cite{W44}, by \cite{poidevin18} and by \cite{ameplanewidesurvey}, and confirms that the QUIJOTE-MFI data are crucial to help breaking the inevitable degeneracy between the AME and the free-free that occurs when only data above 23~GHz are used in regions with AME peak flux densities close to this frequency. From Fig.~\ref{fig:comp_pirxv} (d) it is also clear that the inclusion of QUIJOTE data favours lower AME peak frequencies, which are found to be on average around 4$\,$GHz smaller than in PIRXV. It is also worth stressing that the addition of QUIJOTE data clearly leads to a more precise characterisation of the emission models in the $10-60\,$GHz frequency range. We find on average errors smaller by $\approx 30\,\%$ on $EM$ and $A_{\rm AME}$, by $\approx 70\,\%$ on $W_{\rm AME}$, by $\approx 60\,\%$ on $\nu_{\rm AME}$ and even by $10\,\%$ on $\beta_{\rm dust}$ and $T_{\rm dust}$.

To test that our interpretation of the results is not model-dependent we repeated the analysis described above with the model
proposed by \citet{bonaldi07}. The final plots are very similar to
those displayed in Figure~\ref{fig:comp_pirxv} meaning that the
higher level of detection of AME comes from the addition of the
QUIJOTE maps at 10--20$\,$GHz. In addition to this, our model should provide more reliable estimates of the AME peak frequency thanks to it being fully independent on the AME width.

\section{Regions of AME}\label{ame_regions}

In the following sections we describe the level of detection of AME derived from the modelling analysis of the SEDs (Section~\ref{ame_significance}) and their possible contamination by UC\sc{H\,ii} \rm{regions} (Section~\ref{uchii}). From this analysis we define the final sample of candidate AME sources that will be used for further statistical studies. Additional calculations used to test the robustness and validate this sample are given in Section~\ref{robustness}.

\subsection{Significance of AME detections in our sample} \label{ame_significance}

In order to make a study of the detection of AME in the 52 sources
from our sample we first produced a series of intensity maps at all
available frequencies. The maps were inspected and removed if
some pixels were showing no data 
in the aperture or annulus areas; this process affecting more specifically low frequency maps.

The component separation was operated by including fits for the
free-free, the AME, the thermal dust and the CMB components. The synchrotron component was also included in the six sources indicated in Section~\ref{model_fitting}.
Each SED was then
inspected by eye and it was found that most of the sources were showing the detection 
of a bump in the frequency range \mbox{10--60$\,$GHz}. Some examples of SEDs in
intensity are shown in Figure~\ref{fig:sed_int_subsample}.

\begin{figure}
\begin{center}
\vspace*{2mm}
\hspace*{-0.3cm}
\centering
\includegraphics[width=90mm,angle=0]{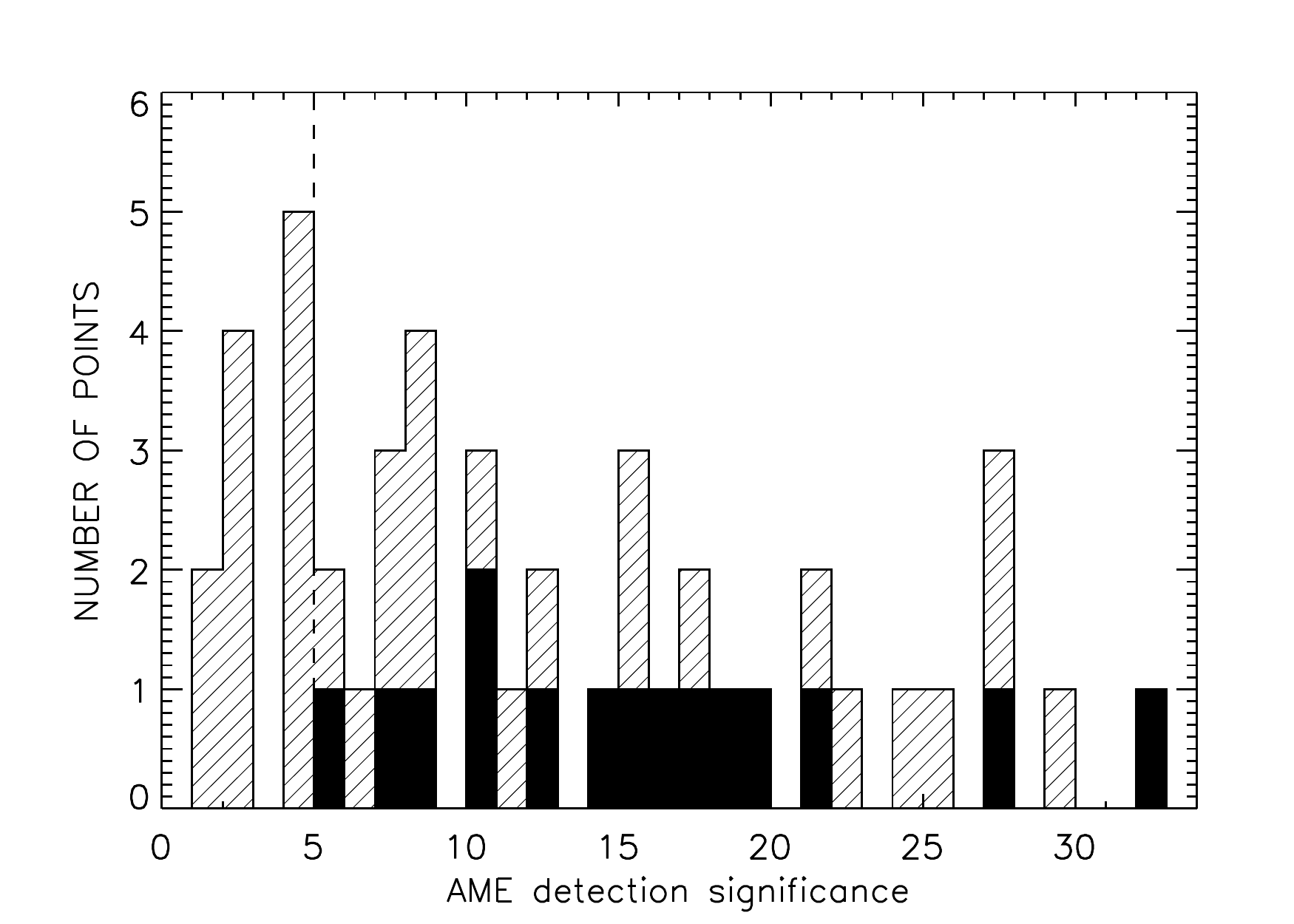}
\vspace*{-0.5cm}
\caption{Histogram of the AME significance 
values ($\sigma_{\rm AME}$) for the sample of 52 sources. 
The 5$\sigma$ limit is shown with
  the vertical dashed line. Sources that are significant and have a maximum contribution from
UC\sc{H\,ii} \rm{} regions, $f^{\rm {UC}\sc{H\,II}}_{\rm max} < 0.25$, are shown as the filled histogram.}
\label{fig:histo_ame_significance}
\end{center}
\end{figure}

The histogram displayed in Figure~\ref{fig:histo_ame_significance}
shows the distribution of the significance of the AME detection,
$\sigma_{\rm AME}$. Following PIRXV we define the sources with $\sigma_{\rm AME}>5$ as ``significant AME sources'', the sources
with $2<\sigma_{\rm AME}<5$ as ``semi significant AME sources'',
and the sources with $\sigma_{\rm AME}<2$ as ``non AME
detections''. 
Some of the ``significant AME sources'' are re-classified as ``semi-significant AME sources'' as will be discussed in the next section. The concerns regarding modelling problems and systematic
errors for a few sources are discussed in Section~\ref{robustness}.

\subsection{Ultra-Compact H\textsc{ii} regions}\label{uchii}

\begin{figure}
\begin{center}
\vspace*{2mm}
\centering
\includegraphics[width=85mm,angle=0]{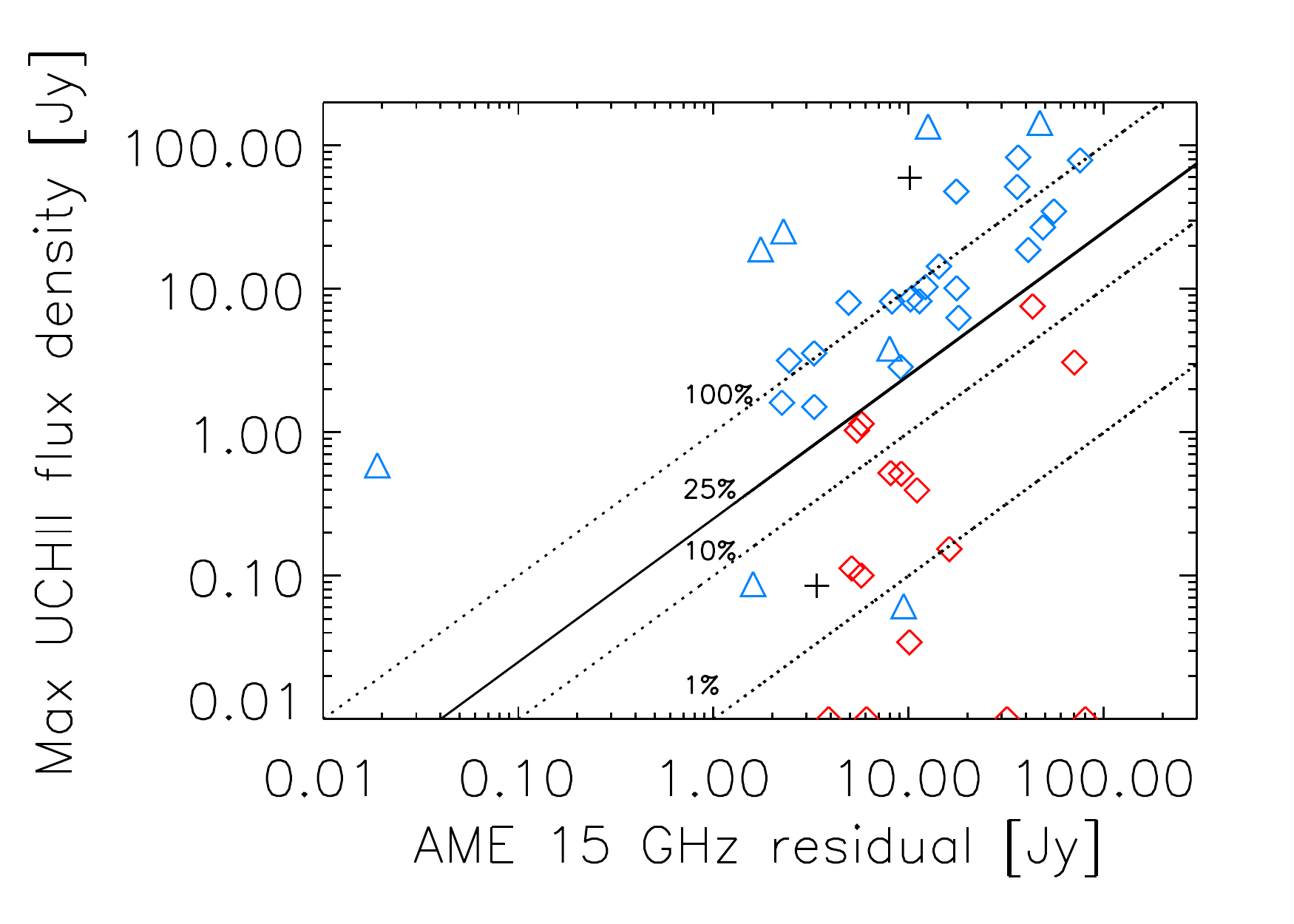}
\vspace*{-0.5cm}
\caption{Estimated maximum contribution from UC\sc{H\,ii} \rm{} regions against
  15$\,$GHz AME residual flux density. 
The most significant AME sources ($\sigma_{\rm AME}>5$ and $S_{\rm
  15}^{\rm residual} >0.25 \times S_{\rm max}^{\rm UCHII}$) are shown
as red diamond symbols, while non-AME regions ($\sigma_{\rm AME}<2$) are
shown as dark cross symbols.
``Semi-significant'' AME sources ($\sigma_{\rm AME} = $2--5) are shown as
blue triangle symbols. ``Significant'' AME regions that have a potentially large
contribution from UC\sc{H\,ii} \rm{} ($S_{\rm
  15}^{\rm residual} <0.25 \times S_{\rm max}^{\rm UCHII}$) are re-classed as ``semi-significant''
and are highlighted by blue diamonds.
The data shown with red diamond symbols are the ``Significant'' AME
regions such that $S_{\rm
  15}^{\rm residual} > 0.25 \times S_{\rm max}^{\rm UCHII}$,
if this information is available.
Regions with no matched UCH II regions are set to 0.01 
for visualization and lie on the
bottom of the plot. The dashed lines correspond to 
different maximum fractions of UC\sc{H\,ii} \rm{} flux 
density: 1, 10, 25 (solid line), and 100$\%$ 
of the 15$\,$GHz residual flux density.}
\label{fig:psc_max_uchii_15ghz}
\end{center}
\end{figure}

Ultra-Compact \sc{H\,ii} \rm{} regions (UC\sc{H\,ii}\rm{}) could bias AME detections and change the free-free typical behaviour. 
It is therefore important to assess their possible impact on our analysis.
UC\sc{H\,ii} \rm{}with EM $\gtrsim 10^{7}$ cm$^{-6}$pc are expected to produce optically thick free-free emission up to 10
GHz or higher \citep[][]{kurtz2002,kurtz2005}. To take into account possible contamination of our sample by emission from arcsec resolution point 
sources \citep[][]{wood1989b} that are not AME in nature we follow the method used in PIRXV as illustrated by their Figure~5.
To this aim we catalog all the \it{IRAS}
\rm points sources retrieved from the \it{IRAS} \rm Point Source Catalog (PSC)\footnote{See the link to the IRAS Faint Source Catalog,
  Version 2.0 in the HEASARC Catalog Resources Index, {\tt https://heasarc.gsfc.nasa.gov/W3Browse/iras/iraspsc.html}}
that lie inside the 2$^{\circ}$ diameter circular apertures of our sample. These sources are classified 
as a function of their colour-colour index defined by the logarithm of flux ratios obtained in several bands. 
The PSC UC\sc{H\,ii} \rm{}potential candidates 
tend to have ratios log$_{10}(S_{60}/S_{12})  \ge 1.30$ and 
log$_{10}(S_{25}/S_{12}) \ge 0.57$ \citep[][]{wood1989a}. They are identified accordingly. 
\cite{kurtz1994} measured the ratio of 100$\,\mu$m to 2 cm (15$\,$GHz) flux
densities and found it lies in the range 1000--400000, with no UC\sc{H\,ii} \rm{}regions having $S_{100 \mu m}/S_{2 \rm cm} < 1000$. Following PIRXV
we use this relation to put limits on the 15 GHz maximum flux
densities that could be emitted by candidate UC\sc{H\,ii} \rm{} regions encountered 
in the apertures used for measuring the flux densities of our sample
of sources. The fluxes at 100 $\mu$m of the PSC sources are summed 
up toward each aperture and then divided by 1000 to get an estimate of the
the maximum UC{\sc H\,ii} flux density at 15$\,$GHz, $S_{\rm max}^{\rm UCHII}$, towards each candidate AME source.
From the multicomponent fits, the flux densities at 15$\,$GHz (or 2\,cm) are
calculated and compared to these maximum UC\sc{H\,ii} \rm{} flux densities.
The distribution is shown in Figure~\ref{fig:psc_max_uchii_15ghz} 
where the maximum UC\sc{H\,ii} \rm{} flux densities are plotted against the
15$\,$GHz flux densities obtained with our analysis. If a candidate
AME source detected with more than 5$\sigma$ has a residual AME flux density at 15$\,$GHz lower 
than 25$\%$ of the maximum UC\sc{H\,ii} \rm{} flux density then it is re-classified as
``semi-significant'', as indicated in Table~\ref{tab:listofclouds}. We believe that this is a very conservative approach, in a way that many of these re-classified sources are actually ``significant'' AME detections. UC\sc{H\,ii} \rm{} contributions to the 30$\,$GHz excess have been recently investigated by \cite{comap6_2021} on a small sample of galactic {\sc H\,ii} regions using data from the 5$\,$GHz CORNISH catalog. The study rejects such regions as the cause of the AME excess.


\begin{figure*}
\begin{center}
\vspace*{2mm}
\centering
\hspace*{0.cm}
\includegraphics[width=80mm,angle=0]{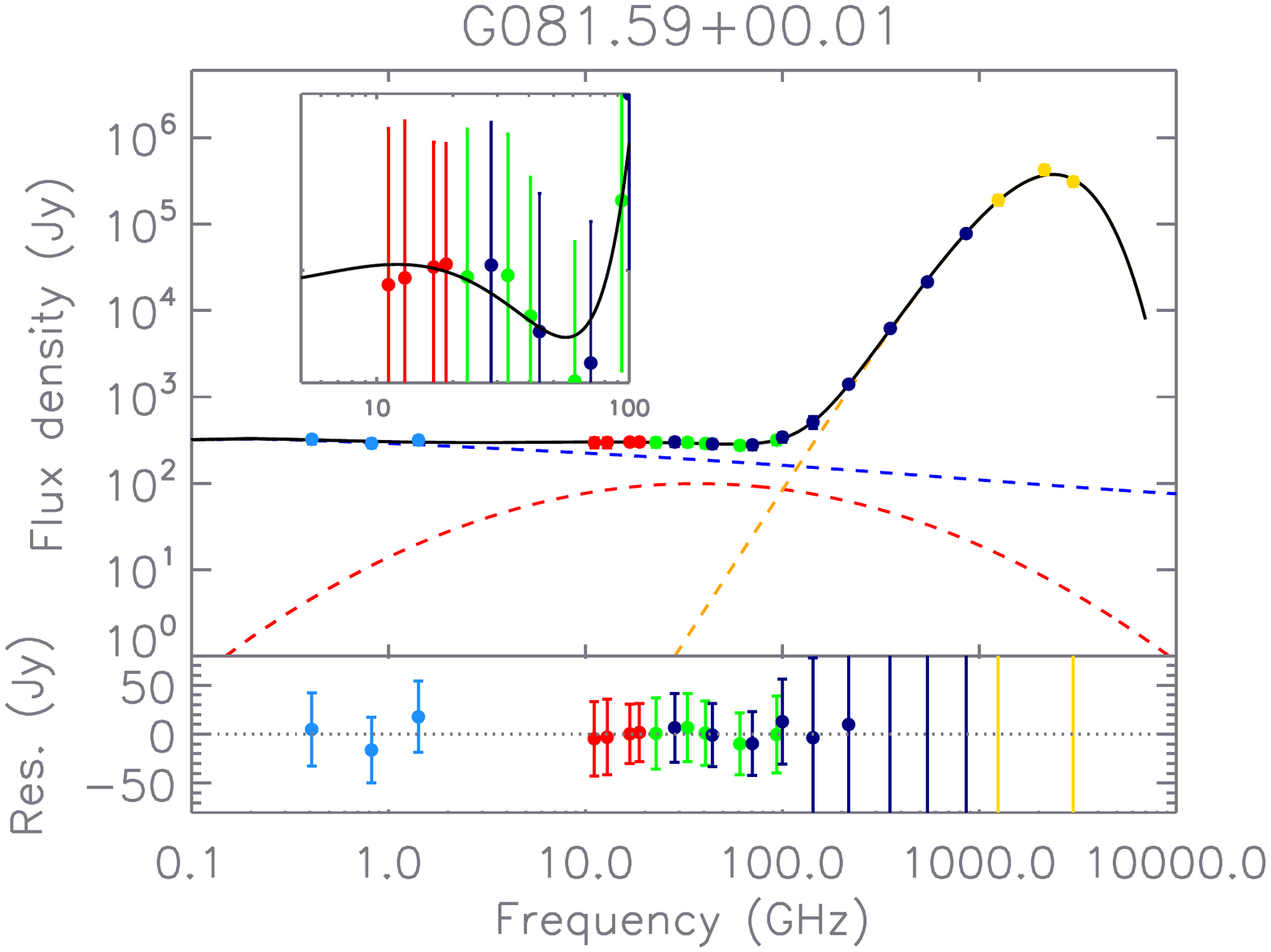}
\hspace*{1.cm}
\vspace*{-3.5cm}
\includegraphics[width=80mm,angle=0]{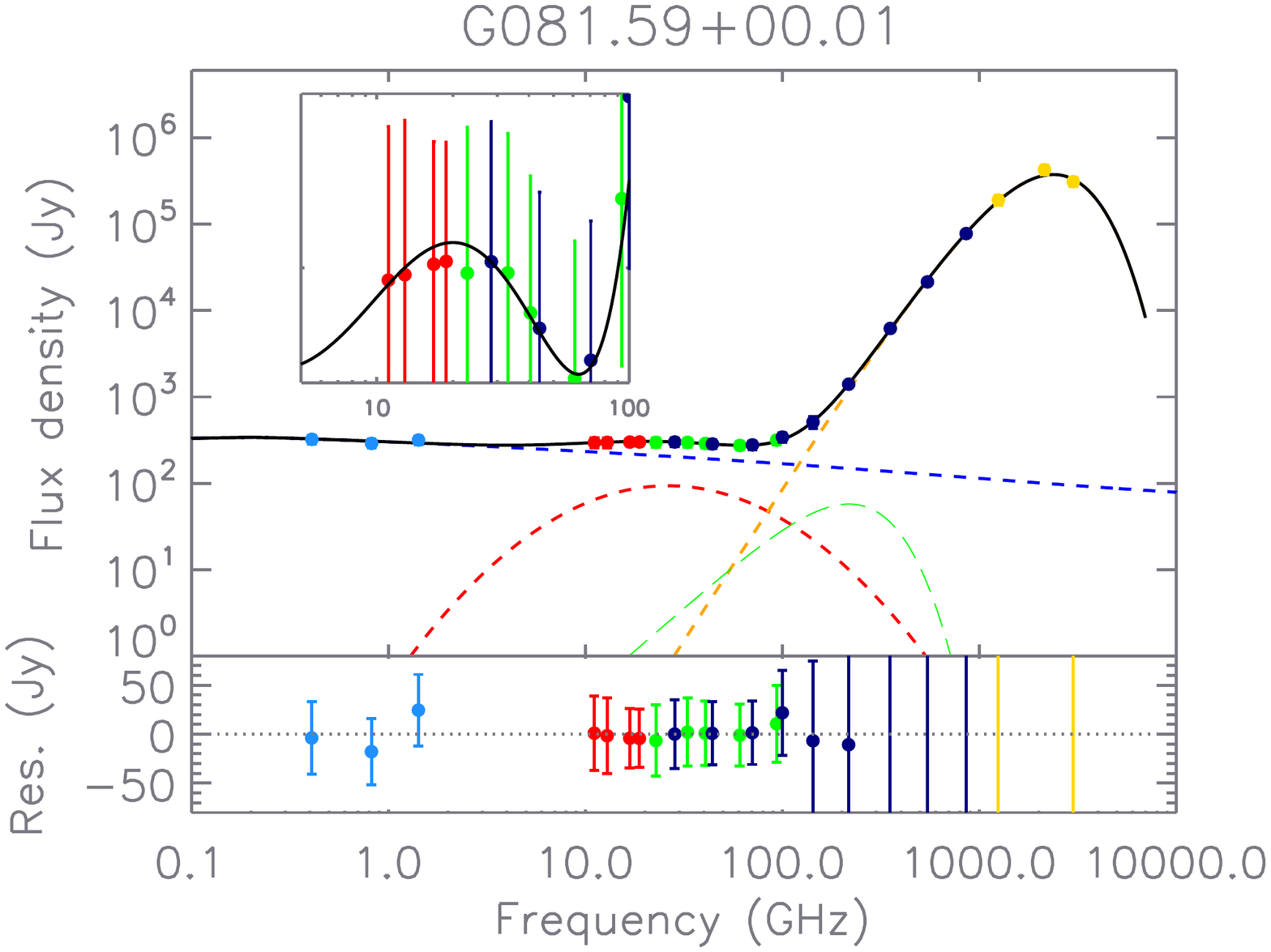}
\hspace*{0.cm}
\vspace*{0.cm}
\caption{Two multicomponent fits of the DR23$/$DR21 region. Colours and symbols definitions are the same as in Figure~\ref{fig:sed_int_subsample}. Left: fits
  obtained with priors on the AME parameters such that $10 < \nu_{\rm
    AME} < 60\,$ GHz, and $0.2 < W_{\rm AME} < 2.5$. Right: Same as
  left but with $0.2 < W_{\rm AME} < 1$. A discussion about the choice
  of priors is given in section~\ref{robustness}. The AME and CMB components
  fit parameters obtained in each case are displayed in Table~\ref{tab:seds_fit_parameters_dr21}.}
\label{fig:dr21}
\end{center}
\end{figure*}

\input{polametex_sedfitparams_table_dr21.txt}

\subsection{Robustness and validation}\label{robustness}

The significance of AME detection, defined by the parameter $\sigma_{\rm AME}$, discussed in 
section~\ref{ame_significance}, is an important indicator 
reflecting the ability of our analysis to detect and fit any excess
of emission observed in the frequency range 10--60$\,$GHz; whether
such a bump is potentially dominated by UC\sc{H\,ii} \rm{} regions or not  
(Section~\ref{uchii}). The significance of AME detection obtained on
each source, though, is also dependent on the
overall accuracy of the multicomponent fit obtained over the full
frequency spectrum considered in the analysis. 

In order to explore the stability of the fitting procedure we made a number of tests to check that our main results are 
not affected by our fitting method and assumptions. This
includes relaxing the assumed calibration uncertainty and changing the sizes of the aperture and annulus radius. 
Overall we were able to fit all the 4 or 5
components on 46 sources from the 52 sources included in the
initial sample, or in other words the multicomponent fit was converging on all the components considered to fit each of the 46 sources. 

The \texttt{SPDust2} models \citep[see][]{ali2009, ali2010} for cold neutral medium, dark cloud, molecular cloud, warm ionized medium and warm neutral medium have widths lying
in the range $[0.4-0.7]$ while in practice slightly wider
distributions could be expected (see discussion in Section~\ref{model_fitting}). To take this into account 
the uniform priors used on the AME parameters are:
$10 < \nu_{\rm AME} < 60\,$ GHz, and $0.2 < W_{\rm AME} < 1.0$. 
Such assumptions on the values allowed to be taken by $W_{\rm AME}$
are important to keep realistic AME detections.
An example of the effect of the priors is shown in Figure~\ref{fig:dr21} where multicomponent fits
obtained on the DR23$/$DR21 maps are displayed. The plot on the left shows
the fit on the AME component with priors on $W_{\rm
  AME}$ such that $0.2 < W_{\rm AME} < 2.5$, while the plot on
the right displays the AME fit component with priors on $W_{\rm
  AME}$ such that $0.2 < W_{\rm AME} < 1.0$. The AME fit
parameters obtained in both cases are given in
Table~\ref{tab:seds_fit_parameters_dr21}.
In the case of loose priors on $W_{\rm AME}$ the AME component
shows an excessively wide looking bump, even if the improvement in the goodness of the fit is marginal (see the values of the $\chi^2_{\rm red}$ in Table~\ref{tab:seds_fit_parameters_dr21}). Such a broad spectrum cannot be reproduced by spinning dust models for environments with reasonable physical parameters, so models like this might be deemed as physically unrealistic. This demonstrates the need for setting realistic priors on the fits to overcome the problem with fit degeneracies. Finally, as it was commented in Section~\ref{model_fitting}, our methodology for error estimation do not properly grasp those parameter degeneracies, leading in some cases to an underestimation of the error (see the too small error of $\nu_{\rm AME}$ in the case of strong prior in Table~\ref{tab:seds_fit_parameters_dr21}).  

As a final test we repeated the analyses with more stringent priors such that $0.4 < W_{\rm AME} < 0.7$ and 16$\,$GHz $ < \nu_{\rm AME} <$ 60$\,$GHz, and found that this does not have a strong impact on the derived results. In particular, we found differences typically smaller than $5\,\%$ in $\nu_{\rm AME}$ and typically smaller than $20\,\%$ in $A_{\rm AME}$.

Our final sample follows the superscript symbols given in the last column in Table~\ref{tab:listofclouds}. A total of 6 sources (labelled as ``BD'') considered as bad detections of AME because of a bad fit of the AME, of the free-free or of the thermal dust component, are not considered on a statistical basis.  
On the other hand, statistics are given for the sample which we refer to as the selected sample (46 sources).
This data set includes sources with low or poor AME detection (2 sources, labelled as ``LD''), with ``semi-significant'' AME detection (29 sources labelled as ``SS'', including 20 ``significant'' AME sources reclassified as "semi-significant AME sources") and with ``significant" AME detections (15 sources labelled as ``S''). 
Statistics are also given on the sample of ``semi-significant'' AME detections and on the sample of ``significant'' AME detections.
The selected sample includes a total of 7 sources with fits reaching the prior upper limit on $W_{\rm AME}$ and such that, the uncertainty on this parameter is, $\sigma_{W_{\rm AME}} = 0$. 
These sources are included in the sample of AME well-detected 44
sources (i.e., the sample including ``semi-significant''  and
``significant'' AME detections).

\section{Statistical study of AME sources}\label{statistics}

Along this section we study the statistical properties
of the physical parameters 
of the sample discussed in the previous section, with the aim 
of better understanding the physical and environmental conditions of the AME sources, as well as to obtain insights
about the nature of the carriers that cause the AME.
The parameter values used to model the components
estimated from the analysis of the SEDs in intensity are
given in Tables~\ref{tab:seds_fit_parameters1} and \ref{tab:seds_fit_parameters2}.  
The method used to calculate the flux densities does not take into
account the effect of the signal integration through the thickness of the clouds as well as across the area sustended by each 
telescope. This limitation will be taken into account, as much as 
possible, in the interpretation of the results.

\subsection{Nature of the sources}

In this section we focus our analysis on the parameters used to model the AME and some of the thermal dust component parameters. This includes the relative strength of the ISRF, which is estimated from the fitted thermal dust parameters.  

\subsubsection{AME fraction at 28.4$\,$GHz} 

\begin{figure}
\begin{center}
\vspace*{2mm}
\hspace*{-0.3cm}
\centering
\includegraphics[width=90mm,angle=0]{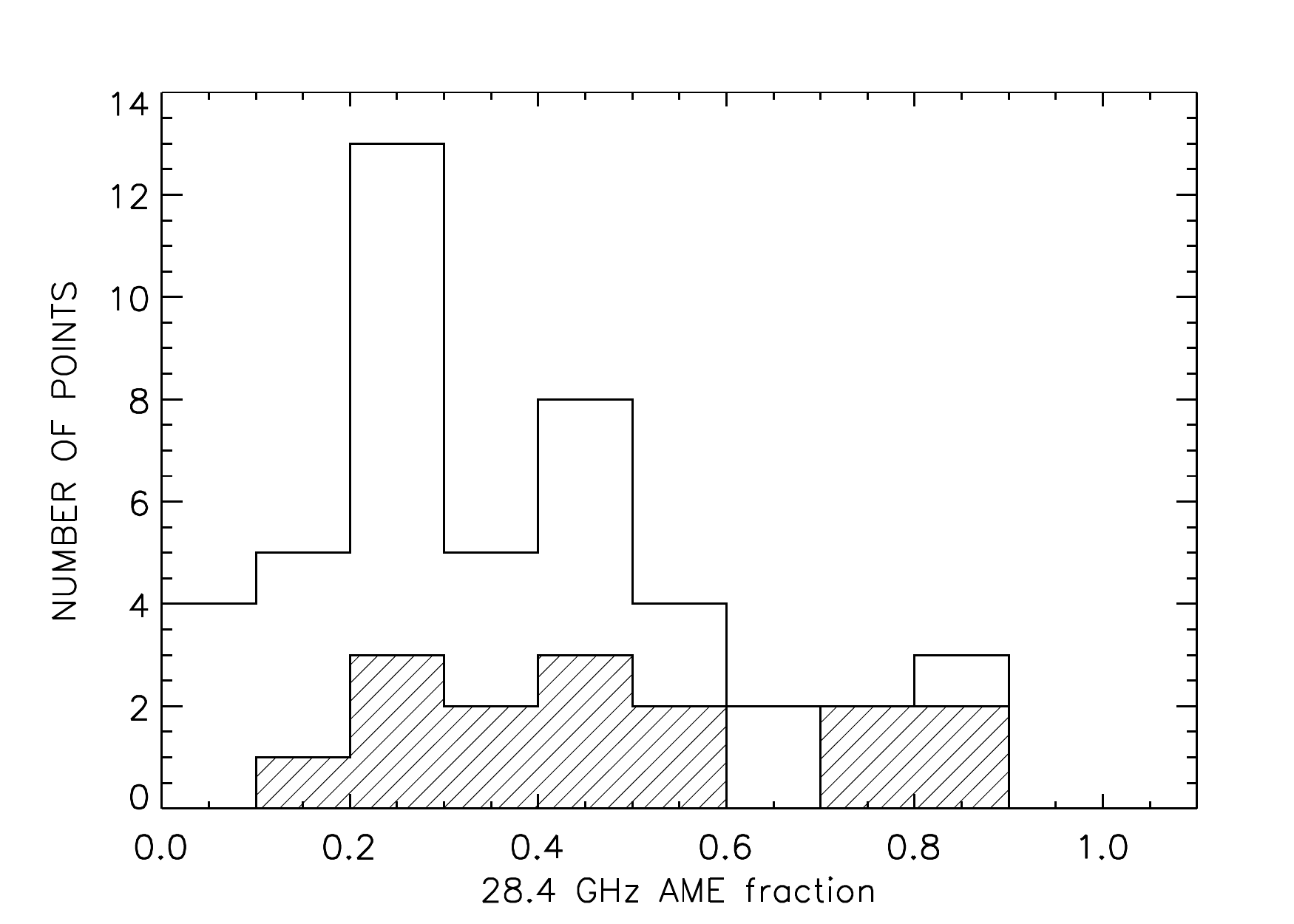}
\caption{Histogram of the AME fraction $S^{28}_{\rm res}/S_{28}$ at
  28.4 GHz. The selected sample is shown as the unfilled histogram.
 The ``significant'' AME detection sample is shown with the hatched area.}
\label{fig:ame_frac_28}
\end{center}
\end{figure}

As a first step we investigate the fraction of the total flux density at
28.4$\,$GHz that is produced by AME under the expectation 
that free-free and AME are the dominant sources of radiation 
at this frequency. For this we calculated the residual AME flux density at $28.4\,$GHz, $S^{28}_{\rm res}$, by
subtracting to the measured flux density at this same frequency
all the other components and propagating their uncertainties.
The histogram of this quantity is plotted in
Figure~\ref{fig:ame_frac_28} and shows that regardless of
whether the sources
are classified as ``significant'' or ``semi-significant'', the
contribution of the AME flux density goes from a few per cent to
almost 100 per cent
of the total flux density. 
This result is different from that obtained by PIRXV who
found that in their sample the sources classified
as ``significant'' AME sources were mainly showing 
$S^{28}_{\rm res}/S^{28} > 50\,\%$, while the remaining 
sources classified as ``semi-significant'' were lying 
in the lower part of the histogram
such that $S^{28}_{\rm res}/S_{28} \lesssim 50\,\%$.
All in all, the majority of the sources in our selected sample show
$S^{28}_{\rm res}/S^{28} < 50\,\%$. This result could come from the AME peak frequency distribution which is found to be about 4$\,$GHz lower than by PIRXV. This result will be presented in Section~\ref{ame_peak_frequency_hist}.

\subsubsection{Dust properties}

\begin{figure}
\begin{center}
\vspace*{2mm}
\hspace*{-1.cm}
\centering
\includegraphics[width=95mm,angle=0]{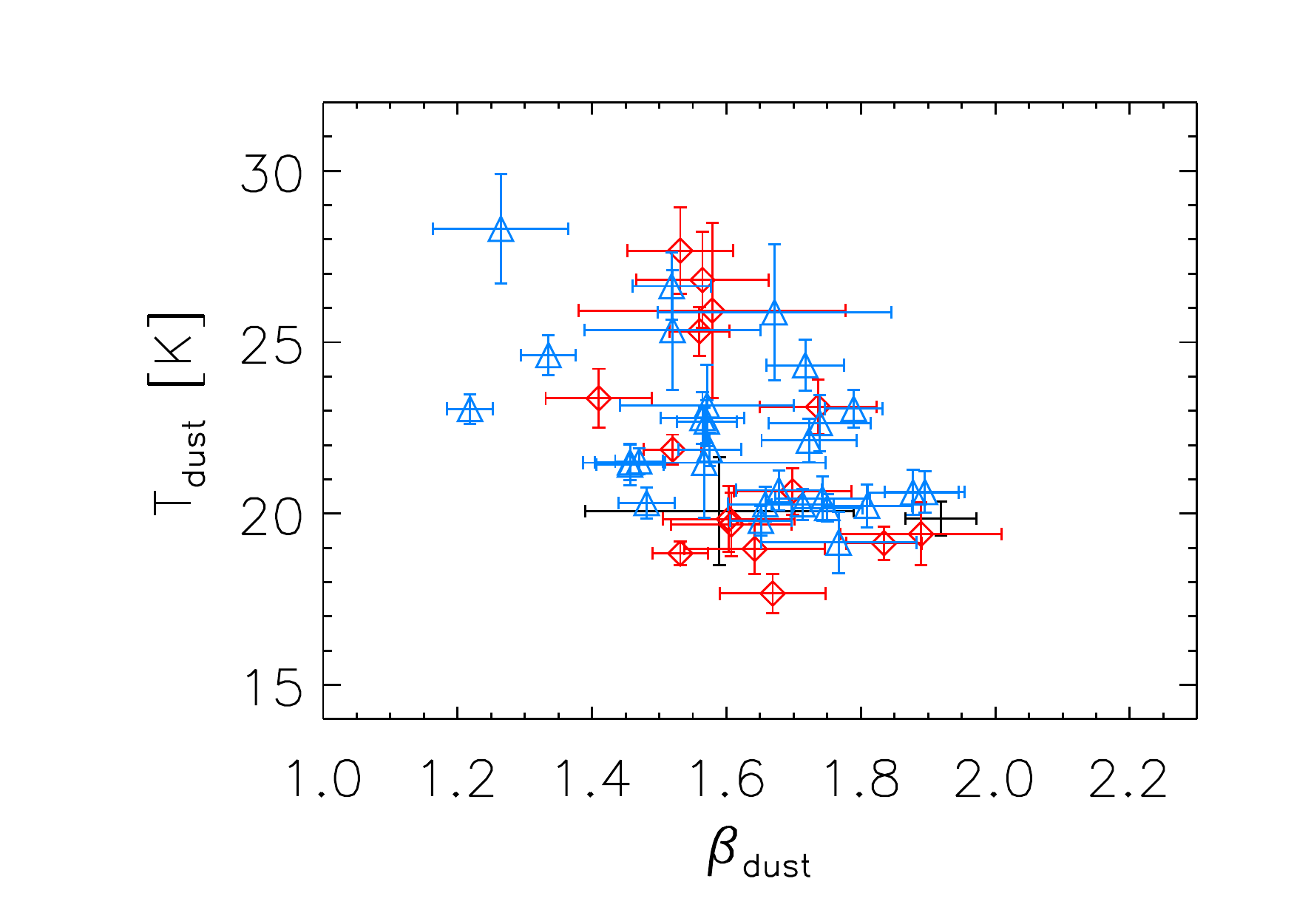}
\vspace*{-0.5cm}
\caption{Distribution of the thermal dust temperature, 
$T_{\rm  dust}$, against the thermal dust emissivisity, $\beta_{\rm dust}$ 
obtained from the SEDs multicomponent fits.
The ``significant'' AME detection sample is shown with red
  diamonds. The ``semi-significant'' AME detection sample is
  shown with blue triangles. Low AME detections are shown in black.}
\label{fig:td_betad}
\end{center}
\end{figure}

The distribution of the thermal dust temperature, 
$T_{\rm  dust}$, against the thermal dust 
emissivisity, $\beta_{\rm dust}$ 
obtained from the SEDs multicomponent fits are displayed in 
Figure~\ref{fig:td_betad}. The expected anti-correlation 
that is discussed and analysed in many works
\citep[e.g.,][]{paradis2014} is also seen in the plot.

\begin{figure}
\begin{center}
\vspace*{2mm}
\hspace*{-1.cm}
\centering
\includegraphics[width=95mm,angle=0]{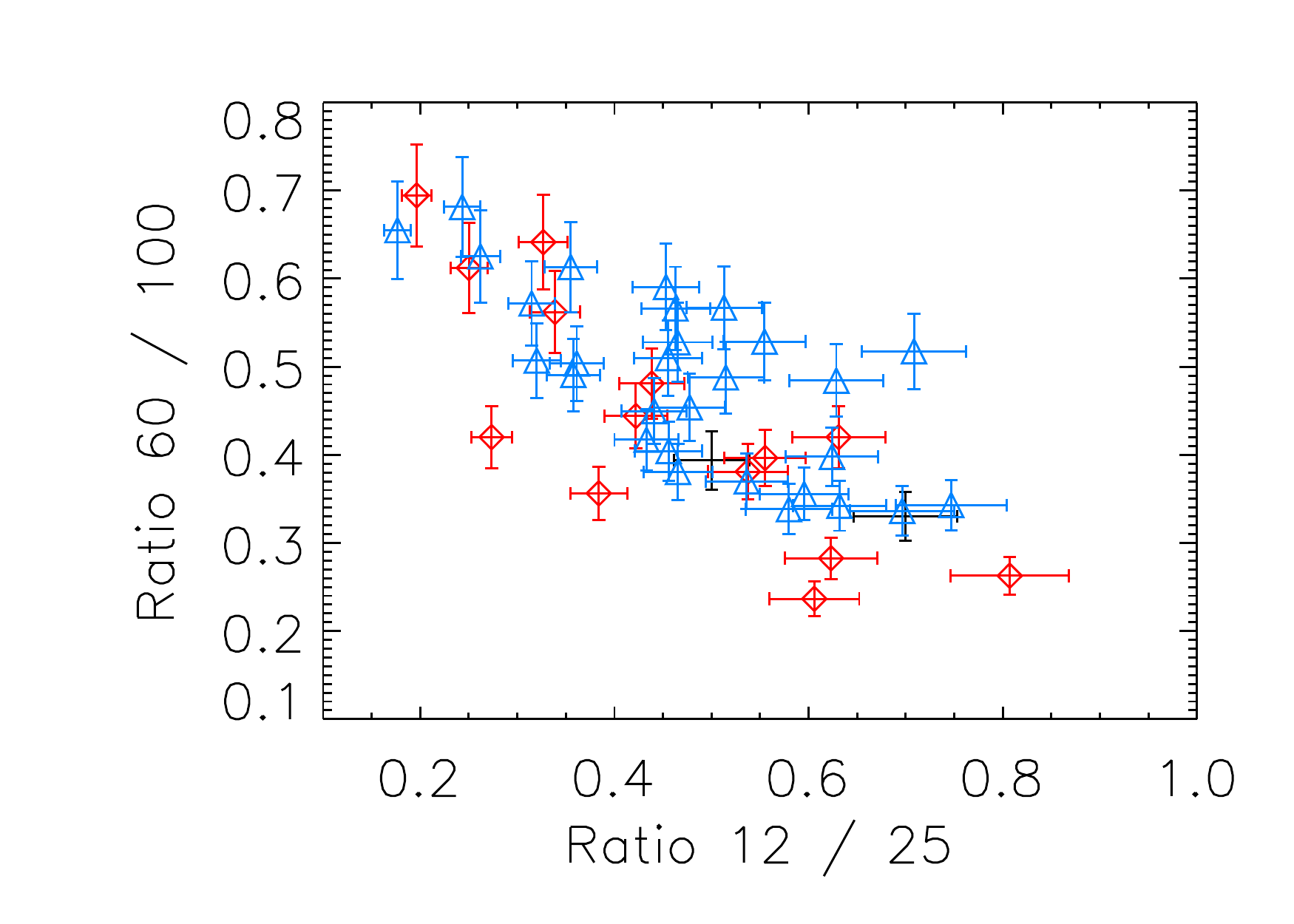}
\vspace*{-0.5cm}
\caption{Colour--colour plot of IRAS $12\mu$m$/25\mu$m against
  $60\mu$m$/100\mu$m for the sample of sources.
Symbols and colours definition are the same as in Figure~\ref{fig:td_betad}.}
\label{fig:ratio_vs_ratio}
\end{center}
\end{figure}

An apparent sequence in the \textit{IRAS} colours given by the
$12\mu$m$/25\mu$m and $60\mu$m$/100\mu$m ratios can also be
expected from previous studies of \sc{H\,ii} \rm{} regions
\citep[][]{chan1995,boulanger1988}, and external galaxies
\citep[][]{helou1986} showing an anti-correlation between the two
ratios. The interpretation relates to the spatial 
distribution of different
grain populations as a function of the Inter-Stellar
Radiation Field (ISRF) intensity.  
This trend was obtained for the sample of sources discussed by 
PIRXV. We find a result similar to their 
analysis but our plots 
shown in Figure~\ref{fig:ratio_vs_ratio} presents 
a lower dynamic range 
of the colour ratio $60\,\mu\text{m}/100\,\mu\text{m}$ 
than the one from their analysis. 
Our sample probes line-of-sights (LOSs) with colour ratios 
$60\,\mu\text{m}/100\,\mu\text{m}$ lying in the 
range 0.2--0.7, which is the range 
in which PIRXV found most of their sources classified 
as ``significant'' AME detections and not expected to be 
dominated by UC\sc{H\,ii} \rm{}region emission.

\subsubsection{Dust optical depth}  \label{dust_optical_depth} 

\begin{figure}
\begin{center}
\vspace*{2mm}
\hspace*{-1.cm}
\centering
\includegraphics[width=95mm,angle=0]{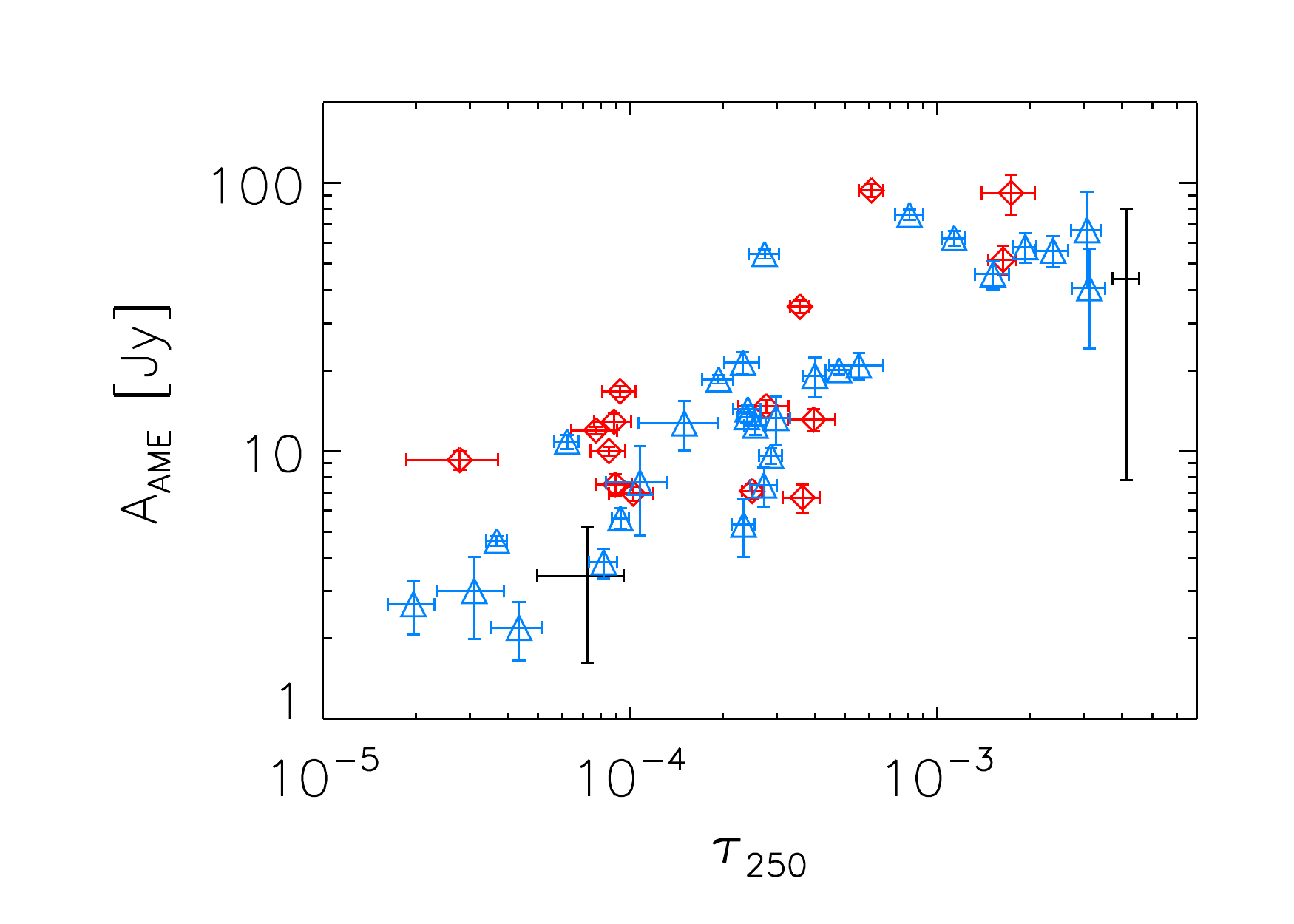}
\vspace*{-0.5cm}
\caption{Distribution of the AME peak flux density A$_{\rm AME}$
  against $\tau_{250}$. All selected data are displayed.
Symbols and colours definition are the same as in Figure~\ref{fig:td_betad}.}
\label{fig:dust_optical_depth}
\end{center}
\end{figure}

The sources of our sample are distributed across regions of different
optical depths. In order to understand how this parameter could help us to
build up a picture of the distribution of the parameters used to fit the
AME components classified as ``semi-significant'' or ``significant'',  
in Figure~\ref{fig:dust_optical_depth} we show the variations of the peak AME flux density, $A_{\rm AME}$, as a
function of the thermal dust optical depth at 250
$\mu$m, $\tau_{250}$, obtained from the fits of the thermal dust components.
One can see a clear trend showing 
an increase of the maximum AME flux density 
with the quantity
of thermal dust matter encountered along the LOSs. 
The Spearman Rank Correlation Coefficient
(SRCC) of that distribution is r$_{\rm s}= 0.80 \pm 0.04$.
This is
not a surprise, as a strong spatial correlation was already observed between the AME
and thermal dust, when AME was first detected
\citep[see][]{kogut1996,leitch1997}, and it is well established that
the interstellar medium is pervaded by a complex non-uniform distribution of
thermal dust material, a fraction of which spatially correlates with
the spiral arms structure of the Galaxy
\citep[e.g.][]{marshall2006,lallement2019}
toward which many sources of our sample are
located (see Figure~\ref{fig:cloudslocation}). 
In addition, no correlation is observed between the AME peak
frequencies and the thermal dust optical depths at 250$\,\mu$m, (see
Figure~\ref{fig:dust_optical_depth2}).
Similarily, no correlation is
observed between the width of the parabola used to fit the AME 
and the thermal dust optical depth
(see Figure~\ref{fig:dust_optical_depth3}). 
One can clearly see in that plot the cases for which the AME width reaches the upper limit of the prior $W_{\rm AME} = 1$. These cases
are not restrained to a specific range of the 
thermal dust optical depth parameter, 
which means that the AME
detections with $W_{\rm AME} = 1$ are not expected to depend on this parameter. 

\subsubsection{The interstellar radiation field: G$_{0}$}

Another important parameter that is useful to describe 
the physics of the several
environments towards AME regions is the relative strength of the ISRF, $G_{0}$ \citep[see][]{mathis1983}. AME
carriers are believed to be tiny particles
lying in
the bottom part of the interstellar dust grain size spectrum
($a \lesssim 1$ nm) (possibly including Polycyclic Aromatic Hydrocarbons or PAHs). Their chemical
properties, physical coherence and total charge could vary over time
and from one environment to another, and therefore
depend on the
relative strength of the ISRF. Therefore, 
having our estimation of $G_{0}$ 
is very useful to explore possible relations with the parameters used
to model the AME component detected at the SED level. 
An estimation of $G_{0}$ can be obtained from the equilibrium dust
temperature of the big dust grains ($T_{\rm BG}$) compared to the
average value of 17.5 K \citep[see][]{mathis1983}, with the relation: 
\begin{equation}
G_{0}= \left ( \frac{T_{\rm BG \it}}{17.5 \rm {K} \it} \right
  )^{4+\beta_{\rm BG}}, 
\label{eq:g0}
\end{equation} 
where $\beta_{\rm BG}$ is the spectral index associated with the opacity of the big grains. In the
following, we assume $T_{\rm BG} \approx T_{\rm dust}$, where $T_{\rm
  dust}$ is the averaged temperature of the thermal dust component
obtained from the fit on each region. As in PIRXV, we also
assume a constant value $\beta_{\rm BG}=2$. We note that using
$\beta_{\rm BG} \approx \beta_{\rm dust}$ could also be considered,
but would not change the conclusions of our analysis.

\begin{figure}
\begin{center}
\vspace*{2mm}
\hspace*{-1.cm}
\centering
\includegraphics[width=95mm,angle=0]{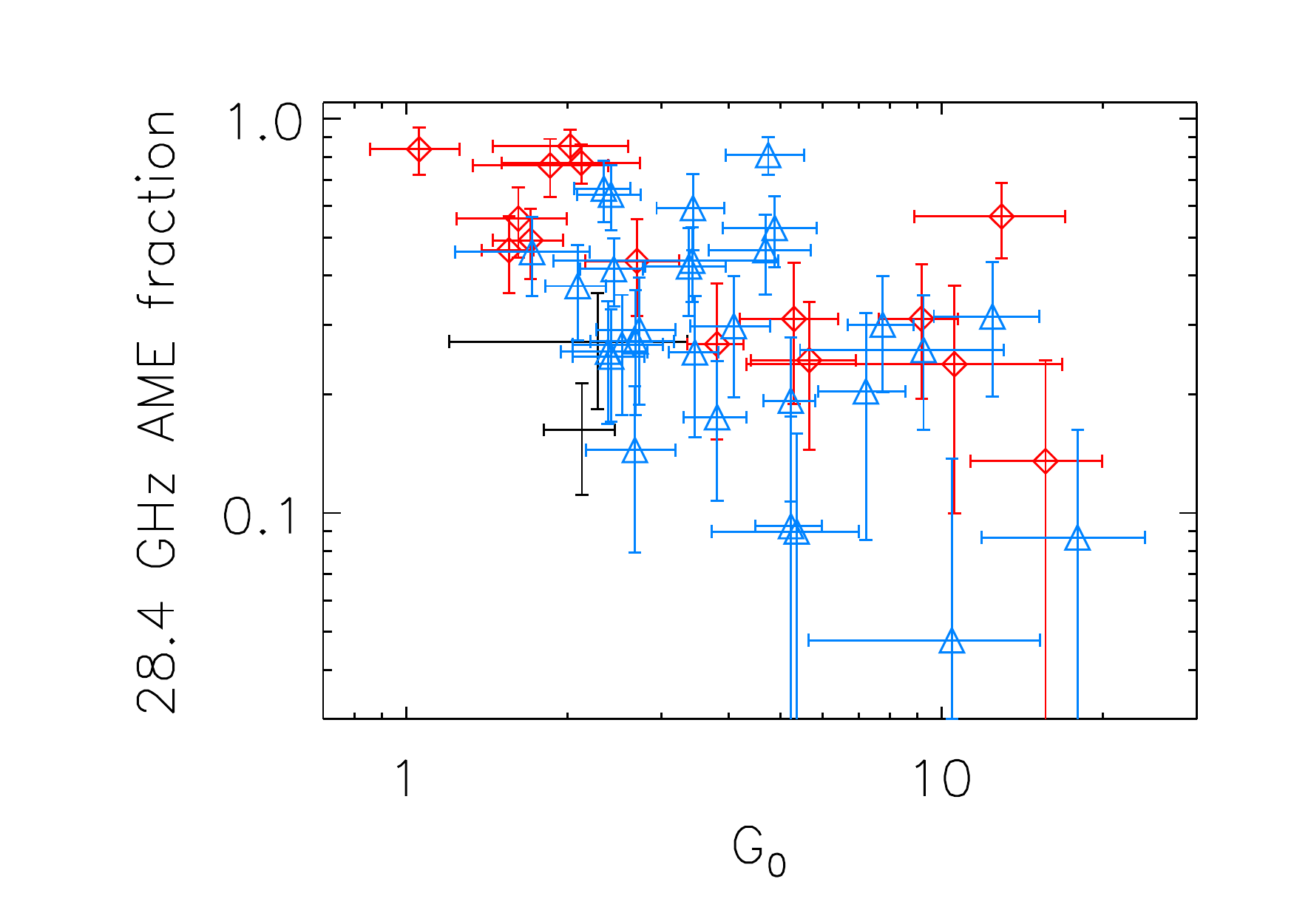}
\vspace*{-0.5cm}
\caption{AME fraction at 28.4$\,$GHz as a function of the estimated
  $G_{0}$. Symbols and colours definition are the same as in Figure~\ref{fig:td_betad}.  }
\label{fig:isrf}
\end{center}
\end{figure}

The correlation between the AME fraction at 28.4$\,$GHz 
(defined as the
residual AME flux density at 28.4$\,$GHz divided
by the total flux density at 28.4$\,$GHz) and G$_{0}$ is shown in
Figure~\ref{fig:isrf}. The data show a decrease of the AME fraction
as a function of G$_{0}$. This trend is similar to the one
obtained by PIRXV in their analysis and seems to be
dependent of the considered subsets. In our analysis
the slope of the ``significant'' AME detection data sample
is of order $\gamma = -0.48$, while the slope of
``semi-significant''  AME detection data sample is of order 
$\gamma = -0.61$. We point out that the uncertainties of
the values of the slopes we estimated are large, $\approx 0.8$ for both ``significant'' and ``semi-significant''  AME detections 
data points, which prevents a full and fair comparison with results
from previous studies. Our slopes, though, can be compared to the
slope of $\gamma = -0.11 \pm 0.04$ obtained by PIRXV on their 
strongest AME sources sample (see their Figure 15 and section
5.1.4), and to the slope of $\gamma = -0.59 \pm 0.11$ obtained 
on their semi-significant AME sources. 
All in all, our results agree with those of PIRXV within the
uncertainties. Differences in the slopes estimates can be 
explained by the different sample sizes (half-sky versus full sky coverage)
and by the introduction of the QUIJOTE data in our analysis.

\subsubsection{Peak frequency of AME} \label{ame_peak_frequency_hist} 
\begin{figure}
\begin{center}
\vspace*{2mm}
\hspace*{-0.5cm}
\centering
\includegraphics[width=90mm,angle=0]{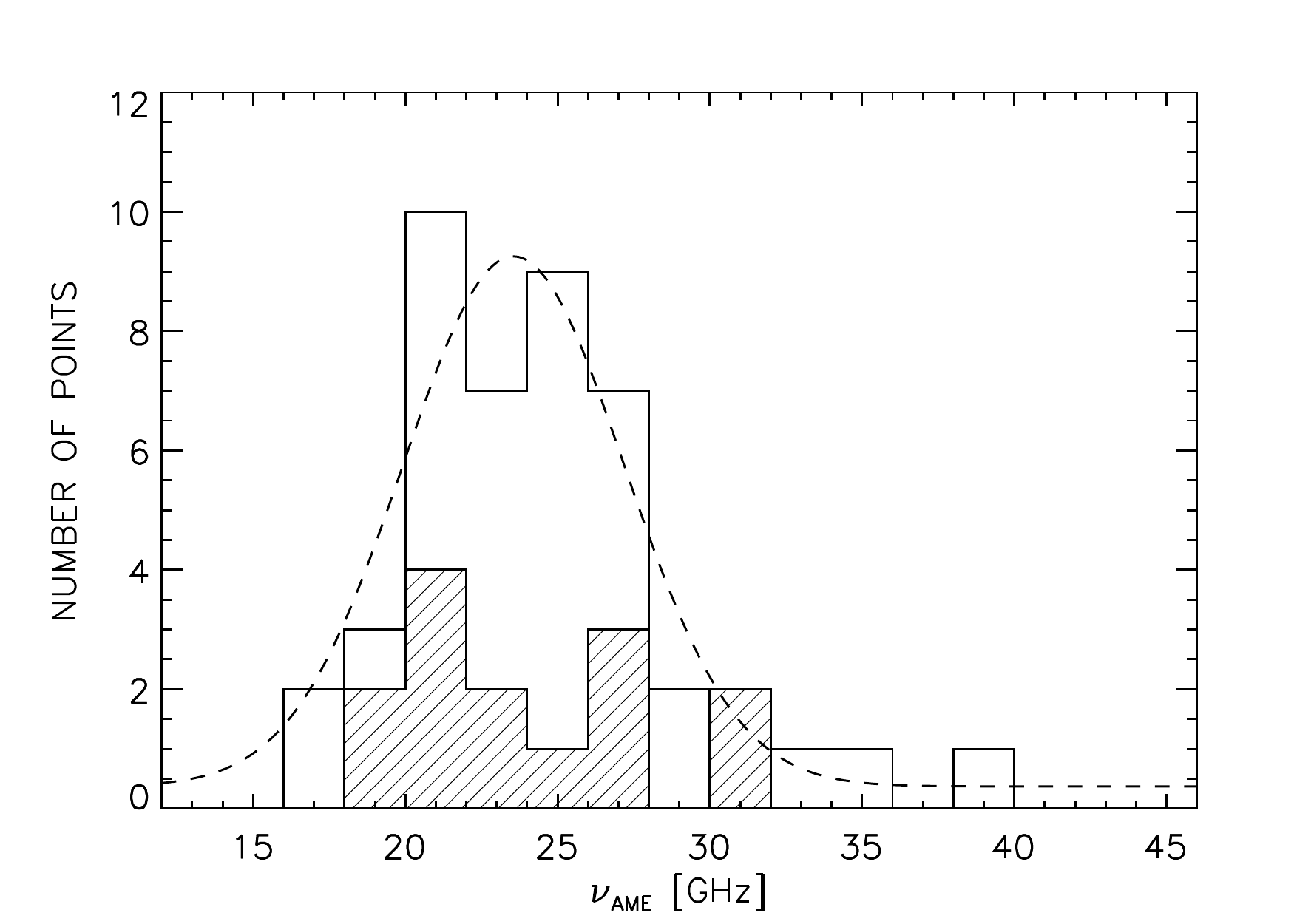}
\caption{Histogram of the AME peak frequency in bins of size 2$\,$GHz. The selected sample is shown as the unfilled histogram.
 The ``significant'' AME detection sample is shown with the hashed area. A Gaussian fit to the histogram is shown with the dashed-line.}
\label{fig:histo_ame_peak_freq}
\end{center}
\end{figure}

Among the three parameters used to fit the AME components in our sample, one is the
peak frequency, which is allowed to vary in the frequency
range 10--60$\,$GHz. Such a degree 
of freedom is important since it allows to get better final fits.
It has also been shown in previous works that one can expect the
frequency of AME to vary from one source to the other, or even within the same region \citep{cepeda-arroita2020}.  
The histogram of the AME peak frequency calculated
for the selected sample is shown in
Figure~\ref{fig:histo_ame_peak_freq}. The Gaussian fit to 
the distribution provides a 
mean frequency and dispersion given by
23.6 $\pm$ 3.6$\,$GHz. The hashed histogram shows 
the distribution of the ``significant'' AME sources sample
peaking around the weighted mean frequency.
PIRXV found their sample of AME sources to peak in 
the range 20--35$\,$GHz, with a weighted mean of 27.9$\,$GHz,
a bit higher than our mean value, the main reason of this difference being that flux densities in the frequency range 
10--20$\,$GHz were not available in their analysis. In fact, the addition of QUIJOTE-MFI data clearly helps reducing the uncertainty in the determination of $\nu_{\rm AME}$, thanks to allowing to trace the down-turn of the AME spectrum at low frequencies. Our average error on $\nu_{\rm AME}$ is 3.4$\,$GHz, and when we repeat our analysis excluding QUIJOTE-MFI data we get an average error of 7.5$\,$GHz (see also discussion in Section~\ref{sec:comparison_other_works}). 
On the other hand our analysis of G160.60-12.05 
(the California nebula/NGC 1499) recovers an AME 
peak frequency at 49.1 $\pm$38.5$\,$GHz, 
which is consistent with values obtained
in previous analyses \citep[][]{per20,pir15}.
The uncertainty on our estimate is quite large because the free-free dominates at $\nu<$100$\,$GHz making the width of the AME bump poorly constrained and the fitted parameters strongly degenerated. On top of that the circular aperture that we use may not be optimal in this case where the emission is elongated and pretty extended.

\subsubsection{Width of the AME bump} \label{ame_width_hist}

\begin{figure}
\begin{center}
\vspace*{2mm}
\hspace*{-0.5cm}
\centering
\includegraphics[width=90mm,angle=0]{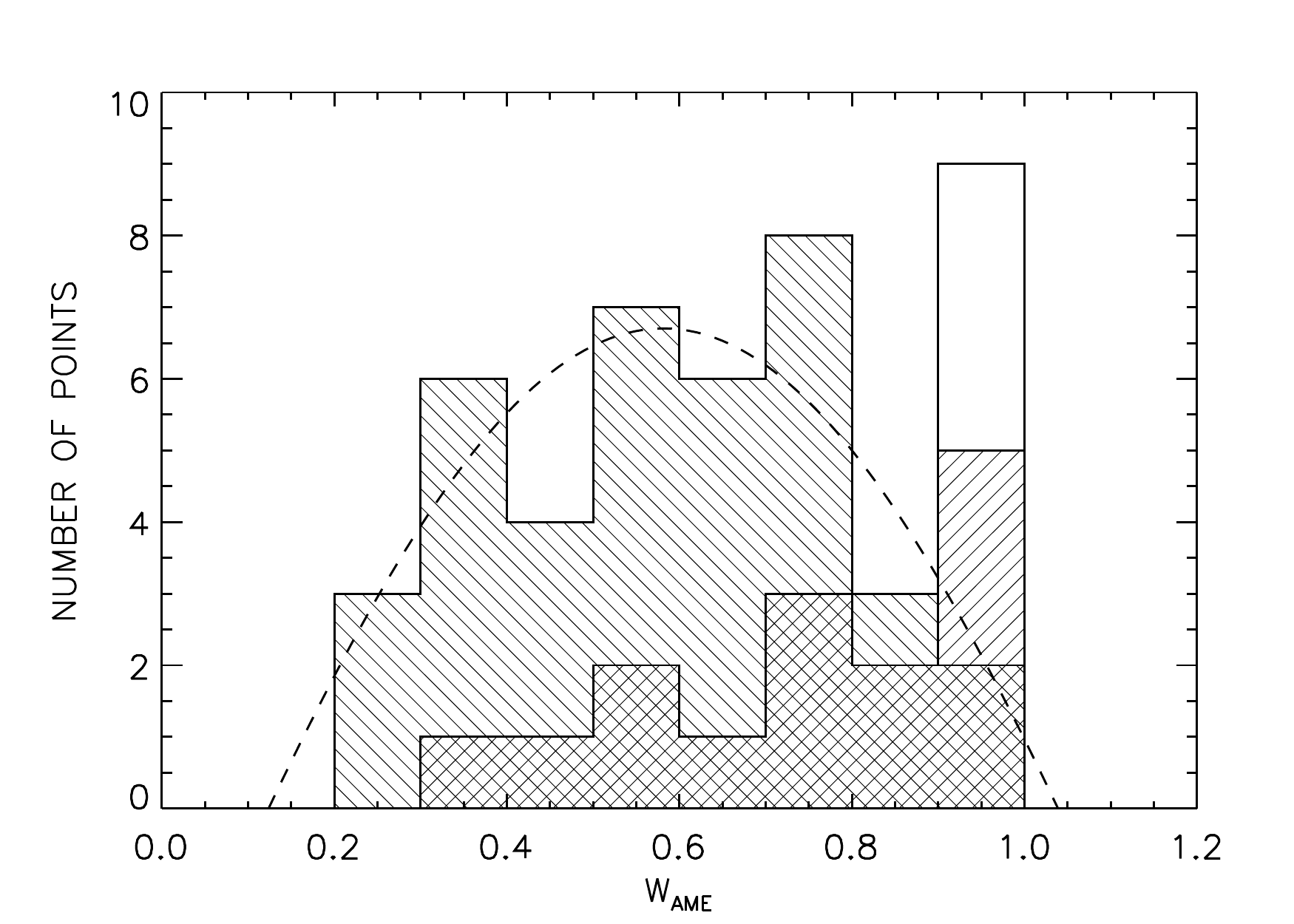}
\caption{Histogram of the width of the AME component parameterized by
  $W_{\rm AME}$ (see Equation~\ref{eq_parabola}) in bins of size 0.1. 
  The selected sample is shown as the whole histogram.
The ``constrained'' AME detections are shown with the unfilled histogram. 
The ``significant'' AME detection sample is shown with the double harshed area. The gaussian fit to the histogram is shown with the dashed-line.}

\label{fig:ame_width}
\end{center}
\end{figure}

In addition to the maximum flux density and peak frequency parameters,
the third parameter used to fit the AME
components is the width of the parabola, $W_{\rm AME}$ (see Equation~\ref{eq_parabola}). 
The allowed range in the fit was 0.2--1
and the initial value was $W_{\rm AME}=0.5$ for all sources,
this value being the expected average value 
from the \texttt{SPDust2} models.
The histogram of our fitted values is displayed in
Figure~\ref{fig:ame_width}. 
As discussed previously, the multicomponent fits
leading to output fit parameters of values
$W_{\rm AME}=1$ and $\sigma_{W_{\rm AME}}=0$
are cases reaching the prior upper limit value, and this artificially leads to a higher number of sources lying in the last bin of the histogram.
The selected sample is shown 
as the whole histogram.
The single-dashed histogram shows the same distribution
without the prior dominated AME detections. 
This distribution has a mean and dispersion given by,
$W_{\rm AME} =0.58 \pm 0.61$.
The distribution looks rather flat, and far from Gaussian, which is reflected in the large error bar of the Gaussian fit. This in fact illustrates that $W_{\rm AME}$ is maybe the worst constrained parameter in our fit, due to large degeneracies with other parameters.

This result is obtained with a bin of size 0.1 and would need a higher
sample for one to drive strong conclusions on a statistical basis. 
Indeed using a bin size of 0.2 the whole histogram looks rather like a 
normal distribution without any clear peak.  
Statistically, we find that $W_{\rm AME}$ does not correlate 
with the free-free component EM parameter. 
Neither do we find any correlation between $W_{\rm AME}$ and 
any of the thermal dust parameters.
On the other hand we observe a mild correlation 
of $W_{\rm AME}$ with the AME emissivity ($A_{\rm AME}/\tau_{250}$).
A detailed definition of the AME emissivity will be given in Section~\ref{amecompchar} where these results will be discussed. 

\subsubsection{Width of AME bump and peak frequency of AME} \label{ame_bump_freq}

The three parameters describing the parabola used to fit the AME flux
density bump (see Equation~\ref{eq_parabola}) are independent from
each other. With this model any correlation found between the AME peak
frequency and the parabola width parameter could therefore be
indicative of the physics underlying the description of the AME carriers.
We checked that neither a negative nor a positive correlation can be 
seen between the two parameters. As shown in Table~\ref{tab:srcc_ame_ame},
all the samples (selected, ``semi-significant'' and ``significant'') 
are showing SRCCs consistent with a null correlation.
These results show that the width
and the peak frequency of the AME component are fully
independent from each other, although this conclusion could be affected by the fact that, in some cases, $W_{\rm AME}$ seems to be poorly constrained in our analysis.

\subsection{Dust correlations}

In this section we focus on the thermal dust component with the aim to better understand its relation with the AME component. We also consider high frequency maps at 100$\,\mu$m,  60$\,\mu$m and  12$\,\mu$m, since these data have the potential to provide information about some of the candidate AME carriers (i.e., spinning dust, PAHs or fullerenes).

\subsubsection{Dust flux densities at 100$\,\mu$m, 60$\,\mu$m, 25$\,\mu$m and 12$\,\mu$m}

\begin{figure*}
\begin{center}
\vspace*{2mm}
\hspace*{-0.5cm}
\centering
\includegraphics[width=60mm,angle=0]{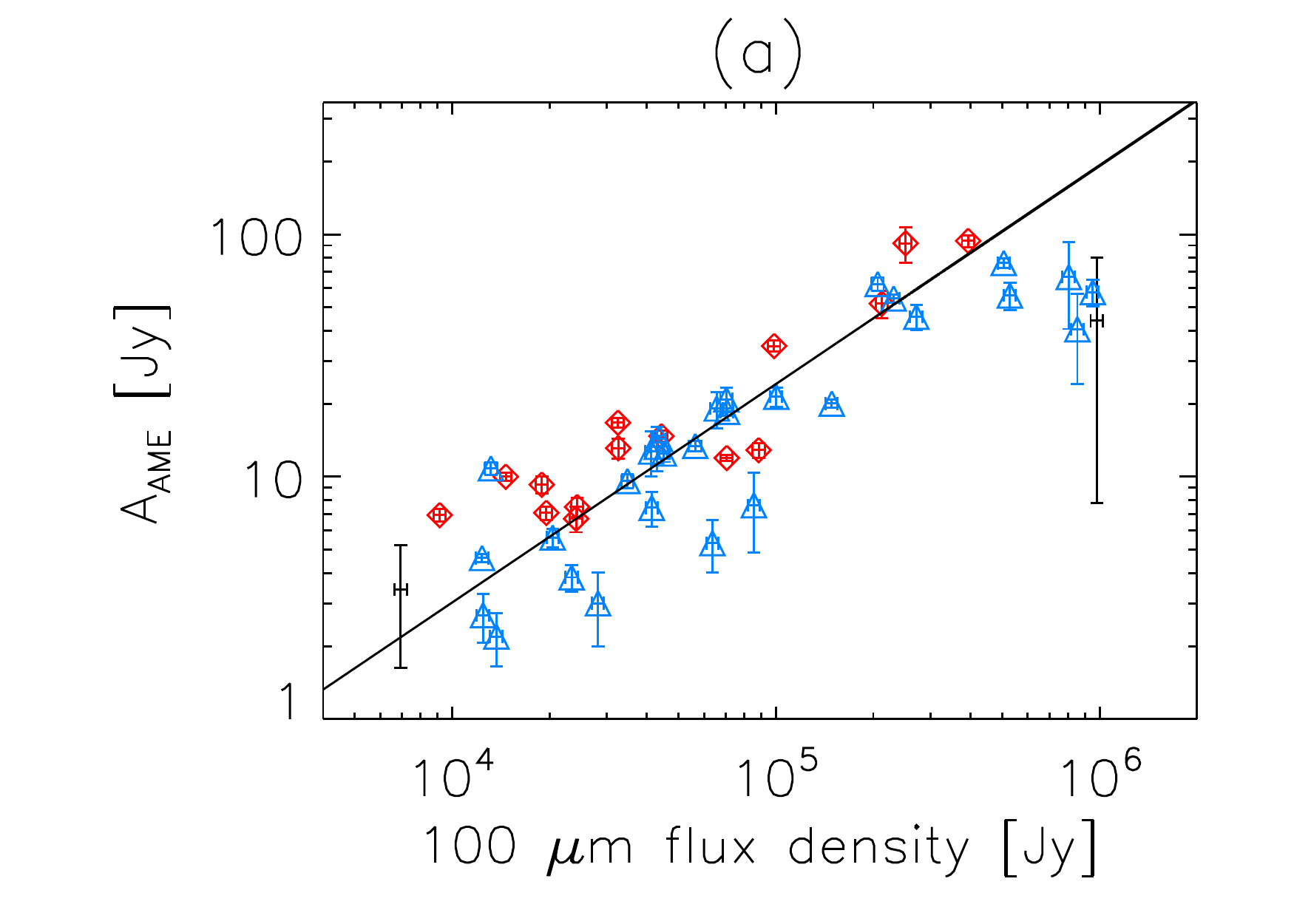}
\includegraphics[width=60mm,angle=0]{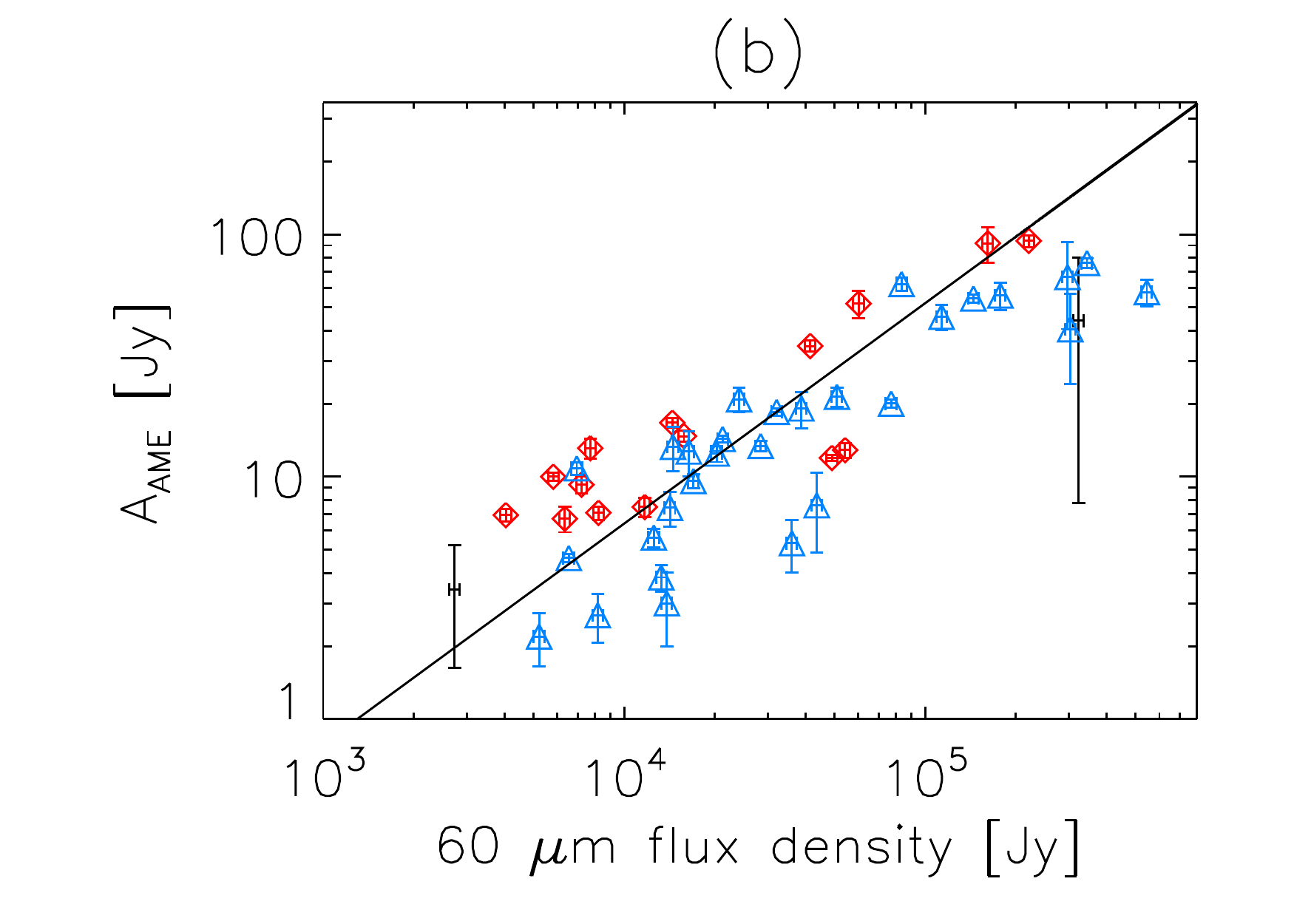}
\includegraphics[width=60mm,angle=0]{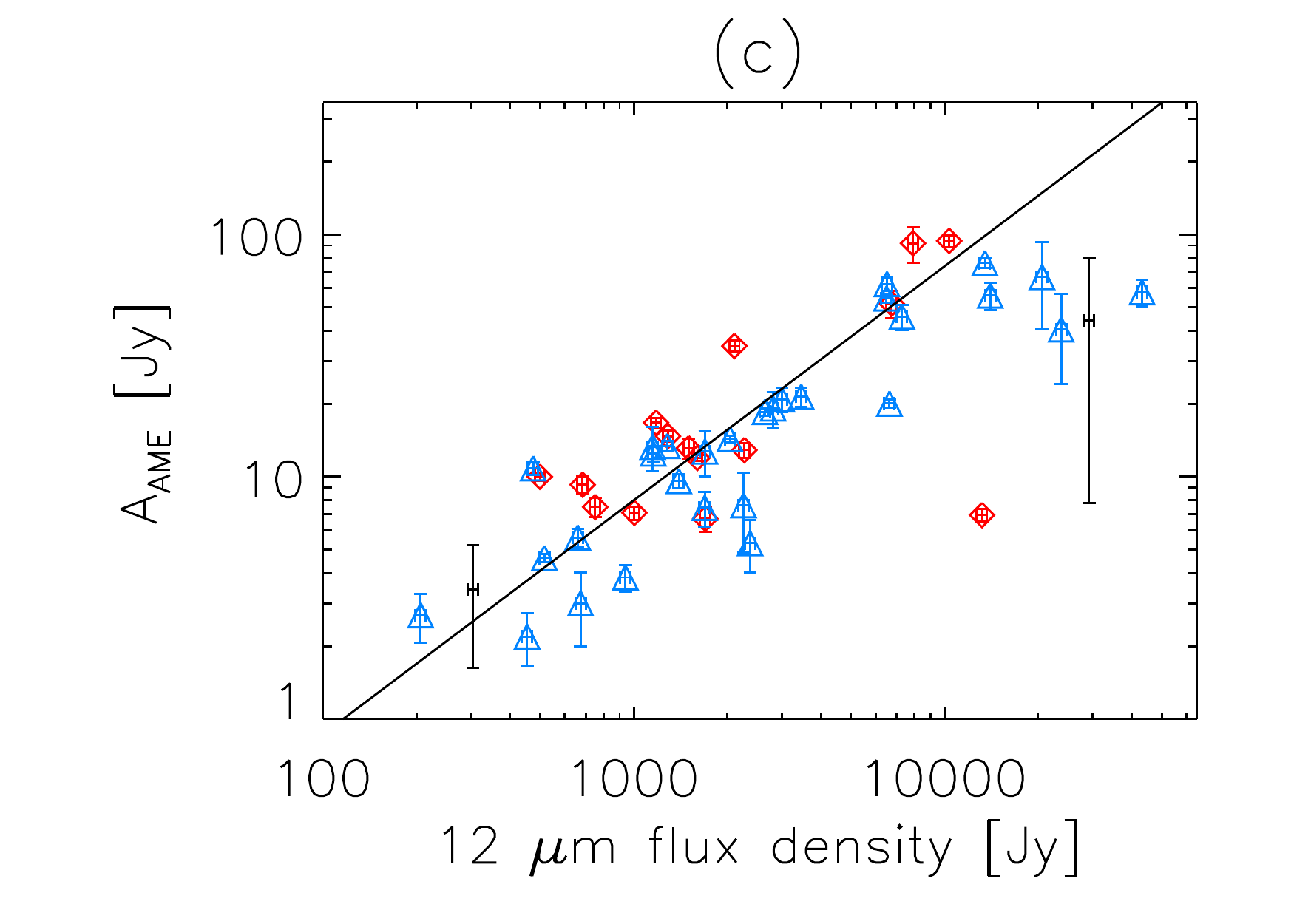}
\hspace*{-0.5cm}
\includegraphics[width=60mm,angle=0]{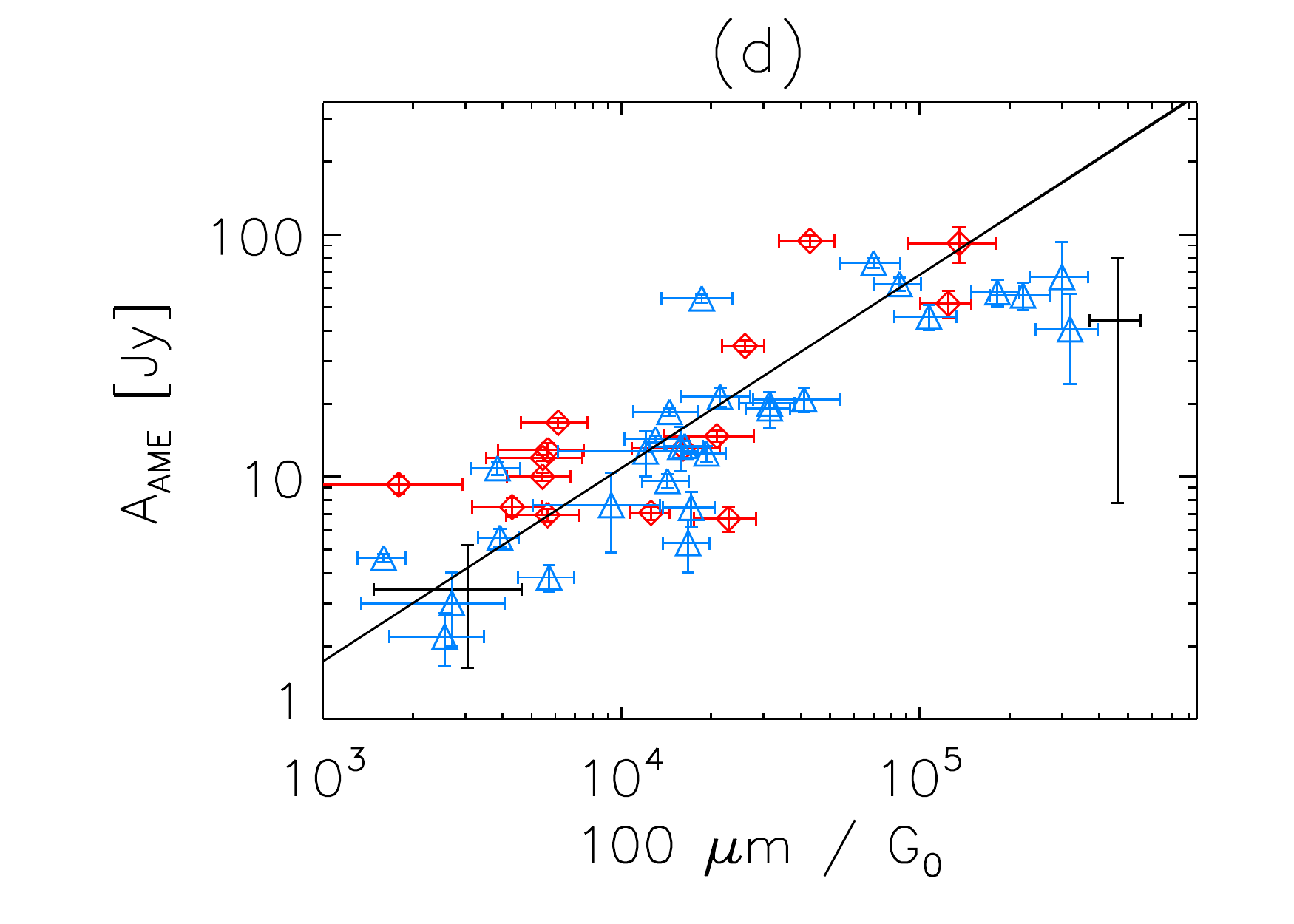}
\includegraphics[width=60mm,angle=0]{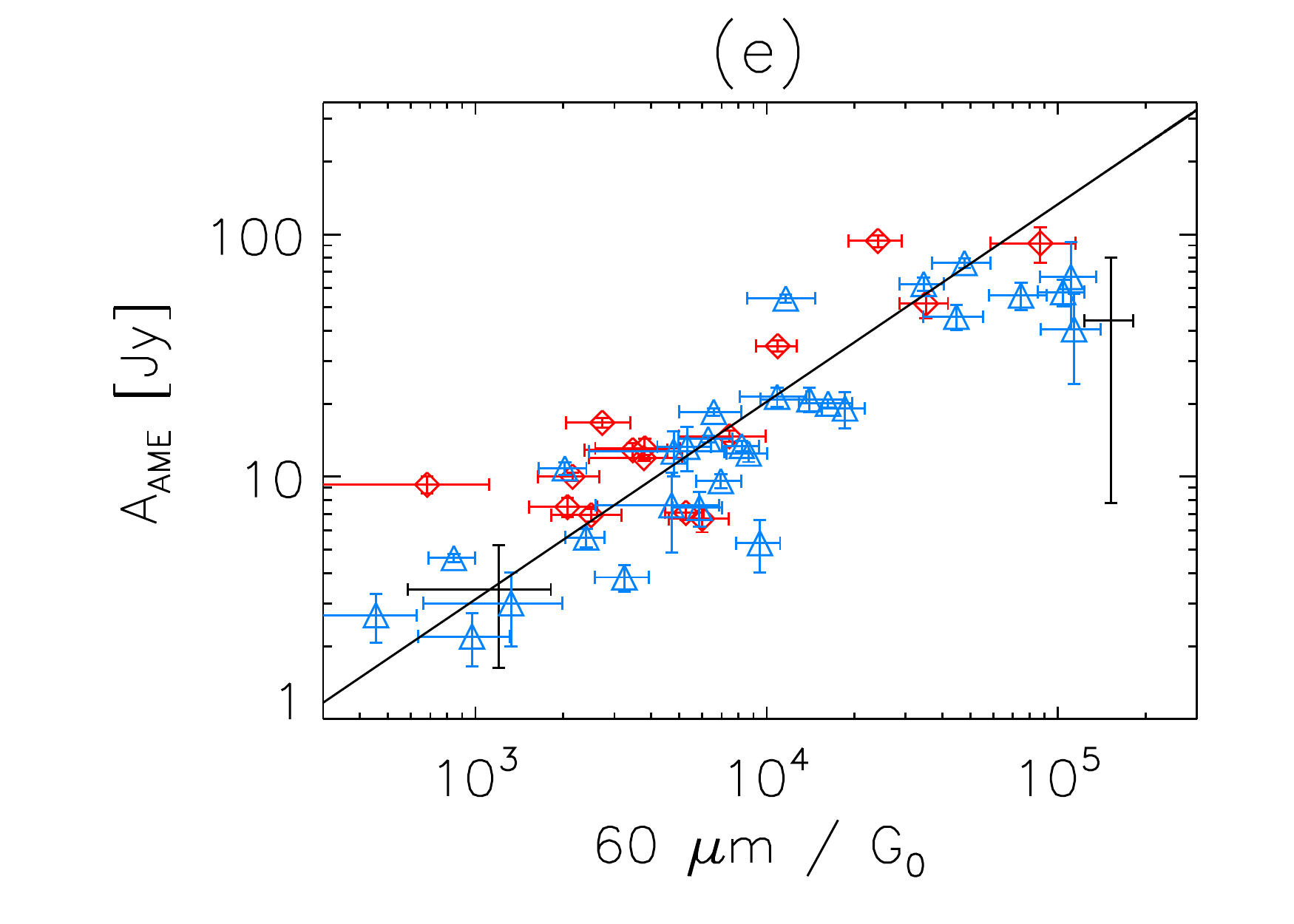}
\includegraphics[width=60mm,angle=0]{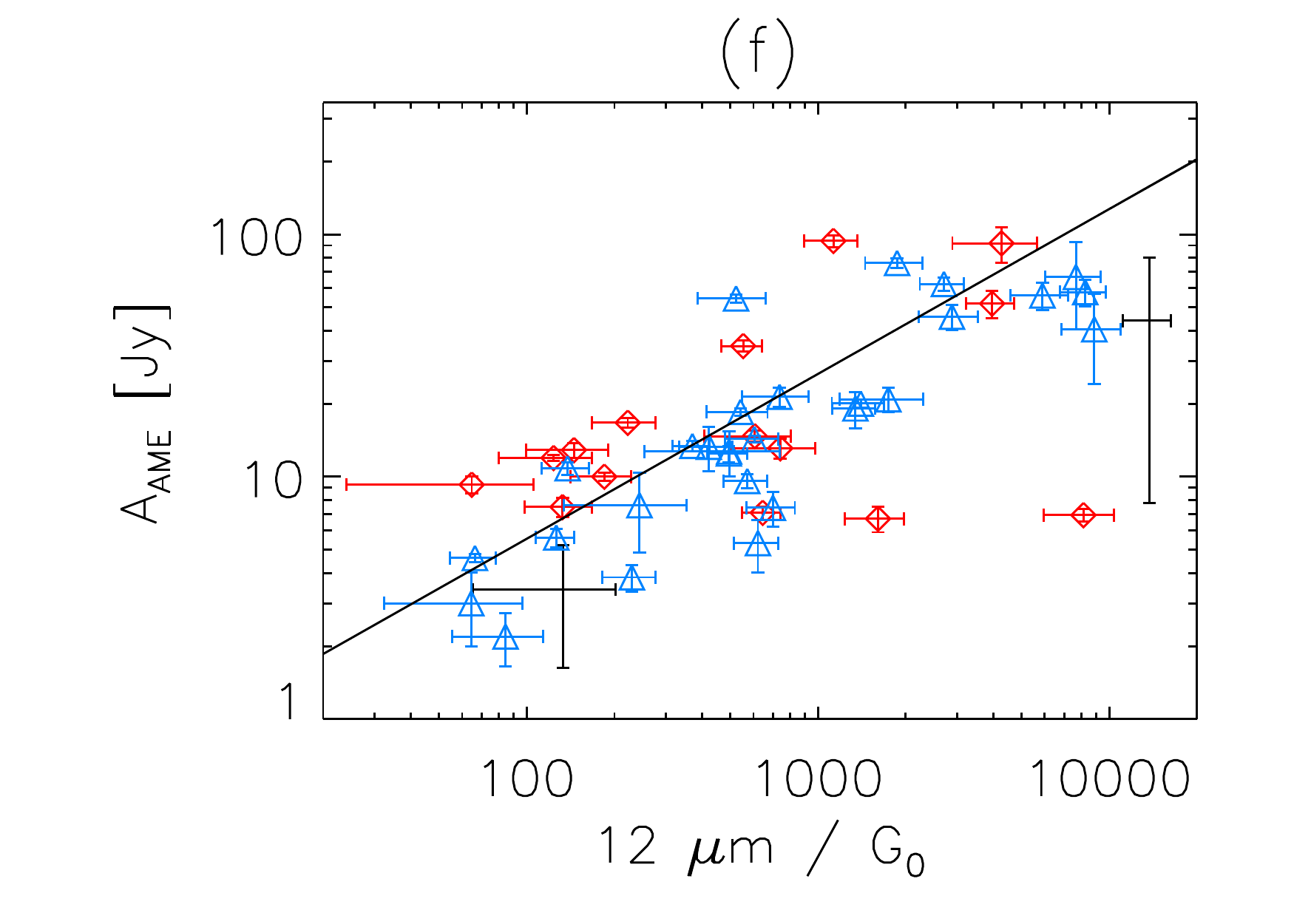}
\vspace*{-0.5cm}
\caption{Top row: AME peak flux density as a function of the
  100$\,\mu$m (panel a),  60$\,\mu$m (panel b) and  12$\,\mu$m (panel
  c) flux density. Bottom row: same as top row but after the infrared
  tracers of dust have been divided by $G_{0}$ (panel d, e and f,
  respectively). Symbols and colours definition are the same as in Figure~\ref{fig:td_betad}. Power-law fits to the full set are shown with
  back solid lines. SRCCs are given in Table~\ref{tab:srcc_dust_12_25_60_100}.}
\label{fig:dust_corr}
\end{center}
\end{figure*}

\input{polametex_table_sources_spearman_DUST_12_25_60_100.txt}

Following the spatial correlation observed between AME and the thermal
dust emission when AME was first discovered, many studies
have explored and discussed the possibility that AME carriers
are spinning dust grains in nature \citep[e.g.,][]{draine98,draine99,ali2009}, 
i.e., possibly a specific subclass of the dust grain population spectrum. 
A look to various dust grain emission templates should therefore be
useful to explore if any specific correlation exists between the
maximum AME flux densities and the flux densities of thermal dust
observed at 100$\,\mu$m, 60$\,\mu$m, 25$\,\mu$m and 12$\,\mu$m.
Such plots are shown in Figure~\ref{fig:dust_corr} (top row) and the strength of the
correlations described by their SRCCs are given in
Table~\ref{tab:srcc_dust_12_25_60_100}. We find very strong correlations
between the AME flux densities and the thermal dust flux densities at
100$\,\mu$m, 60$\,\mu$m, 25$\,\mu$m and 12$\,\mu$m.
This result is consistent with the one obtained by PIRXV from
their analysis.

If the AME carriers are spinning dust grains,
the AME component is expected to be quite insensitive to the ISRF
relative strength, $G_{0}$ \citep[][]{ali2009,ysard10} while 
on the contrary the thermal dust grains population is expected 
to be sensitive to it, mainly because the
UV radiation should control their temperature. If that was true
one would expect better correlations between the maximum
AME flux densities and the flux densities of thermal dust observed
at 100$\,\mu$m, 60$\,\mu$m, 25$\,\mu$m and 12$\,\mu$m,
once they are normalized by $G_{0}$.
This has been discussed in some previous analysis
\citep[e.g.,][]{ysard10}. The plots obtained once the thermal dust
fluxes are normalized by $G_{0}$ are shown in
Figure~\ref{fig:dust_corr} (bottom row) and the strength of 
the correlations described by their SRCCs are given between
parenthesis in Table~\ref{tab:srcc_dust_12_25_60_100}. 
Contrary to what was found on their sample by PIRXV, 
normalizing the thermal dust templates by $G_{0}$ 
leads to less tight correlations. 
These results suggest that the AME carriers could be
coupled to the thermal dust grain components rather than to
a dust grain population relatively insensitive to $G_{0}$. 
On the other hand the dust grain size distribution is very
sensitive to the ISRF, as well as to other parameters such 
as the dipole moments of PAHs \citep[][]{ali2009}, meaning 
that the interpretation of the results obtained with plots 
such as those given in Figure~\ref{fig:dust_corr} may be
complicated. The role of $G_{0}$ will be discussed further 
in Section~\ref{role_of_the_isrf}.

\subsubsection{Thermal Dust peak flux densities}

\begin{figure}
\begin{center}
\vspace*{2mm}
\hspace*{-1.cm}
\centering
\includegraphics[width=95mm,angle=0]{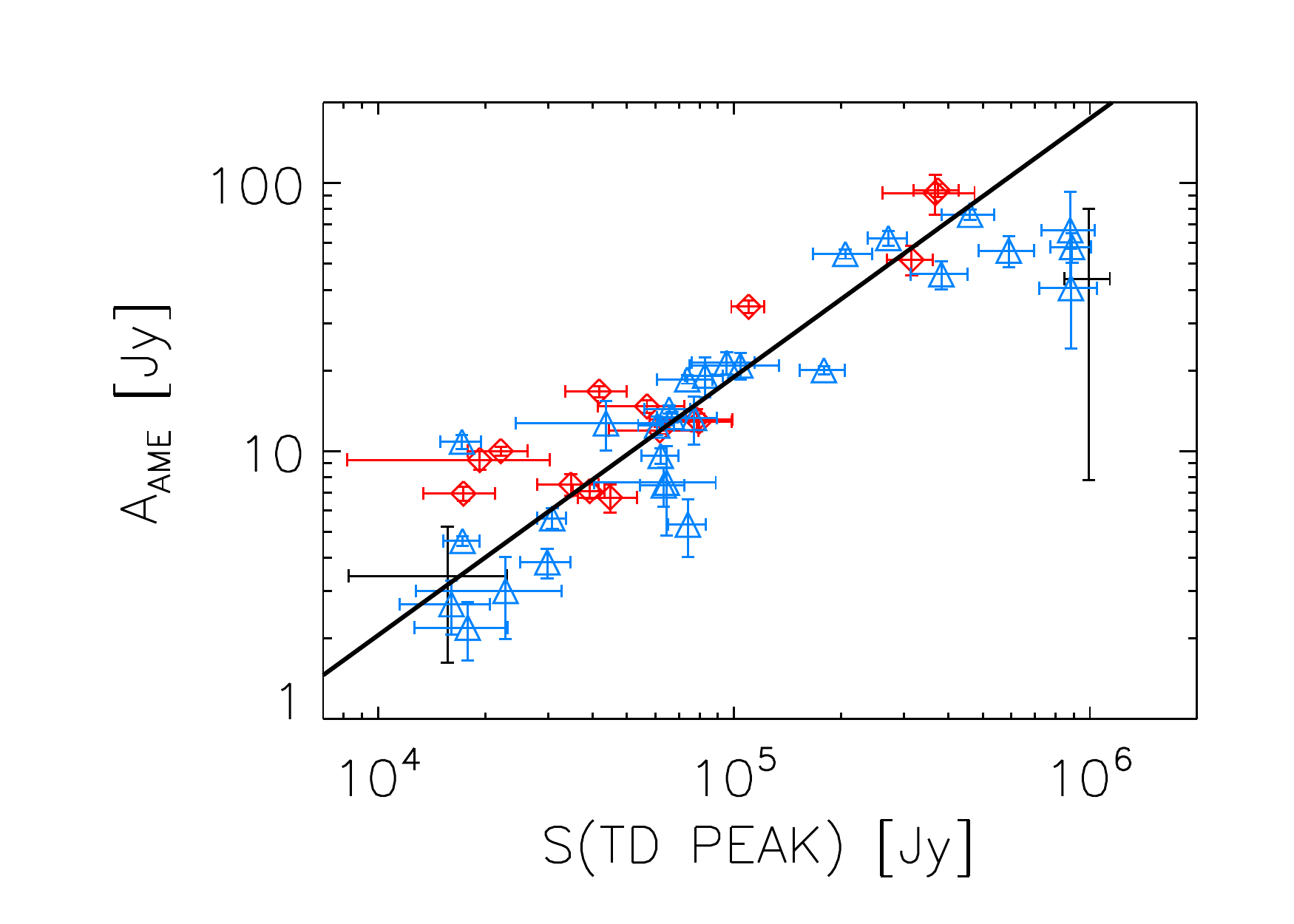}
\vspace*{-0.5cm}
\caption{Maximum AME flux density versus maximum thermal dust flux
  density. Symbols and colours definition are the same as in Figure~\ref{fig:td_betad}. The solid line represents a fitted power-law model to the data.}
\label{fig:ame_peak_dust_peak}
\end{center}
\end{figure}

The size of the aperture used to build the SEDs could
introduce a coupling between some of the thermal dust parameters
$\tau_{250}$, $T_{\rm dust}$ and $\beta_{\rm dust}$ due
to a possible range of degeneracy at the fit level between these
parameters. In order to circumvent this problem, that could mislead the
interpretation of some of the correlations discussed above, 
we looked at the distribution between the flux densities at 
the peak of the AME bumps and at the maximum of the thermal dust
components. This is shown in Figure~\ref{fig:ame_peak_dust_peak} 
where it can be seen a correlation between the two flux components 
at their maximum. The slope of a power-law fit to the selected 
sample is 0.96 and almost consistent with 1 as shown with the dark 
solid line on the plot. The SRRC between the two parameters is 
equal to 0.89 $\pm$ 0.05. 

\subsubsection{Thermal Dust radiance}

\begin{figure}
\begin{center}
\vspace*{2mm}
\hspace*{-1.cm}
\centering
\includegraphics[width=95mm,angle=0]{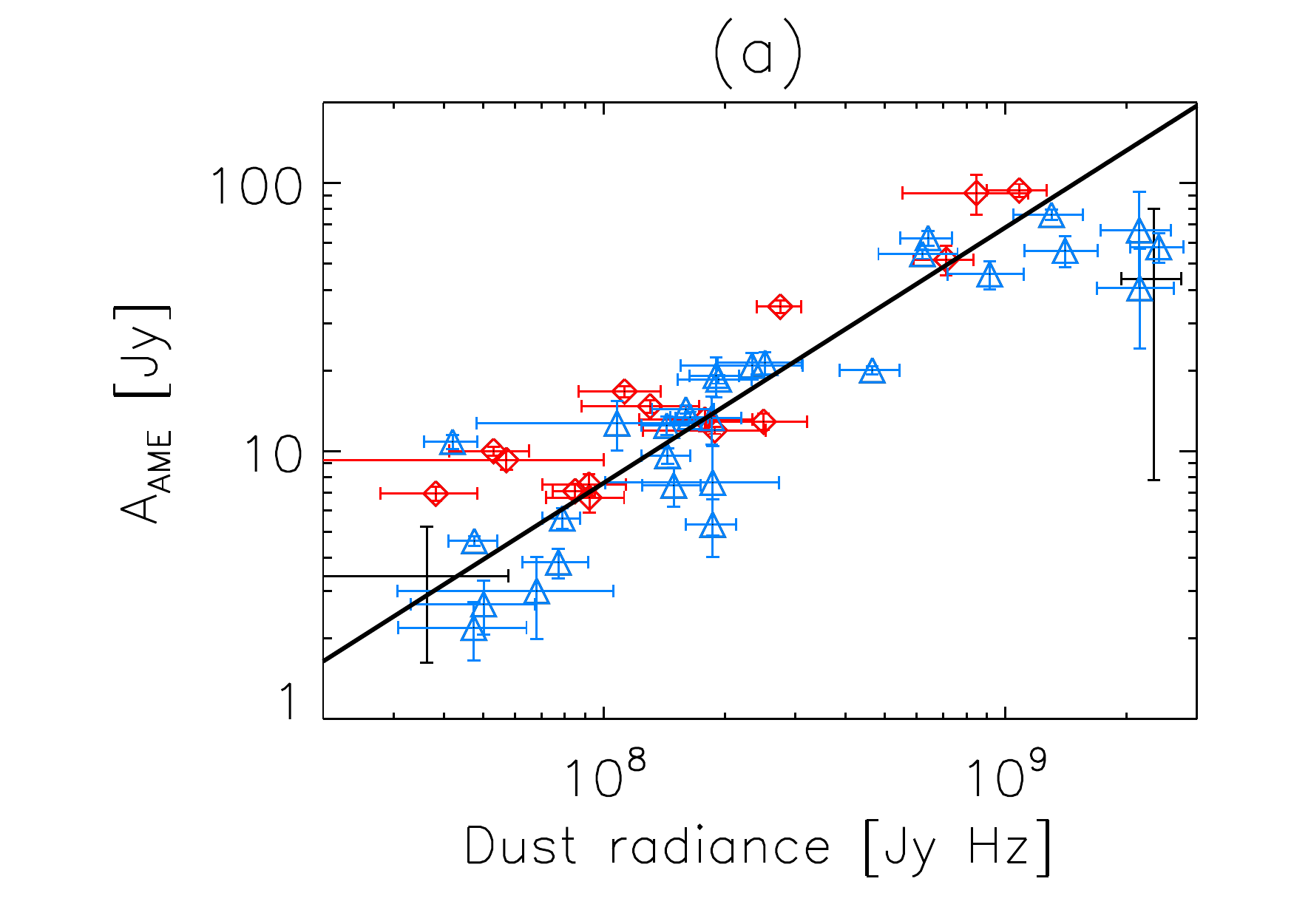}
\includegraphics[width=85mm,angle=0]{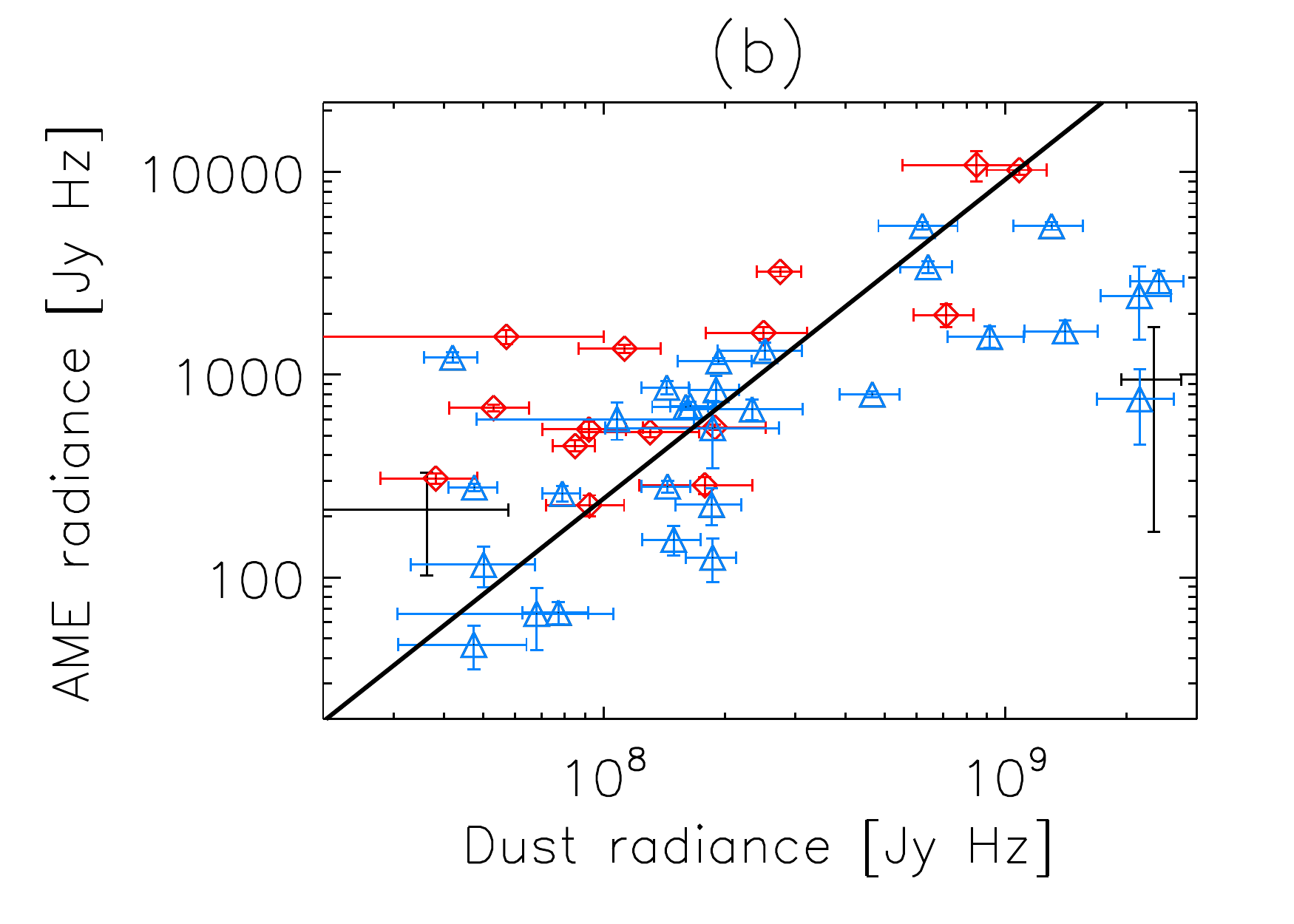}
\vspace*{-0.5cm}
\caption{Top: AME flux density at peak frequency, $A_{\rm AME}$, as a
  function of the thermal dust radiance, $\Re_{\rm td}$. Bottom: AME
  radiance, $\Re_{\rm AME}$, as a function of thermal dust radiance, $\Re_{\rm td}$.
Symbols and colours definition are the same as in Figure~\ref{fig:td_betad}. The solid lines represent fitted power-law models to the data.}
\label{fig:lumbol_ame_lumbol_dust}
\end{center}
\end{figure}

The radiance of a component is defined as the integral of the flux
density of that component over the full spectral range,
$\Re=\int _{-\infty}^{+\infty} S(\nu) d\nu$. 
In this work, all radiances were calculated by integrating the fitted models between 0.4 and 3000\,GHz, which 
is the frequency range where all the maps used in this analysis are
available (see Table~\ref{tab:surveydata}).
Some studies have shown strong correlations between the dust 
radiance and the AME amplitude at the peak frequency
\citep{hensley2016,hensley2017}.
The distribution of both components for our sample is shown 
in Figure~\ref{fig:lumbol_ame_lumbol_dust}, (top). 
A good correlation is observed between the two variables of 
the selected sample, with a SRCC of 0.89 $\pm$ 0.05, and a power-law
slope consistent with 1. This tight correlation suggests a strong
coupling between the big dust grains expected to be the main
contributors to the dust grain radiance considered here (i.e.,
in the wavelength range $\lambda > 100 \mu$m).  
Figure~\ref{fig:lumbol_ame_lumbol_dust} (bottom)
shows the distribution of the AME radiance $\Re_{\rm AME}$ as a
function of the dust radiance $\Re_{\rm td}$. In that case a
lower correlation is observed between the two parameters with a SRCC
of 0.70 $\pm$ 0.06. 

We believe that the reason why the AME amplitude correlates better than the AME radiance is because the latter is quite sensitive to $W_{\rm AME}$, and this parameter has large error bars due to not being very well constrained by our fit (see section~\ref{ame_width_hist}).
This said, these two correlations can be interpreted using two different views.
A first one is that the AME model used to fit the
data and designed to approximate the spectrum of the spinning dust
emission is not fully appropriate to capture the contribution of the
AME carriers, or that in some regions it is
difficult to properly disentangle the AME contribution from the
free-free and thermal dust contributions.  
Another view could be that if the AME model used to fit the data is
good enough to capture the AME components accurately,
then the dust radiance of PAHs and/or Very Small Grains (VSGs) 
could represent a relatively large contribution of the total dust
radiance at wavelengths greater than 100$\,\mu$m. 

\subsection{AME emissivity}

\begin{figure}
\begin{center}
\vspace*{2mm}
\centering
\includegraphics[width=85mm,angle=0]{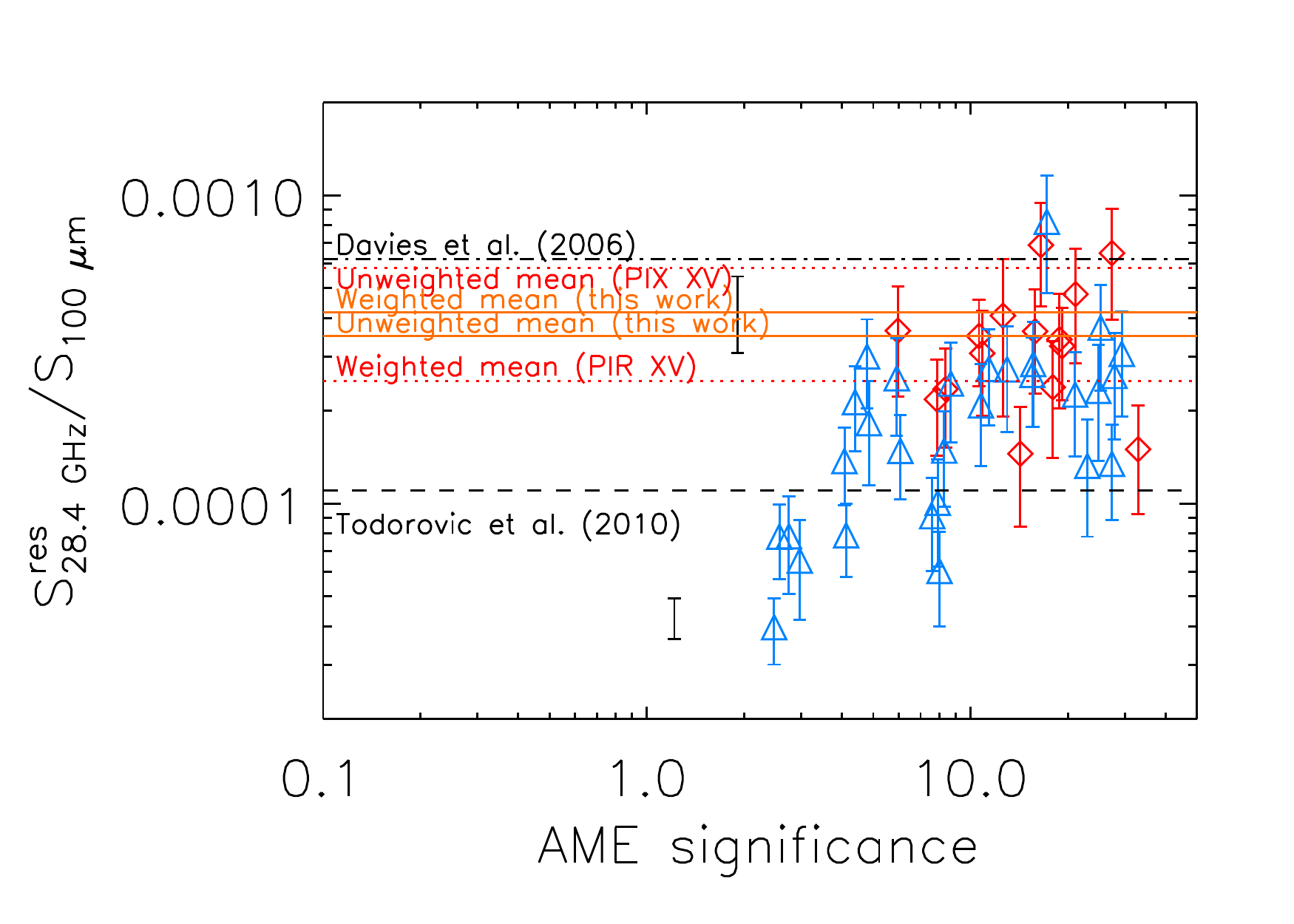}
\vspace*{-0.5cm}
\caption{AME emissivity against AME significance. Symbols and their colours definition are the same as in Figure~\ref{fig:td_betad}.}
\label{fig:60_PARAMETERS_RES28_div_F100mu_vs_sigma_AME}
\end{center}
\end{figure}

As discussed above, strong spatial correlations
were found between the AME emission and thermal dust emission
when AME was first detected \citep[see][]{kogut1996,leitch1997}.
In order to build a picture of the distribution of the AME emission
along the third spatial dimension (i.e., the line-of-sight, LOS), further works have defined the AME emissivity as the ratio between the
AME intensity and the column density, for which the optical depth at a
given wavelength is often used as a proxy \citep[see][and discussion
and references therein]{dickinson18}. In order to make comparisons
with results discussed in the literature we first show in
Figure~\ref{fig:60_PARAMETERS_RES28_div_F100mu_vs_sigma_AME}
the distribution of the AME flux density obtained by subtracting to the measured flux density at
this same frequency all the other components (defined as the residual flux density at 28.4$\,$GHz) normalized
by the 100$\,\mu$m flux density ($S_{28.4 {\rm GHz}}^{\rm res}/S_{100\mu{\rm m}}$), as a function of the AME detection
significance. In this case the 100$\,\mu$m flux density is expected to
be optically thin for a given dust temperature and composition and is
used as a proxy to probe the column density of dust along the LOSs.
$S_{28.4 {\rm GHz}}^{\rm res}/S_{100\mu{\rm m}}$ is in the range $(0.05-9)\times 10^{-4}$ 
with a weighted mean of $(4.2 \pm 0.3) \times 10^{-4}$ and 
an unweighted average of $(3.5 \pm 1.6) \times 10^{-4}$ 
(significant AME sample). These values are 
consistent with each other. They are smaller
than the unweighted average value of $(5.8 \pm 0.7) \times 10^{-4}$
of PIRXV and than the $6.2 \times 10^{-4}$ value of
\cite{davies2006} but are higher than the weighted average
of $(2.5 \pm 0.2) \times 10^{-4}$obtained in PIRXV
and than the value of about $1.1 \times 10^{-4}$ obtained by
\cite{todorovic2010} on a sample of \sc{H\,ii} \rm{} regions.
The differences between our estimates and those obtained by PIRXV could partially come from the different samples used in each study. Our sample only covers the North hemisphere sky while the analysis of PIRXV includes also sources in the Southern hemisphere. Different error treatment may also affect the weighted averages. Regardless of these issues, we have applied a one-to-one comparison between our flux density ratios and those reported in PIRXV in the subsample of 42 common sources. When we represent the former against the latter and fit the data to a straight line we find a slope of $0.76$, meaning that we find $\approx 30\%$ higher emissivities. This is a consequence of the increase of the AME amplitude as a result of the inclusion of QUIJOTE data (see Figure~\ref{fig:comp_pirxv}c and related discussion in section~\ref{sec:comparison_other_works}). A summary of these results is given in Table~\ref{tab:comp_emissivities}.

\begin{table}
\begin{center}
\begin{tabular}{lcc}
\hline\hline
\noalign{\smallskip}
 Sample & \multicolumn{2}{c}{$S_{28.4 {\rm GHz}}^{\rm res}/S_{100\mu{\rm m}}$ [$\times 10^{-4}$]} \\
\cline{2-3}
      & unweighted mean  & weighted mean  \\
\noalign{\smallskip}
\hline
\noalign{\smallskip}
This work - selected sample & 2.5 $\pm$ 1.7 & 3.7 $\pm$ 0.1\\
This work - semi-significant & 2.1 $\pm$ 1.5 & 3.2 $\pm$ 0.1\\
This work - significant &  3.5 $\pm$ 1.6 & 4.2 $\pm$ 0.3 \\
PIRXV - significant & 5.8 $\pm$ 0.7 & 2.5 $\pm$ 0.2 \\
\cite{todorovic2010} & 1.1 $\pm$ - & ... $\pm$ ... \\
\cite{davies2006} & 6.2 $\pm$ - & ... $\pm$ ... \\
\noalign{\smallskip}
\hline\hline
\end{tabular}
\end{center}
\normalsize
\caption{Comparison of the AME flux densities normalized by the 100$\,\mu$m flux densities obtained in this work and in previous studies.}
\label{tab:comp_emissivities}
\end{table}

The small range of values of the ratio of the AME residual flux density at 28.4$\,$GHz to the flux density at
100$\,\mu$m suggests that a power-law index of order 1
could be expected between the two flux density distributions.
This is indeed what the best-fitting power-law confirms as it yields
a power-law index of $1.04\pm 0.21$ in tension with the power-law
index of $0.67\pm 0.03$ obtained by PIRXV on their sample.
Similarily, the best-fitting power-law index between the AME 
residual flux density at 28.4$\,$GHz and the dust optical depth 
at a wavelength of 250 $\mu$m, $\tau_{250}$, yields a power-law 
index of $1.13\pm 0.22$ in agreement with the power-law index
of $1.03\pm 0.03$ obtained by PIRXV. The results obtained by
PIRXV were inferring an AME mainly proportional to the 
column density estimate, i.e., to the amount of
material along the LOS. This is what we find whether we consider
the 100\,$\mu$m map or the $\tau_{250}$ parameters as proxies of the
column density.

\subsection{Role of the ISRF} \label{role_of_the_isrf}

\begin{figure}
\begin{center}
\vspace*{2mm}
\hspace*{-1.cm}
\centering
\includegraphics[width=95mm,angle=0]{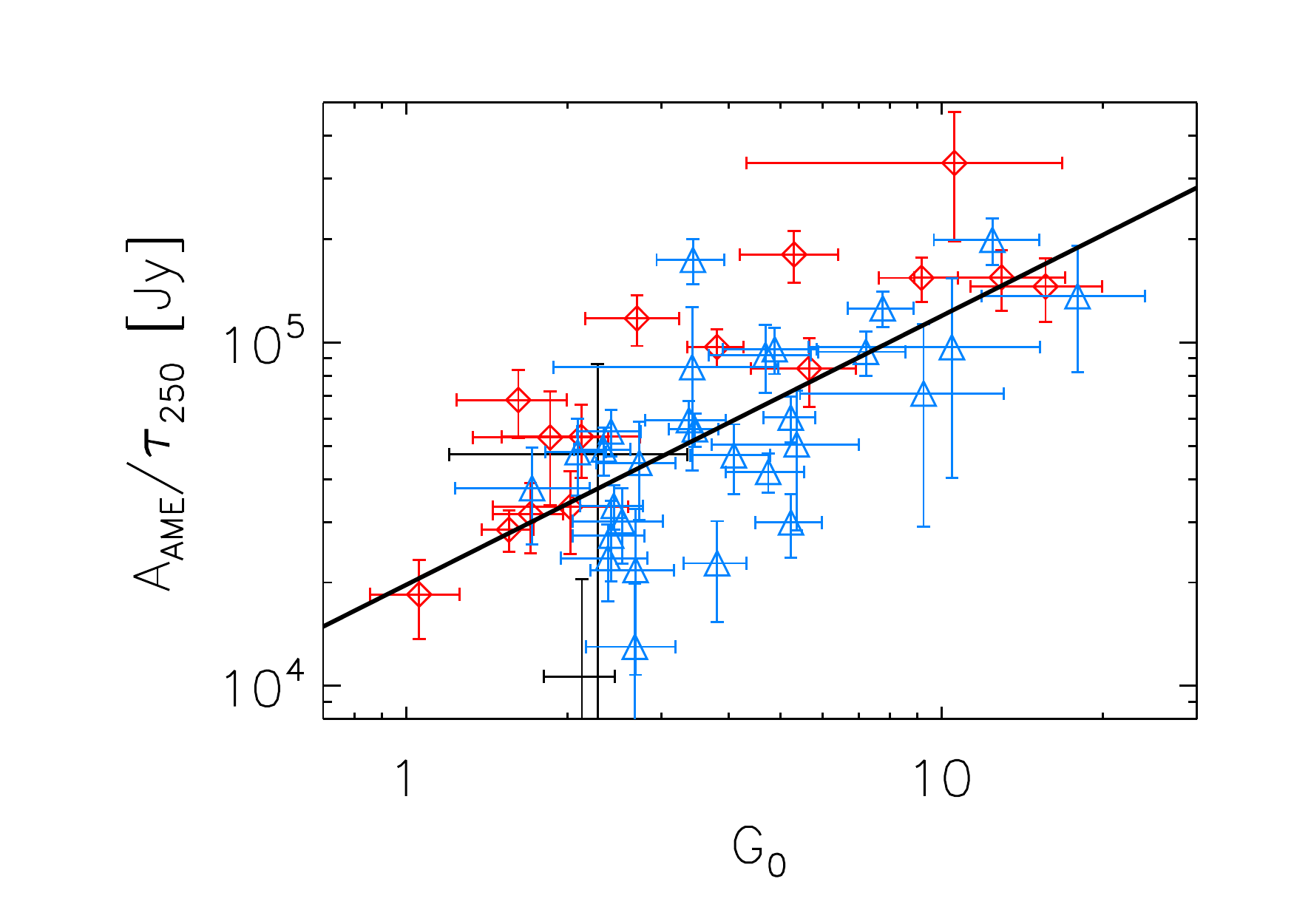}
\vspace*{-0.5cm}
\caption{Variations of the AME emissivity with the
  relative strength of the ISRF, $G_{0}$. Symbols and colours definition are the same as in Figure~\ref{fig:td_betad}. The power-law fit obtained on the selected sample is plotted with
 the black line.}
\label{fig:role_isrf}
\end{center}
\end{figure}

The ISRF is strongly coupled to the nature of the various phases encountered in the
ISM defined in terms of gas temperature and matter density. The UV
light produced by the population of stars pervading the ISM
is absorbed by the dust grain populations and re-radiated in the IR.
The ISRF therefore plays an important dynamic role since it will affect
the chemical composition of the ISM material, the dust grain
distribution as well as the lifetime
of the small dust grain and complex molecule populations
\citep[see][]{jones2013}. It is therefore interesting to
investigate the existence of possible
relationships between the relative strength of the ISRF,
$G_{0}$, and the parameters describing the AME component
derived from the SEDs analysis. For this we looked at the distribution
of the AME emissivity, now defined as $A_{\rm AME}/\tau_{250}$, the AME peak
frequency, $\nu_{\rm AME}$, and the AME bump width parameter,
$W_{\rm AME}$, as a function of $G_{0}$. The plots are
shown in Figures~\ref{fig:role_isrf},
Figure~\ref{fig:role_isrf2} and
Figure~\ref{fig:role_isrf3}, respectively. We find
poor correlations between $G_{0}$ and the AME parameters 
$\nu_{\rm AME}$ and $W_{\rm AME}$. On the other hand, 
we find a SRCC of r$_{\rm s}= 0.68 \pm 0.08$ between
the AME emissivity and $G_{0}$ parameters for the 
selected sample (Figure~\ref{fig:role_isrf}).
This distribution can be fitted by a power-law of index of about
0.8 as shown with the black line in Figure~\ref{fig:role_isrf}.
Since we derived the relative strength of the ISRF, $G_{0}$, by using
the thermal dust grain temperature, $T_{\rm dust}$, obtained from the
SED grey body fits, and by assuming a maximum and constant thermal dust emissivity, $\beta_{\rm dust}=2$ (see equation~\ref{eq:g0}), 
the SRCCs obtained between the
$A_{\rm AME}/\tau_{250}$ and $G_{0}$ parameter distributions
and between the $A_{\rm AME}/\tau_{250}$ and $T_{\rm dust}$ 
parameter distributions are by construction identical.
Similarly, the introduction of the SEDs fit estimates 
of $\beta_{\rm dust}$ in the calculation of $G_{0}$ only 
changes SRCCs values by less than one percent. 
This means that the AME flux densities obtained at the peak
frequency are mainly correlated with the combination of 
the dust optical depth, $\tau_{250}$, and the thermal dust 
temperature $T_{\rm dust}$ parameters. This result is in 
agreement with the strong correlation obtained between the
AME peak flux densities and $\tau_{250}$, and with
the 100$\mu$m thermal dust fluxes discussed in the previous section.

\begin{figure}
\begin{center}
\vspace*{2mm}
\hspace*{-1.cm}
\centering
\includegraphics[width=95mm,angle=0]{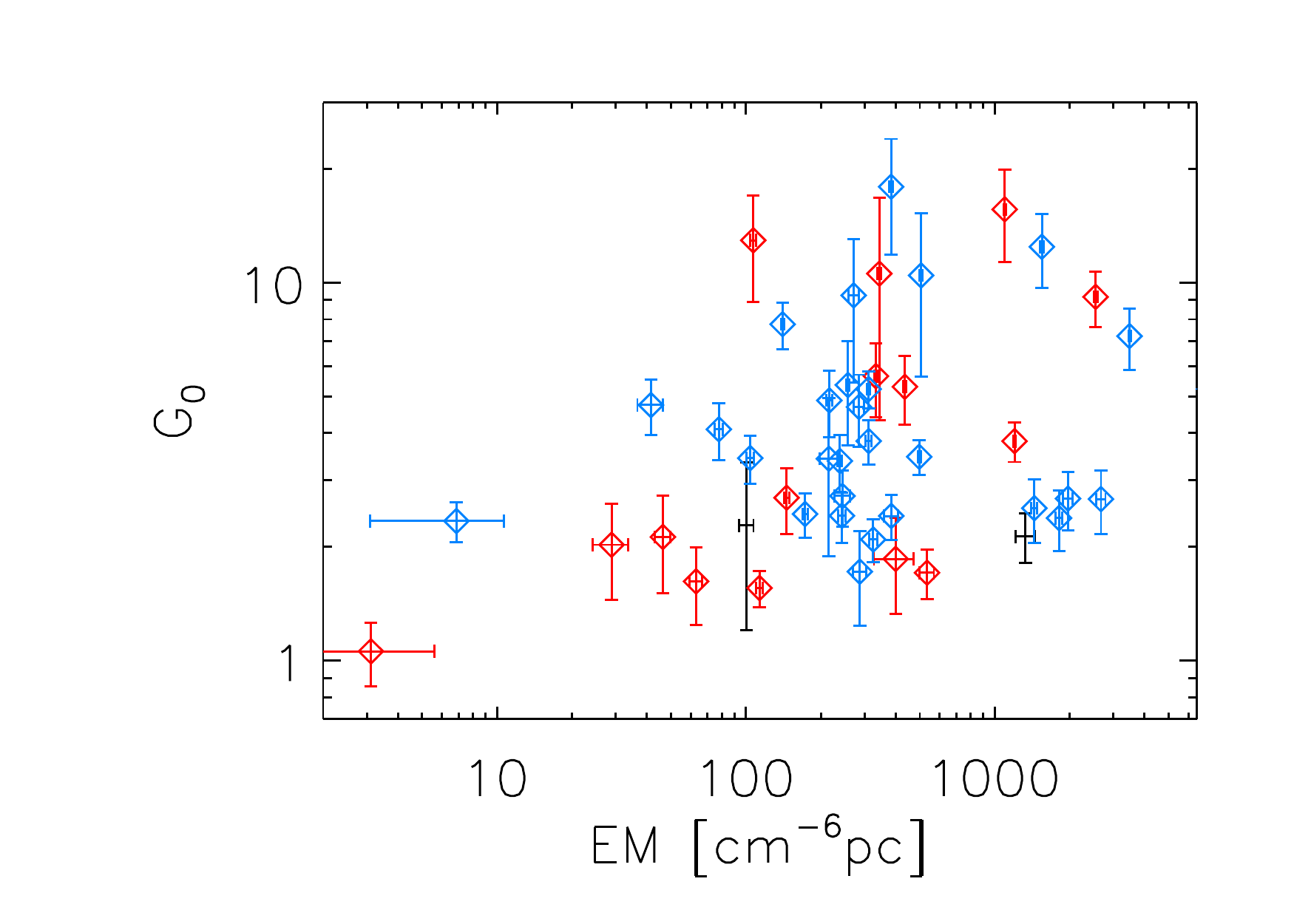}
\vspace*{-0.5cm}
\caption{Free-free Emission Measure (EM) parameter as a function of the relative
strength of the ISRF, $G_{0}$. Symbols and colours definition are the same as in Figure~\ref{fig:td_betad}.}
\label{fig:em_g0}
\end{center}
\end{figure}

In the above we have considered that a good proxy of the relative
strength of the ISRF is given by $G_{0}$, which is a function of the
thermal dust temperature, $T_{\rm dust}$. The EM is another 
interesting parameter associated with hot phases of the ISM, 
i.e., ionized regions. In our sample one can expect 
electron temperatures lying in the range 5\,458--7\,194\,K
as from the electron temperature map provided by \cite{Planck2015results_x}.
Inside molecular clouds, the ionized regions produced by stellar
radiation are expected to represent a fraction of the whole volume
associated with the clouds. Not all the sources displayed in
Table~\ref{tab:listofclouds} are only molecular cloud regions in nature
but they all have thermal dust along their LOSs, which is a component
strongly correlated with the AME component. In this context
we show in Figure~\ref{fig:em_g0} the distribution of the free-free
EM parameter as a function of $G_{0}$. 
The plot shows only a poor correlation between the two parameters, this being also illustrated by the low correlation coefficient, SRCC$=0.30\pm 0.06$, found between the two parameters. This lack of correlation would indicate 
that the AME emissivity does not correlate significantly with the 
EM free-free emission parameter at Galactic scales.

\subsection{Free-Free correlations}

\begin{figure}
\begin{center}
\vspace*{2mm}
\hspace*{-1.cm}
\centering
\includegraphics[width=95mm,angle=0]{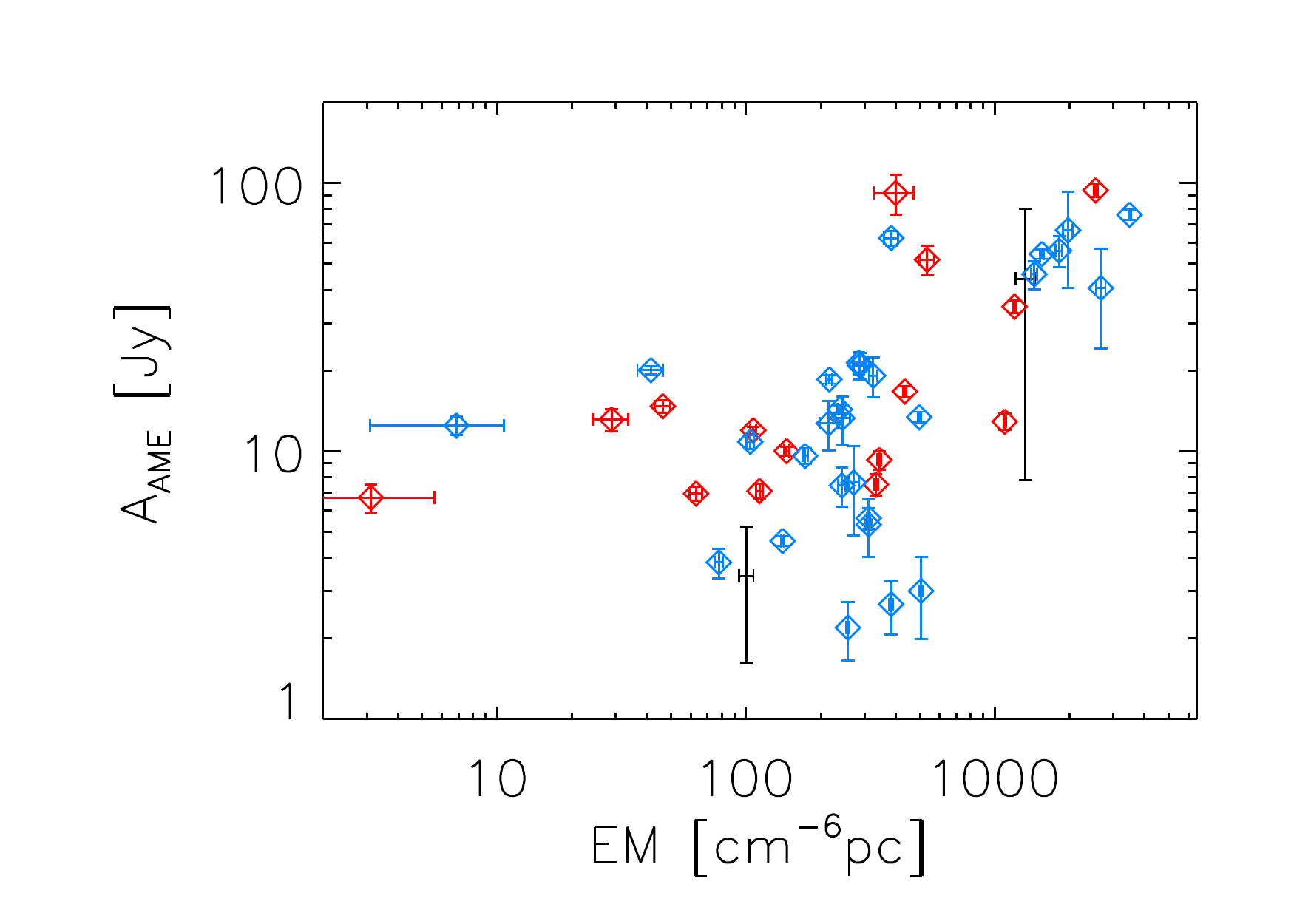}
\includegraphics[width=85mm,angle=0]{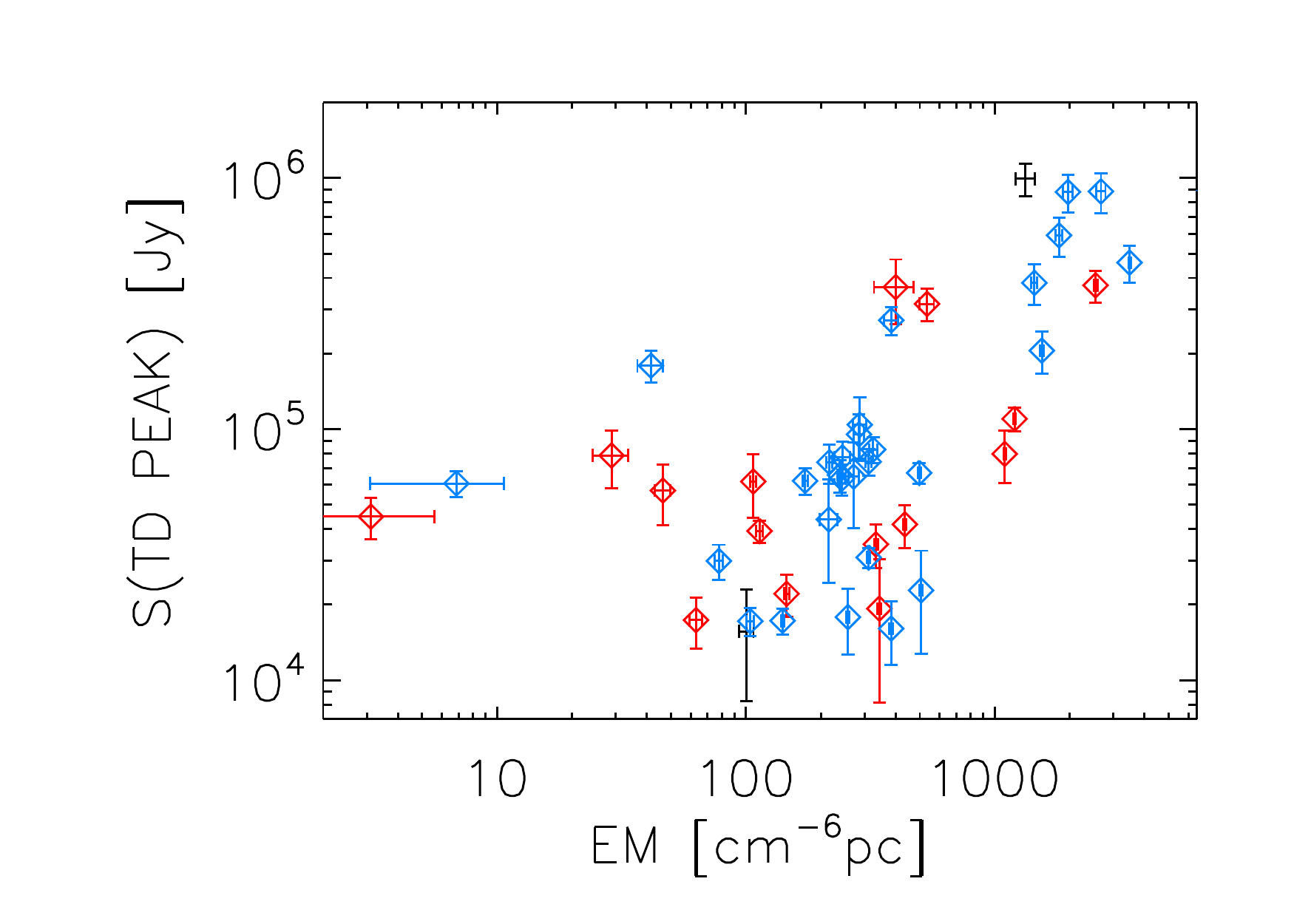}
\caption{Top: AME flux at the peak frequency versus free-free 
emission measure. Bottom: Thermal dust flux at the peak frequency 
versus free-free emission measure. Symbols and colours 
definition are the same as in Figure~\ref{fig:td_betad}.}
\label{fig:ame_max_vs_em}
\end{center}
\end{figure}

\begin{figure}
\begin{center}
\vspace*{2mm}
\centering
\includegraphics[width=105mm,angle=0]{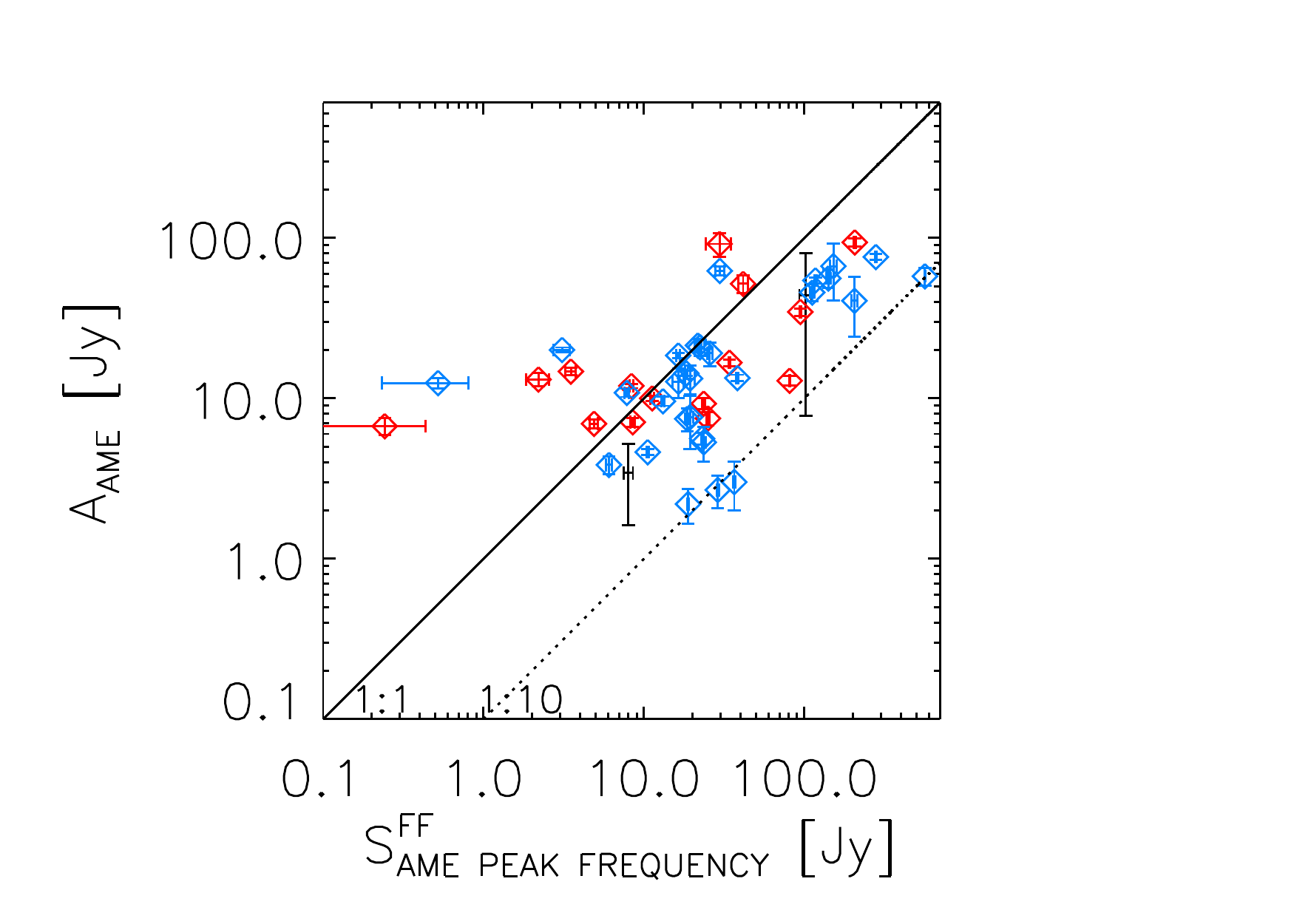}
\vspace*{-0.5cm}
\caption{AME flux at the peak frequency versus free-free flux at the
  AME peak frequency. Symbols and colours definition are the same as in Figure~\ref{fig:td_betad}. The one-to-one relation is displayed by the solid line and
the one-to-ten relation is shown with the dashed line.}
\label{fig:ame_max_vs_ff_at_ame_max}
\end{center}
\end{figure}

In our study the EM of the free-free does not correlate with the AME
emissivity estimated by $A_{\rm AME}/\tau_{250}$. On the other hand,
a mild correlation is observed between the amplitude of the AME at the
peak frequency, $A_{\rm AME}$, and the EM. This is shown on the plot
displayed at the top panel in Figure~\ref{fig:ame_max_vs_em}, with 
a SRCC between the two parameters of 0.66 $\pm$ 0.0.5. Since a
strong correlation is observed between $A_{\rm AME}$ and the
emission of the thermal dust at the peak frequency, 
$S_{\rm TD,  PEAK}$, this also means
that a correlation can be expected between EM
and, $S_{\rm TD, PEAK}$. This is shown in the plot displayed in
the bottom panel of Figure~\ref{fig:ame_max_vs_em}. 
In that case the SRCC between
the two parameters of the selected dataset is 0.64 $\pm$ 0.04.

In the interpretation of these results it must be taken into account that our EMs are estimated directly from integrated flux densities, and given the non-linear dependency between the two, those estimates could not be representative of the real averaged EMs of each region, as it was already commented in section~\ref{model_fitting}. This could indeed contribute to smear out any underlying real correlation. In addition, the fact that the correlation in the top panel of Figure~\ref{fig:ame_max_vs_em} is only seen for the sources with highest AME amplitudes could be a hint that there could be a selection effect, in such a way that when the free-free is high the AME can only be detected when it is also very high. In order to better understand this, in Figure~\ref{fig:ame_max_vs_ff_at_ame_max} we plot $A_{\rm AME}$ as a function of the flux density of 
the free-free at $\nu_{\rm AME}$. The one-to-one relation is 
displayed by the solid line while the one-to-ten relation is shown with the dashed line. Given that calibration uncertainties are of order $5-10\%$ the lack of sources below the one-to-ten line could in fact tell that the AME cannot be separated when it is less than $10\%$ of the free-free. On the contrary, the plot also shows that there are a few regions (like the Perseus and $\rho$ oph molecular clouds, respectively G160.26-18.62 and G353.05+16.90) with more AME than free-free. 

It must also be taken into account that our SED multicomponent fit is subject to an anti-correlation between the AME and free-free amplitudes which may contribute to worsening the correlation observed in Figure~\ref{fig:ame_max_vs_ff_at_ame_max}. This parameter degeneracy, which upcoming 5$\,$GHz data from the C-BASS experiment \citep[][]{cbass2018} will help to break, is clearly seen in MCMC analyses like those presented in \cite{cepeda-arroita2020} and in \cite{ameplanewidesurvey}.

\section{Discussion}\label{discussion}

In this section we summarize our results suggesting that the AME carriers may be preferentially located in cold rather than in hot phases of the ISM. Some limitations of our modelling of the AME component are then discussed, followed by a comparisons of our results with those from previous works.

\subsection{Does AME originate from the Cold ISM Phase ?}

In the last sections we searched for correlations between some of the
parameters obtained from the multicomponent fits of the AME
component and ISM tracers including the flux densities obtained
at 12, 25, 60 and 100$\,\mu$m. Interestingly, we find that the
flux densities obtained at the peak frequency of the AME bumps
show strong correlation with the flux densities
at 100, 60 and 25 $\,\mu$m, with a small loss
of correlation with the flux densities at 12 $\,\mu$m.
On the other hand, once these four flux densities tracers
are normalized by the relative strength of the ISRF, $G_{0}$, the correlations with $A_{\rm AME}$ are found to be about a few to ten percent lower in the high frequency bands.
These results could discard tiny dust particles (PAHs or VSGs in
nature) as AME carriers, if such particles are poorly sensitive
to the relative strength of the ISRF. For this reason we explored in
more detail possible relationships between the AME component parameters 
with dust modelling parameters, with $G_{0}$, as well as with the
free-free component parameters. Table~\ref{tab:srcc_main} gives a
summary of some of the most relevant SRCCs obtained from
the previous analysis in this respect. They could help to shed light
on some existing physical relationships between the astrophysical components.

From spectral energy distribution analysis of the sample
of 46 good candidate AME sources the strongest correlation 
is found between the maximum flux density of the thermal dust, 
$S_{\rm TD,peak}$, and of the AME peak, $A_{\rm AME}$
(Figure~\ref{fig:ame_peak_dust_peak}). A lower 
correlation is found between the AME emissivity, 
$A_{\rm AME}/\tau_{250}$, and the interstellar radiation field 
relative strength, $G_{0}$ (Figure~\ref{fig:role_isrf}), 
and a mild correlation is obtained between $A_{\rm AME}$ and the 
free-free EM (Figure~\ref{fig:ame_max_vs_em}, top). 
On the other hand no correlation is found between 
$A_{\rm AME}/\tau_{250}$ and $EM$ (see end of
Section~\ref{role_of_the_isrf}), 
and neither between the AME peak frequency, $\nu_{\rm AME}$, 
and $G_{0}$ (Figure~\ref{fig:role_isrf2}). 
As discussed in the previous section, averaging effects in our estimates of EM, as well as a selection effect associated with only the brightest AME sources being detected above very high free-free amplitudes, could have an impact on the tentative correlation seen between EM and $A_{\rm AME}$.
On the other
hand the correlation found between $A_{\rm AME}$ and $S_{\rm TD,peak}$
is expected to be real since these two components are associated with
distinct wavelength ranges with poor overlap between each other. Since
there is a null correlation between $A_{\rm AME}/\tau_{250}$ and $EM$,
this means that $A_{\rm AME}/\tau_{250}$, which also correlates with the 
dust grain emissivity, $S_{\rm TD,peak}/\tau_{250}$, is rather driven by 
$G_{0}$, which in turn is a function of the thermal dust temperature 
approximated by $T_{\rm dust}$ obtained from the modelling. 
In other words the interstellar radiation field still could be the 
main driver of the AME in terms of spinning dust excitation mechanisms,
but the spinning dust could be more likely associated with cold phases 
of the ISM rather than to hot phases associated with free-free radiation.

\begin{table*}
\begin{center}
\begin{tabular}{llccccl}
\hline\hline
\noalign{\smallskip}
Variable 1 & Variable 2 & SRCC & SRCC & SRCC & Power-Law Slope$^{\rm(a)}$ & Figure\\
 &                & selected sample & AME significant & AME
                                                        semi-significant
  &selected sample&\\
\noalign{\smallskip}
\hline
\noalign{\smallskip}
  $A_{\rm AME}[\rm Jy]$          & $S_{\rm  TD,peak}[\rm Jy]$  & 0.88$\pm$ 0.05  & 0.82$\pm$ 0.07  &   0.91$\pm$ 0.04 & 0.96$\pm$1.56 &\ref{fig:ame_peak_dust_peak}\\
  $A_{\rm AME}[\rm Jy]$          & $\Re_{\rm Dust}$            & 0.88$\pm$ 0.05  & 0.85$\pm$ 0.08  &   0.90$\pm$ 0.05 & 0.95$\pm$2.37 &\ref{fig:lumbol_ame_lumbol_dust} (top)\\
  $W_{\rm AME}$    & $A_{\rm AME}/\tau_{250}[\rm Jy]$          & 0.66$\pm$ 0.12  & 0.64$\pm$ 0.18  &   0.57$\pm$ 0.15 &...&\ref{fig:63_PARAMETERS_AME_PEAK_div_TAU250_vs_WAME}\\
$A_{\rm AME}/\tau_{250}[\rm Jy]$ &  $G_{0}$ or T$_{\rm td}$    & 0.68$\pm$ 0.08  & 0.87$\pm$ 0.07  &   0.62$\pm$ 0.11 & 0.78$\pm$0.94 &\ref{fig:role_isrf}\\
   $S_{\rm TD,peak}[\rm Jy]$     & EM [cm$^{-6}$ pc]           & 0.64$\pm$0.03   & 0.50$\pm$ 0.08  &   0.64$\pm$ 0.04 & 0.43$\pm$0.16 &\ref{fig:ame_max_vs_em} (bottom)\\
 $A_{\rm AME}[\rm Jy]$           & EM [cm$^{-6}$ pc]           & 0.59$\pm$ 0.05  & 0.65$\pm$ 0.11  &   0.55$\pm$ 0.03 & 1.42$\pm$0.89 &\ref{fig:ame_max_vs_em} (top)\\
$\Re_{\rm AME}$                  & $\Re_{\rm td} x 10^{-4}$    & 0.70$\pm$ 0.14  & 0.66$\pm$ 0.23  &   0.73$\pm$ 0.17 & 1.57$\pm$4.32 &\ref{fig:lumbol_ame_lumbol_dust} (bottom)\\
$\nu_{\rm AME}[\rm GHz]$         & $G_{0}$ or T$_{\rm td}$     & 0.40$\pm$ 0.12  & 0.21$\pm$ 0.22  &   0.60$\pm$ 0.15 & -0.05$\pm$0.56 &\ref{fig:role_isrf2}\\
 $G_{0}$ or T$_{\rm td}$         & EM [cm$^{-6}$ pc]           & 0.30$\pm$ 0.06  & 0.49$\pm$ 0.13  &   0.09$\pm$ 0.07 &...            & \ref{fig:em_g0}\\
   $W_{\rm AME}$                  & $G_{0}$ or T$_{\rm td}$     & 0.23$\pm$ 0.12  & 0.57$\pm$ 0.18  &   0.07$\pm$ 0.15 &...            &\ref{fig:role_isrf3}\\
$\nu_{\rm AME}[\rm GHz]$         & EM [cm$^{-6}$ pc]           & 0.06$\pm$ 0.11  & 0.27$\pm$ 0.20  &  -0.24$\pm$ 0.14 &...            & ...\\
$A_{\rm AME}/\tau_{250}[\rm Jy]$& EM [cm$^{-6}$ pc]           & 0.01$\pm$ 0.07  & 0.46$\pm$ 0.14  &  -0.21$\pm$ 0.07 &...            &...\\
  $A_{\rm AME}[\rm Jy]$          & $G_{0}$ or T$_{\rm td}$     &-0.16$\pm$ 0.06  & 0.26$\pm$ 0.10  &  -0.33$\pm$ 0.07 &...            &...\\
\noalign{\smallskip}
\hline\hline
\end{tabular}
\end{center}
\normalsize
\caption{Selection of Spearman rank correlation coefficients (SRCCs) between
  several model parameters in decreasing strength for the selected sample. $^{\rm(a)}$: Slopes obtained from linear fits in log--log space.}
\label{tab:srcc_main}
\end{table*}

\begin{figure}
\begin{center}
\vspace*{2mm}
\hspace*{-1.cm}
\centering
\includegraphics[width=95mm,angle=0]{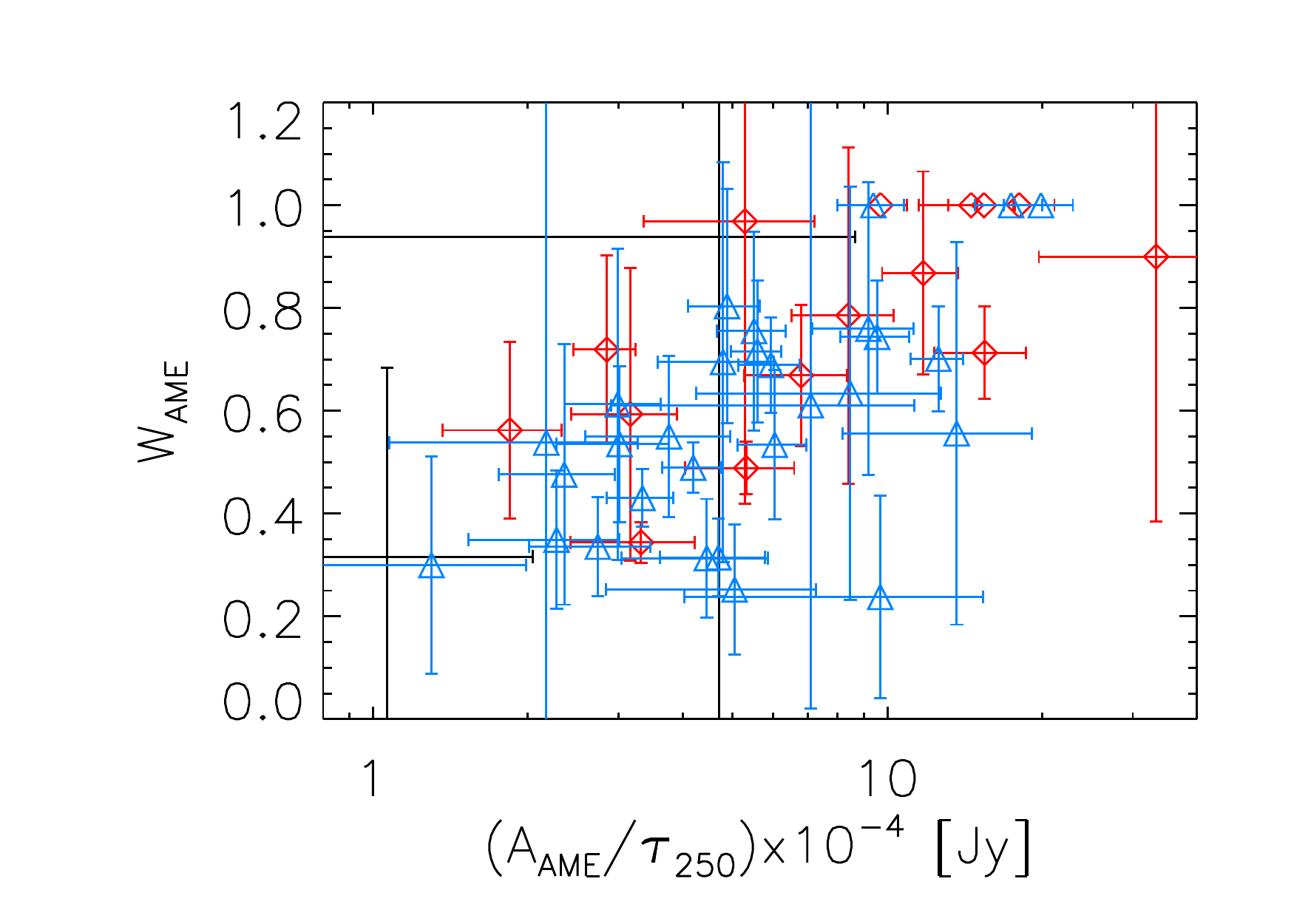}
\vspace*{-0.5cm}
\caption{AME emissivity against the width of the AME parabola model,
  $W_{\rm AME}$. Symbols and colours definition are the same as in Figure~\ref{fig:ame_nu_and_width}.}
\label{fig:63_PARAMETERS_AME_PEAK_div_TAU250_vs_WAME}
\end{center}
\end{figure}

\subsection{AME components characterization} \label{amecompchar}

From the results obtained with the multicomponent fit analysis we tested the level of independency between the parameters used to fit the AME. This model is the analytical approximation of the spectrum of spinning dust emission proposed by \cite{stevenson2014}. Indeed, we find null or very low correlations between parameters, $A_{\rm AME}$ and $\nu_{\rm AME}$, $\nu_{\rm AME}$ and $W_{\rm AME}$, and $W_{\rm AME}$ and $A_{\rm AME}$. On the other hand, we find a small correlation between $W_{\rm AME}$ and $A_{\rm AME}/\tau_{250}$. The distribution of these two parameters is shown in  Figure~\ref{fig:63_PARAMETERS_AME_PEAK_div_TAU250_vs_WAME}. By definition, the AME emissivity depends on the total amount of material along the LOS as estimated by $\tau_{250}$, and this correlation means that, on average, $A_{\rm AME}/W_{\rm AME}$ is not directly proportional to $\tau_{250}$. Testing this result using a physical AME modelling is out of the scope of this work, but could be investigated in future analyses. On the other hand, in a previous section we discussed the strong correlation obtained between $A_{\rm AME}$ and the dust radiance, $\Re_{\rm Dust}$. Put all together these results favor a strong coupling between the peak AME flux densities and the total amount of dust probed at 250$\,\mu$m, but only a fraction of the total amount of material would be at the origin of the AME radiance.

\subsection{Comparison with previous works}\label{sec:comparison_other_works}

The main differences found in this work with respect 
to the results discussed in PIRXV have been 
discussed along the previous sections. Below we compare 
and discuss our results with those from other works.

Using hierarchical Bayesian inference and full dust 
spectral energy distribution (SED) 
modelling, \cite{bell2019} argue
that, on angular scales of approximately 1$^\circ$, AME 
in $\lambda$ Orionis correlates more strongly with PAH mass
than with total dust mass, giving support for a spinning PAH
hypothesis within this region. 
Here, on similar angular scales, we find a better correlation 
with the 100$\,\mu$m dust template than with the 12$\,\mu$m 
dust template giving hints that, on Galactic scale, the dust 
grain components producing AME are more likely
associated with the cold ISM. 
This hypothesis is also supported by the strong
correlation we find between the maximum flux density of the AME
components with the dust radiance obtained from the integration 
of the dust flux models at wavelengths lower than 100$\,\mu$m.
This result may suffer the lack of modelling, in 
this work, at wavelengths shorter
than 100$\,\mu$m though, but it suggests that the AME carriers are
spatially closely associated with the thermal dust components. 

\cite{cepeda-arroita2020} discuss AME spectral variations in the $\lambda$ Orionis region with mild
correlation between the AME peak frequency and the free-free emission
measure, and strong correlation between the thermal dust temperature and the free-free emission measure.
Their results obtained at 1$^{\circ}$-angular scale give an overall picture consistent with spinning 
dust where the local radiation field plays a key role. In our analysis we find mild and null correlations 
between the AME peak frequency distribution and the thermal dust temperature, or the free-free emission measure, respectively. 
At face value, our result obtained at similar angular scale tends to discard the free-free emission as the main driver of the
excitation of the AME carriers. On the other hand, our analysis is
obtained on a sample of sources distributed on a
Galactic scale. This makes direct comparisons with results obtained on individual regions quite difficult. 
One should also bear in mind that some of the correlations obtained at low angular resolutions break down on finer angular scales. E.g. 
\citet{casassus2006} discuss 31\,GHz Cosmic Background Imager (CBI) observations of LDN 1622;  \citet{casassus2008} discuss similar observations of the $\rho$ Oph molecular cloud;  \cite{arce-tord2020} discuss $\rho$ Oph 4.5 arcmin resolution observations at 31\,GHz with CBI 2; and \cite{casassus2021} discuss ATCA high resolution observations of the
$\rho$ Oph West photo-dissociation region suggesting spectral variations that could be explained with two different cut offs 
on PAHs populations with the \texttt{SPDust} model. Actually, these studies demonstrate that finer angular resolution observations are important to identify the physical regions where spectral variations occur.

From another perspective, \cite{bernstein2020} discuss fullerenes based modelling of AME in 14 different regions. The models are calibrated using the well studied LDN\,1622 dark cloud physical conditions. The rotational temperatures are of the order of the dust grains temperatures for most of the regions, suggesting that in this scenario the AME carriers are associated with cold ISM phases. This result could support our discussion above (i.e. that AME emissivity correlates slightly with the dust temperature while not with EM). Our study is focused on high column density regions pervaded by molecular clouds, i.e., including cold neutral medium (CNM) phases, mainly located along the Galactic plane. Using a completely different method, \cite{hensley2021} investigated the relationship between the CNM, the AME and the abundance of PAHs over large areas associated with diffuse ISM regions ($N_{\rm HI} < 4 \times 10^{20}$ cm$^{-2}$) at high Galactic latitudes ($\mid b \mid > 30^{\circ}$). Their study shows that the CNM fraction strongly correlates with the fraction of dust in PAHs, and that PAHs preferentially reside in cold and relatively dense phases of the gas. If PAHs are indeed at the origin of the AME probed in our work, they could also preferentially be associated with cold phases of the ISM, i.e., with the CNM.

Finally, we point out that AME has been detected in other galaxies.
The first detection of AME in another galaxy, namely, NGC\,6946, 
was reported by \cite{murphy2010}. Detection of AME has also been
reported by \cite{murphy2018} in NGC\,4725B using VLA data.
In a following work \cite{murphy2020} discussed complementary ALMA
observations on NGC\,4725B that show discrepancy with expected 
thermal dust component making the interpretation of the results quite
puzzling. In our study we sampled the AME component over several AME
candidate regions in our Galaxy. The results show a 
distribution of peak frequencies close to 25$\ $GHz which is consistent with the average 
peak frequency observed by \cite{battistelli2019} on M31. Here, the relatively low resolution used in our study allows to sample our galaxy at about kiloparsec scales or lower. This is an asset allowing more straightforward comparisons with results obtained on close-by galaxies sampled at kiloparsec scales \citep[see for example Figure 1 in][for comparison with our Figure~\ref{fig:cloudslocation}]{murphy2010}.

\section{Summary}\label{summary}

In this work we revisited the approach proposed by PIRXV and their analysis of the multicomponent parameters obtained on Galactic candidates AME sources on the full sky at 1$^\circ$-angular scales. The main difference with their work comes from the inclusion of flux densities provided by the QUIJOTE-MFI wide survey maps at 11, 13, 17 and 19$\,$GHz covering the northern hemisphere. These maps allow generally improved detections, a better separation of the AME and the free-free components and a
better characterizations of the AME spectra observed between 10$\,$GHz and 60$\,$GHz on a sample of 46 sources. From our analysis we find the following:

\begin{itemize}
 
    \item The distribution of the AME peak frequency has a weighted mean
    frequency and dispersion of \mbox{23.6 $\pm$ 3.6$\,$GHz}, about
    4$\,$GHz  lower than the mean value obtained by PIRXV on their full-sky
    sample. Our result demonstrates the importance of using low frequency
    data in the range 10--20$\,$GHz to properly characterize the AME bump
    turnover. The value is in agreement with estimates obtained on nearby spiral galaxies. 
    \item The strongest correlations, of order 88$\%$, are found
  between the thermal dust peak flux density, and of the AME peak flux
  density, and between the AME peak flux density and the thermal dust
  radiance.
  \item Mild correlation coefficients of order 66-68 per cent are found between the AME emissivity (defined as $A_{\rm AME}/\tau_{250}$) and the width of the AME component, as well as between the AME emissivity and the interstellar radiation field relative strength.
    \item A mild correlation of order 59$\%$ is found between the AME peak flux density and the free-free EM, but this could be affected by averaging effects in the calculation of EM, as well as by the fact that only very bright AME sources would be clearly detected above strong free-free emission, whose determination is subject to uncertainties associated with calibration errors of order $10\%$.
    \item No correlation is found between the AME emissivity, $A_{\rm AME}/\tau_{250}$, and the free-free radiation EM.
    \item No significant correlation is observed between the peak frequencies of
  the AME and the thermal dust components as it has been reported 
  in the case of Lambda Orionis in a previous study by \cite{cepeda-arroita2020}.

\end{itemize}

  From our analysis we conclude that the interstellar radiation field still
  can be the main driver of the intensity of the AME toward spinning dust
  excitation mechanisms, but it is not clear 
  whether spinning dust would be most likely associated with 
  cold phases of the interstellar medium rather than with hot 
  phases dominated by free-free radiation. 
Future data over large sky fractions coming from projects currently under development
like C-BASS \citep[][]{cbass2018}, 
TFGI (\citet{RubinoSPIE12}, and see also the introduction in \citet{mfiwidesurvey}) and 
MFI2 \citep[][]{MFI2} should help to clarify these aspects and to further refine similar statistical analyses.

\section*{Acknowledgements}

\input{quijote_acknow}

FP acknowledges the European Commission under the Marie 
Sklodowska-Curie Actions within the 
\textit{European Union's Horizon 2020}  research 
and innovation programme under Grant Agreement number
658499 (PolAME). FP acknowledges support from the 
Spanish State Research Agency (AEI) under grant numbers
PID2019-105552RB-C43. 
FG acknowledges funding from the European Research 
Council (ERC) under the \textit{European Union's Horizon 2020} 
research and innovation programme (grant agreement 
No 101001897). 
EdlH acknowledge partial financial support from 
the \textit{Concepci\'on Arenal Programme} of the 
Universidad de Cantabria. 
BR-G acknowledges ASI-INFN Agreement 2014-037-R.0.
DT acknowledges the support from the Chinese Academy 
of Sciences President's International Fellowship 
Initiative, Grant N. 2020PM0042.
We acknowledge the use of data from the 
\textit{Planck} /ESA mission, downloaded
from the \textit{Planck} Legacy Archive, and of 
the Legacy Archive for
Microwave Background Data Analysis (LAMBDA). 
Support for LAMBDA is
provided by the NASA Office of Space Science. 
Some of the results in
this paper have been derived using the 
HEALP{\sc ix} (G{\'o}rski et al. 2005) package.

\section*{Data Availability}

The QUIJOTE nominal mode maps, are expected to be made publicly available in the first QUIJOTE data release. Other ancillary data employed in this work are publicly available and can be accessed online as detailed in the paper text.




\bibliographystyle{mnras}
\bibliography{polame,quijote}



\clearpage
    \appendix
    
    \noindent


\appendix

\section{Additional Plots}  \label{append_additional_plots}

A few more Figures, all showing a lack of correlation between some of the modelling parameters, are displayed in this appendix for the interested reader.

\begin{figure}
\begin{center}
\vspace*{2mm}
\hspace*{-1.cm}
\centering
\includegraphics[width=95mm,angle=0]{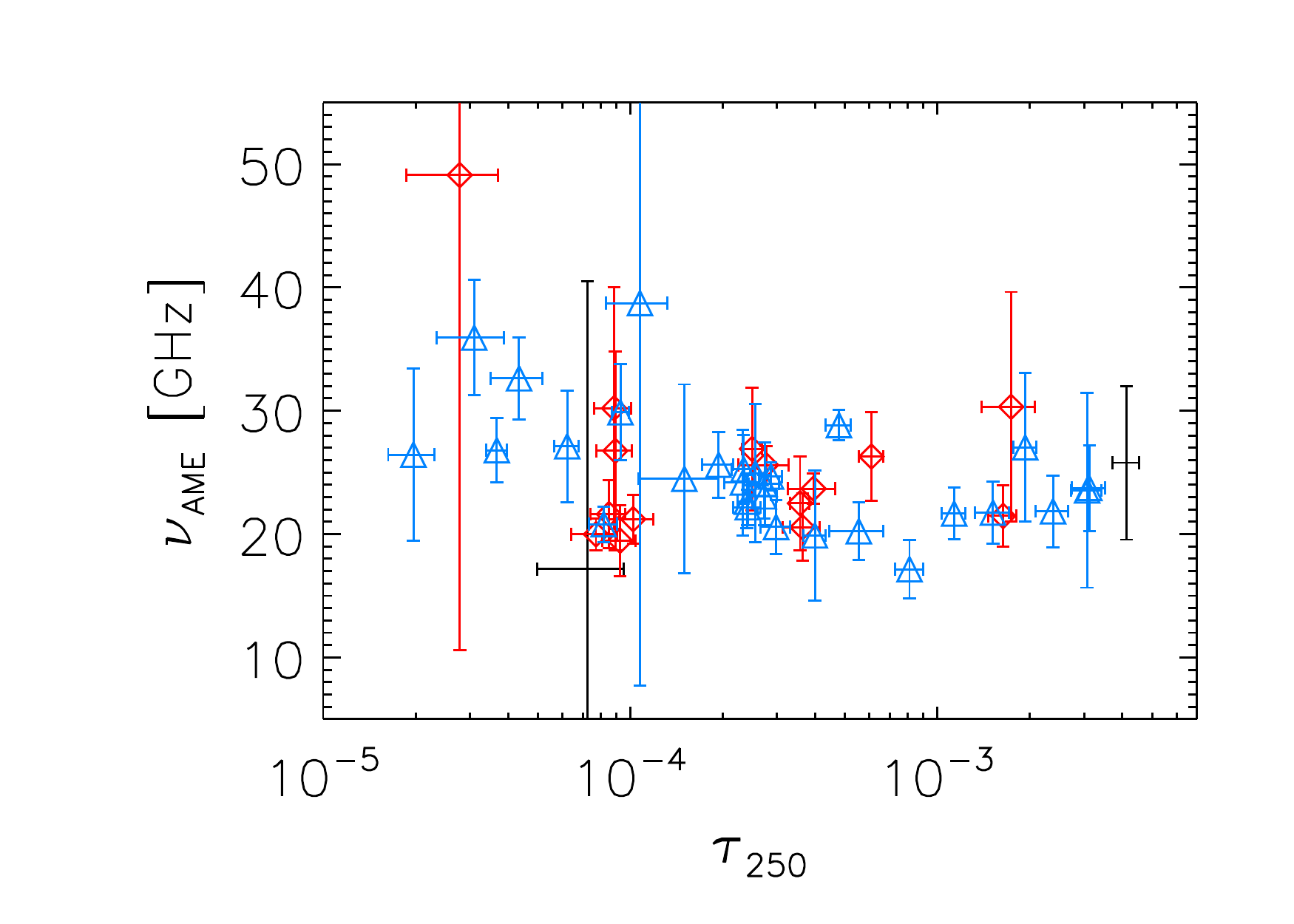}
\vspace*{-0.5cm}
\caption{Distribution of the 
AME peak frequency $\nu_{\rm AME}$ against $\tau_{250}$.
All selected data are displayed.
Symbols and colours definition are the same as in Figure~\ref{fig:td_betad}.}
\label{fig:dust_optical_depth2}
\end{center}
\end{figure}

\begin{figure}
\begin{center}
\vspace*{2mm}
\hspace*{-1.cm}
\centering
\includegraphics[width=95mm,angle=0]{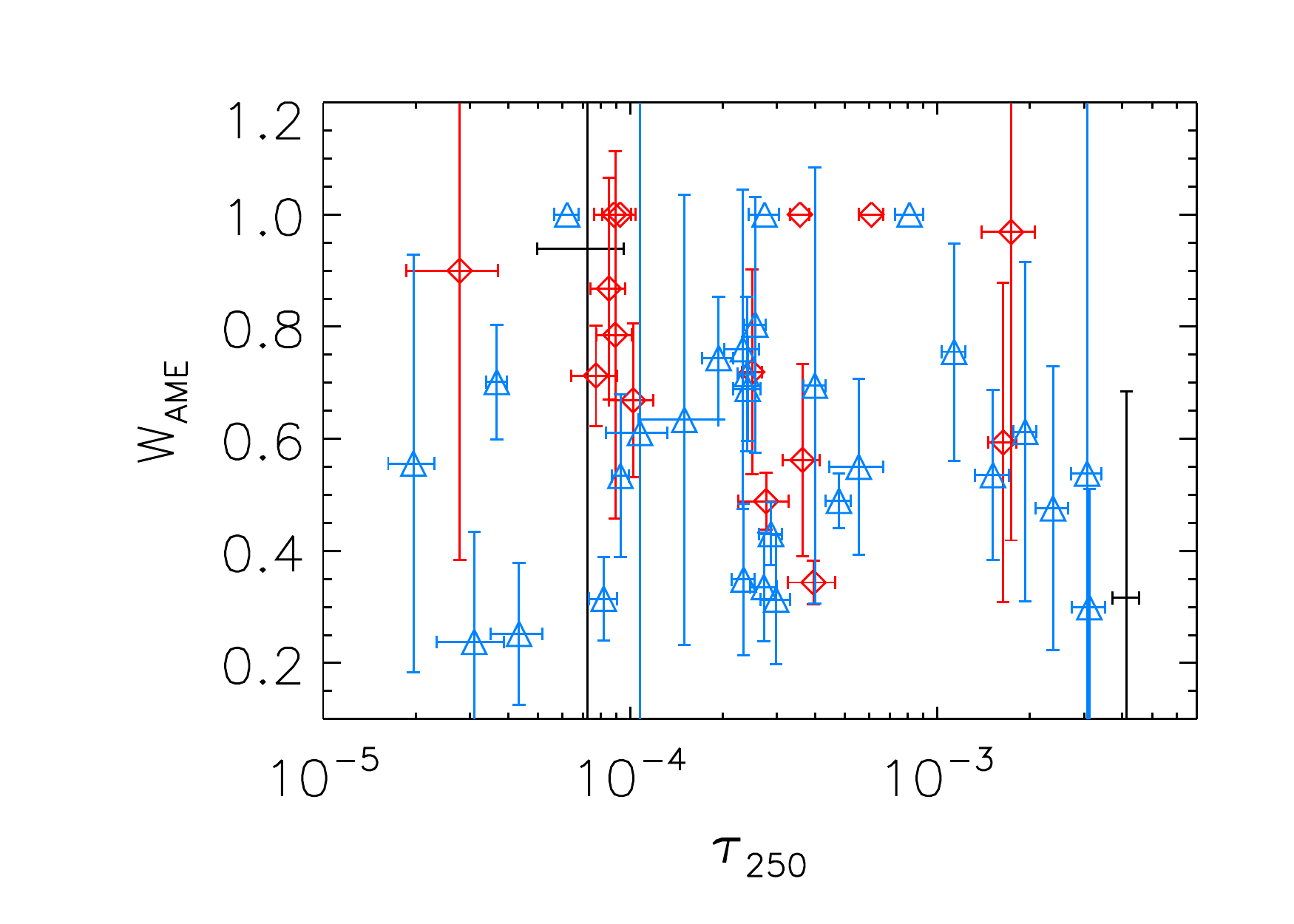}
\vspace*{-0.5cm}
\caption{Distribution of the
width of the AME $\rm W_{\rm AME}$  against $\tau_{250}$.
All selected data are displayed.
Symbols and colours definition are the same as in Figure~\ref{fig:td_betad}.}
\label{fig:dust_optical_depth3}
\end{center}
\end{figure}

\begin{figure}
\begin{center}
\vspace*{2mm}
\hspace*{-1.cm}
\centering
\includegraphics[width=95mm,angle=0]{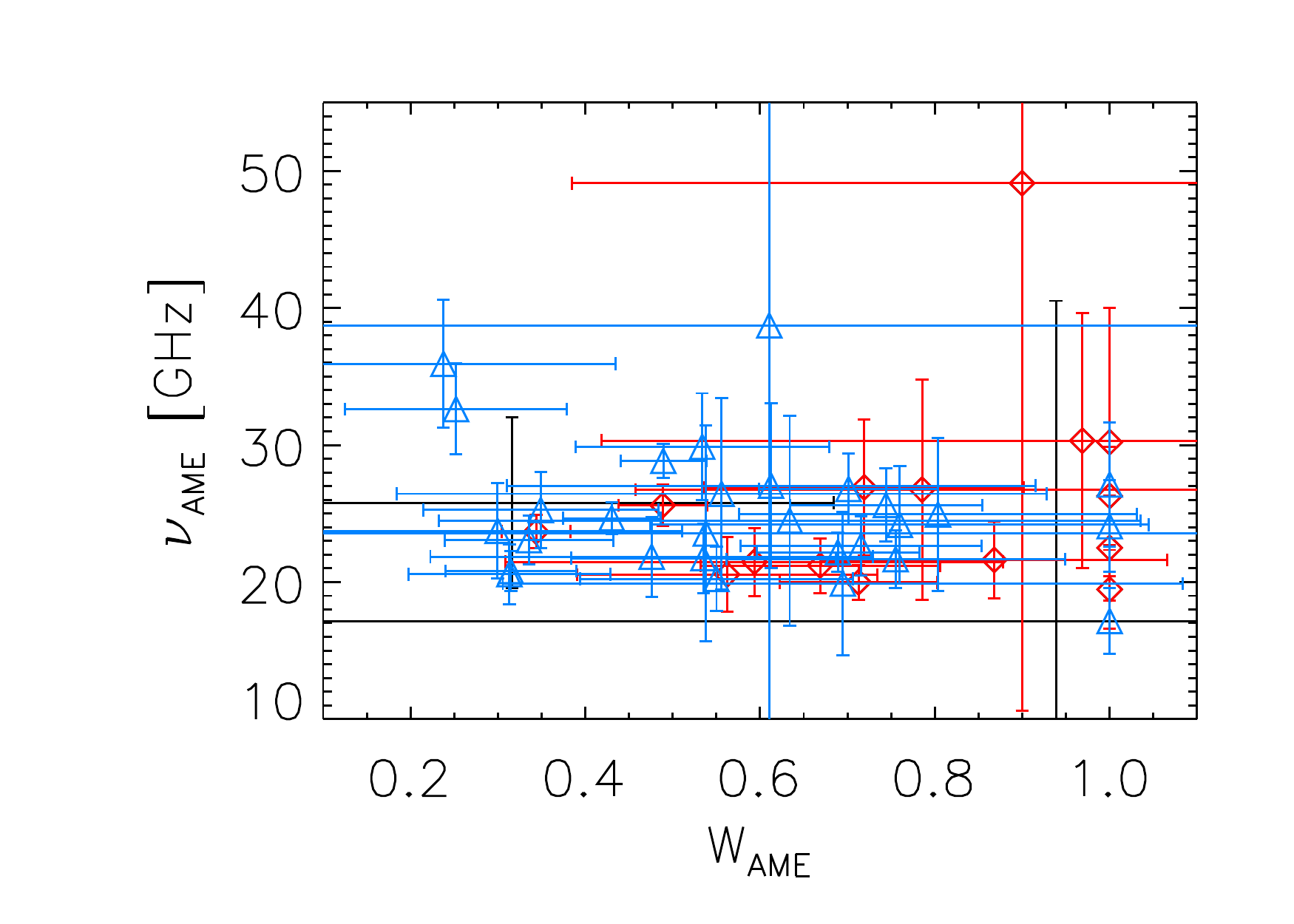}
\vspace*{-0.5cm}
\caption{Variations of the AME peak frequency $\nu_{\rm AME}$ as a
  function of  the width of the parabola, $\sigma_{\rm AME, \it
    \nu}$. Symbols and colours definition are the same as in Figure~\ref{fig:td_betad}.}
\label{fig:ame_nu_and_width}
\end{center}
\end{figure}

\begin{figure}
\begin{center}
\vspace*{2mm}
\hspace*{-1.cm}
\centering
\includegraphics[width=95mm,angle=0]{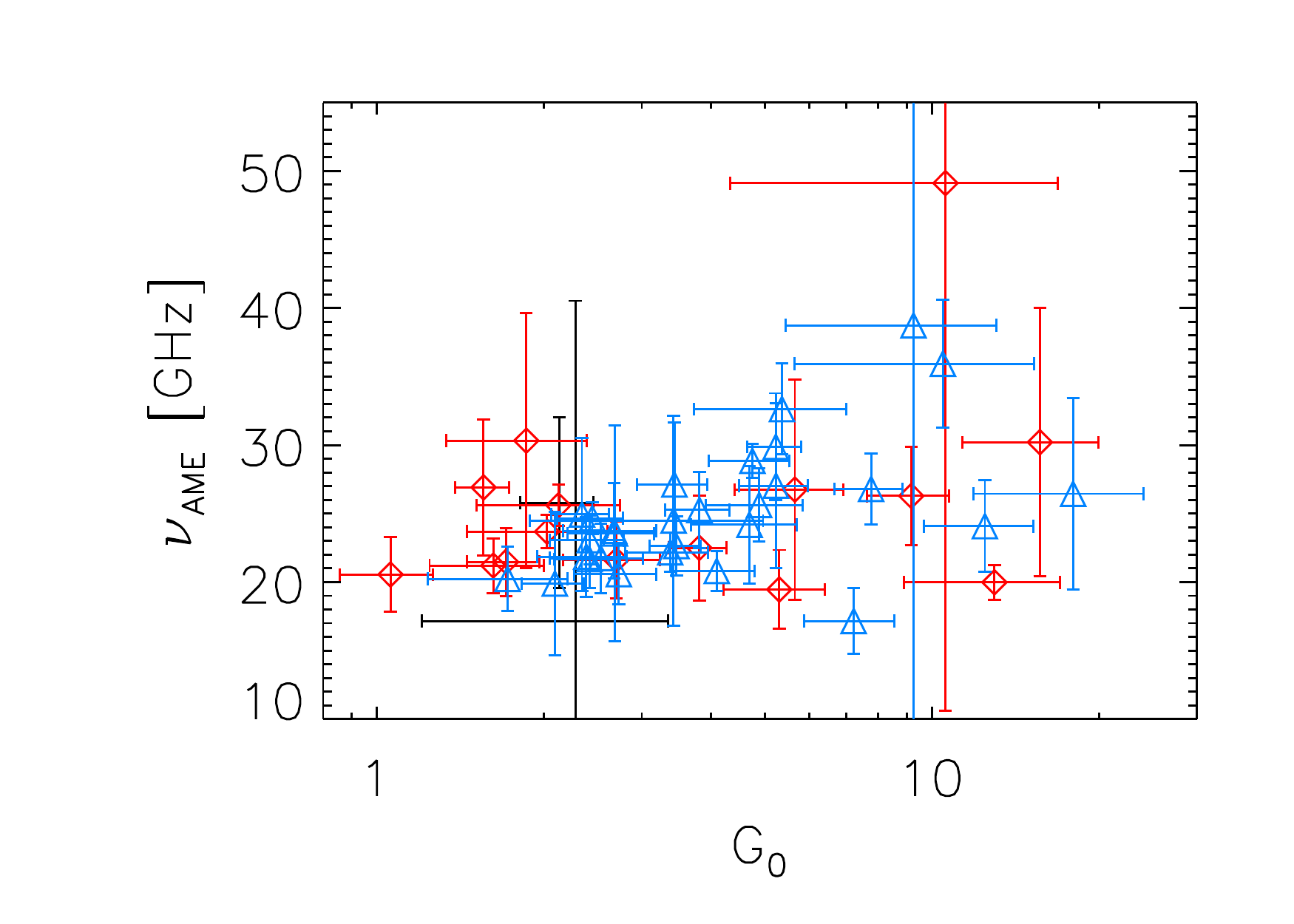}
\vspace*{-0.5cm}
\caption{Variations of the AME peak frequency with the
  relative strength of the ISRF, $G_{0}$. Symbols and 
  colours definition are the same as in Figure~\ref{fig:td_betad}. 
  }
\label{fig:role_isrf2}
\end{center}
\end{figure}

\begin{figure}
\begin{center}
\vspace*{2mm}
\hspace*{-1.cm}
\centering
\includegraphics[width=95mm,angle=0]{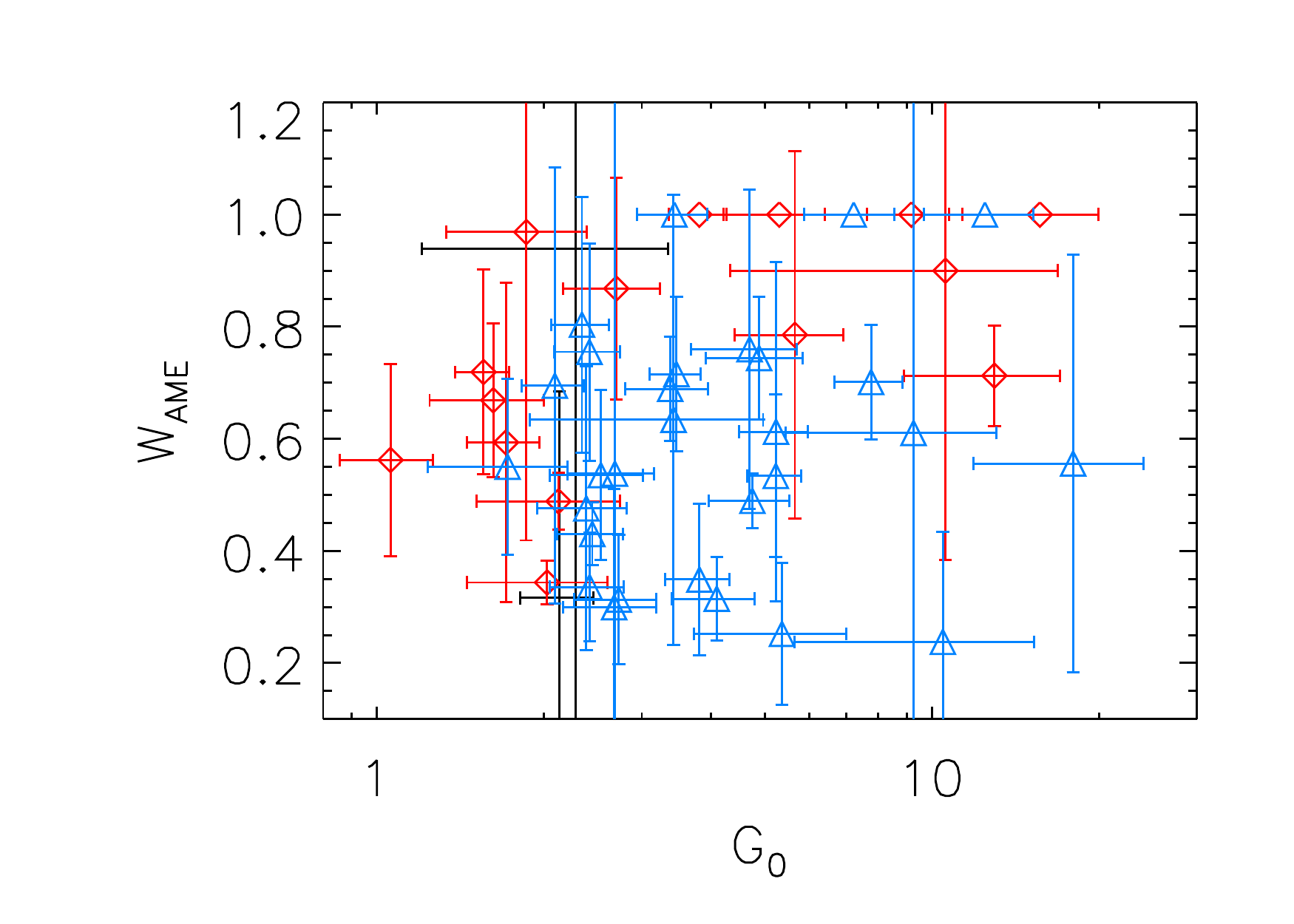}
\vspace*{-0.5cm}
\caption{Variations of the AME parabola width parameter 
with the relative strength of the ISRF, $G_{0}$. Symbols 
and colours definition are the same as in Figure~\ref{fig:td_betad}.}
\label{fig:role_isrf3}
\end{center}
\end{figure}

\begin{itemize}
 
    \item Figure~\ref{fig:dust_optical_depth2} shows the variations of the AME peak frequency, $\nu_{\rm AME}$, as a function of the proxy of the thermal dust material, $\tau_{250}$, as discussed in section~\ref{dust_optical_depth}. The distribution of $\nu_{\rm AME}$ seems to be independent of the quantity of matter along the LOSs. 
 
    \item Figure~\ref{fig:dust_optical_depth3} shows the Variations of the AME characteristic width, $W_{\rm AME}$, as a function of the proxy of the thermal dust material, $\tau_{250}$, as discussed in section~\ref{dust_optical_depth}. This AME parameter also seems to be independent of the quantity of matter along the LOSs.

    \item  Figure~\ref{fig:ame_nu_and_width} shows the
    Variations of the AME peak frequency, $\nu_{\rm AME}$, 
    as a function of  the width of the 
    parabola, $W_{\rm AME}$, as discussed in
    section~\ref{ame_bump_freq}. As expected from 
    the formalism used to model AME \citep[see][]{stevenson2014}, this plot 
    confirms that the two parameters are 
    independent from each other.

    \item  Figure~\ref{fig:role_isrf2} and 
    Figure~\ref{fig:role_isrf3} show
    the distribution of the AME peak frequency and
    of the AME parabola width parameter with the
    relative strength of the ISRF, $G_{0}$, respectively, 
    as discussed in section~\ref{role_of_the_isrf}. 
    The two AME parameters show no dependence on $G_{0}$.

\end{itemize}
 
\section{SED MULTICOMPONENT FIT PARAMETERS} \label{append_seds_params}

\input{polametex_sedfitparams_tables_1_cmb.txt}

\input{polametex_sedfitparams_tables_2_parabola.txt}

The parameters obtained on each source from the multicomponent analysis are displayed in Table~\ref{tab:seds_fit_parameters1} and Table~\ref{tab:seds_fit_parameters2}. The name of the sources are given in column one from each Table. The parameters used to model the synchrotron component, and the free-free component, are given in columns 2 and 3 from Table~\ref{tab:seds_fit_parameters1}, respectively. Columns 4 to 6 of the same table give the parameters used to model the thermal dust grain component, while column 7 gives the relative strength of the ISRF derived using $T_{\rm dust}$ displayed in column 5. The parameter used to model the CMB component is found in the last column of Table~\ref{tab:seds_fit_parameters1}. All the parameters related to the modelling of AME are given in Table~\ref{tab:seds_fit_parameters2}. The maximum fraction of emission that could be attributed to UC\sc{H\,ii} \rm{}regions is given in column 8 of that table, while the reduced $\chi^{2}$ of the multicomponent fits are given in the last column.

\section{SEDs of the full sample} \label{append_seds}

The plots of the SEDs obtained on each of the 52 candidate AME
sources of the sample are displayed in Figures C1--C7.
In each plot, the QUIJOTE intensity flux densities
are shown with red filled circles, and the \textit{WMAP},
\textit{Planck}, and DIRBE intensity flux densities are 
shown with green, blue, and yellow opened diamonds,
respectively. The low frequency points used to fit the models are 
shown in pale blue, and in blue if they were not included in the 
fitting procedure. The result of the multicomponent fit is 
illustrated by the continuous black curve. 
The fit to the AME component
is shown with the dashed red line. The fit to the 
free--free component is shown with the dashed blue line. 
The fit to the thermal dust component is shown with
the dashed yellow line. The fit to the CMB component is 
shown with the dashed green line.

\begin{figure*}
\begin{center}
\vspace*{0mm}
\centering
\includegraphics[width=77mm,angle=0]{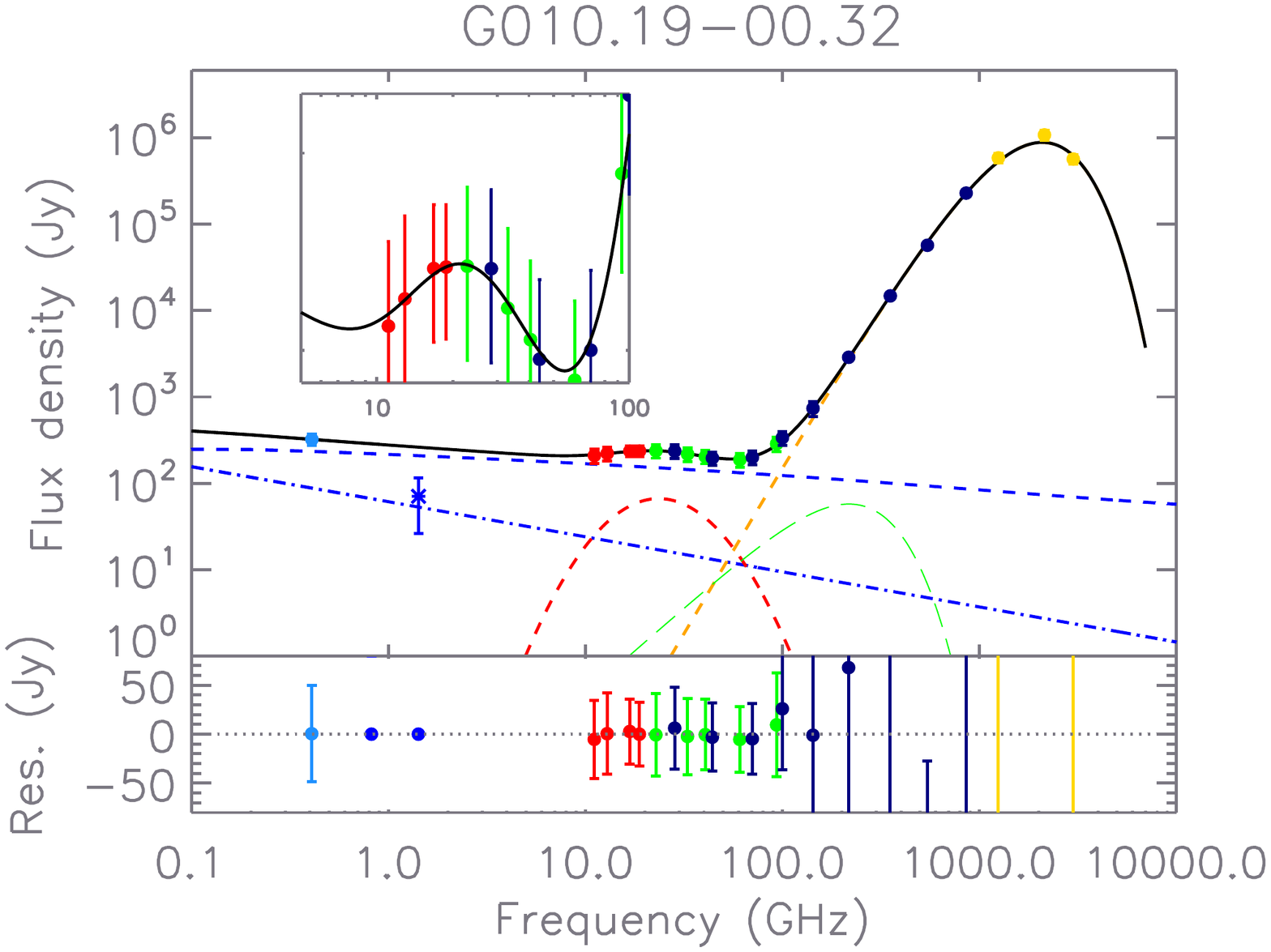}
\vspace*{-4.5cm}
\hspace*{10mm}
\includegraphics[width=77mm,angle=0]{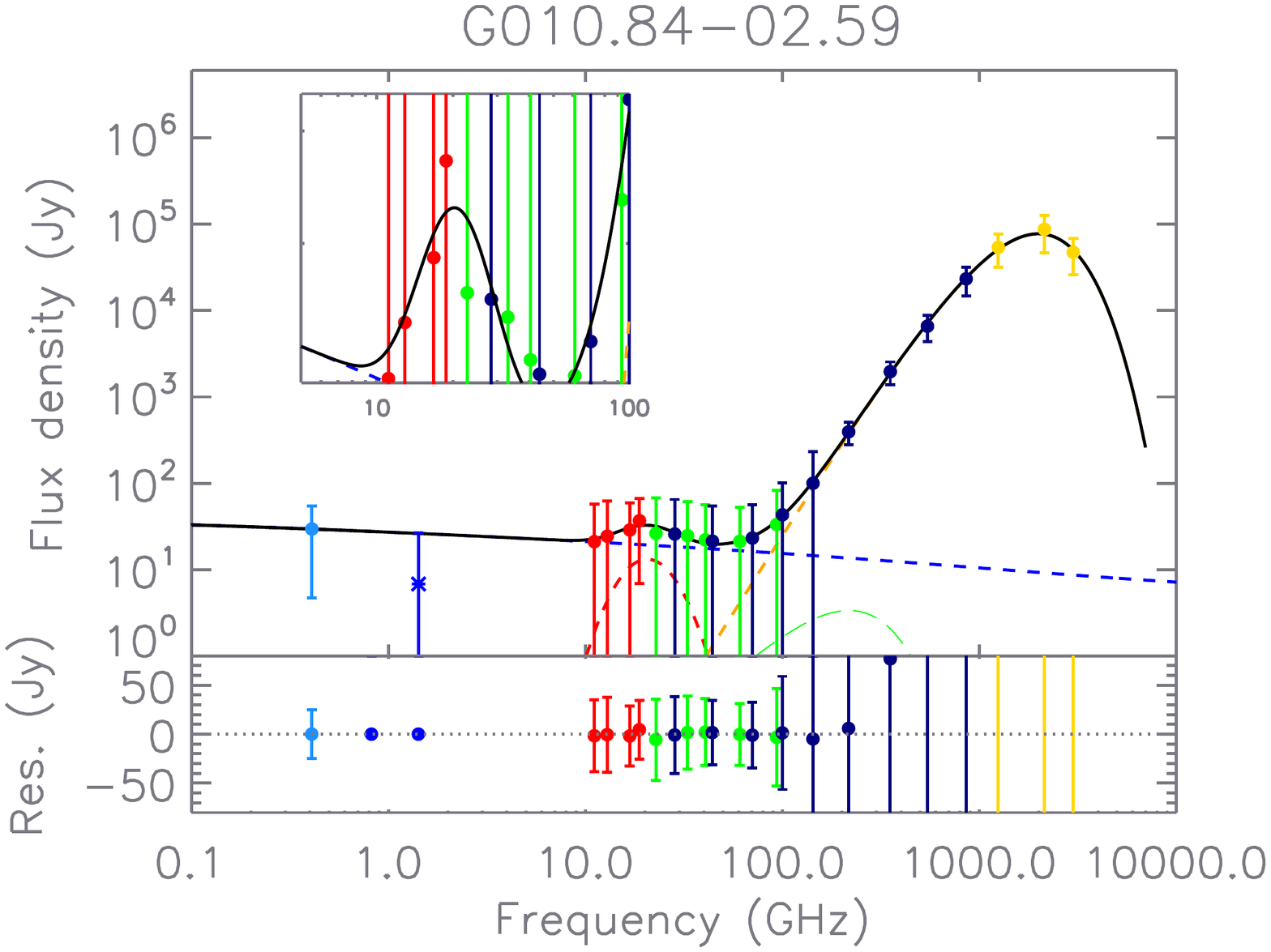}
\vspace*{-4.5cm}
\includegraphics[width=77mm,angle=0]{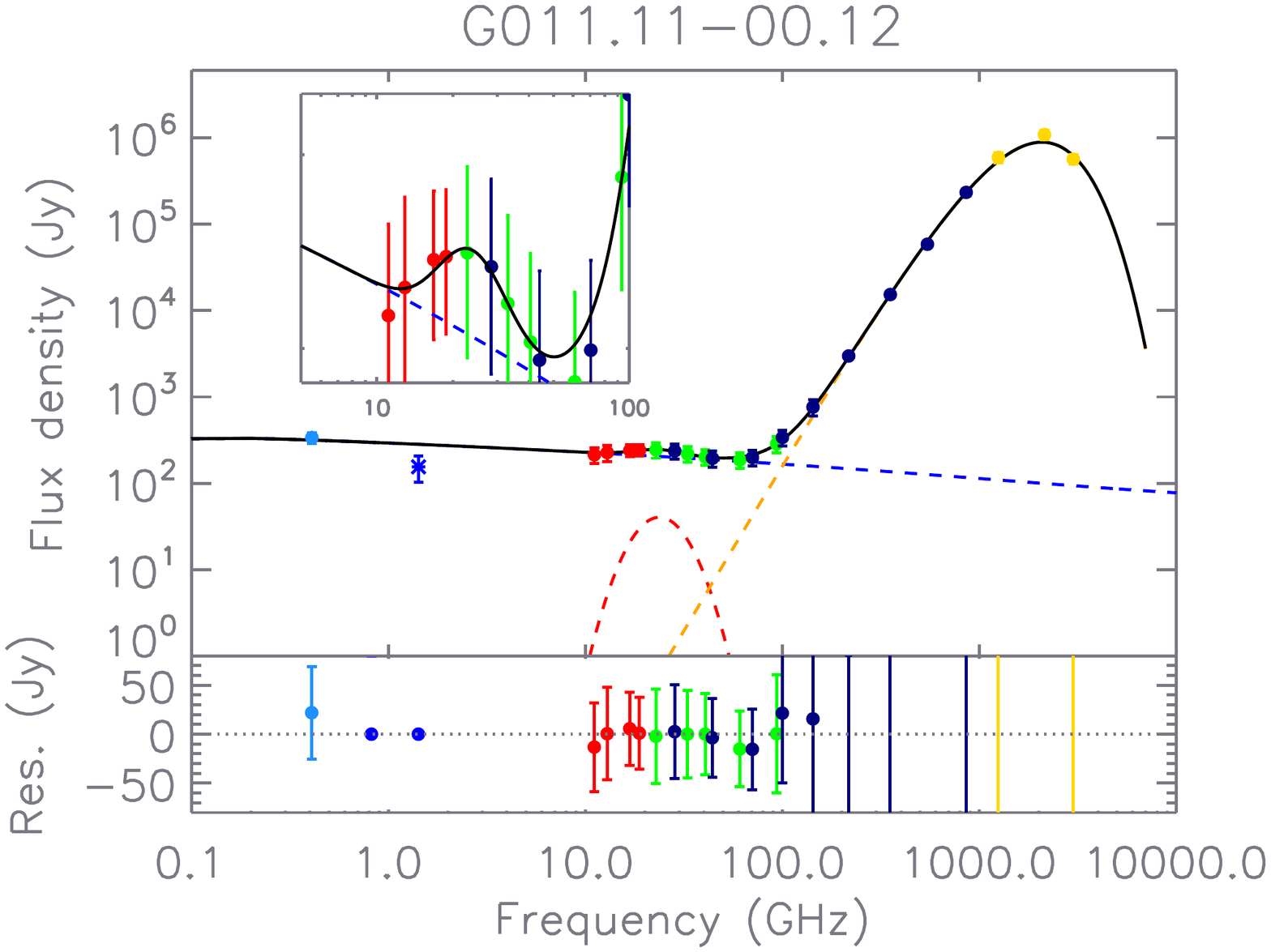}
\hspace*{10mm}
\includegraphics[width=77mm,angle=0]{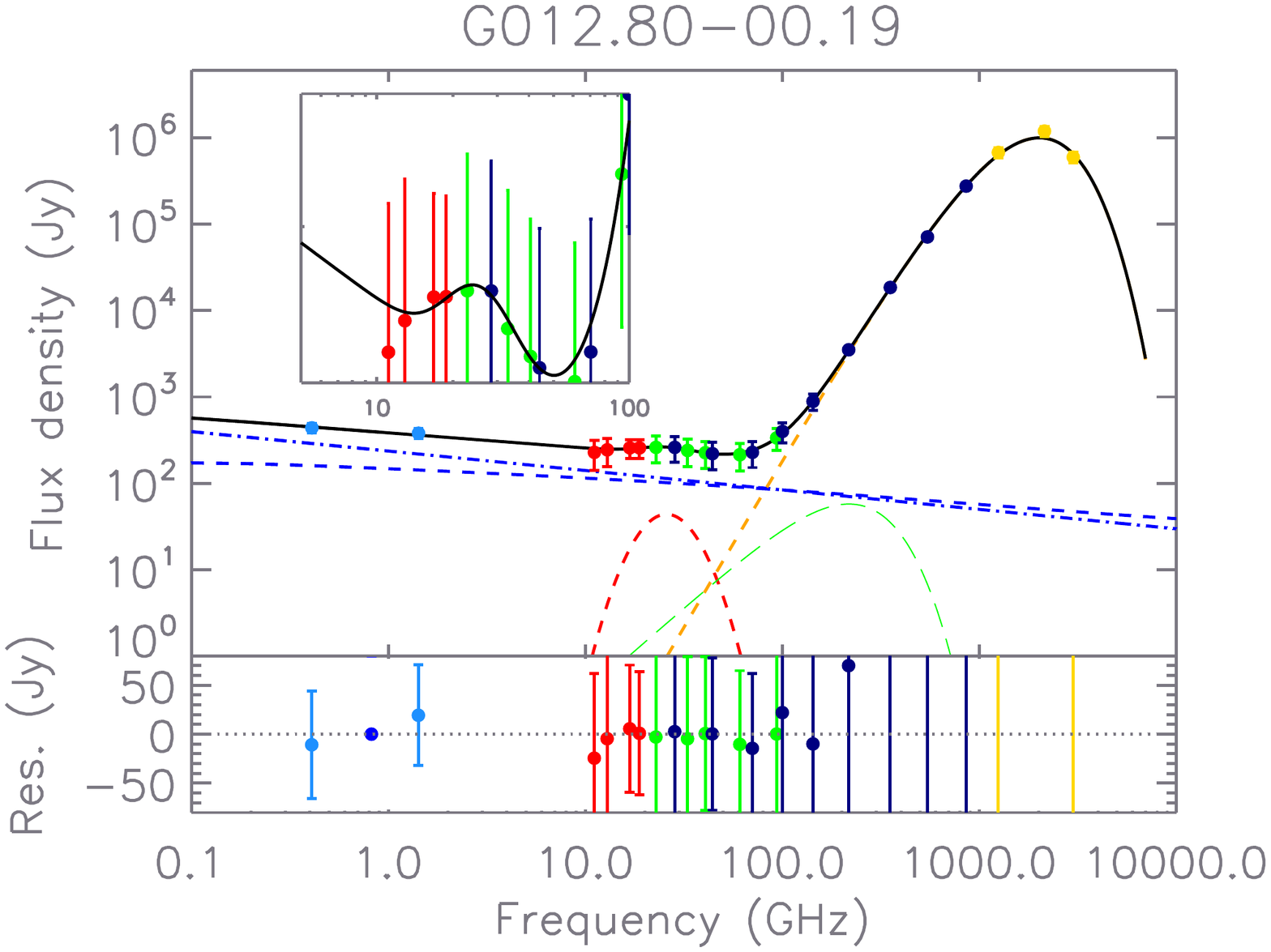}
\vspace*{-4.5cm}
\includegraphics[width=77mm,angle=0]{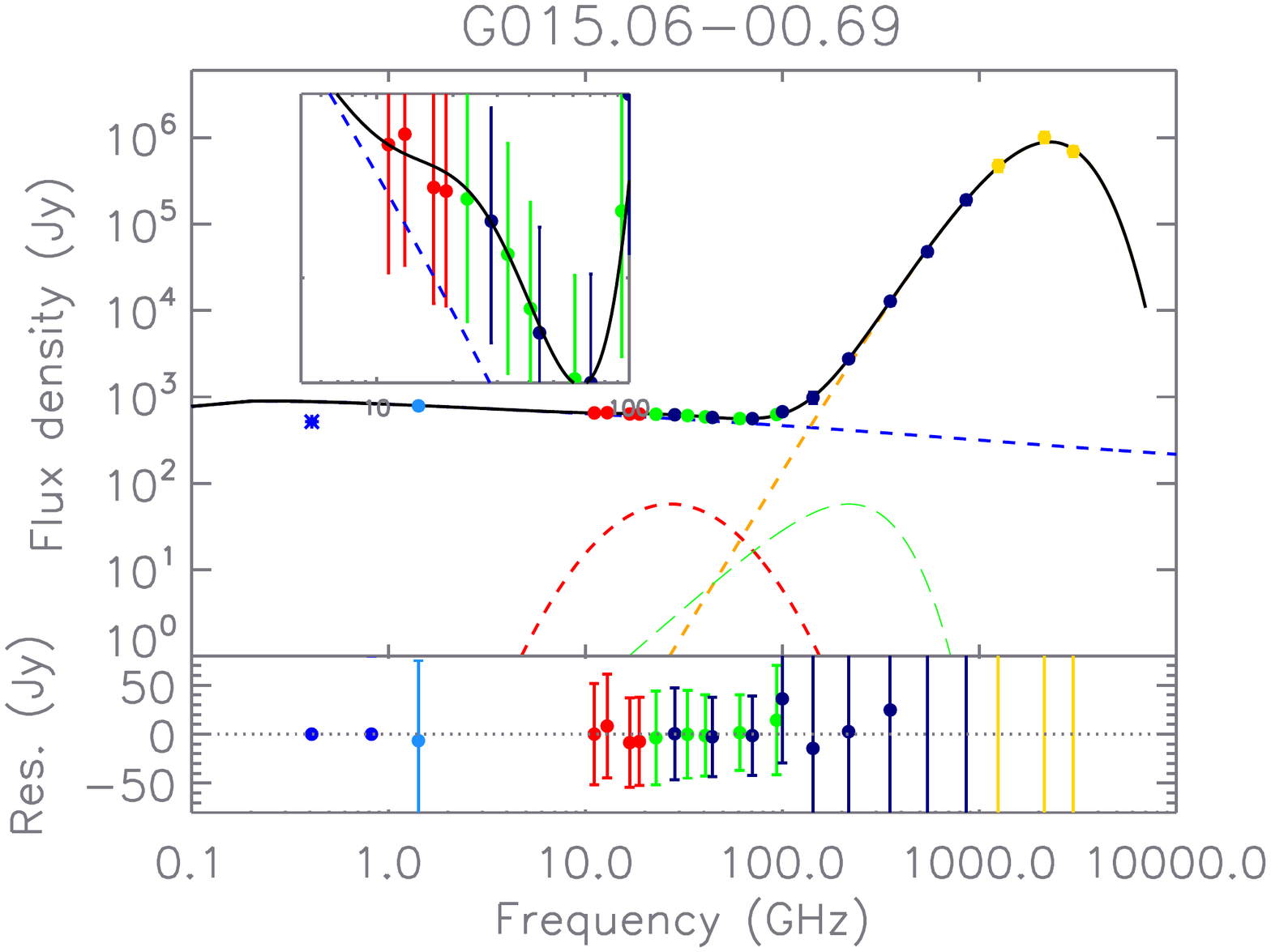}
\hspace*{10mm}
\includegraphics[width=77mm,angle=0]{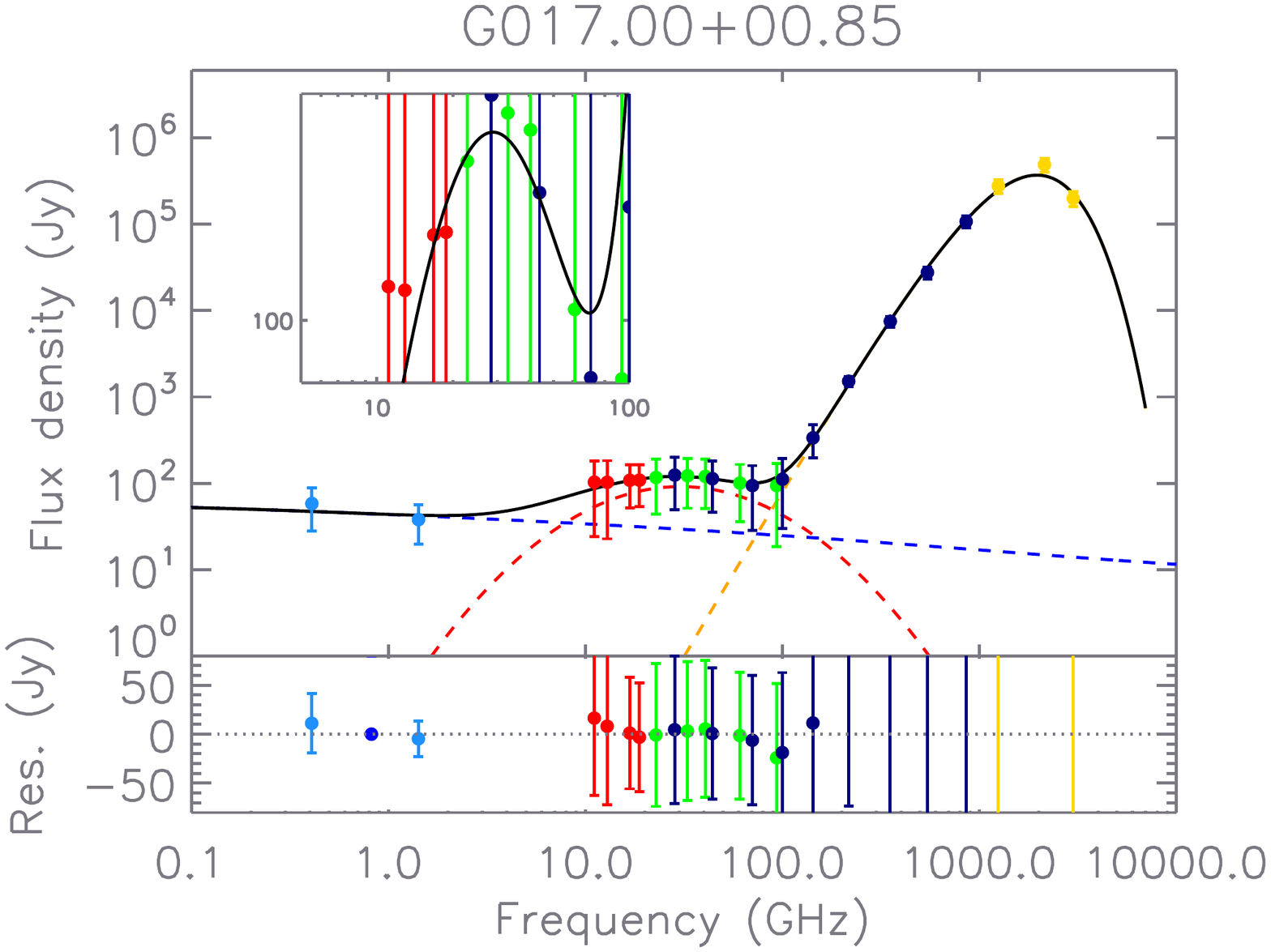}
\vspace*{-4.5cm}
\includegraphics[width=77mm,angle=0]{G037.79-00.11_SED_INT.pdf}
\hspace*{10mm}
\includegraphics[width=77mm,angle=0]{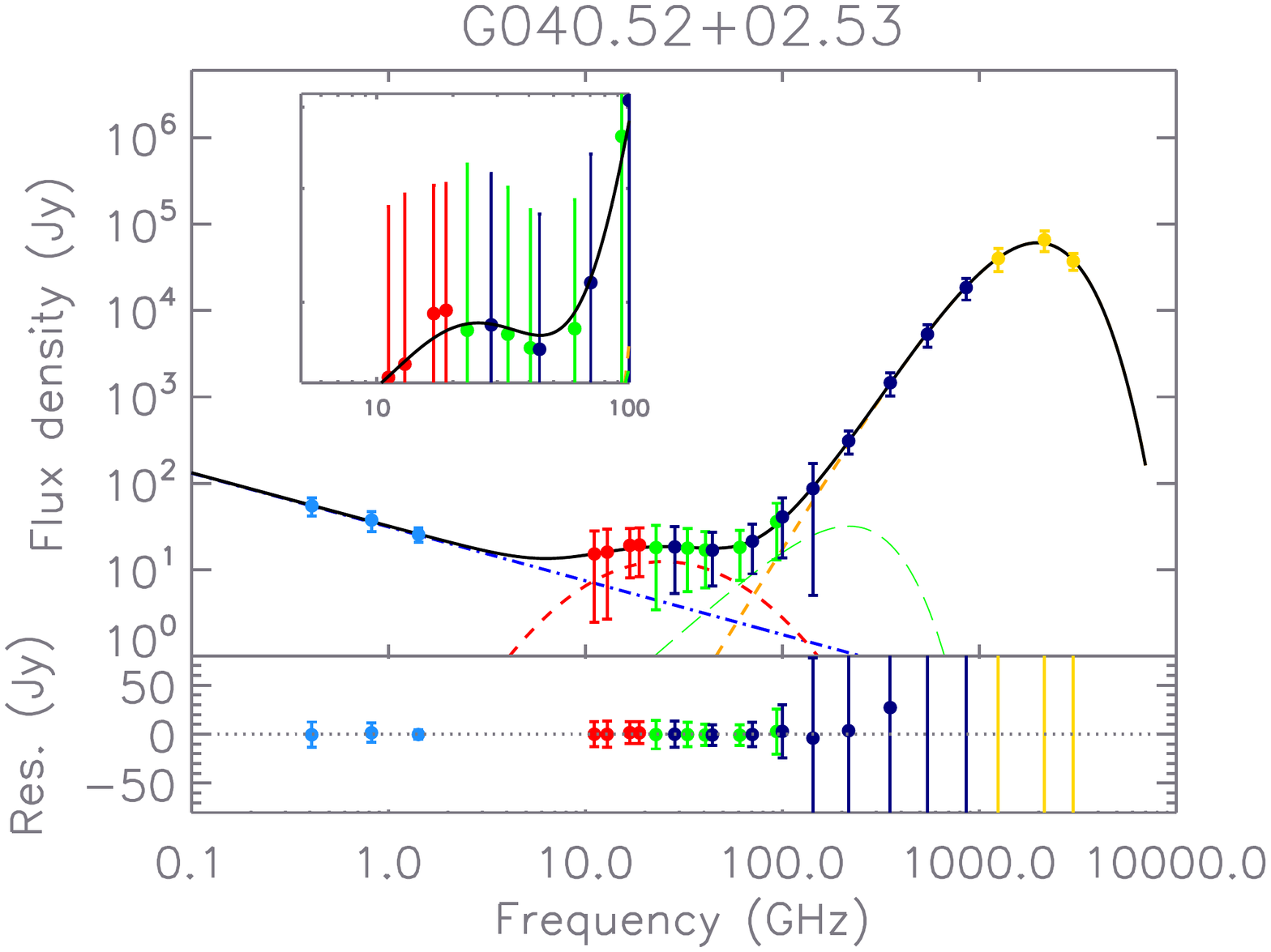}
\vspace*{1cm}
\caption{SED of the sample of regions discussed in this work.
See caption of Figure~\ref{fig:sed_int_subsample} for 
symbols, lines and colours conventions.}
\label{fig:sed_int1}
\end{center}
\end{figure*}

\begin{figure*}
\begin{center}
\vspace*{0mm}
\centering
\includegraphics[width=77mm,angle=0]{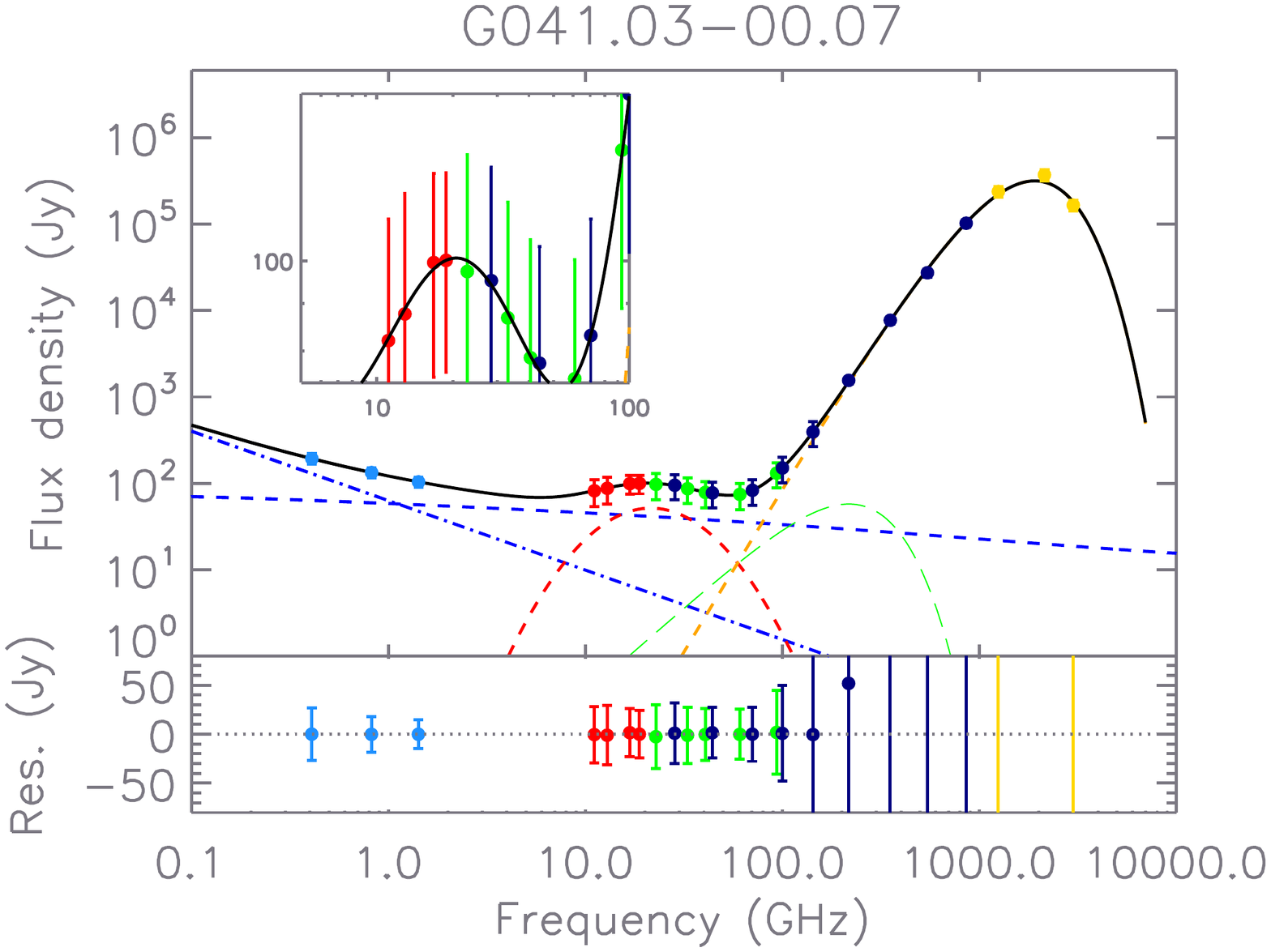}
\vspace*{-4.5cm}
\hspace*{10mm}
\includegraphics[width=77mm,angle=0]{G043.20-00.10_SED_INT.pdf}
\vspace*{-4.5cm}
\includegraphics[width=77mm,angle=0]{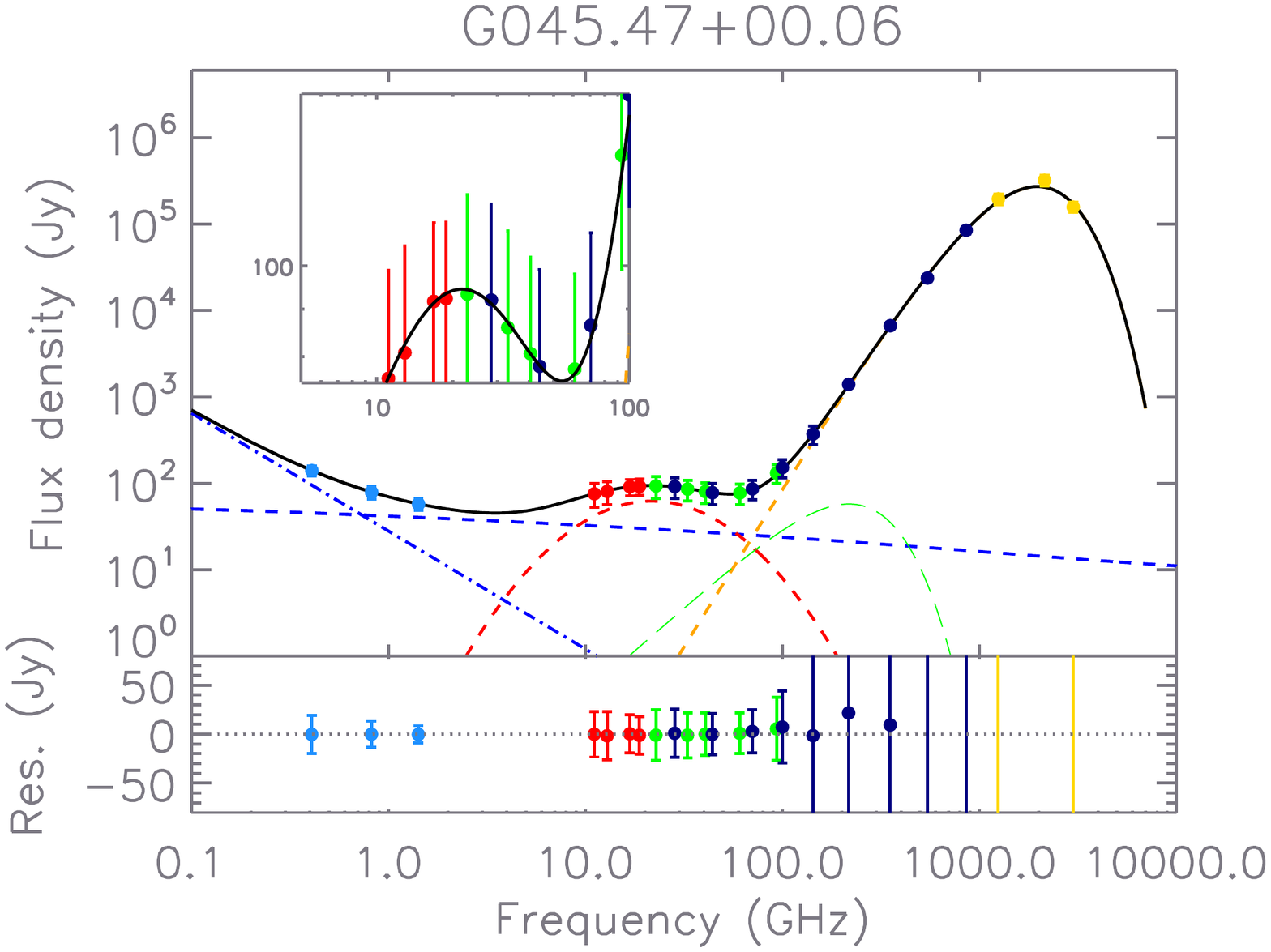}
\hspace*{10mm}
\includegraphics[width=77mm,angle=0]{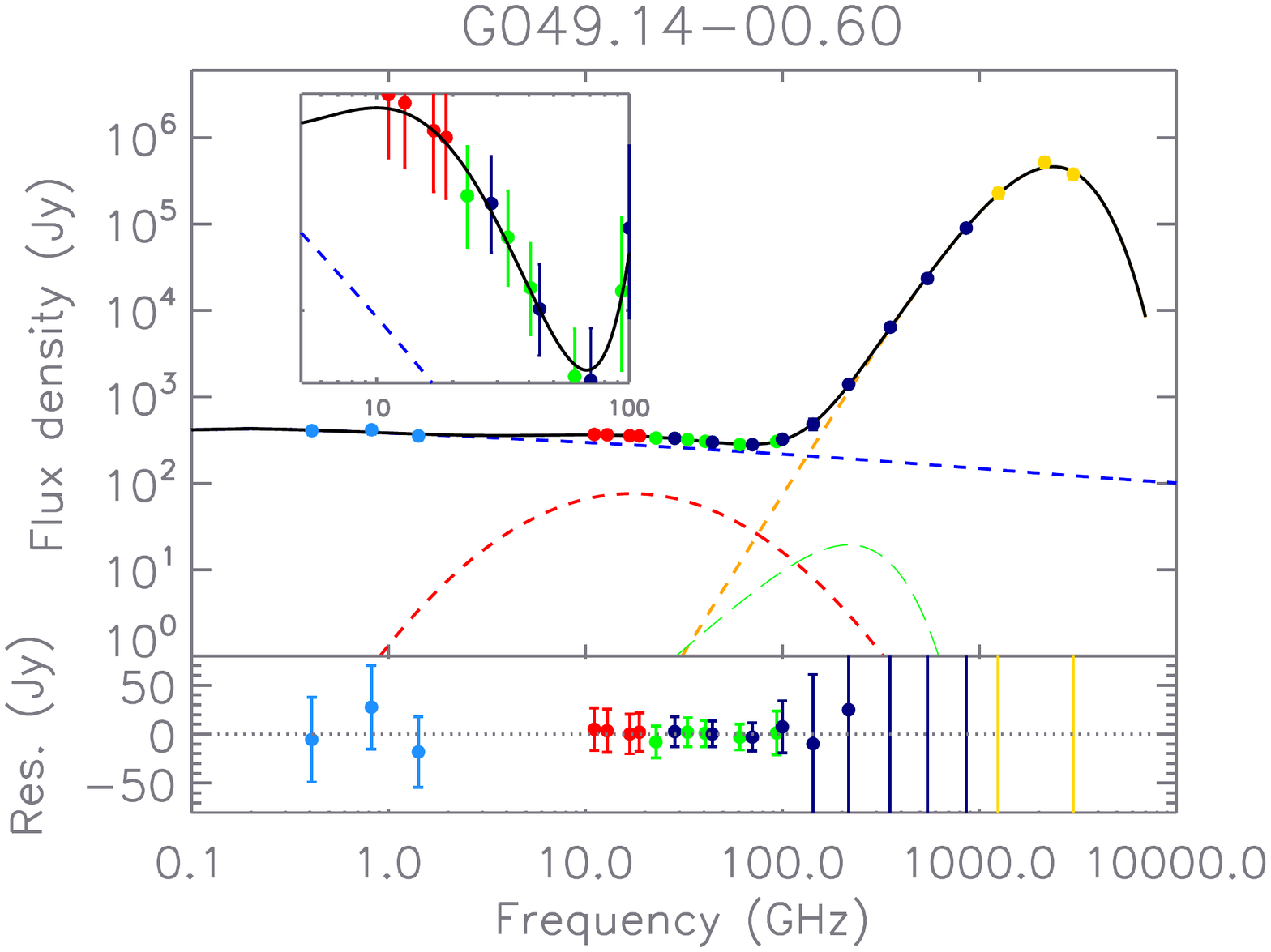}
\vspace*{-4.5cm}
\includegraphics[width=77mm,angle=0]{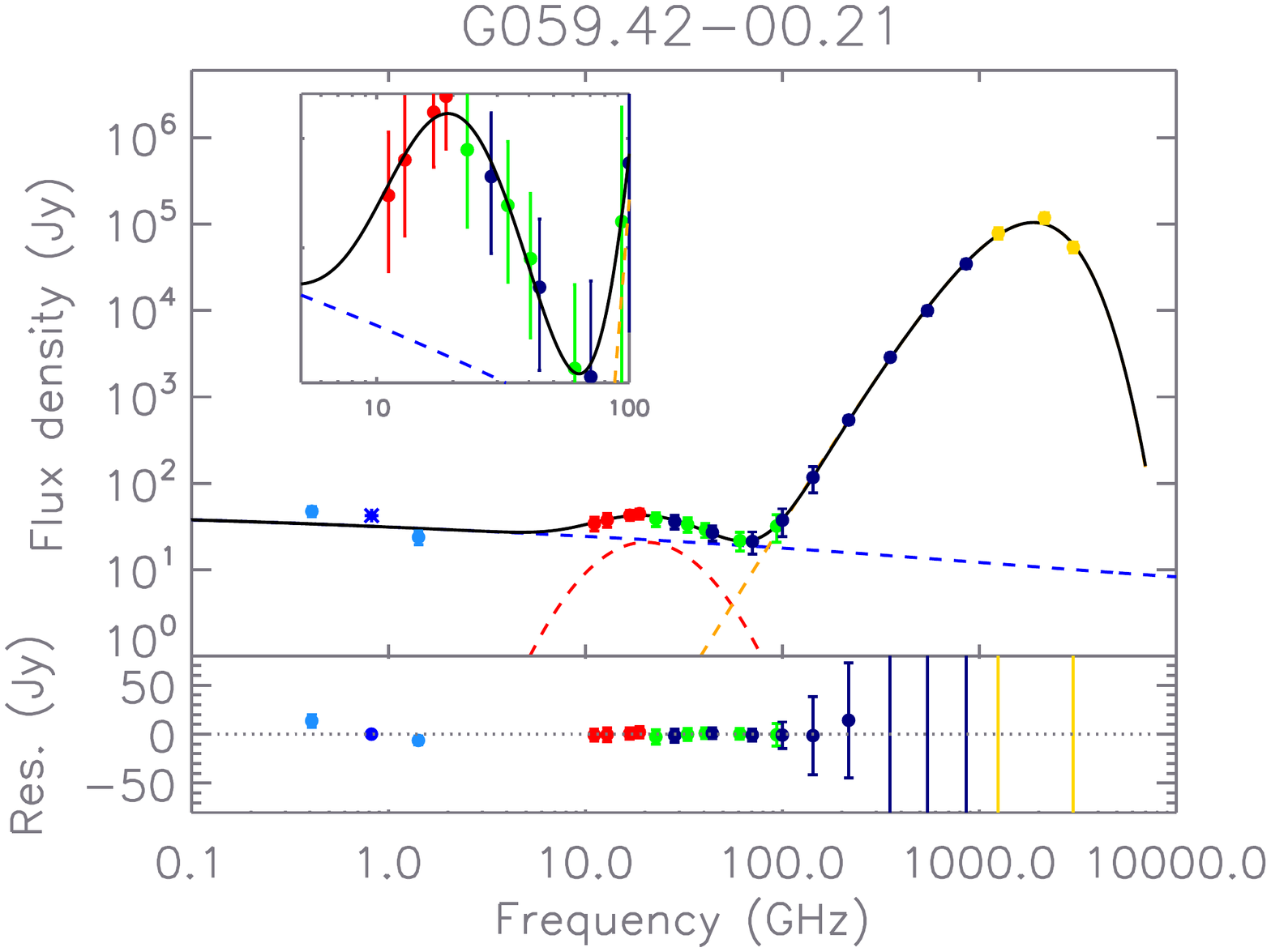}
\hspace*{10mm}
\includegraphics[width=77mm,angle=0]{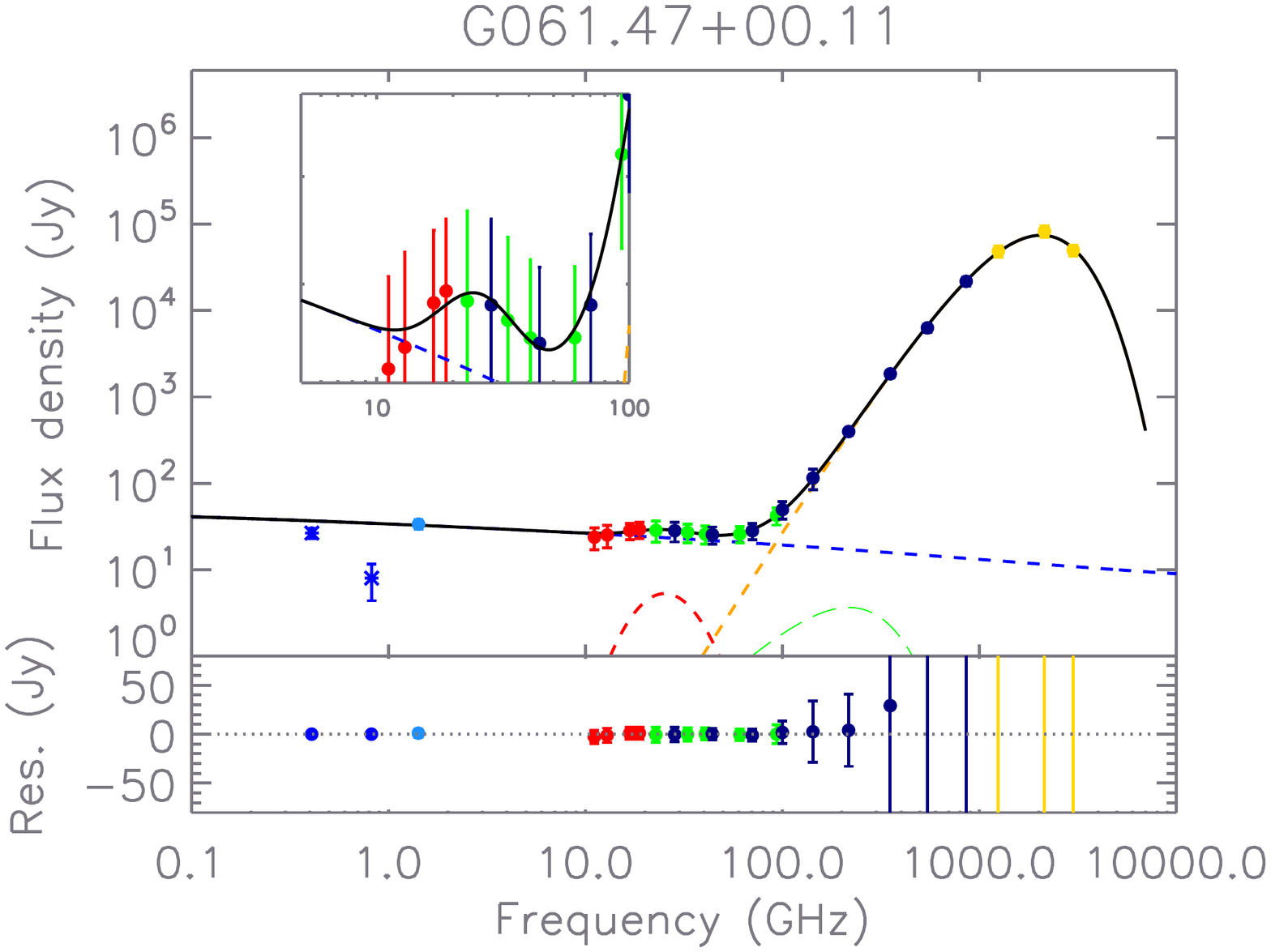}
\vspace*{-4.5cm}
\includegraphics[width=77mm,angle=0]{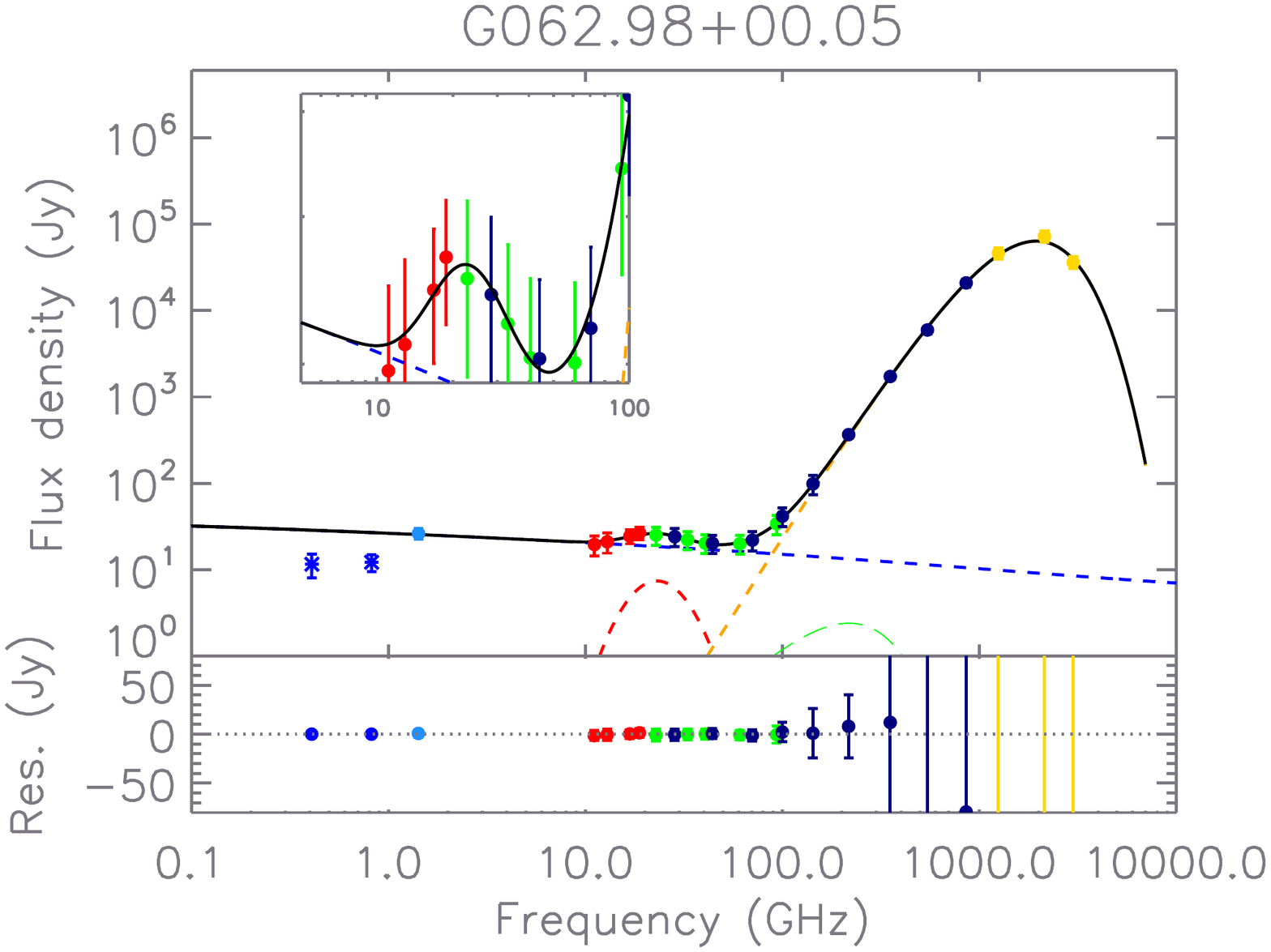}
\hspace*{10mm}
\includegraphics[width=77mm,angle=0]{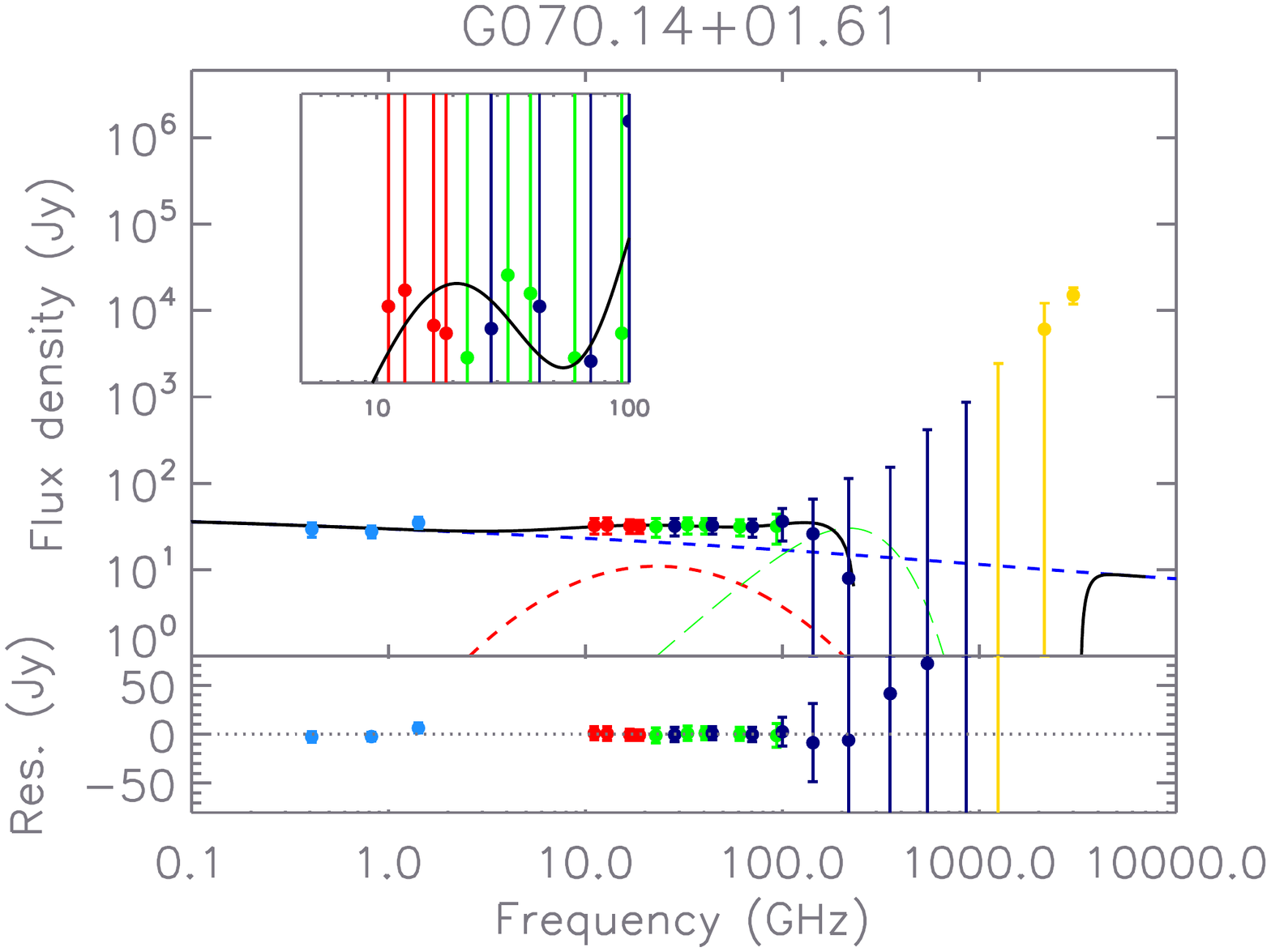}
\vspace*{1cm}
\caption{Same as Figure~\ref{fig:sed_int1}.}
\label{fig:sed_int2}
\end{center}
\end{figure*}

\begin{figure*}
\begin{center}
\vspace*{0mm}
\centering
\clearpage
\includegraphics[width=77mm,angle=0]{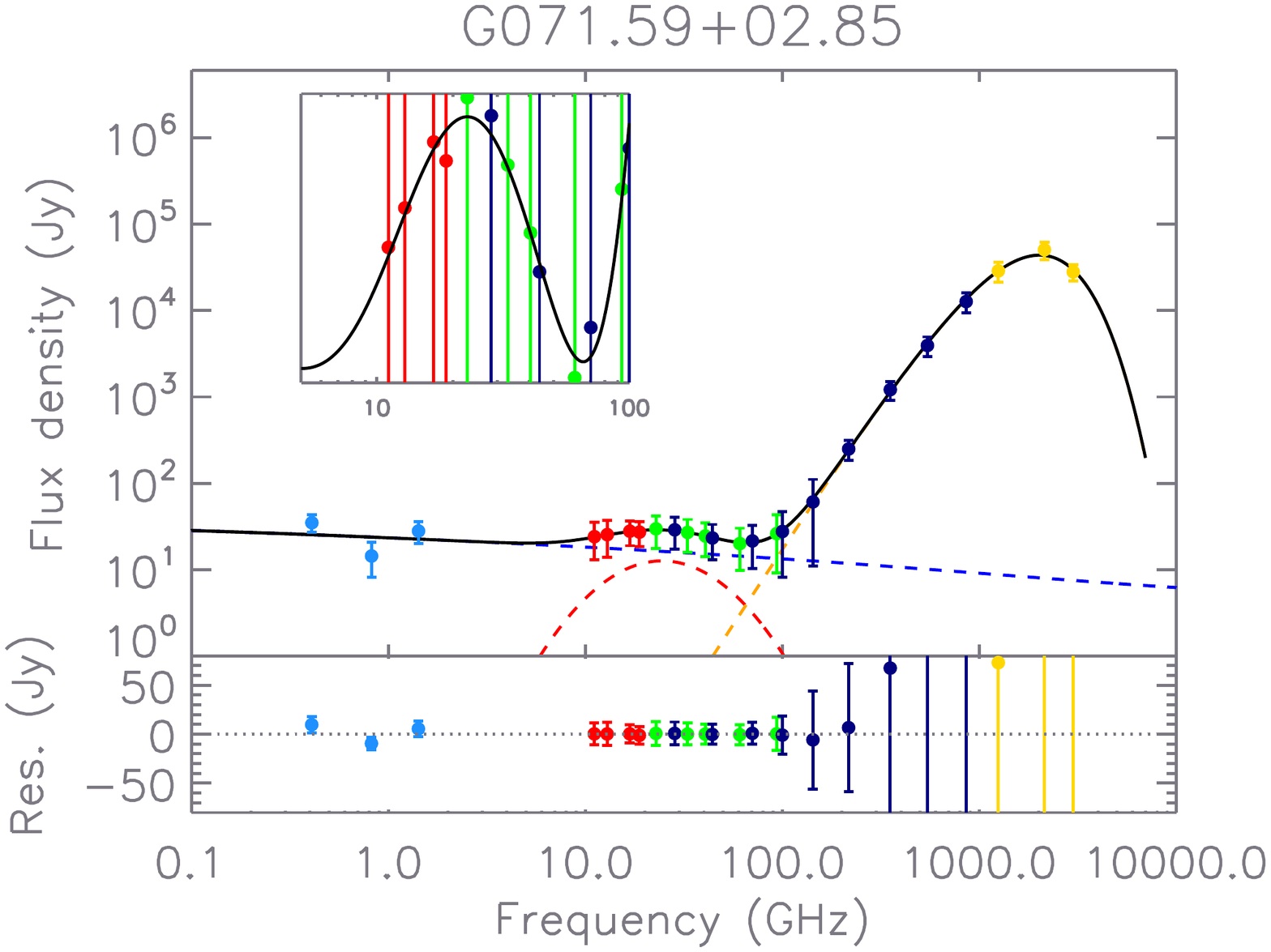}
\vspace*{-4.5cm}
\hspace*{10mm}
\includegraphics[width=77mm,angle=0]{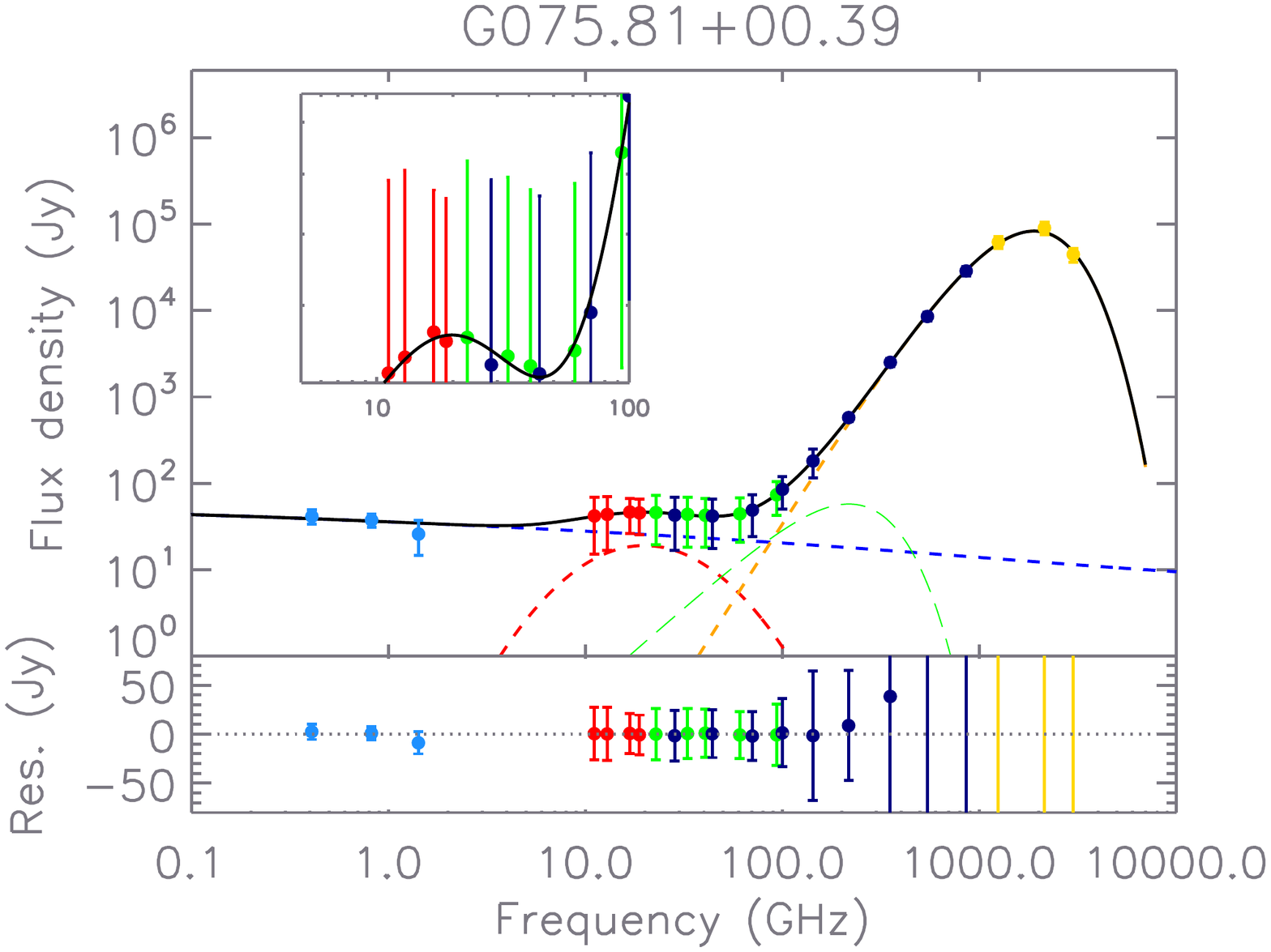}
\vspace*{-4.5cm}
\includegraphics[width=77mm,angle=0]{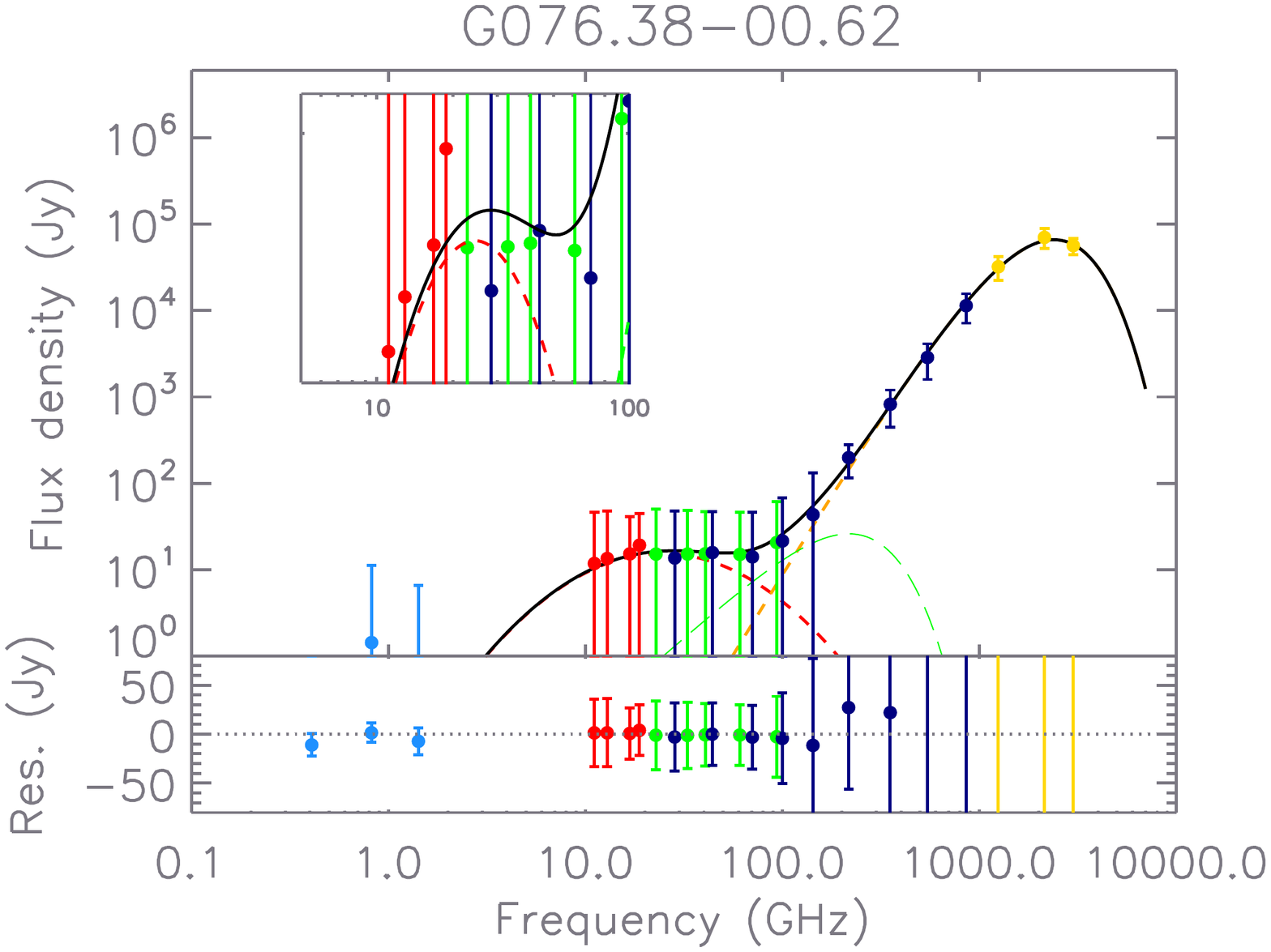}
\hspace*{10mm}
\includegraphics[width=77mm,angle=0]{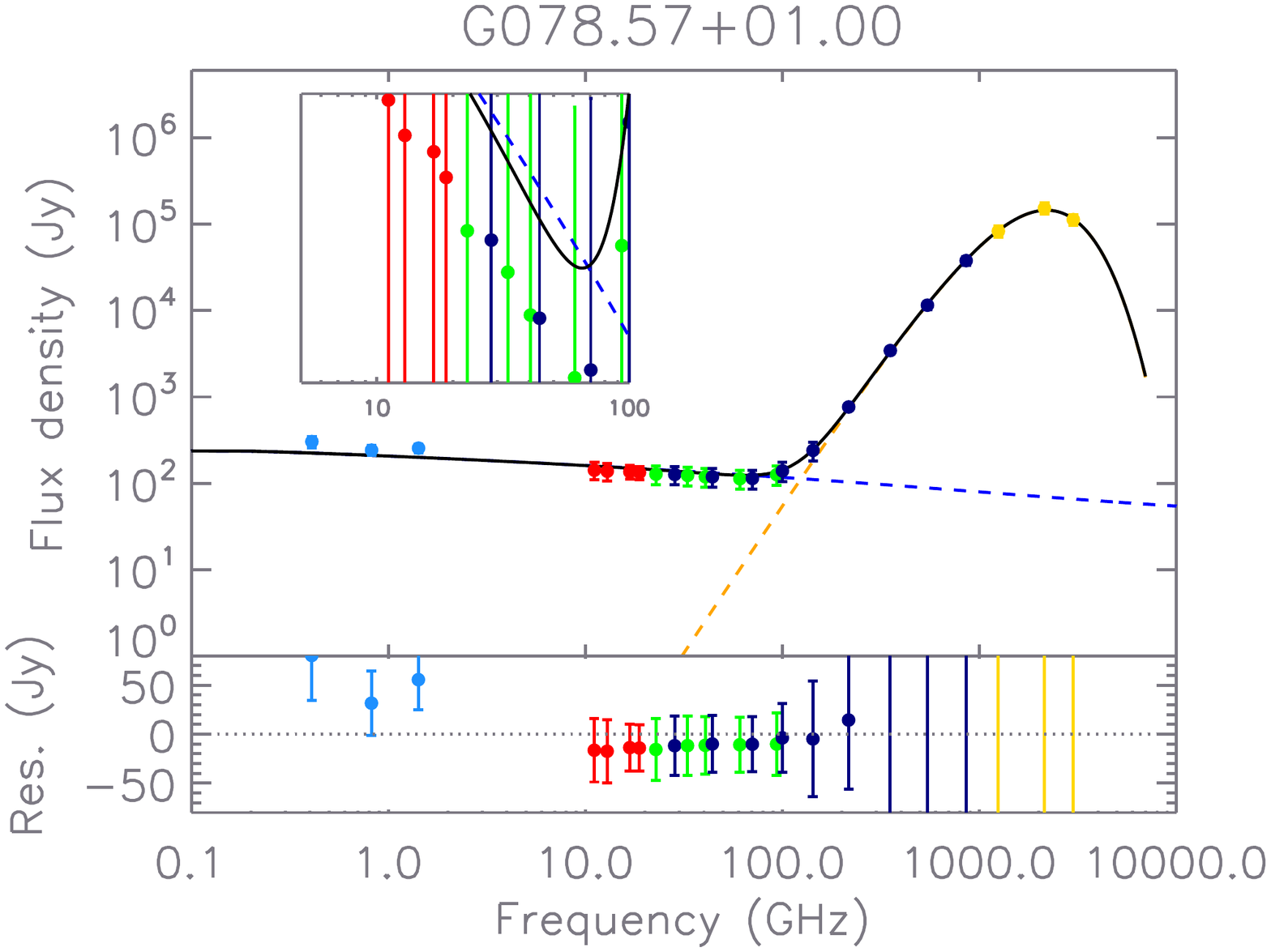}
\vspace*{-4.5cm}
\includegraphics[width=77mm,angle=0]{G081.59+00.01_SED_INT.pdf}
\hspace*{10mm}
\includegraphics[width=77mm,angle=0]{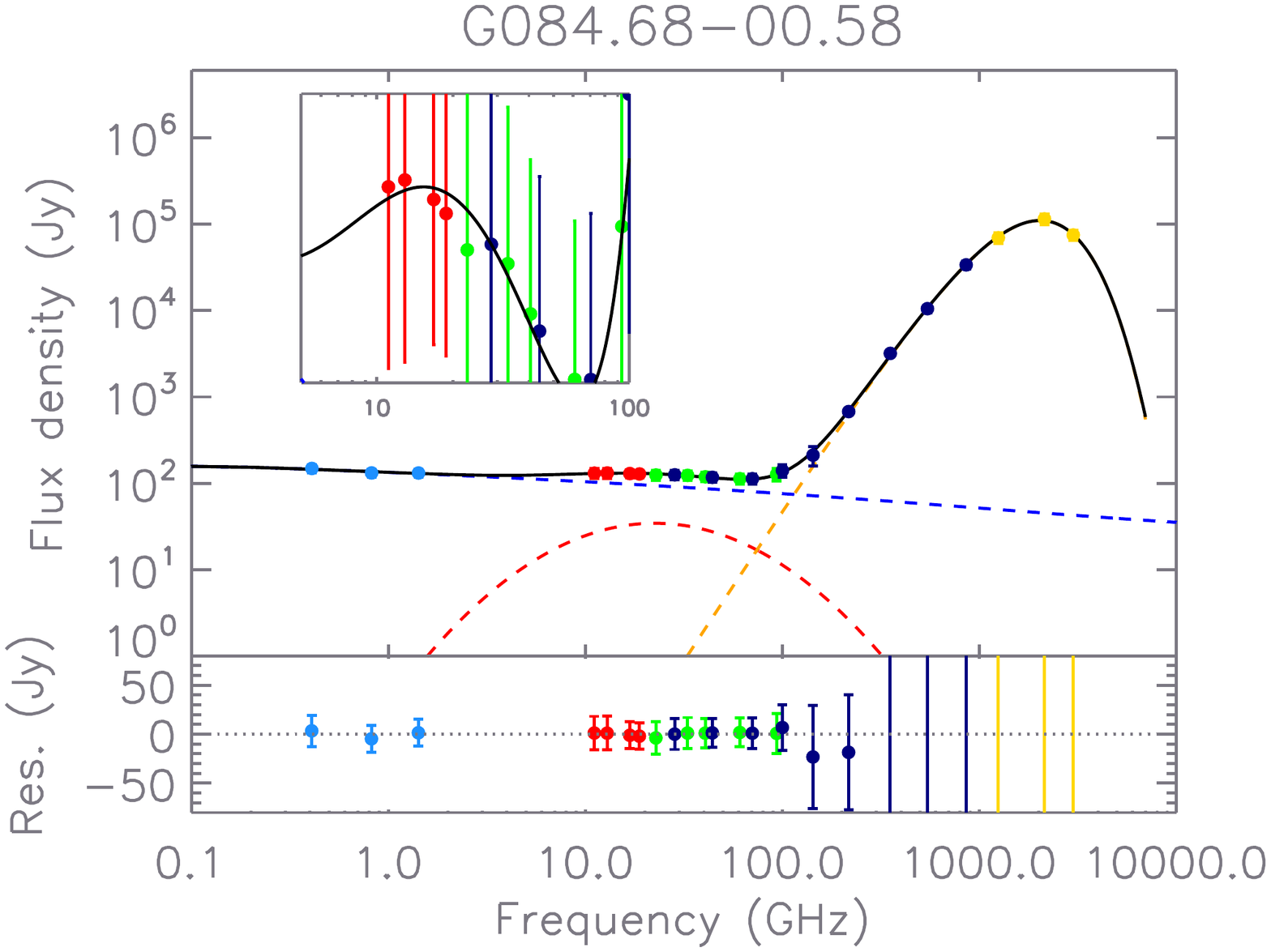}
\vspace*{-4.5cm}
\includegraphics[width=77mm,angle=0]{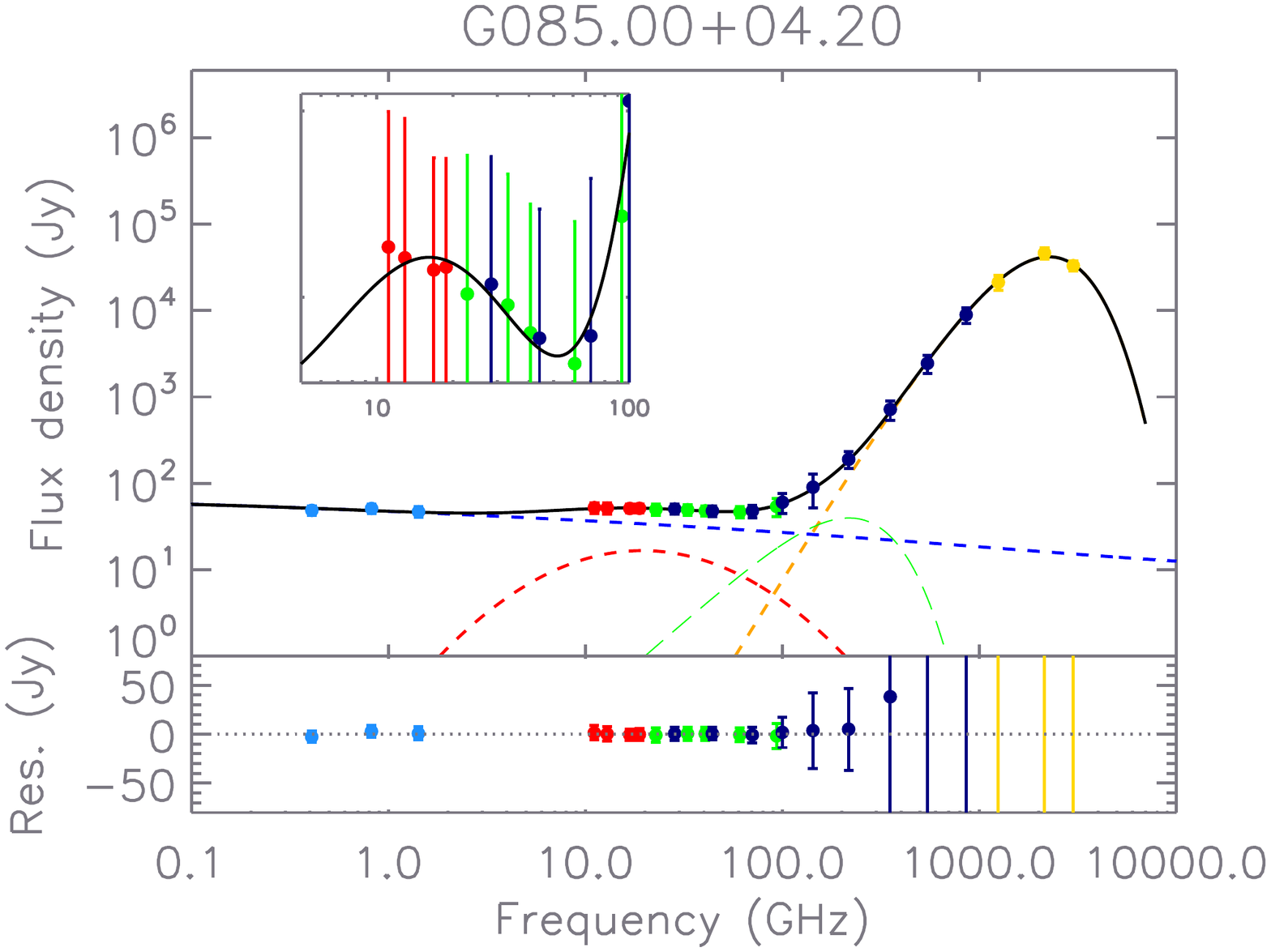}
\hspace*{10mm}
\includegraphics[width=77mm,angle=0]{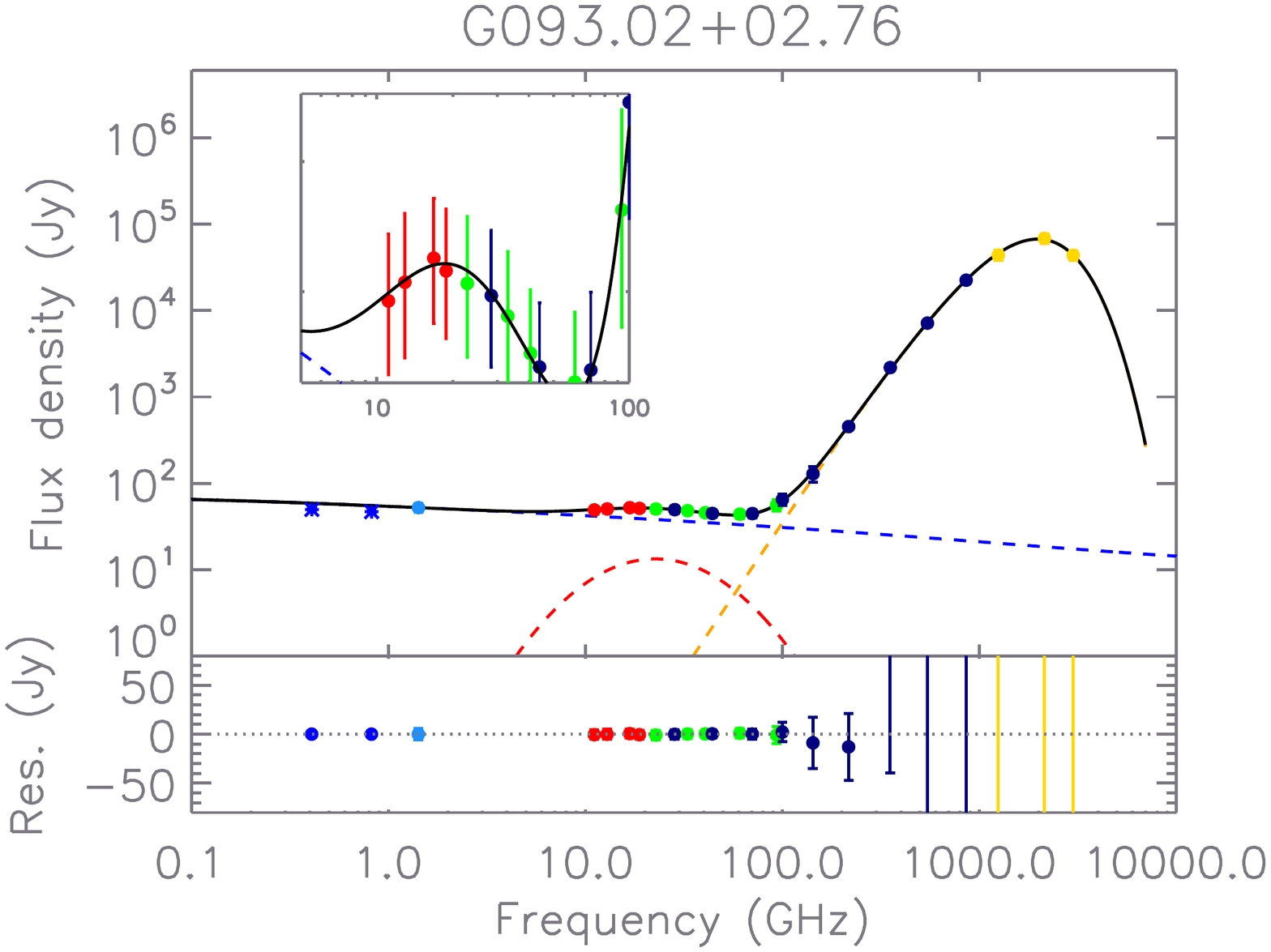}
\vspace*{1cm}
\caption{Same as Figure~\ref{fig:sed_int1}.}
\label{fig:sed_int3}
\end{center}
\end{figure*}

\begin{figure*}
\begin{center}
\vspace*{0mm}
\centering
\clearpage
\includegraphics[width=77mm,angle=0]{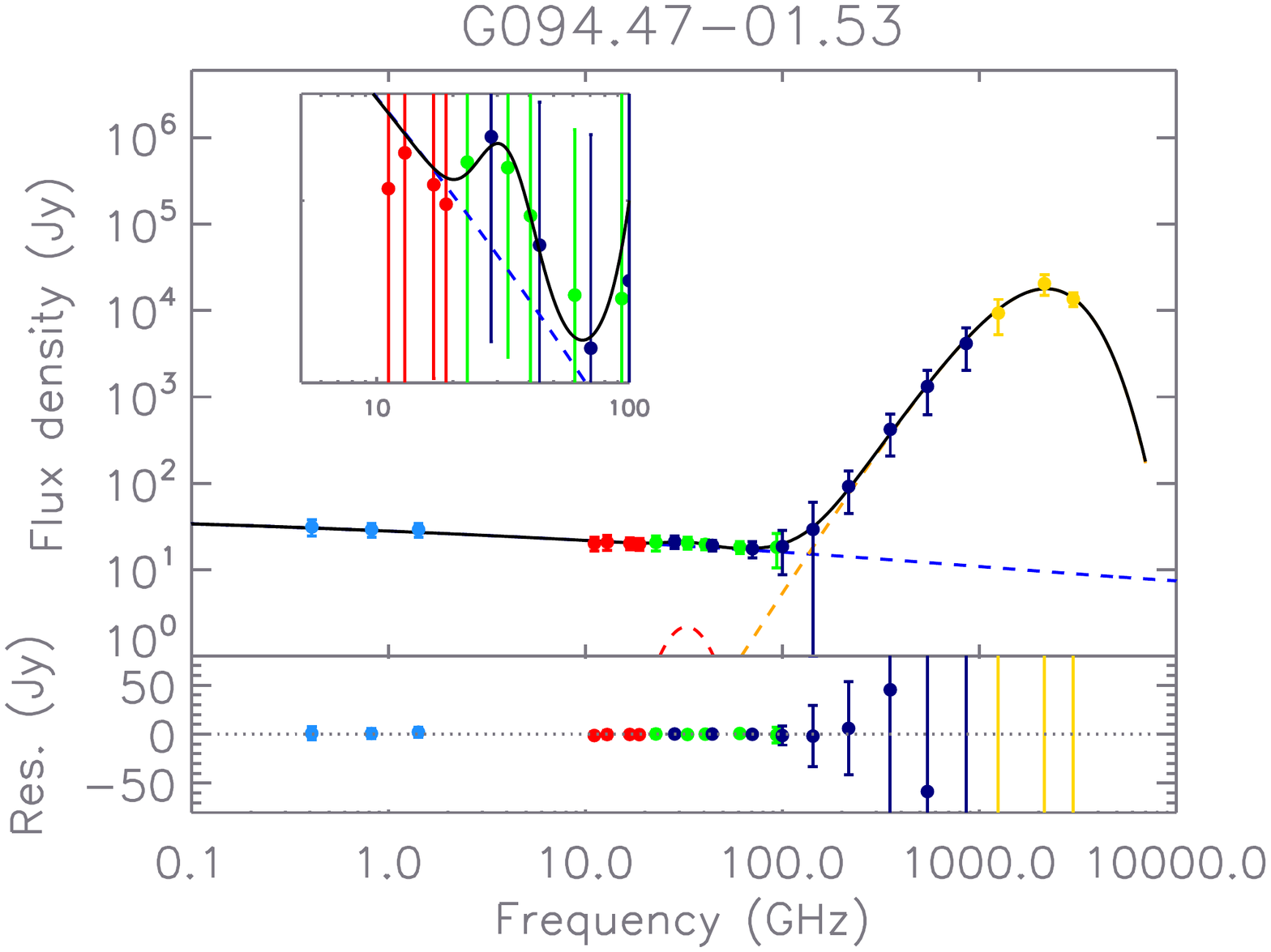}
\vspace*{-4.5cm}
\hspace*{10mm}
\includegraphics[width=77mm,angle=0]{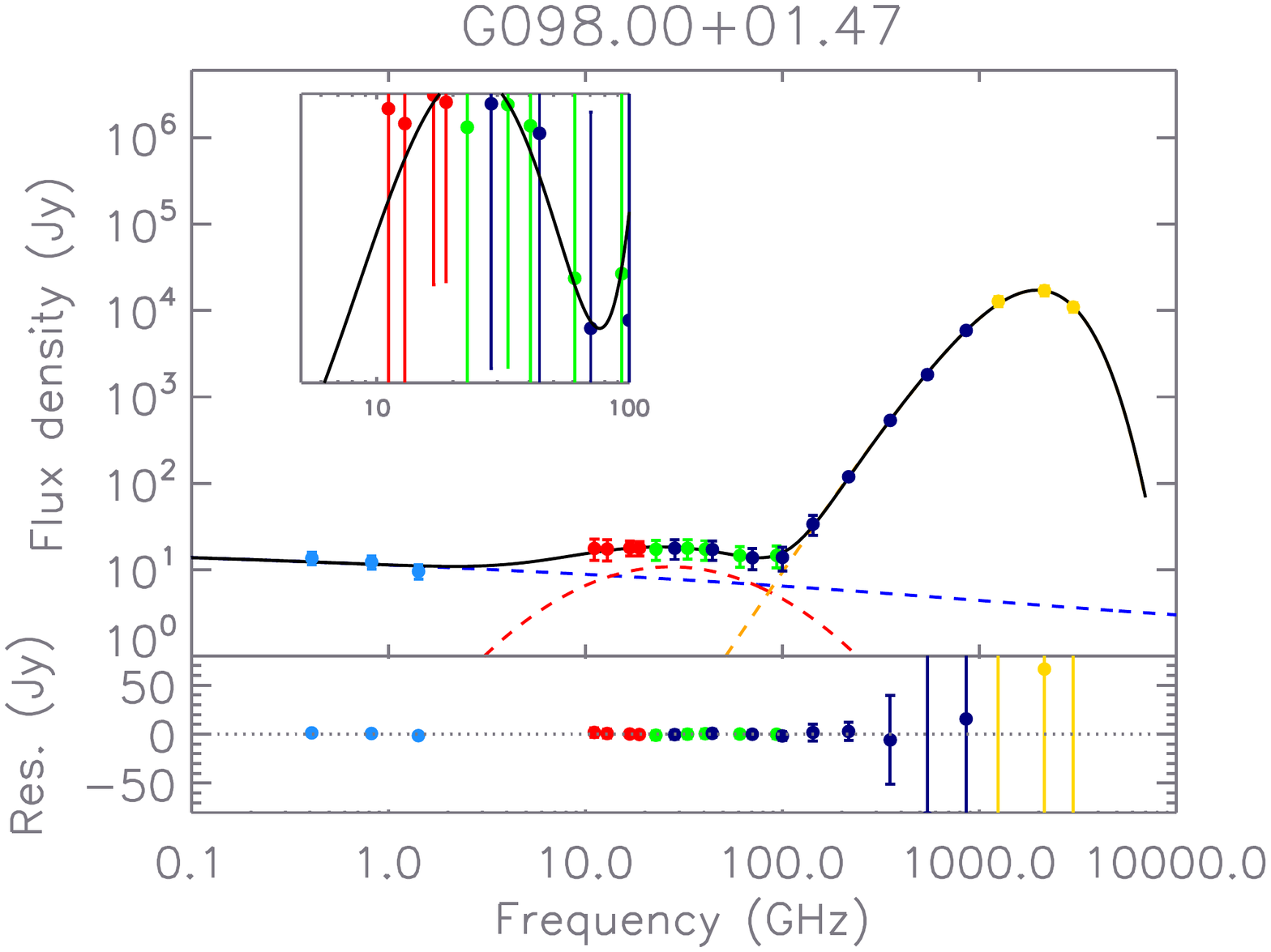}
\vspace*{-4.5cm}
\includegraphics[width=77mm,angle=0]{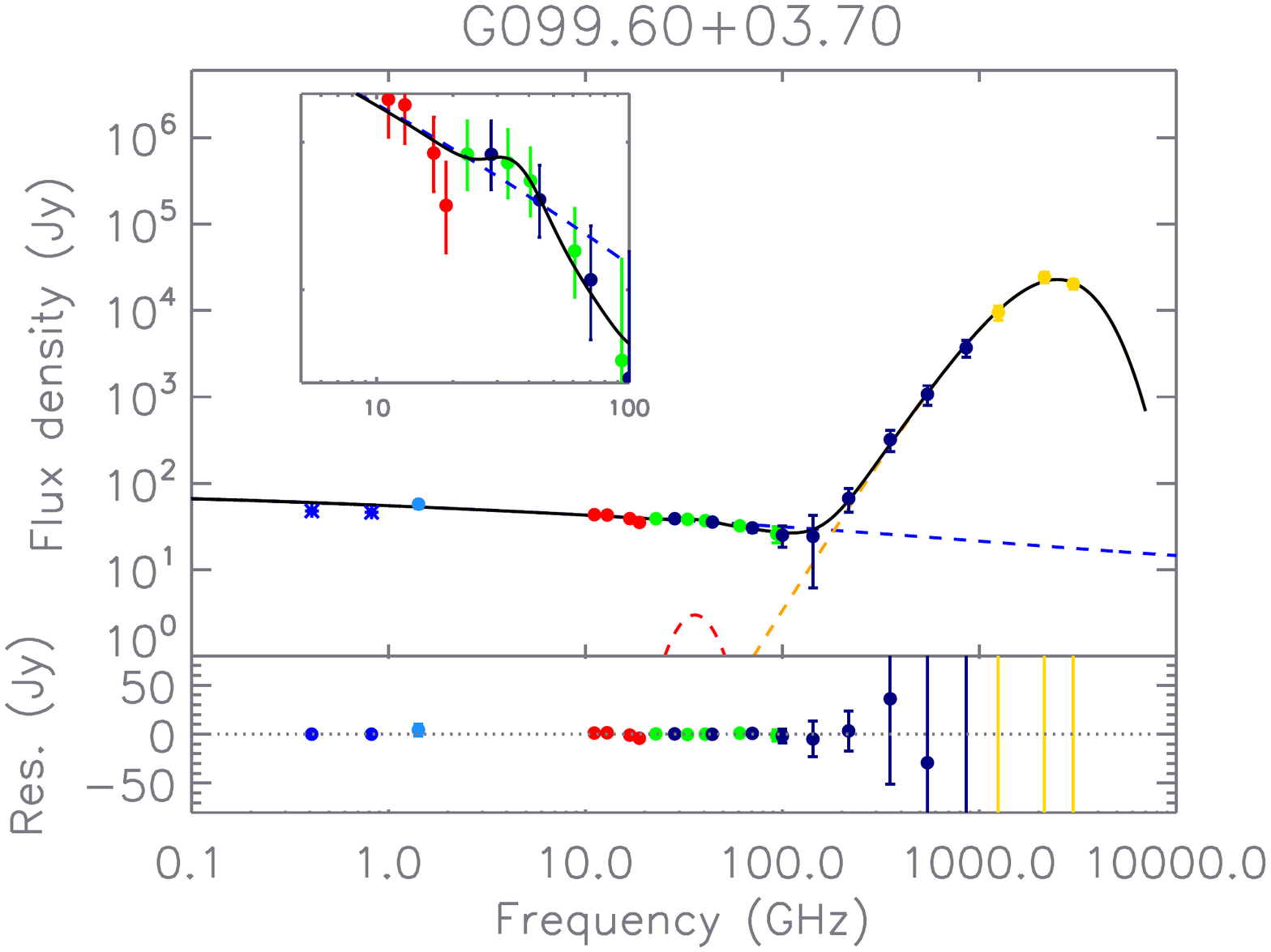}
\hspace*{10mm}
\includegraphics[width=77mm,angle=0]{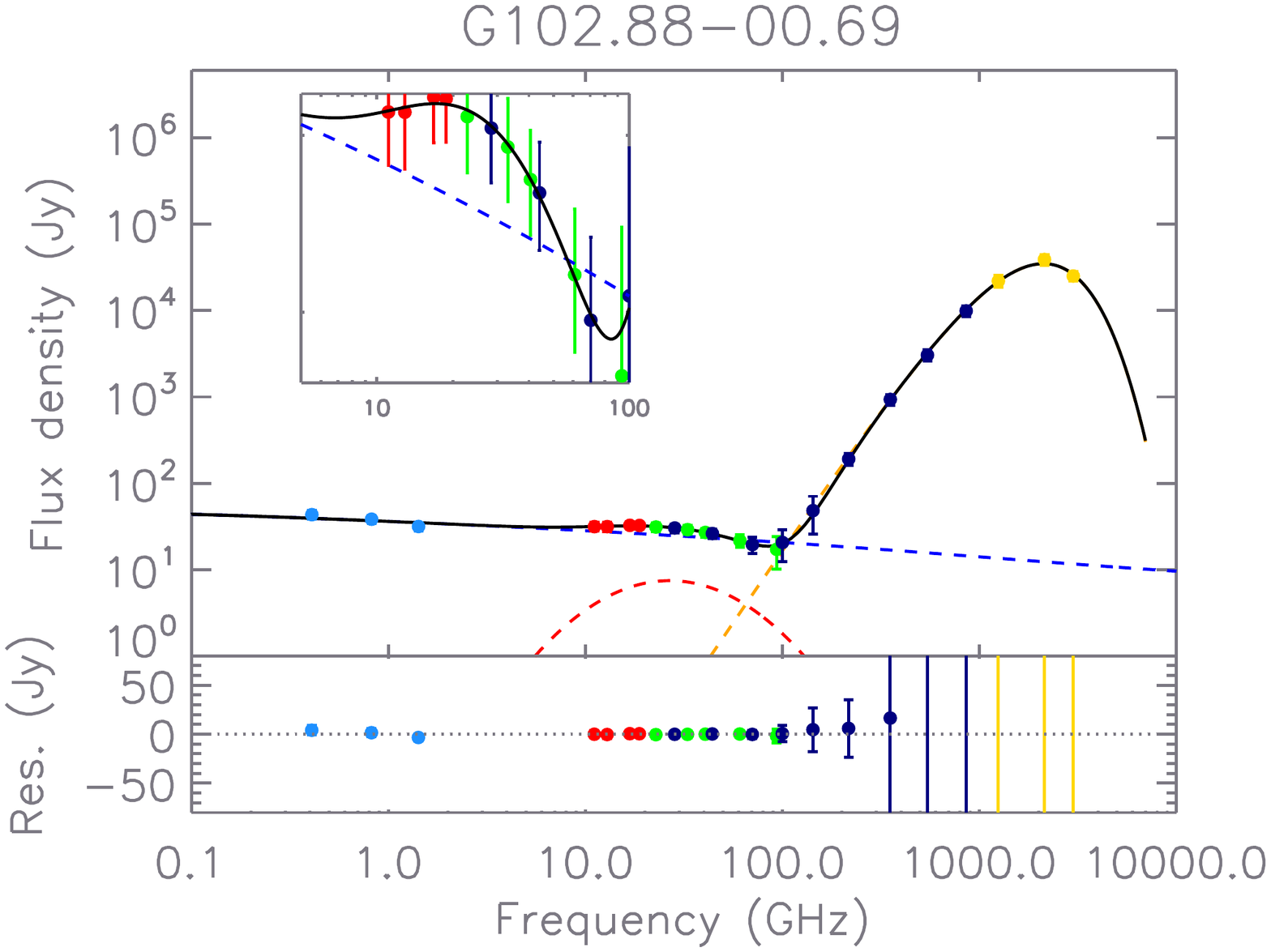}
\vspace*{-4.5cm}
\includegraphics[width=77mm,angle=0]{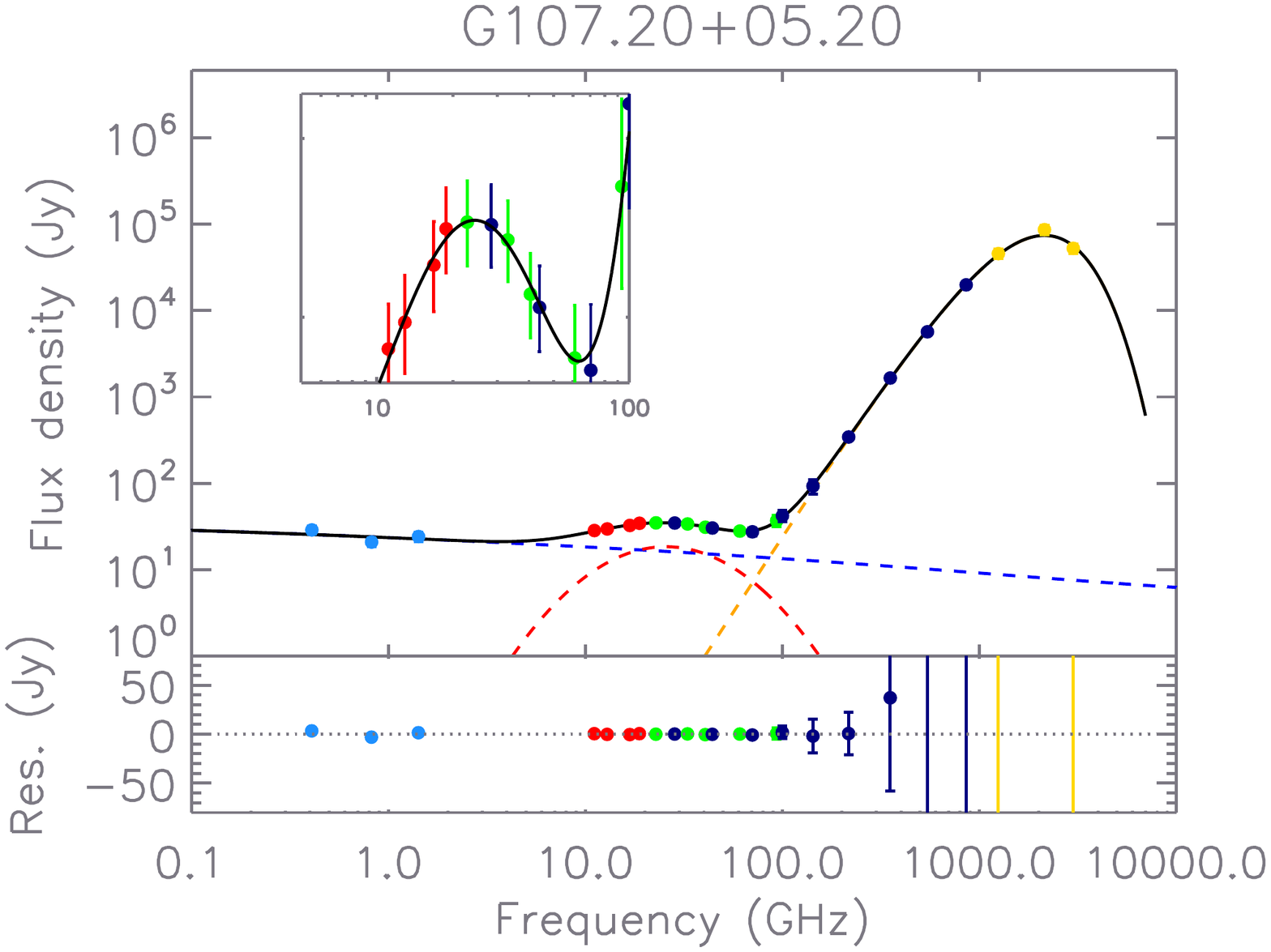}
\hspace*{10mm}
\includegraphics[width=77mm,angle=0]{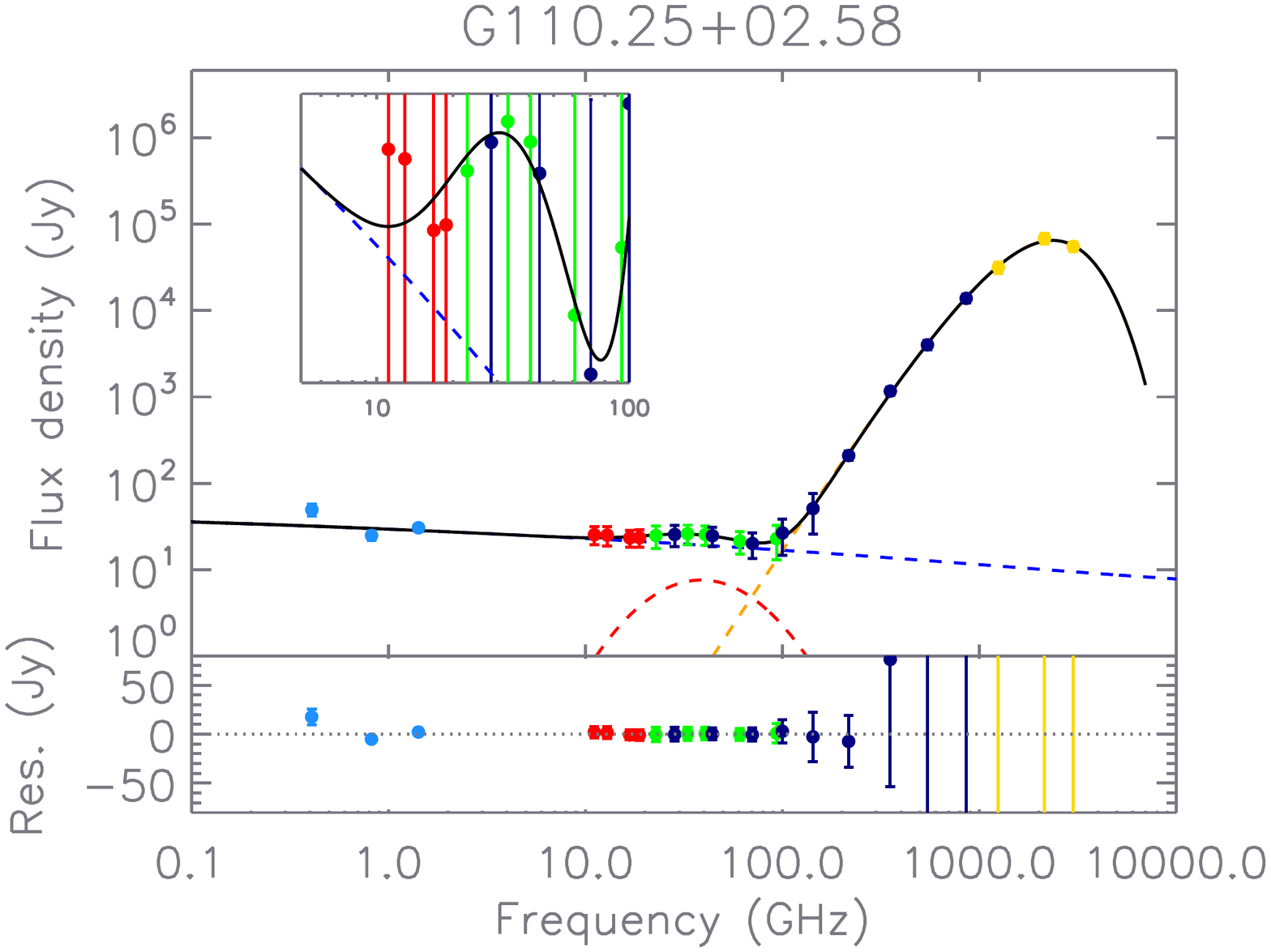}
\vspace*{-4.5cm}
\includegraphics[width=77mm,angle=0]{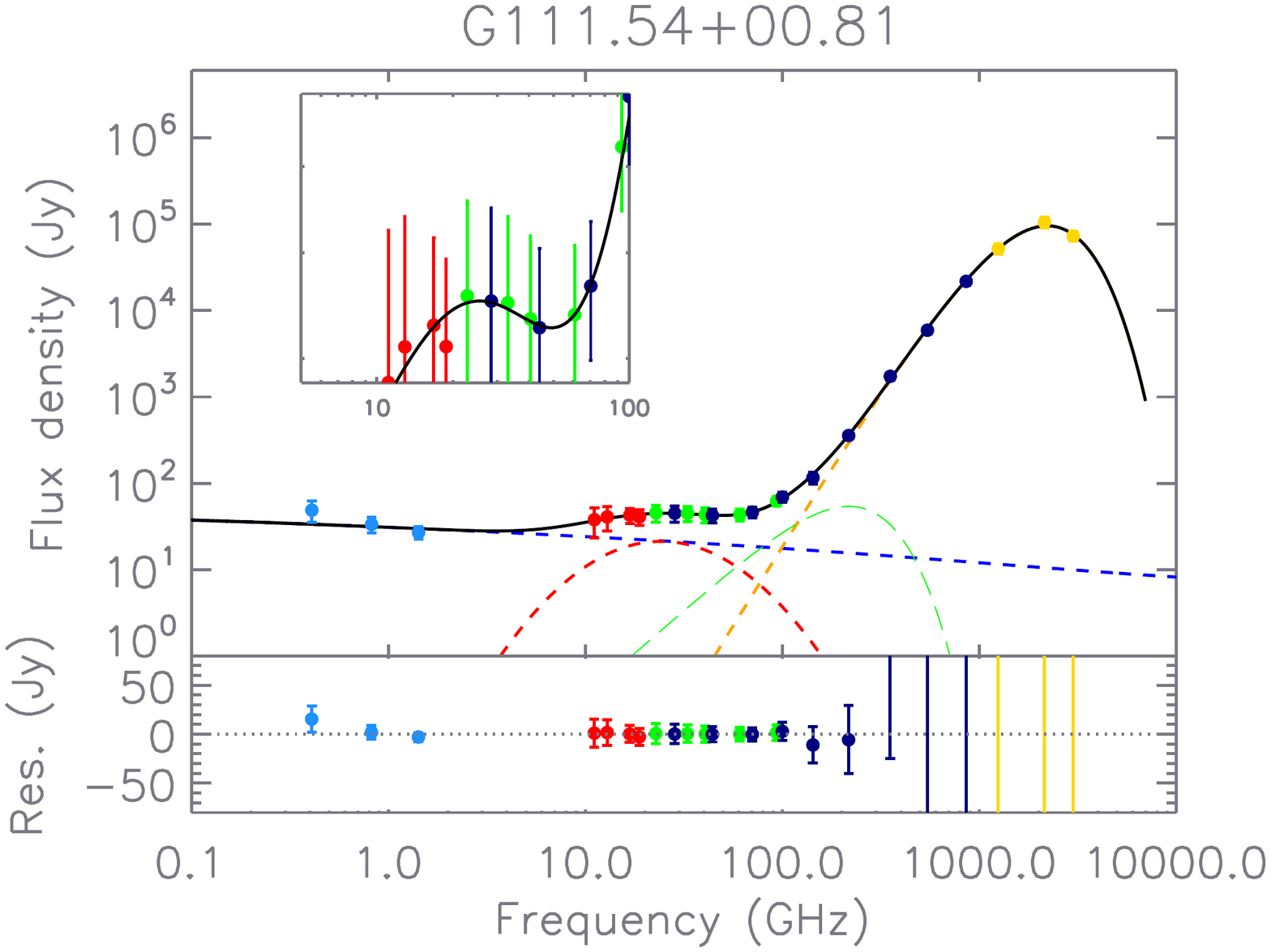}
\hspace*{10mm}
\includegraphics[width=77mm,angle=0]{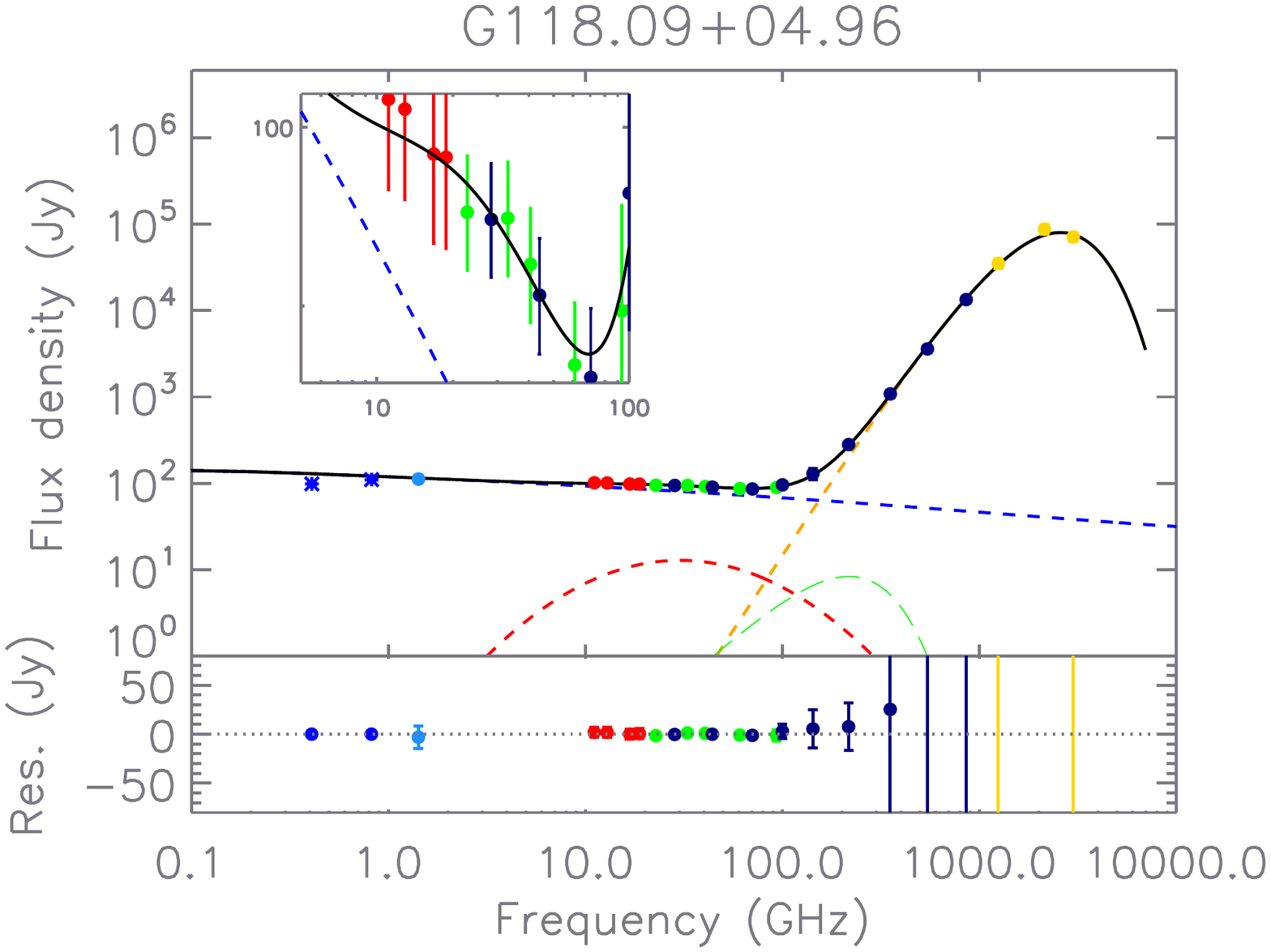}
\vspace*{1cm}
\caption{Same as Figure~\ref{fig:sed_int1}.}
\label{fig:sed_int4}
\end{center}
\end{figure*}

\begin{figure*}
\begin{center}
\vspace*{0mm}
\centering
\clearpage
\includegraphics[width=77mm,angle=0]{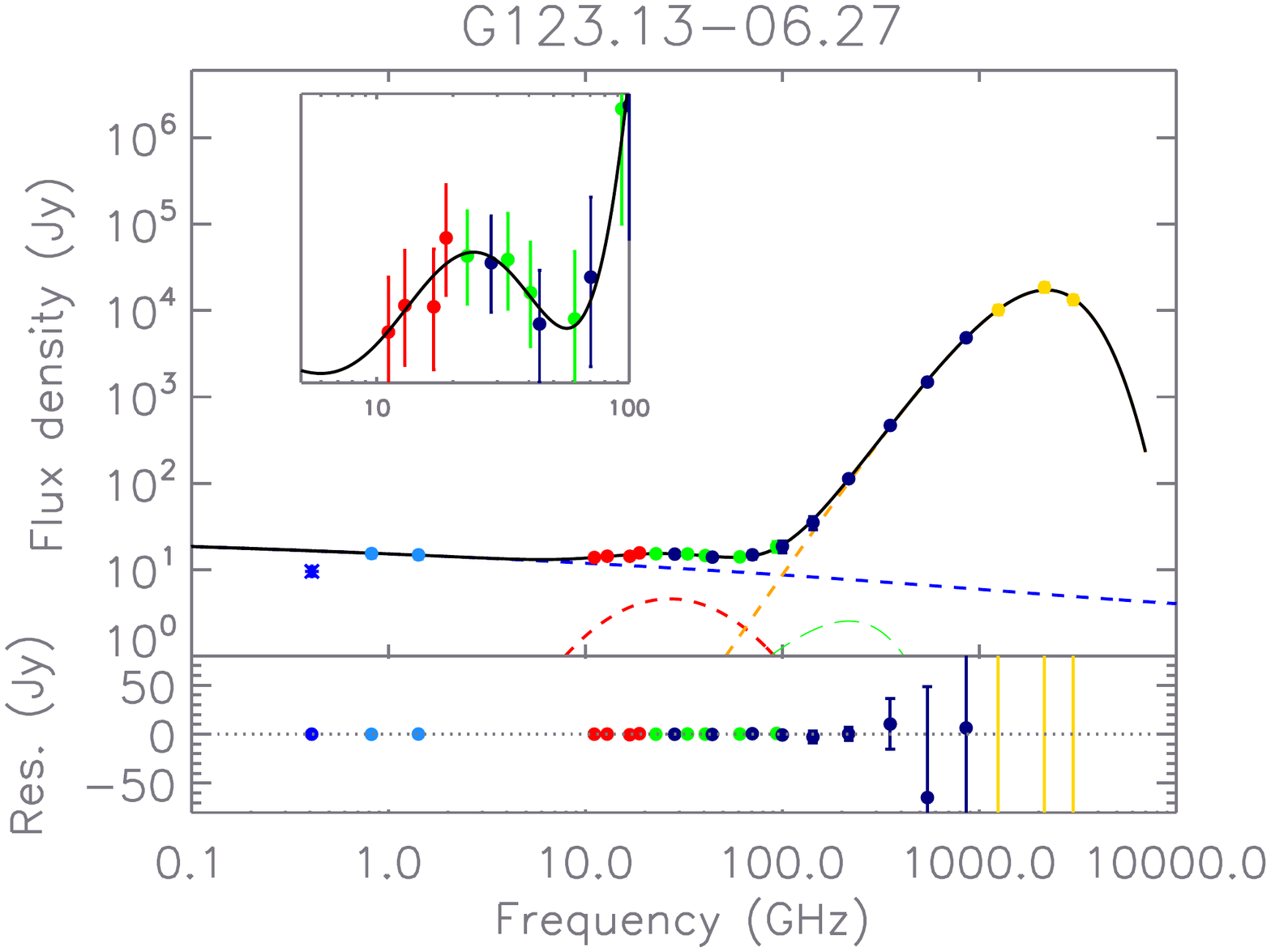}
\vspace*{-4.5cm}
\hspace*{10mm}
\includegraphics[width=77mm,angle=0]{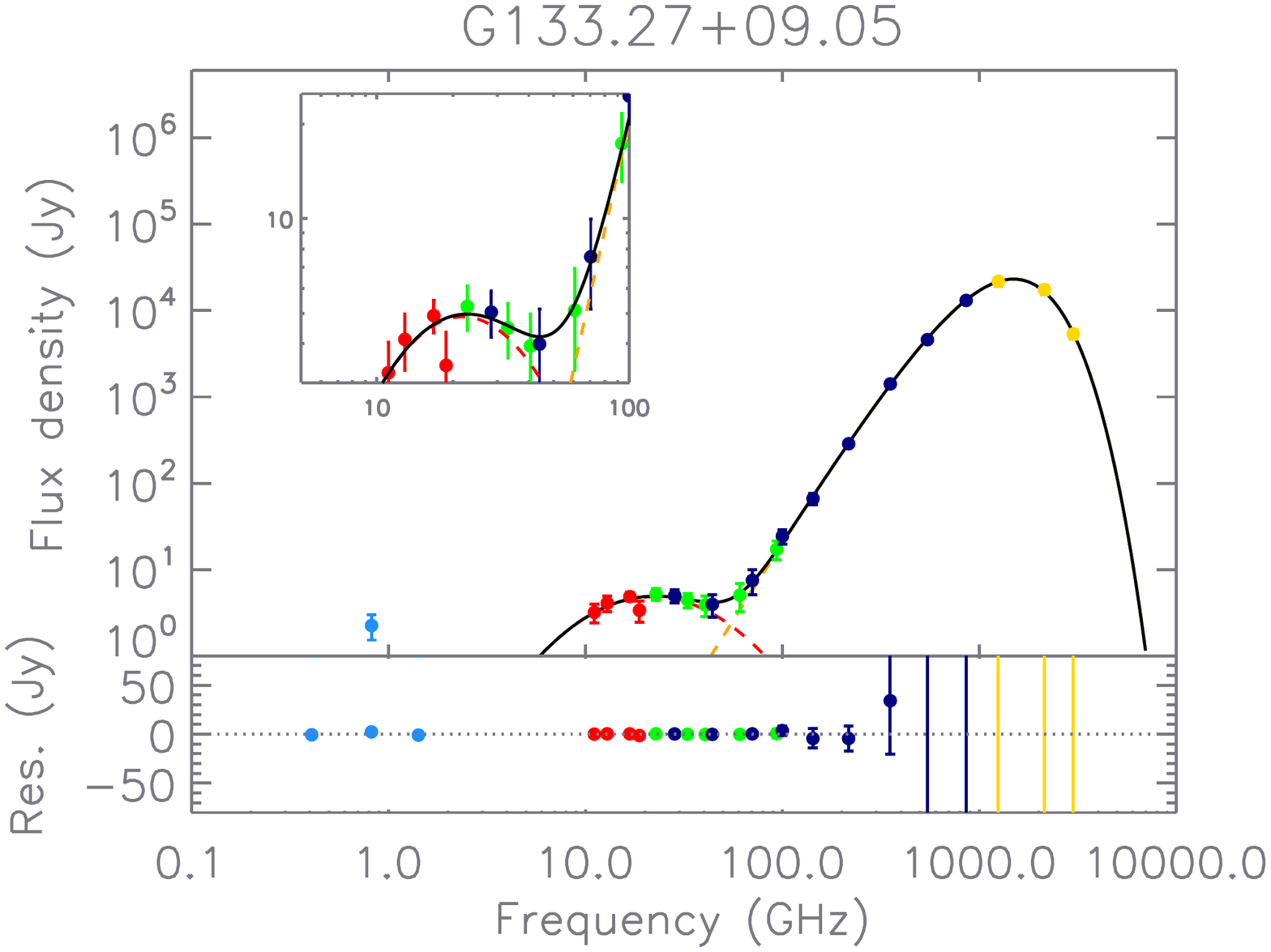}
\vspace*{-4.5cm}
\includegraphics[width=77mm,angle=0]{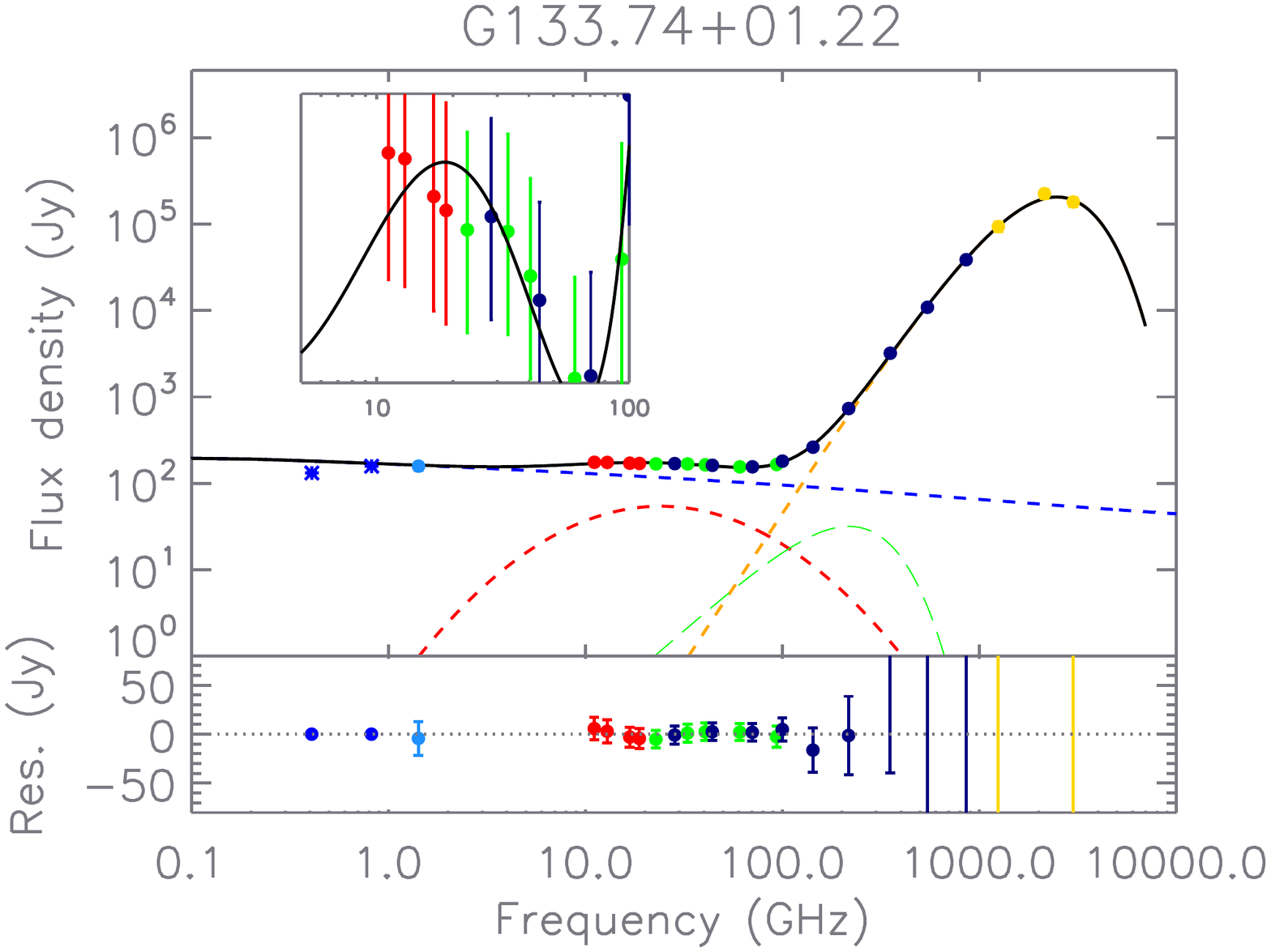}
\hspace*{10mm}
\includegraphics[width=77mm,angle=0]{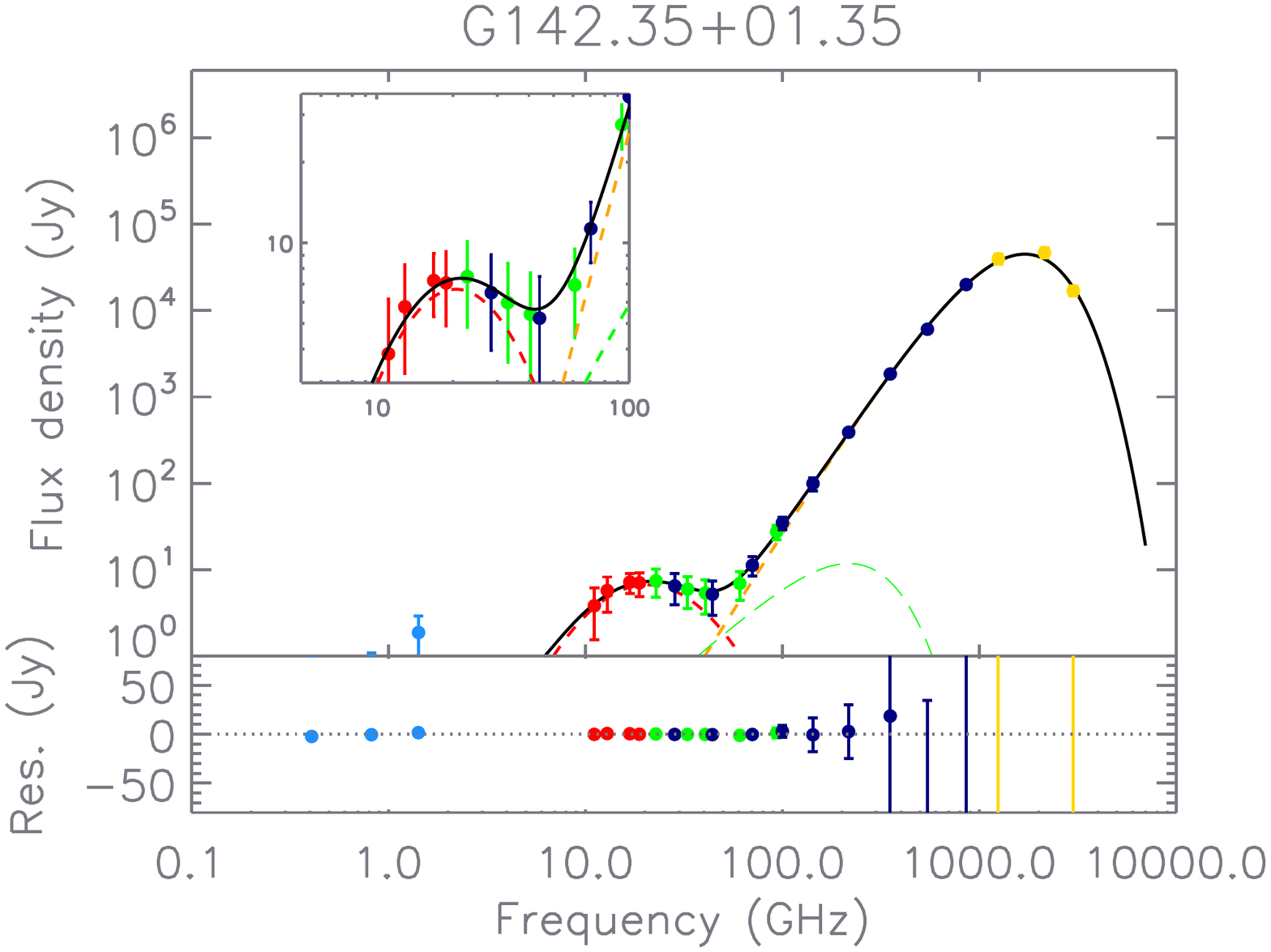}
\vspace*{-4.5cm}
\includegraphics[width=77mm,angle=0]{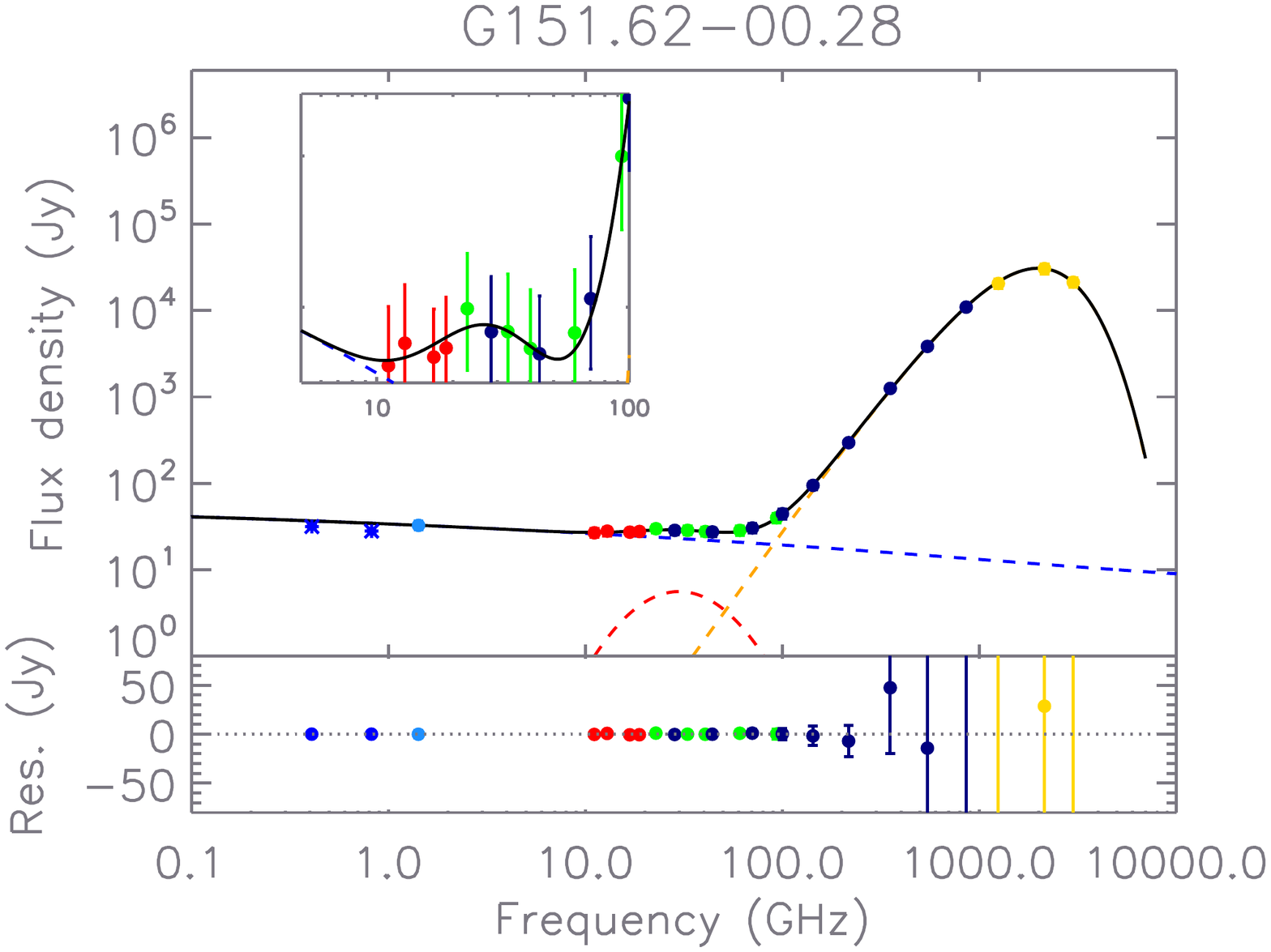}
\hspace*{10mm}
\includegraphics[width=77mm,angle=0]{G160.26-18.62_SED_INT.pdf}
\vspace*{-4.5cm}
\includegraphics[width=77mm,angle=0]{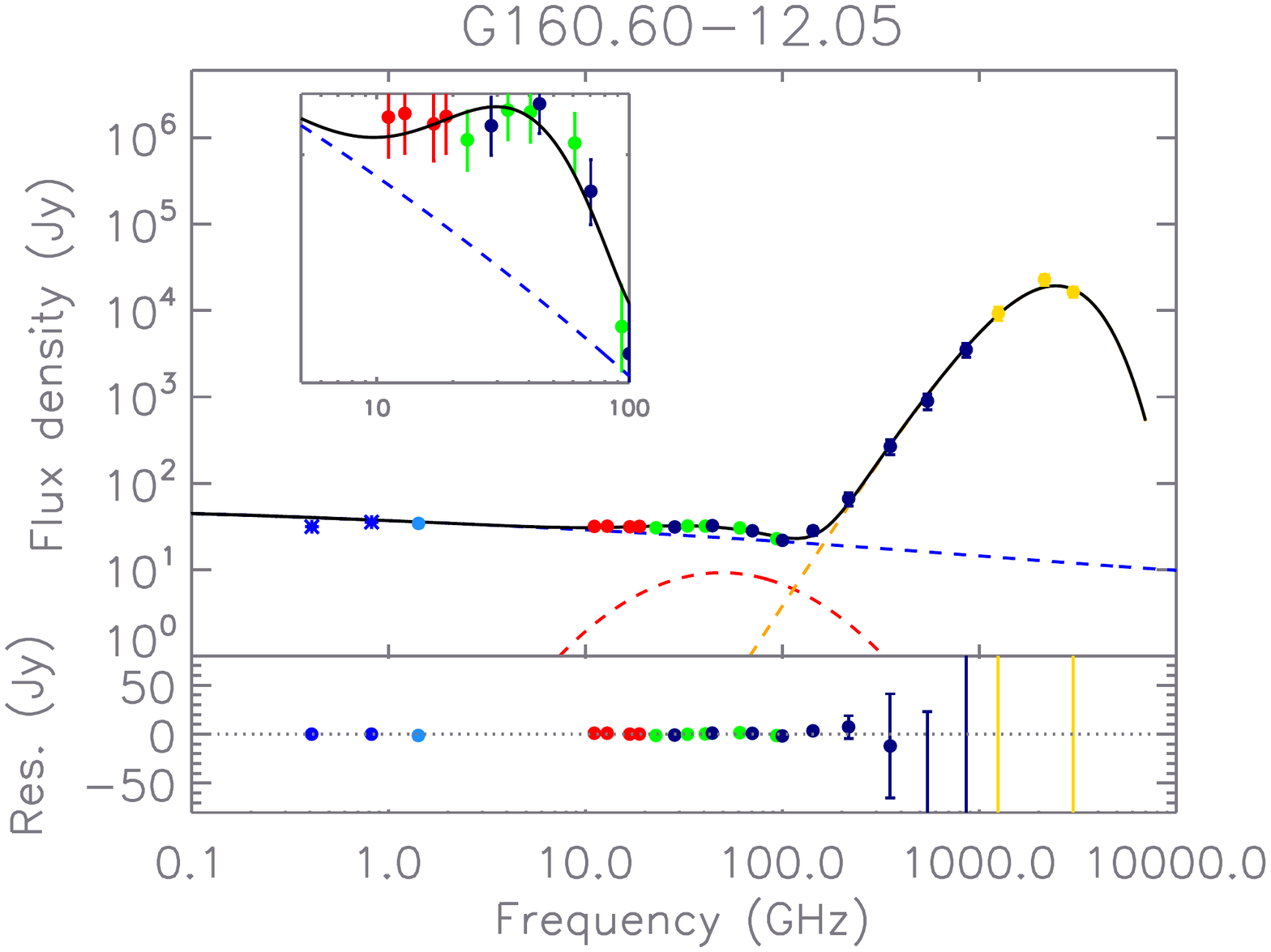}
\hspace*{10mm}
\includegraphics[width=77mm,angle=0]{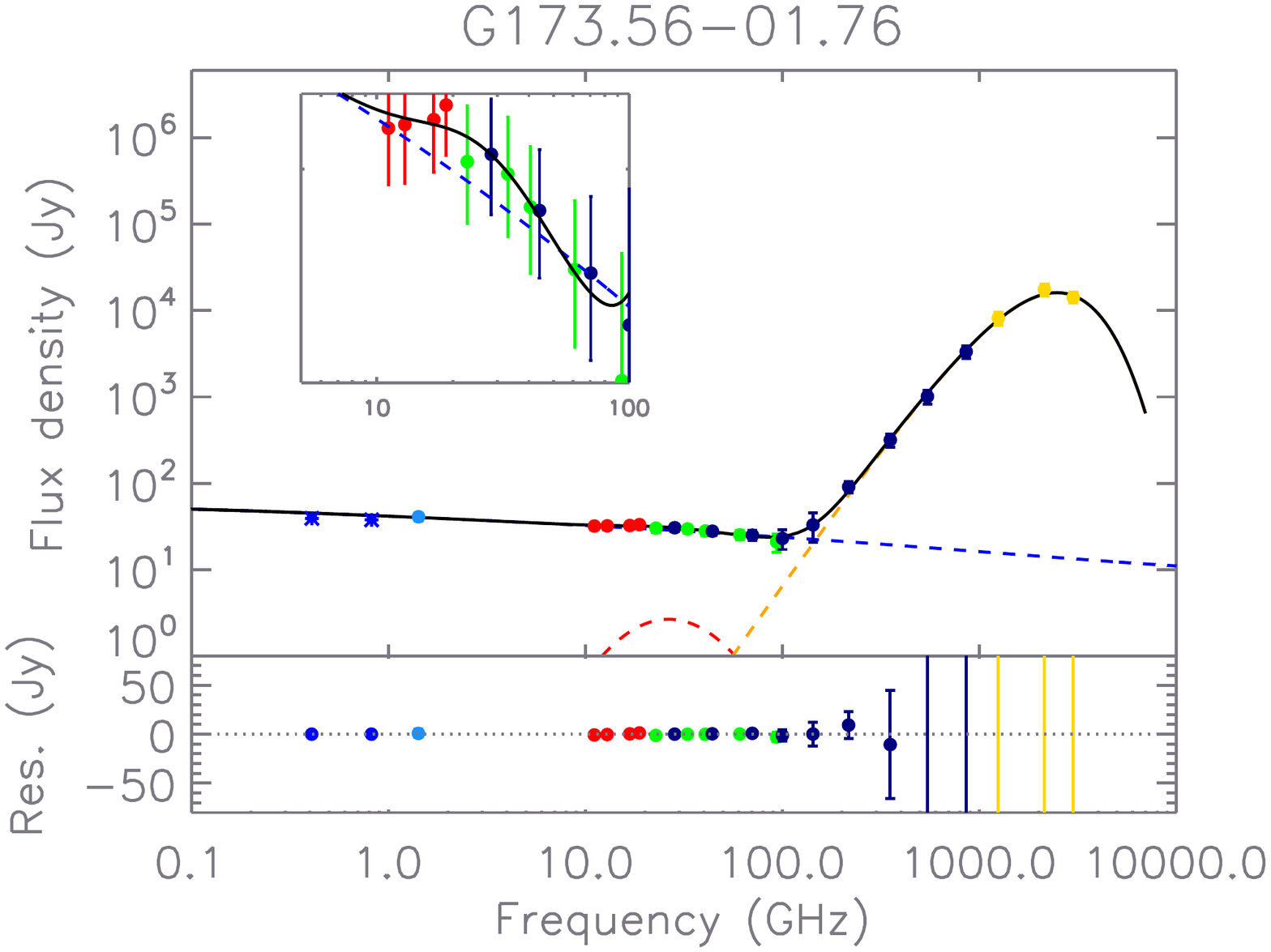}
\vspace*{1cm}
\caption{Same as Figure~\ref{fig:sed_int1}.}
\label{fig:sed_int5}
\end{center}
\end{figure*}

\begin{figure*}
\begin{center}
\vspace*{0mm}
\centering
\clearpage
\includegraphics[width=77mm,angle=0]{G173.62+02.79_SED_INT.pdf}
\vspace*{-4.5cm}
\hspace*{10mm}
\includegraphics[width=77mm,angle=0]{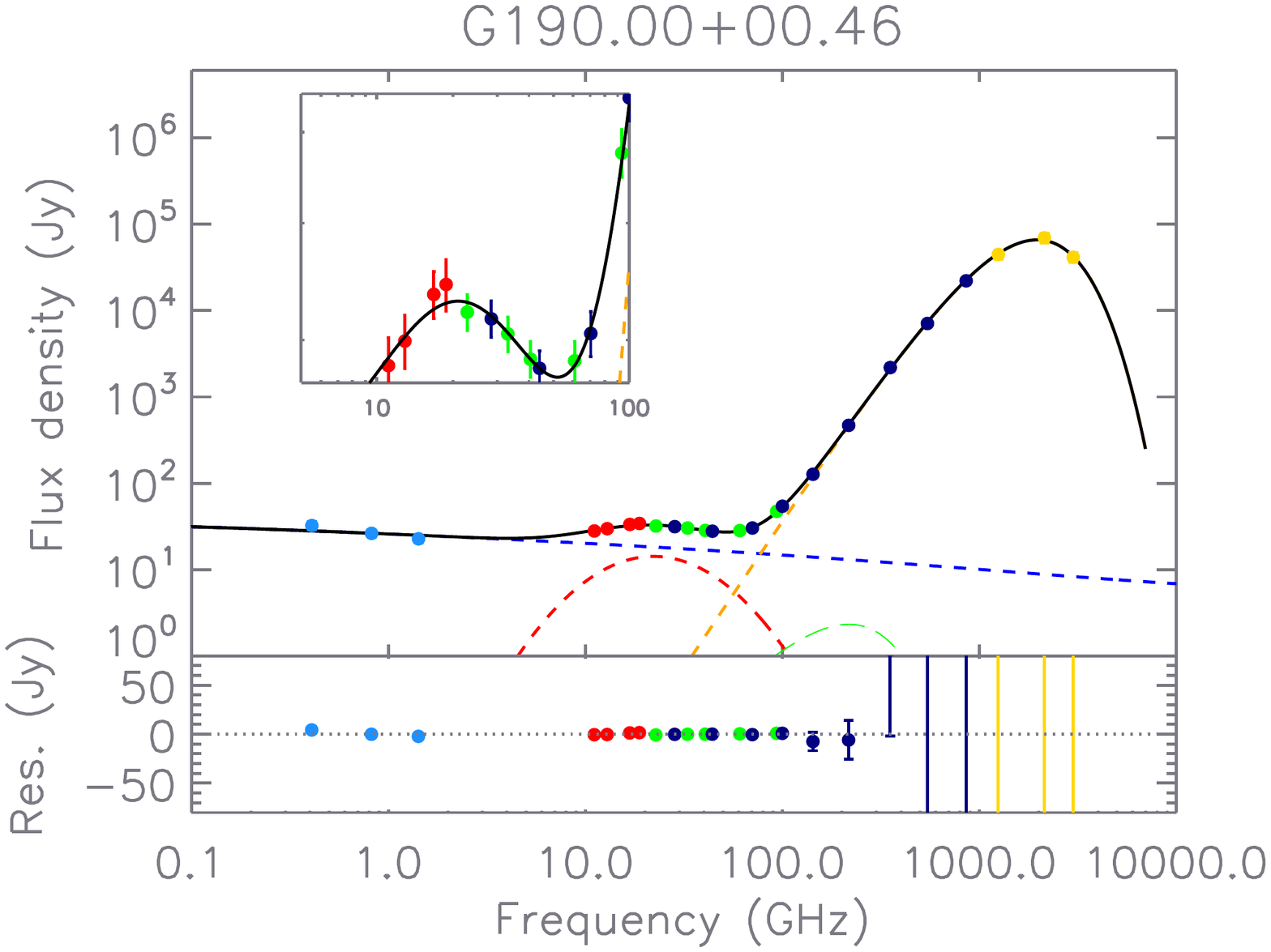}
\vspace*{-4.5cm}
\includegraphics[width=77mm,angle=0]{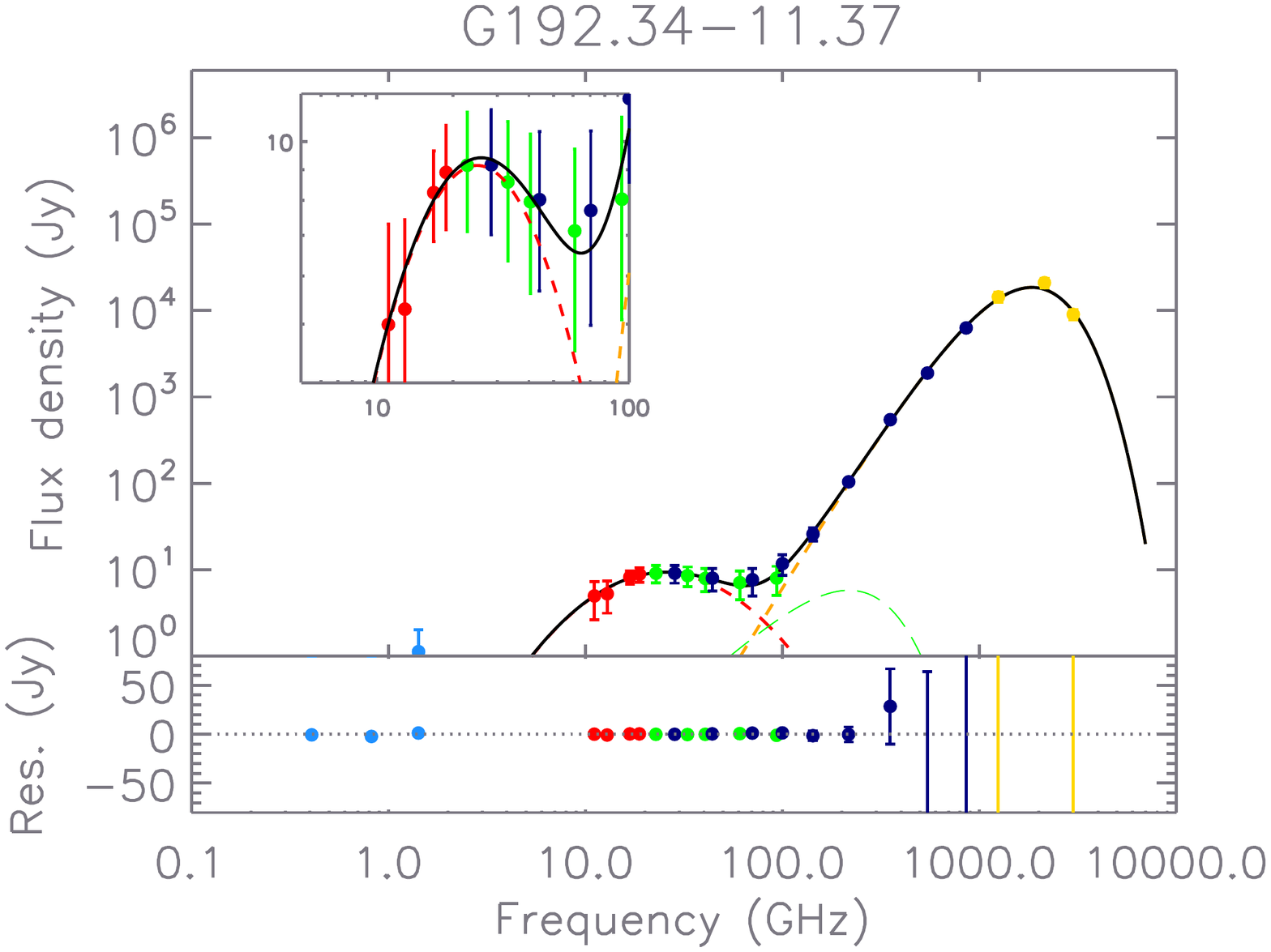}
\hspace*{10mm}
\includegraphics[width=77mm,angle=0]{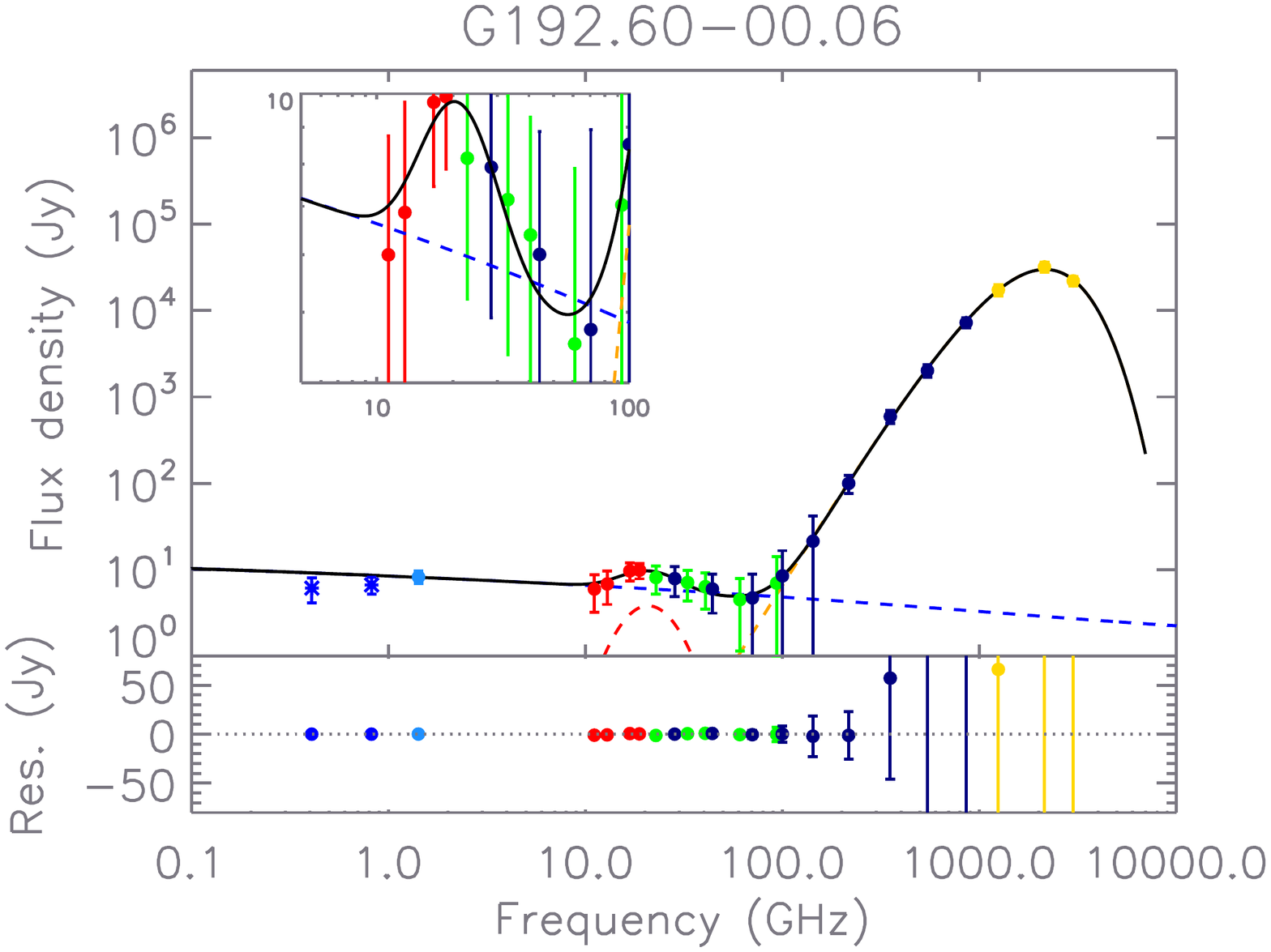}
\vspace*{-4.5cm}
\includegraphics[width=77mm,angle=0]{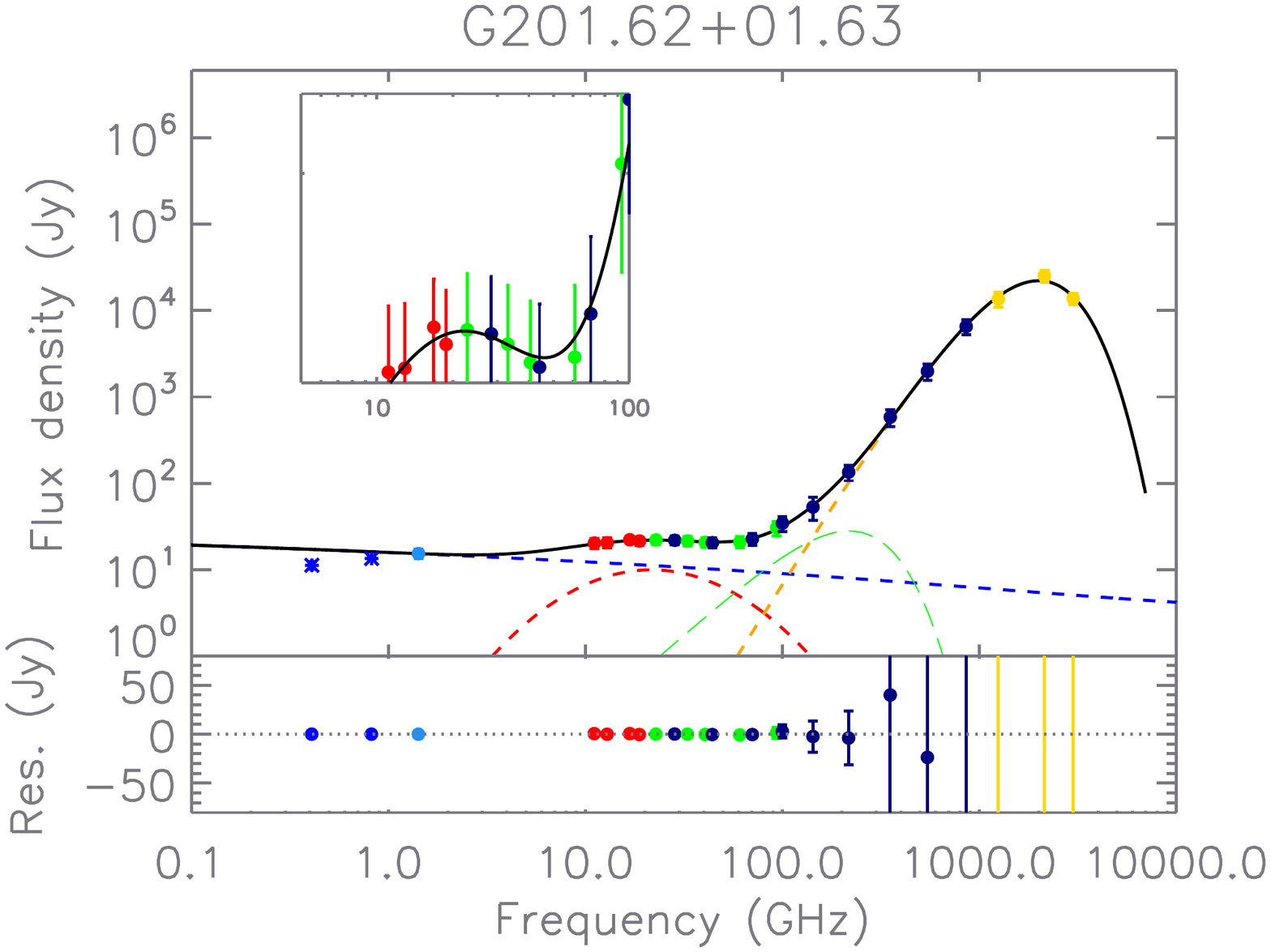}
\hspace*{10mm}
\includegraphics[width=77mm,angle=0]{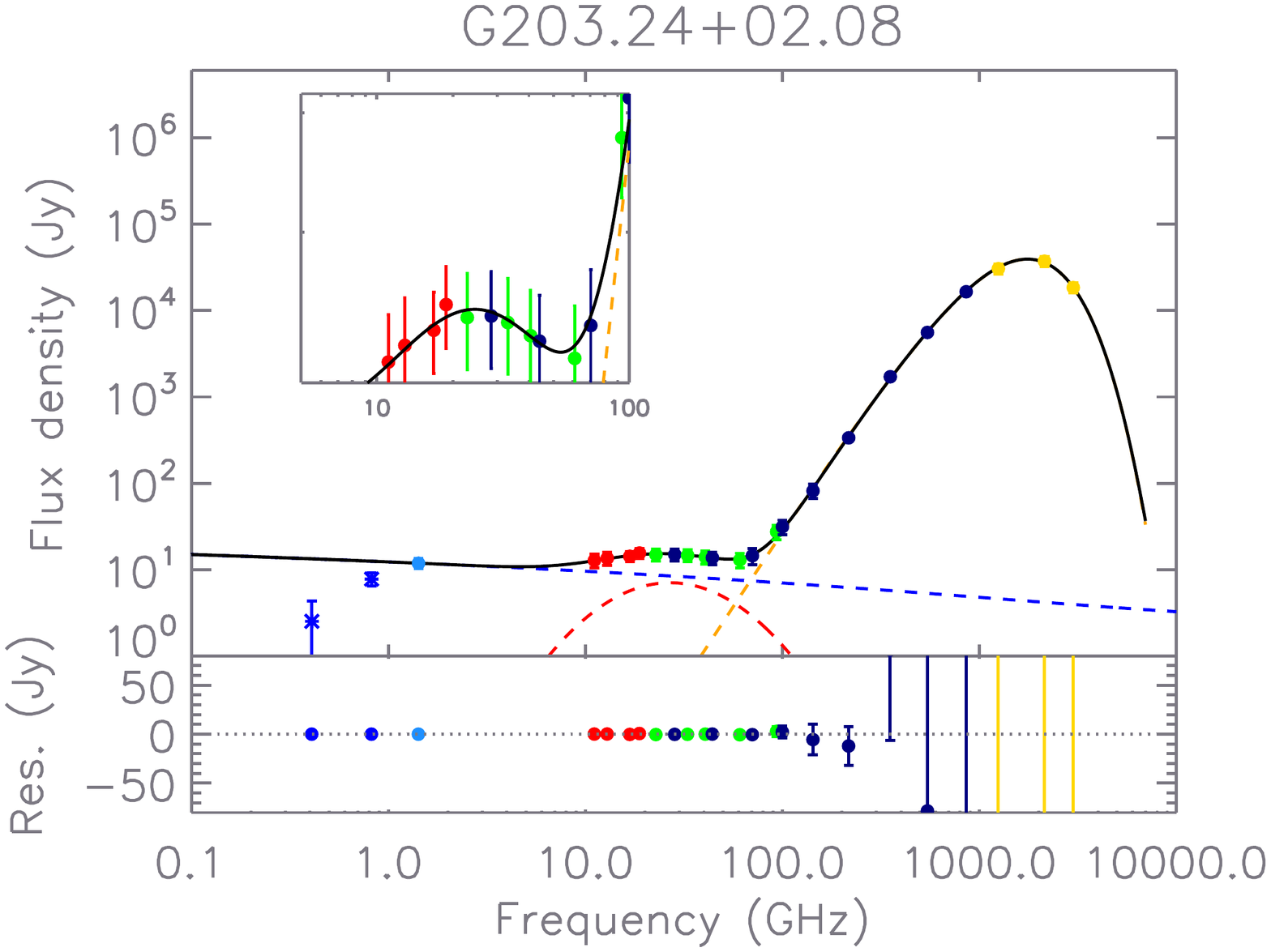}
\vspace*{-4.5cm}
\includegraphics[width=77mm,angle=0]{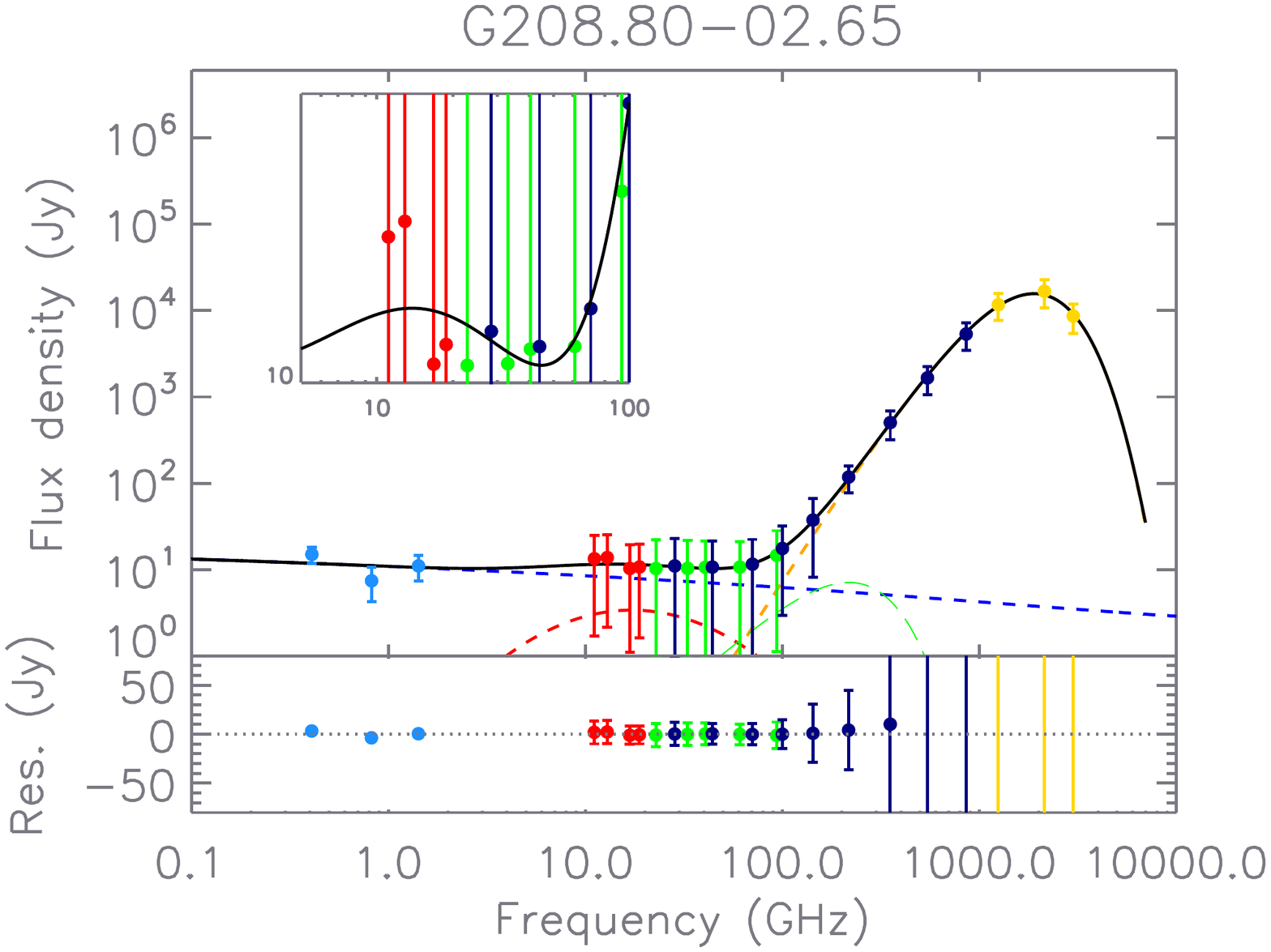}
\hspace*{10mm}
\includegraphics[width=77mm,angle=0]{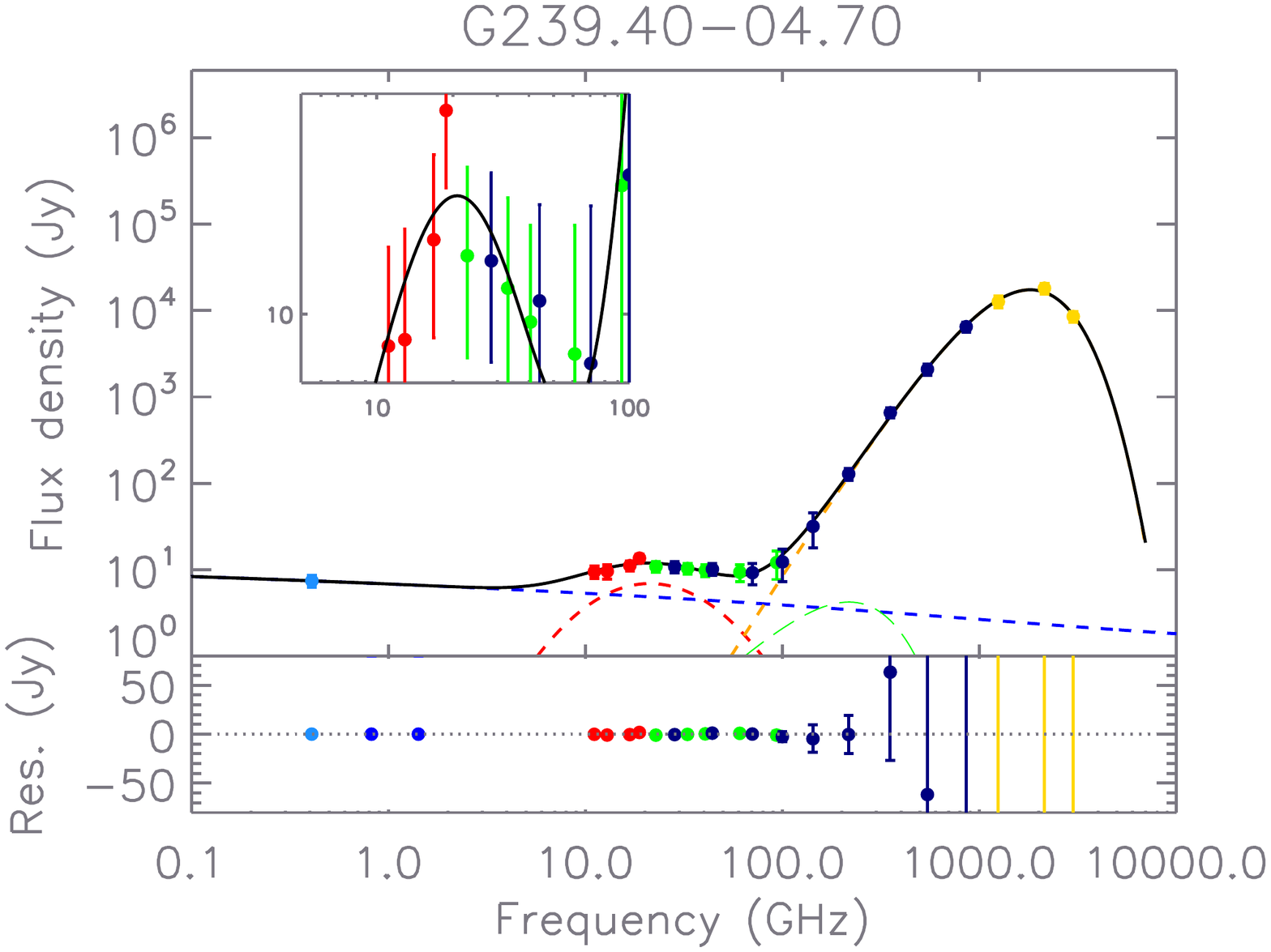}
\vspace*{1cm}
\caption{Same as Figure~\ref{fig:sed_int1}.}
\label{fig:sed_int6}
\end{center}
\end{figure*}

\begin{figure*}
\begin{center}
\vspace*{0mm}
\centering
\clearpage
\includegraphics[width=77mm,angle=0]{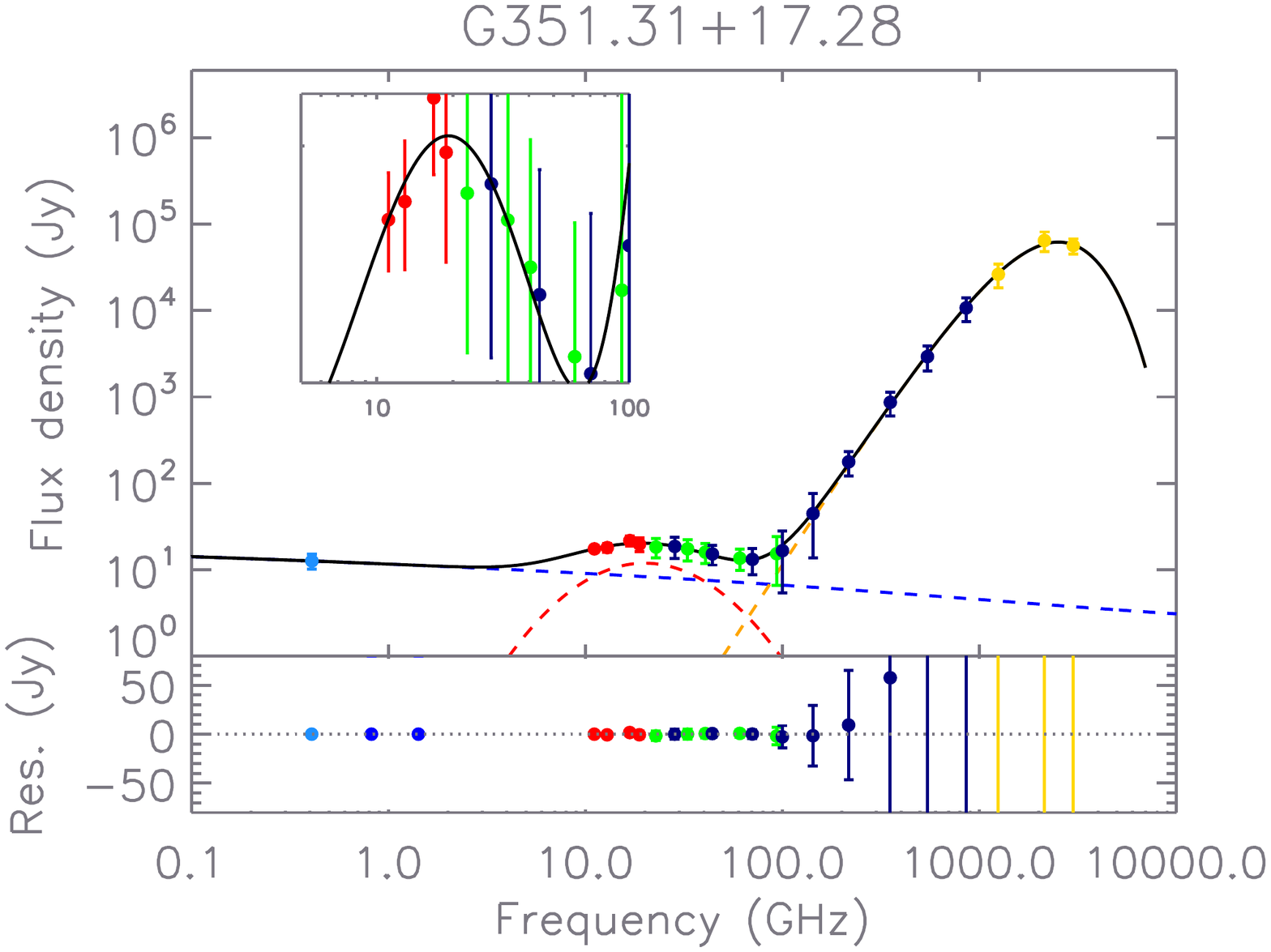}
\vspace*{-4.5cm}
\hspace*{10mm}
\includegraphics[width=77mm,angle=0]{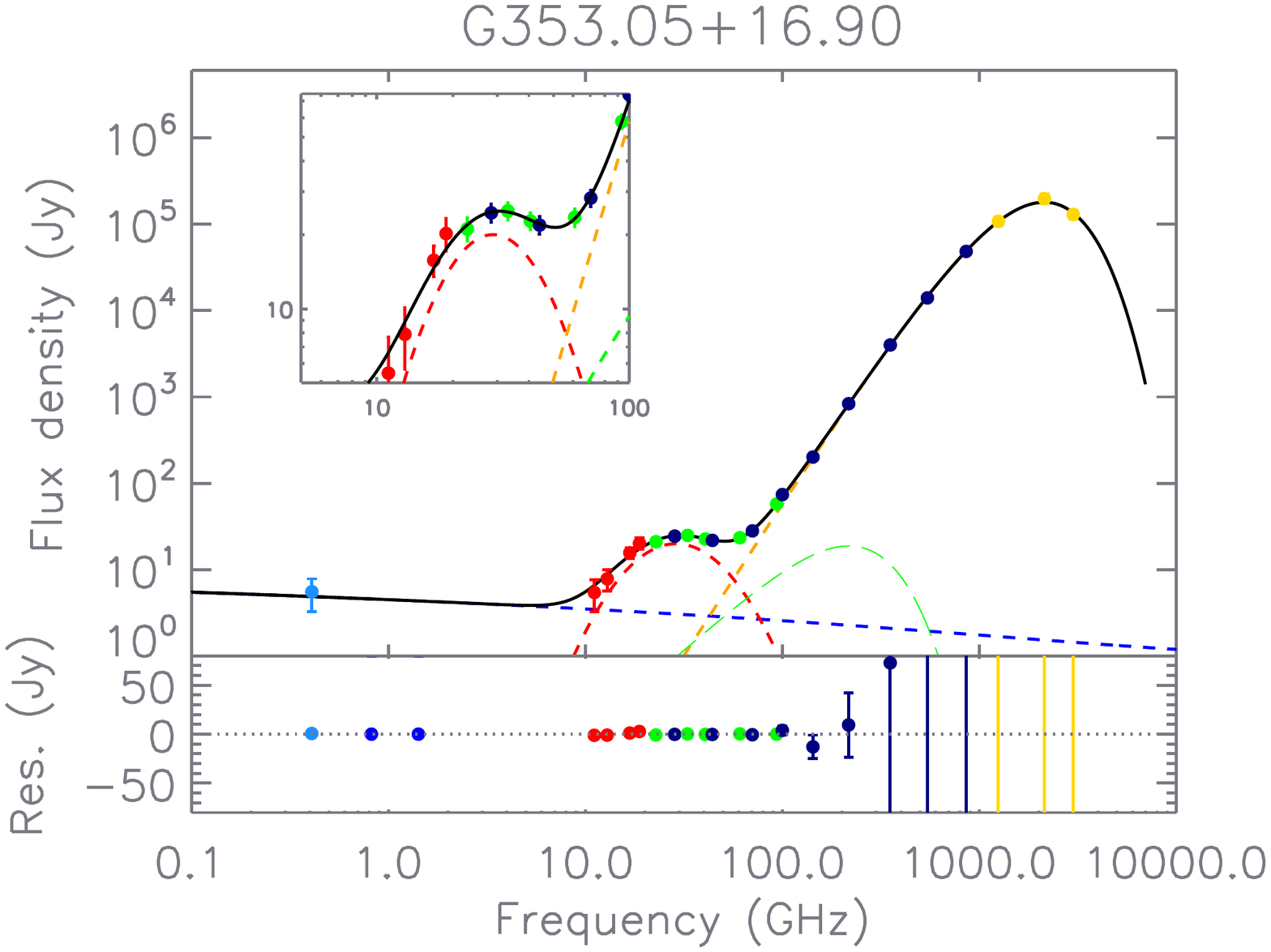}
\vspace*{-4.5cm}
\includegraphics[width=77mm,angle=0]{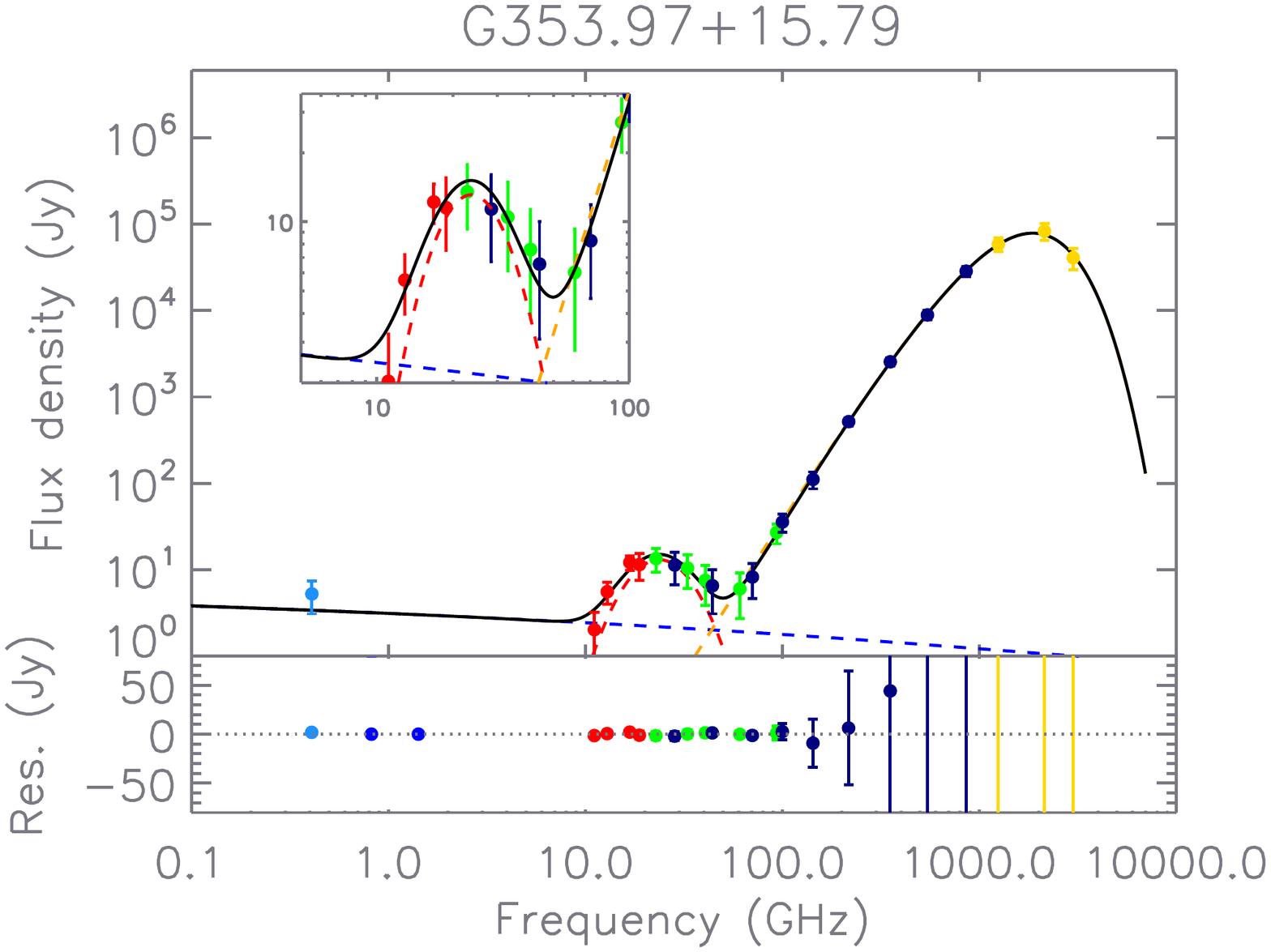}
\hspace*{10mm}
\includegraphics[width=77mm,angle=0]{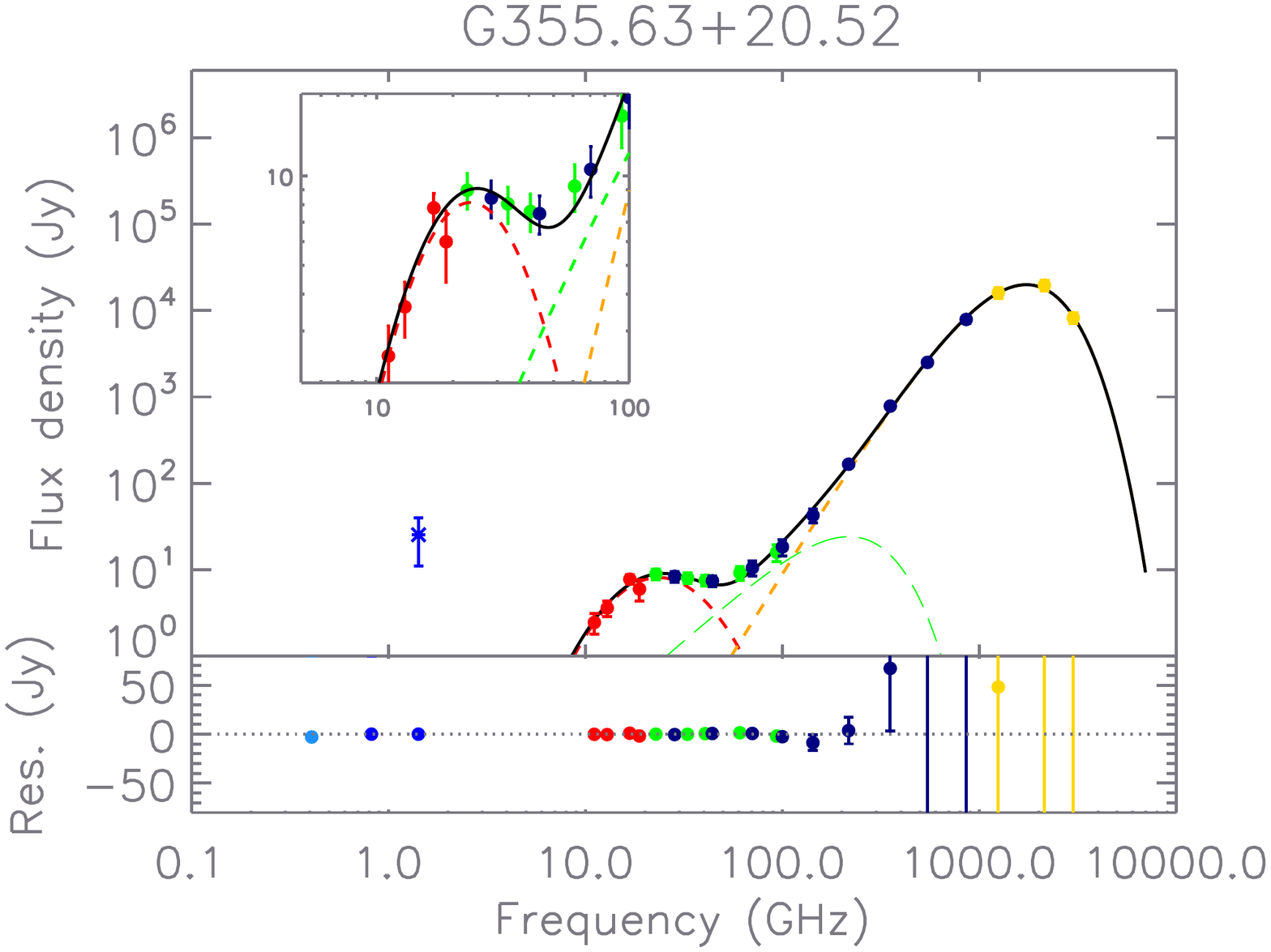}
\vspace*{1cm}
\caption{Same as Figure~\ref{fig:sed_int1}.}
\label{fig:sed_int7}
\end{center}
\end{figure*}

\section{Additional TABLES}  \label{append_tables}

All the SRCCs obtained between the several parameters used to model the free-free, AME, and thermal dust components are displayed in Tables D1--D5. In additional to several plots used to study correlations between some of the parameters, these tables have been used to systematically identify the parameters showing the most meaningful correlation factors, and guide our study.

\input{polametex_table_sources_spearman_AME_AME.txt}

\input{polametex_table_sources_spearman_AME_TD.txt}

\input{polametex_table_sources_spearman_AME_FF.txt}

\input{polametex_table_sources_spearman_TD_FF.txt}

\input{polametex_table_sources_spearman_TD_TD.txt}


\bsp	
\label{lastpage}
\end{document}

%% file: polame_authors_and_institutes.tex
\author[F. Poidevin et al.]{F. Poidevin$^{1,2}$,\thanks{E-mail: fpoidevin@iac.es}
R.T. G\'{e}nova-Santos$^{1,2}$,\thanks{E-mail: rgs@iac.es}, 
J.A. Rubi{\~n}o-Mart{\'{\i}}n$^{1,2}$,
C.~H. L\'opez-Caraballo,$^{1,2}$
\newauthor
R. A. Watson$^{3}$,
E. Artal,$^{4}$
M. Ashdown,$^{5,6}$
R.~B. Barreiro,$^{7}$
F.~J. Casas,$^{7}$
\newauthor
E. de la Hoz,$^{7,8}$
M. Fern\'andez-Torreiro,$^{1,2}$
F. Guidi,$^{1,2,9}$
D. Herranz,$^{7}$
R.~J. Hoyland,$^{1,2}$
\newauthor
A.~N. Lasenby,$^{5,6}$
E. Martinez-Gonzalez,$^{7}$
M.~W. Peel,$^{1,2}$
L. Piccirillo,$^{3}$
R. Rebolo,$^{1,2,10}$
\newauthor
B. Ruiz-Granados,$^{1,2,11}$
D. Tramonte,$^{12,13,1,2}$
F. Vansyngel$^{1,2}$
P. Vielva,$^{7}$
\\
$^{1}$Instituto de Astrof\'{\i}sica de Canarias, E-38205 La Laguna, Tenerife, Spain\\
$^{2}$Departamento de Astrof\'{\i}sica, Universidad de La Laguna,
E-38206 La Laguna, Tenerife, Spain\\
$^{3}$Jodrell Bank Centre for Astrophysics, Alan Turing Building, 
Department of Physics and Astronomy, School of Natural Sciences, The 
University of Manchester,\\ Oxford Road, Manchester M13 9PL, Manchester, UK\\
$^{4}$Departamento de Ingenieria de COMunicaciones (DICOM),
Laboratorios de I+D de Telecomunicaciones, Plaza de la Ciencia s/n, E-39005 Santander, Spain\\ 
$^{5}$Astrophysics Group, Cavendish Laboratory, University of Cambridge, J J Thomson Avenue, Cambridge CB3 0HE, UK\\
$^{6}$Kavli Institute for Cosmology, University of Cambridge, Madingley Road, Cambridge CB3 0HA, UK\\
$^{7}$Instituto de F\'{\i}sica de Cantabria (IFCA), CSIC-Univ. de Cantabria, Avda. los Castros, s/n, E-39005 Santander, Spain\\
$^{8}$Departamento de F\'{\i}sica Moderna, Universidad de Cantabria,
Avda. de los Castros s/n, 39005 Santander, Spain\\
$^{9}$Institut d'Astrophysique de Paris, UMR 7095, CNRS \& Sorbonne Universit\'e, 98 bis boulevard Arago, 75014 Paris, France\\
$^{10}$Consejo Superior de Investigaciones Cientificas, E-28006 Madrid, Spain\\
$^{11}$Departamento de F\'{\i}sica. Facultad de Ciencias. Universidad de C\'ordoba.  Campus de Rabanales, Edif. C2. Planta Baja.  E-14071 C\'ordoba, Spain\\
$^{12}$Purple Mountain Observatory, CAS, No.10 Yuanhua Road, Qixia District, Nanjing 210034, China \\
$^{13}$NAOC-UKZN Computational Astrophysics Center (NUCAC), University of Kwazulu-Natal, Durban 4000, South Africa
}

%% file: polametex_table_sources_2.txt
\begin{table*}
\begin{center}
\begin{tabular}{lccllcll}
\hline\hline
\noalign{\smallskip}
Source Name & Glon  & Glat   &Region Type & Other Name$^{(a)}$ & References & $\sigma_{\rm AME}$ & $\sigma_{\rm AME}$ \\
& $[^{\circ}]$ & $[^{\circ}]$ & &&& PIRXV & This Work \\
\noalign{\smallskip}
\hline
\noalign{\smallskip}
 G010.19$-$00.32  &  10.19  &  $-$0.32  &                           SNR  &          Kes62. Synch. SNR9.9$-$0.8  &                   1 &\textbf{3.4}$^{S}$& 2.6$^{SS}$\\
 G010.84$-$02.59  &  10.84  &  $-$2.59  &                            MC  &                                            GGD 27  &                   2 &...& 4.8$^{SS}$\\
 G011.11$-$00.12  &  10.60  &  $-$0.12  &                            MC  &                                      G011.11$-$0.12  &                   2 &...& 2.5$^{SS}$\\
 G012.80$-$00.19  &  12.80  &  $-$0.19  &                           SNR  &                                               W33  &                   1 &\textbf{2.7}& 1.2$^{LD}$\\
 G015.06$-$00.69  &  15.06  &  $-$0.69  &                            MC  &                                               M17  &                   1 &1.9& 8.0$^{S \rightarrow SS}$\\
 G017.00+00.85  &  17.00  &   0.85  &                            MC  &                                               M16  &                 1,2 &5.3& 6.0$^{S}$\\
 G037.79$-$00.11  &  37.79  &  $-$0.11  &                           SNR  &                                               W47  &                   1 &3.4& 7.6$^{S \rightarrow SS}$\\
 G040.52+02.53  &  40.52  &   2.53  &                        MC/HII  &                                               W45  &                   1 &0.2& 12.9$^{S \rightarrow SS}$\\
 G041.03$-$00.07  &  41.03  &  $-$0.07  &                            MC  &                                 SDC G41.003$-$0.097  &                   4 &...& 7.9$^{S}$\\
 G043.20$-$00.10  &  43.20  &  $-$0.10  &                            MC  &                                               W49  &                 1,3 &5.3&  8.3$^{S \rightarrow SS}$ \\
 G045.47+00.06  &  45.47  &   0.06  &                           SNR  &                                           NRAO601  &                   1 &5.9& 15.6$^{S \rightarrow SS}$\\
 G049.14$-$00.60  &  49.14  &  $-$0.60  &                        MC/HII  &                                               W51  &                   2 &...& 22.9$^{S \rightarrow SS}$\\
 G059.42$-$00.21  &  59.42  &  $-$0.21  &                        MC/HII  &                                               W55  &                   1 &7.0& 8.7$^{S \rightarrow SS}$\\
 G061.47+00.11  &  61.47  &   0.11  &                        MC/HII  &                       HII LBN061.50+00.29. SH2$-$88  &                   1 &1.9& 4.1$^{SS}$\\
 G062.98+00.05  &  62.98  &   0.05  &                            MC  &                                               S89  &                   1 &\textbf{7.5}& 6.1$^{S \rightarrow SS}$\\
 G070.14+01.61  &  70.14  &   1.61  &                       Cluster  &                                          NGC 6857  &                   4 &...&3.1$^{BD}$\\
 G071.59+02.85  &  71.59  &   2.85  &                        MC/HII  &                                              s101  &                   1 &1.8& 4.8$^{SS}$\\
 G075.81+00.39  &  75.81  &   0.39  &                        MC/HII  &              HII GAL075.84+00.40. SH2$-$105. Cyg 2N  &                   1 &2.5& 5.9$^{S \rightarrow SS}$\\
 G076.38$-$00.62  &  76.38  &  $-$0.62  &                        MC/HII  &                                              S106  &                 1,3 &...& 3.9$^{BD}$\\
 G078.57+01.00  &  78.57  &   1.00  &                        MC/HII  &                                           LDN 889  &                 2,3 &...& ...$^{BD}$\\
 G081.59+00.01  &  81.59  &   0.01  &                        MC/HII  &                                         DR23/DR21  &                 1,2 &1.3& 17.9$^{S}$\\
 G084.68$-$00.58  &  84.68  &  $-$0.58  &                            MC  &                                      DOBASHI 2732  &                   4 &...& 18.8$^{S}$\\
 G085.00+04.20  &  84.90  &   3.80  &                        MC/HII  &                                  LBN 084.97+04.21  &                   4 &...& 21.1$^{S}$\\
 G093.02+02.76  &  93.02  &   2.76  &                        MC/HII  &                                HII GAL093.06+2.81  &                   1 &1.6& 21.0$^{S \rightarrow SS}$\\
 G094.47$-$01.53  &  94.47  &  $-$1.53  &                        MC/HII  &                                          LDN 1059  &                   1 &0.6& 4.1$^{SS}$\\
 G098.00+01.47  &  98.00  &   1.47  &                        MC/HII  &                          RNe GM1-12, DNe TGU H582  &                   1 &6.1& 17.2$^{S \rightarrow SS}$\\
 G099.60+03.70  &  99.60  &   3.70  &                            MC  &                                           LDN1111  &                   1 &0.6& 3.0$^{SS}$\\
 G102.88$-$00.69  & 102.88  &  $-$0.69  &                        MC/HII  &                                      LDN1161/1163  &                   1 &2.5& 10.9$^{S}$\\
 G107.20+05.20  & 107.20  &   5.20  &                            MC  &                                              S140  &                 1,2 &9.9& 27.8$^{S \rightarrow SS}$\\
 G110.25+02.58  & 110.25  &   2.58  &                        MC/HII  &                  HII G110.2+02.5. LBN110.11+02.44  &                   1 &\textbf{3.4}& 2.7$^{SS}$\\
 G111.54+00.81  & 111.54  &   0.81  &                  Open Cluster  &                                          NGC 7538  &                   2 &...& 10.8$^{S \rightarrow SS}$\\
 G118.09+04.96  & 118.09  &   4.96  &                           SNR  &                                          NGC 7822  &                   1 &...& 14.2$^{S}$\\
 G123.13$-$06.27  & 123.13  &  $-$6.27  &                        MC/HII  &                                              S184  &                   2 &...& 25.2$^{S \rightarrow SS}$\\
 G133.27+09.05  & 133.27  &   9.05  &                            MC  &                                LDN 1358/1355/1357  &                   1 &8.5$^{S}$& 11.1$^{BD}$\\
 G133.74+01.22  & 133.74  &   1.22  &                            MC  &                                                W3  &                   1 &1.5& 24.8$^{S \rightarrow SS}$\\
 G142.35+01.35  & 142.35  &   1.35  &                            MC  &                        DNe TGU H942, DOBASHI 3984  &                   1 &\textbf{9.5}$^{S}$& 8.4$^{S}$\\
 G151.62$-$00.28  & 151.62  &  $-$0.28  &                        MC/HII  &                                       HII SH2$-$209  &                   1 &1.5& 11.4$^{S \rightarrow SS}$\\
 G160.26-18.62  & 160.26  & $-$18.62  &                            MC  &                                           Perseus  &                 1,2 &17.4$^{S}$& 19.2$^{S}$\\
 G160.60$-$12.05  & 160.60  & $-$12.05  &                            MC  &                      NGC 1499 (California nebula)  &                   1 &5.1$^{S}$& 12.6$^{S}$\\
 G173.56$-$01.76  & 173.56  &  $-$1.76  &                  Open Cluster  &                                          NGC 1893  &                   1 &0.8& 4.4$^{SS}$\\
 G173.62+02.79  & 173.62  &   2.79  &                       Cluster  &                                              S235  &                   1 &5.6& 15.5$^{S \rightarrow SS}$\\
 G190.00+00.46  & 190.00  &   0.46  &                        MC/HII  &                                     NGC 2174/2175  &                   1 &7.4& 29.3$^{S \rightarrow SS}$\\
 G192.34$-$11.37  & 192.34  & $-$11.37  &                            MC  &                                     LDN 1582/1584  &                   1 &12.3$^{S}$& 12.5$^{BD}$\\
 G192.60$-$00.06  & 192.60  &  $-$0.06  &                       Cluster  &                                              S255  &                   1 &4.3& 7.9$^{S \rightarrow SS}$\\
 G201.62+01.63  & 201.62  &   1.63  &                            MC  &                                     LDN 1608/1609  &                   1 &7.4$^{S}$& 27.3$^{S}$\\
 G203.24+02.08  & 203.24  &   2.08  &                        MC/HII  &                                          LDN 1613  &           1,2 &8.3$^{S}$& 15.8$^{S}$\\
 G208.80$-$02.65  & 208.80  &  $-$2.65  &                        MC/HII  &                                     S280--LBN 970  &                   1 &\textbf{2.0}& 1.9$^{LD}$\\
 G239.40$-$04.70  & 239.40  &  $-$4.70  &                            MC  &                   LDN 1667, HII LBN1059, V VY Cma  &             1 &9.9$^{S}$& 16.5$^{S}$\\
 G351.31+17.28  & 351.31  &  17.28  &                        MC/HII  &                                  HII LBN1105/1104  &                       1 & 5.3$^{S}$& 32.9$^{S}$\\
 G353.05+16.90  & 353.05  &  16.90  &                            MC  &                  Rho Ophiuchi, AME-G353.05+16.901  &                 1,3 &\textbf{29.8}$^{S}$& 27.3$^{S \rightarrow SS}$\\
 G353.97+15.79  & 353.97  &  15.79  &                            MC  &                                      In Ophiuchus  &                   1 &\textbf{10.9}$^{S}$& 10.6$^{S}$\\
 G355.63+20.52  & 355.63  &  20.52  &                            MC  &                                  In Rho Ophiuchus  &                   1 &13.3$^{S}$& 17.0$^{BD}$\\
\noalign{\smallskip}
\hline\hline
\end{tabular}
\end{center}
\normalsize
\caption{List of sources. References: 1: \citet{pir15} (PIRXV), 2: \citet{matthews2009}, 3: \citet{crutcher1999}, 4: \citet{lee2016}.
Note: $^{(a)}$ information retrieved from the Simbad database ({\tt http://simbad.u-strasbg.fr/simbad//}). Sources such that $\sigma_{\rm AME}$ from PIRXV are greater than $\sigma_{\rm AME}$ from this work are shown in bold. 
Superscript symbols in last two columns are $^{S}$ for ``significant'' AME detection, $^{SS}$ for ``semi-significant'' AME detection, $^{S \rightarrow SS}$ for ``significant'' AME detection reclassified as ``semi-significant'' AME detection (see text for details), $^{LD}$ for low detection of AME and, $^{BD}$ for bad detection because of a bad fit of the AME, of the free-free or of the thermal dust component. See section~\ref{robustness} for details.}
\label{tab:listofclouds}
\end{table*}

%% file: polametex_sedfitparams_table_dr21.txt
\begin{table*}
\begin{center}
\begin{tabular}{lccccccccc}
\hline\hline
\noalign{\smallskip}
 DR23/DR21 & $A_{\rm AME}$& $\sigma_{\rm AME}$ &   $\nu_{\rm AME}$ &   $W_{\rm AME}$  & $\Delta T_{\rm CMB}$ & $A_{\rm AME}$ priors & $\nu_{\rm AME}$ priors & $W_{\rm AME}$ priors  & $\chi^2_{\rm red}$\\
\noalign{\smallskip}
\hline
 [G081.59+00.01]        &  [Jy]              &                    &  [GHz]                 &      &       [Jy] &       [Jy]  &  [GHz]    &       &   \\
\noalign{\smallskip}
\hline
\noalign{\smallskip}
See plot on Figure~\ref{fig:dr21}, left  &  99.4 $\pm$ 5.7 & 17.4  &   36.8 $\pm$40.5  &   1.8 $\pm$ 1.2    &  -1.2 $\pm$   69.6  &     [0 , 300] &     [10 , 60] &     [0.2 , 2.5]  &   0.18   \\ 
See plot on Figure~\ref{fig:dr21}, right  &  94.0 $\pm$ 5.2 & 18.1  &   26.3 $\pm$ 3.6  &       1.0  &  125.0 &     [0 , 300] &     [10 , 60]   &      [0.2 , 1.0]  &     0.20\\  
\noalign{\smallskip}
\hline\hline
\end{tabular}
\end{center}
\normalsize
\caption{\small Fit parameters of the AME and CMB components obtained with different priors on the AME width parameter, $W_{\rm AME}$. Note that in the case of stronger priors the best-fit values for $W_{\rm AME}$ and $\Delta T_{\rm CMB}$ are found in the border of the prior.
The corresponding plots are shown in Figure~\ref{fig:dr21}.
}
\label{tab:seds_fit_parameters_dr21}
\end{table*}

%% file: polametex_table_sources_spearman_DUST_12_25_60_100.txt
\begin{table*}
\begin{center}
\begin{tabular}{lrrr}
\hline\hline
\noalign{\smallskip}
 Wavelentgh & SRCC & SRCC & SRCC \\
     & selected sample & AME significant & AME semi-significant \\
\noalign{\smallskip}
\hline
\noalign{\smallskip}
                             100$\,\mu$m  &   0.87$\pm$ 0.04  (   0.84$\pm$ 0.05  )&   0.86$\pm$ 0.02  (   0.65$\pm$ 0.08 )&   0.89$\pm$ 0.03  (   0.88$\pm$ 0.05)\\
                              60$\,\mu$m  &   0.84$\pm$ 0.04  (   0.86$\pm$ 0.05  )&   0.82$\pm$ 0.03  (   0.65$\pm$ 0.08 )&   0.88$\pm$ 0.03  (   0.90$\pm$ 0.05)\\
                              25$\,\mu$m  &   0.85$\pm$ 0.04  (   0.81$\pm$ 0.05  )&   0.65$\pm$ 0.03  (   0.43$\pm$ 0.07 )&   0.90$\pm$ 0.03  (   0.90$\pm$ 0.05)\\
                              12$\,\mu$m  &   0.80$\pm$ 0.04  (   0.70$\pm$ 0.05  )&   0.39$\pm$ 0.04  (   0.19$\pm$ 0.07 )&   0.89$\pm$ 0.03  (   0.85$\pm$ 0.06)\\
\noalign{\smallskip}
\hline\hline
\end{tabular}
\end{center}
\normalsize
\caption{Spearman rank correlation coefficients (SRCCs) between the AME maximum flux densities
and the IR/submm flux densities. The values displayed between parentheses are the SRCCs obtained
once the IR/submm flux densities are divided by the IRSFs estimates G$_{0}$.}
\label{tab:srcc_dust_12_25_60_100}
\end{table*}

%% file: quijote_acknow.tex
We thank the staff of the Teide Observatory for invaluable assistance in the commissioning and operation of QUIJOTE.
The {\it QUIJOTE} experiment is being developed by the Instituto de Astrofisica de Canarias (IAC),
the Instituto de Fisica de Cantabria (IFCA), and the Universities of Cantabria, Manchester and Cambridge.
Partial financial support was provided by the Spanish Ministry of Science and Innovation 
under the projects AYA2007-68058-C03-01, AYA2007-68058-C03-02,
AYA2010-21766-C03-01, AYA2010-21766-C03-02, AYA2014-60438-P,
ESP2015-70646-C2-1-R, AYA2017-84185-P, ESP2017-83921-C2-1-R,
AYA2017-90675-REDC (co-funded with EU FEDER funds),
PGC2018-101814-B-I00, 
PID2019-110610RB-C21, PID2020-120514GB-I00, IACA13-3E-2336, IACA15-BE-3707, EQC2018-004918-P, the Severo Ochoa Programs SEV-2015-0548 and CEX2019-000920-S, the
Maria de Maeztu Program MDM-2017-0765, and by the Consolider-Ingenio project CSD2010-00064 (EPI: Exploring
the Physics of Inflation). We acknowledge support from the ACIISI, Consejeria de Economia, Conocimiento y 
Empleo del Gobierno de Canarias and the European Regional Development Fund (ERDF) under grant with reference ProID2020010108.
This project has received funding from the European Union's Horizon 2020 research and innovation program under
grant agreement number 687312 (RADIOFOREGROUNDS).

%% file: polametex_sedfitparams_tables_1_cmb.txt
%

\begin{table*}
\begin{center}
\begin{tabular}{llllllll}
\hline\hline
\noalign{\smallskip}
Source Name     &$\alpha_{\rm synch, int}$   &EM               &$\tau_{250} $    &$T_{\rm dust}$ &$\beta_{\rm dust}$ & G$_{0}$ & $\Delta T_{\rm CMB}$\\
\noalign{\smallskip}
\hline
                &                            & [cm$^{-6}$ pc]  &[$\times 10^{5}$] & [K]           &                  &     & [$\mu$K]    \\
 \\
\noalign{\smallskip}
\hline
\noalign{\smallskip}
  G010.19-00.32  & -0.41 $\pm$  0.27  &    1969.0 $\pm$      86.8  & 307.0 $\pm$  35.2  &  20.6 $\pm$   0.6  & 1.90 $\pm$ 0.06  & 2.68 $\pm$ 0.48  &  125.0 $^{(\rm PUL)}$ \\
  G010.84-02.59  &  ...               &     244.2 $\pm$      17.1  &  29.8 $\pm$   3.3  &  20.7 $\pm$   0.6  & 1.68 $\pm$ 0.06  & 2.73 $\pm$ 0.46  &    7.3 $\pm$   20.5 \\
  G011.11-00.12  &  ...               &    2671.4 $\pm$     111.7  & 312.3 $\pm$  38.6  &  20.6 $\pm$   0.7  & 1.88 $\pm$ 0.07  & 2.67 $\pm$ 0.51  &  -28.2 $\pm$  115.6 \\
  G012.80-00.19  & -0.22 $\pm$  0.05  &    1328.8 $\pm$     118.4  & 413.0 $\pm$  41.1  &  19.9 $\pm$   0.5  & 1.92 $\pm$ 0.05  & 2.13 $\pm$ 0.32  &  125.0 $^{(\rm PUL)}$ \\
  G015.06-00.69  &  ...               &    7265.6 $\pm$      63.8  & 193.1 $\pm$  16.6  &  23.1 $\pm$   0.5  & 1.79 $\pm$ 0.04  & 5.23 $\pm$ 0.74  &  125.0 $^{(\rm PUL)}$ \\
  G017.00+00.85  &  ...               &     399.9 $\pm$      72.9  & 173.7 $\pm$  34.3  &  19.4 $\pm$   0.9  & 1.89 $\pm$ 0.12  & 1.86 $\pm$ 0.53  &  -74.7 $\pm$  151.9 \\
  G037.79-00.11  & -1.33 $\pm$  0.51  &    1812.5 $\pm$      55.6  & 237.8 $\pm$  28.9  &  20.2 $\pm$   0.6  & 1.81 $\pm$ 0.07  & 2.38 $\pm$ 0.44  &   69.5 $\pm$   72.9 \\
  G040.52+02.53  & -0.62 $\pm$  0.03  &       6.9 $\pm$       3.8  &  25.5 $\pm$   2.1  &  20.2 $\pm$   0.4  & 1.75 $\pm$ 0.05  & 2.34 $\pm$ 0.28  &   69.3 $\pm$   10.3 \\
  G041.03-00.07  & -0.80 $\pm$  0.14  &     534.3 $\pm$      34.8  & 163.6 $\pm$  17.1  &  19.1 $\pm$   0.5  & 1.83 $\pm$ 0.06  & 1.71 $\pm$ 0.26  &  125.0 $^{(\rm PUL)}$ \\
  G043.20-00.10  &  ...               &    1442.5 $\pm$      40.2  & 151.6 $\pm$  19.0  &  20.4 $\pm$   0.7  & 1.74 $\pm$ 0.07  & 2.53 $\pm$ 0.48  &   67.5 $\pm$   62.3 \\
  G045.47+00.06  & -1.37 $\pm$  0.18  &     383.9 $\pm$      24.0  & 113.3 $\pm$  10.1  &  20.3 $\pm$   0.5  & 1.71 $\pm$ 0.05  & 2.41 $\pm$ 0.33  &  125.0 $^{(\rm PUL)}$    \\
  G049.14-00.60  &  ...               &    3479.9 $\pm$      35.5  &  81.1 $\pm$   8.5  &  24.3 $\pm$   0.8  & 1.72 $\pm$ 0.06  & 7.22 $\pm$ 1.34  &   42.2 $\pm$   38.1 \\
  G059.42-00.21  &  ...               &     286.9 $\pm$      18.4  &  55.4 $\pm$  11.0  &  19.2 $\pm$   0.9  & 1.77 $\pm$ 0.12  & 1.72 $\pm$ 0.48  &  -60.4 $\pm$   29.1 \\
  G061.47+00.11  &  ...               &     311.4 $\pm$       8.1  &  23.4 $\pm$   2.0  &  21.9 $\pm$   0.5  & 1.57 $\pm$ 0.05  & 3.81 $\pm$ 0.51  &    8.0 $\pm$   10.6 \\
  G062.98+00.05  &  ...               &     243.1 $\pm$       8.2  &  27.2 $\pm$   2.7  &  20.3 $\pm$   0.5  & 1.66 $\pm$ 0.06  & 2.42 $\pm$ 0.37  &    5.2 $\pm$   11.4 \\
  G070.14+01.61  &  ...               &     272.4 $\pm$      26.6  & ...  &  10.0 $^{(\rm PLL)}$  & 2.11 $\pm$ 2.99  & 0.03 $^{(\rm PLL)}$  &   65.8 $\pm$   54.9 \\
  G071.59+02.85  &  ...               &     215.4 $\pm$      17.1  &  15.0 $\pm$   4.3  &  21.5 $\pm$   1.6  & 1.57 $\pm$ 0.18  & 3.42 $\pm$ 1.54  &  -12.8 $\pm$   32.9 \\
  G075.81+00.39  &  ...               &     324.7 $\pm$      12.5  &  39.9 $\pm$   3.4  &  19.8 $\pm$   0.4  & 1.65 $\pm$ 0.05  & 2.09 $\pm$ 0.27  &  125.0 $^{(\rm PUL)}$ \\
  G076.38-00.62  &  ...               &       ...                  &  11.3 $\pm$   2.7  &  24.3 $\pm$   1.7  & 1.78 $\pm$ 0.17  & 7.09 $\pm$ 2.93  &   56.5 $\pm$   38.5 \\
  G078.57+01.00  &  ...               &    1835.0 $\pm$      74.2  &  32.4 $\pm$   6.2  &  23.9 $\pm$   1.3  & 1.46 $\pm$ 0.09  & 6.53 $\pm$ 2.19  & -125.0 $^{(\rm PLL)}$ \\
  G081.59+00.01  &  ...               &    2541.5 $\pm$      44.1  &  60.9 $\pm$   5.6  &  25.3 $\pm$   0.7  & 1.56 $\pm$ 0.04  & 9.17 $\pm$ 1.54  &  125.0 $^{(\rm PUL)}$ \\
  G084.68-00.58  &  ...               &    1201.3 $\pm$      15.4  &  35.7 $\pm$   2.6  &  21.9 $\pm$   0.4  & 1.52 $\pm$ 0.04  & 3.81 $\pm$ 0.45  &   -1.3 $\pm$   23.2 \\
  G085.00+04.20  &  ...               &     435.2 $\pm$       6.6  &   9.3 $\pm$   1.1  &  23.1 $\pm$   0.8  & 1.74 $\pm$ 0.09  & 5.30 $\pm$ 1.10  &   86.2 $\pm$   11.0 \\
  G093.02+02.76  &  ...               &     497.2 $\pm$       6.6  &  24.0 $\pm$   1.6  &  21.5 $\pm$   0.4  & 1.47 $\pm$ 0.04  & 3.46 $\pm$ 0.37  &  -19.6 $\pm$    9.7 \\
  G094.47-01.53  &  ...               &     257.0 $\pm$       3.7  &   4.3 $\pm$   0.8  &  23.2 $\pm$   1.2  & 1.57 $\pm$ 0.13  & 5.36 $\pm$ 1.64  &   -6.1 $\pm$    5.6 \\
  G098.00+01.47  &  ...               &     104.1 $\pm$       3.5  &   6.2 $\pm$   0.6  &  21.5 $\pm$   0.5  & 1.46 $\pm$ 0.05  & 3.43 $\pm$ 0.50  &  -20.9 $\pm$    6.4 \\
  G099.60+03.70  &  ...               &     506.0 $\pm$       7.6  &   3.1 $\pm$   0.8  &  25.9 $\pm$   2.0  & 1.67 $\pm$ 0.17  &10.45 $\pm$ 4.80  &  -34.5 $\pm$    9.1 \\
  G102.88-00.69  &  ...               &     332.9 $\pm$       5.9  &   8.9 $\pm$   1.2  &  23.4 $\pm$   0.9  & 1.41 $\pm$ 0.08  & 5.66 $\pm$ 1.25  &  -83.2 $\pm$   11.7 \\
  G107.20+05.20  &  ...               &     216.5 $\pm$       5.6  &  19.4 $\pm$   2.3  &  22.8 $\pm$   0.8  & 1.56 $\pm$ 0.06  & 4.88 $\pm$ 0.97  &   -1.7 $\pm$   11.8 \\
  G110.25+02.58  &  ...               &     271.0 $\pm$      12.9  &  10.8 $\pm$   2.4  &  25.4 $\pm$   1.7  & 1.52 $\pm$ 0.13  & 9.26 $\pm$ 3.81  &  -52.8 $\pm$   30.4 \\
  G111.54+00.81  &  ...               &     284.8 $\pm$      13.8  &  23.3 $\pm$   3.0  &  22.6 $\pm$   0.8  & 1.74 $\pm$ 0.08  & 4.69 $\pm$ 1.02  &  117.3 $\pm$   16.2 \\
  G118.09+04.96  &  ...               &    1096.0 $\pm$      10.6  &   8.9 $\pm$   1.2  &  27.7 $\pm$   1.3  & 1.53 $\pm$ 0.08  &15.63 $\pm$ 4.30  &   18.2 $\pm$   13.0 \\
  G123.13-06.27  &  ...               &     140.4 $\pm$       1.7  &   3.7 $\pm$   0.3  &  24.6 $\pm$   0.6  & 1.34 $\pm$ 0.04  & 7.76 $\pm$ 1.09  &    5.5 $\pm$    3.4 \\
  G133.27+09.05  &  ...               &       ...                  &  35.5 $\pm$   6.1  &  15.4 $\pm$   0.6  & 1.69 $\pm$ 0.09  & 0.46 $\pm$ 0.10  &    0.2 $\pm$   13.1 \\
  G133.74+01.22  &  ...               &    1546.5 $\pm$      25.3  &  27.4 $\pm$   3.1  &  26.6 $\pm$   1.0  & 1.52 $\pm$ 0.06  &12.44 $\pm$ 2.77  &   68.9 $\pm$   25.9 \\
  G142.35+01.35  &  ...               &       3.1 $\pm$       2.5  &  36.4 $\pm$   5.1  &  17.7 $\pm$   0.6  & 1.67 $\pm$ 0.08  & 1.06 $\pm$ 0.20  &   25.6 $\pm$   13.1 \\
  G151.62-00.28  &  ...               &     311.3 $\pm$       4.2  &   9.3 $\pm$   0.6  &  23.1 $\pm$   0.4  & 1.22 $\pm$ 0.03  & 5.23 $\pm$ 0.59  &  -12.5 $\pm$    7.3 \\
  G160.26-18.62  &  ...               &      46.4 $\pm$       3.4  &  27.7 $\pm$   5.2  &  19.8 $\pm$   1.0  & 1.60 $\pm$ 0.10  & 2.12 $\pm$ 0.61  &   19.8 $\pm$   16.1 \\
  G160.60-12.05  &  ...               &     343.8 $\pm$       6.4  &   2.8 $\pm$   0.9  &  25.9 $\pm$   2.6  & 1.58 $\pm$ 0.20  &10.57 $\pm$ 6.25  &  -35.3 $\pm$    8.5 \\
  G173.56-01.76  &  ...               &     383.3 $\pm$       5.9  &   2.0 $\pm$   0.3  &  28.3 $\pm$   1.6  & 1.26 $\pm$ 0.10  &17.94 $\pm$ 6.07  &  -26.6 $\pm$    7.6 \\
  G173.62+02.79  &  ...               &     172.9 $\pm$       4.3  &  28.7 $\pm$   2.4  &  20.3 $\pm$   0.5  & 1.48 $\pm$ 0.04  & 2.44 $\pm$ 0.33  &  -26.6 $\pm$    8.6 \\
  G190.00+00.46  &  ...               &     238.6 $\pm$       4.7  &  24.1 $\pm$   2.5  &  21.4 $\pm$   0.6  & 1.46 $\pm$ 0.05  & 3.37 $\pm$ 0.58  &    5.1 $\pm$    9.5 \\
  G192.34-11.37  &  ...               &       ...                  &  11.0 $\pm$   2.0  &  18.7 $\pm$   0.8  & 1.79 $\pm$ 0.11  & 1.47 $\pm$ 0.39  &   12.6 $\pm$    8.7 \\
  G192.60-00.06  &  ...               &      77.9 $\pm$       3.1  &   8.2 $\pm$   0.9  &  22.1 $\pm$   0.6  & 1.72 $\pm$ 0.07  & 4.09 $\pm$ 0.70  &  -13.4 $\pm$    6.0 \\
  G201.62+01.63  &  ...               &     145.7 $\pm$       3.5  &   8.5 $\pm$   1.1  &  20.6 $\pm$   0.7  & 1.70 $\pm$ 0.09  & 2.70 $\pm$ 0.53  &   60.9 $\pm$    6.4 \\
  G203.24+02.08  &  ...               &     113.4 $\pm$       3.6  &  24.9 $\pm$   1.9  &  18.8 $\pm$   0.3  & 1.53 $\pm$ 0.04  & 1.56 $\pm$ 0.17  &  -26.4 $\pm$    8.1 \\
  G208.80-02.65  &  ...               &     100.5 $\pm$       6.7  &   7.3 $\pm$   2.3  &  20.1 $\pm$   1.6  & 1.59 $\pm$ 0.20  & 2.28 $\pm$ 1.07  &   15.5 $\pm$   19.2 \\
  G239.40-04.70  &  ...               &      63.0 $\pm$       3.7  &  10.2 $\pm$   1.7  &  19.0 $\pm$   0.7  & 1.64 $\pm$ 0.10  & 1.62 $\pm$ 0.38  &    9.2 $\pm$    7.7 \\
  G351.31+17.28  &  ...               &     107.1 $\pm$       3.1  &   7.7 $\pm$   1.3  &  26.8 $\pm$   1.4  & 1.56 $\pm$ 0.10  &12.95 $\pm$ 4.05  &    1.7 $\pm$    8.2 \\
  G353.05+16.90  &  ...               &      41.5 $\pm$       4.9  &  47.8 $\pm$   4.5  &  22.7 $\pm$   0.6  & 1.57 $\pm$ 0.04  & 4.75 $\pm$ 0.79  &   40.9 $\pm$   12.7 \\
  G353.97+15.79  &  ...               &      28.8 $\pm$       4.7  &  39.5 $\pm$   7.0  &  19.7 $\pm$   0.9  & 1.61 $\pm$ 0.09  & 2.02 $\pm$ 0.57  &  -25.4 $\pm$   17.1 \\
  G355.63+20.52  &  ...               &       ...                  &  15.7 $\pm$   2.8  &  17.6 $\pm$   0.7  & 1.76 $\pm$ 0.11  & 1.05 $\pm$ 0.26  &   52.5 $\pm$    8.1 \\
\noalign{\smallskip}
\hline\hline
\end{tabular}
\end{center}
\normalsize
\caption{\small SEDs Fit parameters. $^{(\rm PUL)}$: prior upper limit. $^{(\rm PLL)}$: prior lower limit.
}
\label{tab:seds_fit_parameters1}
\end{table*}

%% file: polametex_sedfitparams_tables_2_parabola.txt
%

\begin{table*}
\begin{center}
\begin{tabular}{lllllllcc}
\hline\hline
\noalign{\smallskip}
Source Name & $A_{\rm AME}$& $\sigma_{\rm AME}$ &   $\nu_{\rm AME}$ &   $W_{\rm AME}$  &$S_{\rm resid}^{28.4}$  &$ \frac{S_{\rm resid}^{28.4}}{S_{100 \mu \rm m}}$ & $f_{\rm max}^{\rm UCHII} [15GHz]$  & $\chi^2_{\rm red}$\\
\noalign{\smallskip}
\hline
            &  [Jy]              &                    &  [GHz]                 &  [log[GHz]]    &       [Jy]    &       &   \\
 \\
\noalign{\smallskip}
\hline
\noalign{\smallskip}
  G010.19-00.32  &  66.7 $\pm$ 26.0  &       2.6     &      23.5 $\pm$       7.9  &       0.5 $\pm$       0.7  &      62.8 $\pm$      14.2  &       0.8 $\pm$       0.2  &     3.073  &       0.4 \\
  G010.84-02.59  &  13.3 $\pm$  2.7  &       4.8     &      20.6 $\pm$       2.2  &       0.3 $\pm$       0.1  &       7.8 $\pm$       2.6  &       1.8 $\pm$       0.7  &     0.482  &       0.0 \\
  G011.11-00.12  &  40.6 $\pm$ 16.4  &       2.5     &      23.7 $\pm$       3.5  &       0.3 $\pm$       0.2  &      33.9 $\pm$       6.8  &       0.4 $\pm$       0.1  &    10.818  &       0.4 \\
  G012.80-00.19  &  44.0 $\pm$ 36.2  &       1.2     &      25.8 $\pm$       6.2  &       0.3 $\pm$       0.4  &      42.0 $\pm$       4.5  &       0.4 $\pm$       0.1  &     5.904  &       0.3 \\
  G015.06-00.69  &  57.7 $\pm$  7.2  &       8.0     &      27.0 $\pm$       6.0  &       0.6 $\pm$       0.3  &      57.5 $\pm$      17.0  &       0.6 $\pm$       0.2  &     2.275  &       0.2 \\
  G017.00+00.85  &  91.8 $\pm$ 15.4  &       6.0     &      30.3 $\pm$       9.3  &       1.0 $\pm$       0.5  &      91.6 $\pm$      31.7  &       3.6 $\pm$       1.4  &     0.044  &       0.3 \\
  G037.79-00.11  &  56.0 $\pm$  7.4  &       7.6     &      21.8 $\pm$       2.9  &       0.5 $\pm$       0.3  &      48.1 $\pm$      14.0  &       0.9 $\pm$       0.3  &     0.458  &       0.4 \\
  G040.52+02.53  &  12.5 $\pm$  1.0  &      12.9     &      24.9 $\pm$       5.6  &       0.8 $\pm$       0.2  &      12.3 $\pm$       4.1  &       2.7 $\pm$       1.0  &     0.831  &       0.0 \\
  G041.03-00.07  &  51.8 $\pm$  6.6  &       7.9     &      21.5 $\pm$       2.5  &       0.6 $\pm$       0.3  &      46.3 $\pm$      14.0  &       2.2 $\pm$       0.8  &     0.175  &       0.2 \\
  G043.20-00.10  &  45.7 $\pm$  5.5  &       8.3     &      21.7 $\pm$       2.5  &       0.5 $\pm$       0.2  &      40.4 $\pm$      12.0  &       1.5 $\pm$       0.5  &     1.437  &       0.4 \\
  G045.47+00.06  &  62.4 $\pm$  4.0  &      15.6     &      21.7 $\pm$       2.1  &       0.8 $\pm$       0.2  &      58.5 $\pm$      19.3  &       2.8 $\pm$       1.1  &     0.628  &       0.2 \\
  G049.14-00.60  &  76.2 $\pm$  3.3  &      22.9     &      17.2 $\pm$       2.4  &       1.0 $^{(\rm PUL)}$  &      67.1 $\pm$      24.6  &       1.3 $\pm$       0.5  &     1.047  &       0.2 \\
  G059.42-00.21  &  20.9 $\pm$  2.4  &       8.7     &      20.2 $\pm$       2.3  &       0.6 $\pm$       0.2  &      17.2 $\pm$       5.4  &       2.5 $\pm$       0.9  &     0.350  &       0.7 \\
  G061.47+00.11  &   5.3 $\pm$  1.3  &       4.1     &      25.3 $\pm$       2.8  &       0.3 $\pm$       0.1  &       5.0 $\pm$       1.1  &       0.8 $\pm$       0.2  &    10.855  &       0.1 \\
  G062.98+00.05  &   7.4 $\pm$  1.2  &       6.1     &      23.1 $\pm$       1.8  &       0.3 $\pm$       0.1  &       6.1 $\pm$       1.6  &       1.5 $\pm$       0.5  &     1.087  &       0.2 \\
  G070.14+01.61  &  11.0 $\pm$  3.6  &       3.1     &      23.0 $\pm$      29.3  &       1.0 $^{(\rm PUL)}$  &      10.8 $\pm$       3.1  &       3.8 $\pm$       1.2  &     2.820  &       1.7 \\
  G071.59+02.85  &  12.7 $\pm$  2.7  &       4.8     &      24.5 $\pm$       7.7  &       0.6 $\pm$       0.4  &      12.4 $\pm$       3.4  &       3.0 $\pm$       1.0  &     0.007  &       0.3 \\
  G075.81+00.39  &  19.1 $\pm$  3.2  &       5.9     &      19.9 $\pm$       5.3  &       0.7 $\pm$       0.4  &      16.8 $\pm$       5.1  &       2.6 $\pm$       0.9  &     0.575  &       0.1 \\
  G076.38-00.62  &  15.5 $\pm$  4.0  &       3.9     &      24.5 $\pm$      14.3  &       0.9 $\pm$       0.9  &      15.2 $\pm$       4.5  &       2.3 $\pm$       0.8  &     1.195  &       0.1 \\
  G078.57+01.00  & ...  &  ...     &      ...  &        ...  &       ...  &       ...  &  ... &       0.7 \\
  G081.59+00.01  &  94.0 $\pm$  5.2  &      17.9     &      26.3 $\pm$       3.6  &       1.0 $^{(\rm PUL)}$  &      93.7 $\pm$      34.3  &       2.4 $\pm$       1.0  &     0.000  &       0.2 \\
  G084.68-00.58  &  34.6 $\pm$  1.8  &      18.8     &      22.5 $\pm$       3.8  &       1.0 $^{(\rm PUL)}$   &      33.7 $\pm$      12.1  &       3.4 $\pm$       1.4  &     0.000  &       0.1 \\
  G085.00+04.20  &  16.7 $\pm$  0.8  &      21.1     &      19.5 $\pm$       2.9  &       1.0 $^{(\rm PUL)}$  &      15.5 $\pm$       5.6  &       4.8 $\pm$       1.9  &     0.010  &       0.1 \\
  G093.02+02.76  &  13.4 $\pm$  0.6  &      21.0     &      22.6 $\pm$       2.1  &       0.7 $\pm$       0.1  &      12.7 $\pm$       4.2  &       2.3 $\pm$       0.8  &     0.724  &       0.1 \\
  G094.47-01.53  &   2.2 $\pm$  0.5  &       4.1     &      32.6 $\pm$       3.3  &       0.3 $\pm$       0.1  &       1.9 $\pm$       0.5  &       1.4 $\pm$       0.4  &    31.097  &       0.1 \\
  G098.00+01.47  &  10.8 $\pm$  0.6  &      17.2     &      27.1 $\pm$       4.5  &       1.0 $^{(\rm PUL)}$  &      10.8 $\pm$       4.0  &       8.2 $\pm$       3.4  &     0.314  &       0.2 \\
  G099.60+03.70  &   3.0 $\pm$  1.0  &       3.0     &      35.9 $\pm$       4.7  &       0.2 $\pm$       0.2  &       1.8 $\pm$       0.6  &       0.7 $\pm$       0.2  &   133.803  &       0.3 \\
  G102.88-00.69  &   7.5 $\pm$  0.7  &      10.9     &      26.7 $\pm$       8.1  &       0.8 $\pm$       0.3  &       7.5 $\pm$       2.5  &       3.1 $\pm$       1.1  &     0.018  &       0.2 \\
  G107.20+05.20  &  18.5 $\pm$  0.7  &      27.8     &      25.6 $\pm$       2.7  &       0.7 $\pm$       0.1  &      18.3 $\pm$       6.1  &       2.6 $\pm$       1.0  &     1.008  &       0.4 \\
  G110.25+02.58  &   7.6 $\pm$  2.8  &       2.7     &      38.7 $\pm$      31.0  &       0.6 $\pm$       0.6  &       6.7 $\pm$       2.1  &       0.8 $\pm$       0.3  &    11.067  &       0.6 \\
  G111.54+00.81  &  21.4 $\pm$  2.0  &      10.8     &      24.2 $\pm$       4.3  &       0.8 $\pm$       0.3  &      20.9 $\pm$       6.7  &       2.1 $\pm$       0.8  &     2.730  &       0.3 \\
  G118.09+04.96  &  12.9 $\pm$  0.9  &      14.2     &      30.2 $\pm$       9.8  &       1.0 $^{(\rm PUL)}$  &      12.9 $\pm$       4.8  &       1.5 $\pm$       0.6  &     0.003  &       0.3 \\
  G123.13-06.27  &   4.6 $\pm$  0.2  &      25.2     &      26.8 $\pm$       2.6  &       0.7 $\pm$       0.1  &       4.6 $\pm$       1.5  &       3.7 $\pm$       1.4  &     0.459  &       0.1 \\
  G133.27+09.05  &   4.9 $\pm$  0.4  &      11.1     &      21.9 $\pm$       6.2  &       0.7 $\pm$       0.3  &       4.6 $\pm$       1.5  &       8.4 $\pm$       3.1  &     0.055  &       1.3 \\
  G133.74+01.22  &  54.4 $\pm$  2.2  &      24.8     &      24.1 $\pm$       3.3  &       1.0 $^{(\rm PUL)}$  &      53.7 $\pm$      19.6  &       2.3 $\pm$       0.9  &     0.553  &       0.3 \\
  G142.35+01.35  &   6.7 $\pm$  0.8  &       8.4     &      20.6 $\pm$       2.7  &       0.6 $\pm$       0.2  &       5.7 $\pm$       1.8  &       2.4 $\pm$       0.8  &     0.200  &       0.6 \\
  G151.62-00.28  &   5.6 $\pm$  0.5  &      11.4     &      29.9 $\pm$       3.9  &       0.5 $\pm$       0.1  &       5.6 $\pm$       1.7  &       2.7 $\pm$       0.9  &     1.301  &       0.1 \\
  G160.26-18.62  &  14.7 $\pm$  0.8  &      19.2     &      25.6 $\pm$       1.5  &       0.5 $\pm$       0.1  &      14.4 $\pm$       4.2  &       3.2 $\pm$       1.1  &     0.064  &       1.3 \\
  G160.60-12.05  &   9.3 $\pm$  0.7  &      12.6     &      49.1 $\pm$      38.5  &       0.9 $\pm$       0.5  &       7.7 $\pm$       3.7  &       4.1 $\pm$       2.1  &     0.000  &       0.8 \\
  G173.56-01.76  &   2.7 $\pm$  0.6  &       4.4     &      26.4 $\pm$       7.0  &       0.6 $\pm$       0.4  &       2.7 $\pm$       0.7  &       2.1 $\pm$       0.7  &     0.055  &       0.2 \\
  G173.62+02.79  &   9.6 $\pm$  0.6  &      15.5     &      24.7 $\pm$       1.1  &       0.4 $\pm$       0.1  &       9.1 $\pm$       2.5  &       2.6 $\pm$       0.8  &     1.631  &       0.3 \\
  G190.00+00.46  &  14.3 $\pm$  0.5  &      29.3     &      22.2 $\pm$       1.4  &       0.7 $\pm$       0.1  &      13.4 $\pm$       4.4  &       3.1 $\pm$       1.1  &     0.848  &       0.4 \\
  G192.34-11.37  &   9.1 $\pm$  0.7  &      12.5     &      24.9 $\pm$       4.0  &       0.7 $\pm$       0.2  &       9.0 $\pm$       2.9  &       9.1 $\pm$       3.3  &     0.000  &       0.8 \\
  G192.60-00.06  &   3.8 $\pm$  0.5  &       7.9     &      20.8 $\pm$       1.4  &       0.3 $\pm$       0.1  &       2.4 $\pm$       0.8  &       1.0 $\pm$       0.4  &     0.719  &       0.1 \\
  G201.62+01.63  &  10.0 $\pm$  0.4  &      27.3     &      21.6 $\pm$       2.8  &       0.9 $\pm$       0.2  &       9.5 $\pm$       3.3  &       6.5 $\pm$       2.5  &     0.056  &       0.1 \\
  G203.24+02.08  &   7.1 $\pm$  0.4  &      15.8     &      26.9 $\pm$       5.0  &       0.7 $\pm$       0.2  &       7.1 $\pm$       2.3  &       3.6 $\pm$       1.3  &     0.022  &       0.2 \\
  G208.80-02.65  &   3.4 $\pm$  1.8  &       1.9     &      17.2 $\pm$      23.4  &       0.9 $\pm$       1.7  &       3.0 $\pm$       0.7  &       4.3 $\pm$       1.2  &     0.025  &       0.2 \\
  G239.40-04.70  &   6.9 $\pm$  0.4  &      16.5     &      21.2 $\pm$       2.0  &       0.7 $\pm$       0.1  &       6.3 $\pm$       2.1  &       6.9 $\pm$       2.5  &     0.000  &       0.3 \\
  G351.31+17.28  &  12.0 $\pm$  0.4  &      32.9     &      20.0 $\pm$       1.3  &       0.7 $\pm$       0.1  &      10.6 $\pm$       3.6  &       1.5 $\pm$       0.6  &     0.036  &       0.1 \\
  G353.05+16.90  &  20.0 $\pm$  0.7  &      27.3     &      28.8 $\pm$       1.2  &       0.5 $\pm$       0.1  &      20.0 $\pm$       6.0  &       1.3 $\pm$       0.5  &     0.997  &       0.4 \\
  G353.97+15.79  &  13.1 $\pm$  1.2  &      10.6     &      23.7 $\pm$       1.2  &       0.3 $\pm$       0.1  &      11.4 $\pm$       3.1  &       3.5 $\pm$       1.1  &     0.190  &       0.3 \\
  G355.63+20.52  &   8.1 $\pm$  0.5  &      17.0     &      23.5 $\pm$       1.2  &       0.5 $\pm$       0.1  &       7.5 $\pm$       2.2  &       7.7 $\pm$       2.6  &     0.000  &       0.7 \\
\noalign{\smallskip}
\hline\hline
\end{tabular}
\end{center}
\normalsize
\caption{\small SEDs Fit parameters. $^{(\rm PUL)}$: prior upper limit. 
}
\label{tab:seds_fit_parameters2}
\end{table*}

%% file: polametex_table_sources_spearman_AME_AME.txt
\begin{table*}
\begin{center}
\begin{tabular}{llrrr}
\hline\hline
\noalign{\smallskip}
Variable 1 & Variable 2 & SRCC & SRCC & SRCC \\
 &                & selected sample & AME significant & AME semi-significant \\
\noalign{\smallskip}
\hline
\noalign{\smallskip}
                   $A_{\rm AME}[\rm Jy]$  &                $\nu_{\rm AME}[\rm GHz]$  &  $-$0.20$\pm$ 0.11  &   0.10$\pm$ 0.22  &  $-$0.52$\pm$ 0.13 \\
                   $A_{\rm AME}[\rm Jy]$  &                           $W_{\rm AME}$  &   0.27$\pm$ 0.12  &   0.39$\pm$ 0.20  &   0.40$\pm$ 0.14 \\
                   $A_{\rm AME}[\rm Jy]$  &        $A_{\rm AME}/\tau_{250}[\rm Jy]$  &  $-$0.11$\pm$ 0.08  &   0.19$\pm$ 0.09  &  $-$0.27$\pm$ 0.10 \\
                   $A_{\rm AME}[\rm Jy]$  &                         $\Re_{\rm AME}$  &   0.88$\pm$ 0.15  &   0.77$\pm$ 0.27  &   0.91$\pm$ 0.22 \\
                $\nu_{\rm AME}[\rm GHz]$  &                           $W_{\rm AME}$  &  $-$0.06$\pm$ 0.14  &   0.27$\pm$ 0.26  &  $-$0.12$\pm$ 0.17 \\
                $\nu_{\rm AME}[\rm GHz]$  &        $A_{\rm AME}/\tau_{250}[\rm Jy]$  &   0.18$\pm$ 0.13  &   0.02$\pm$ 0.22  &   0.31$\pm$ 0.16 \\
                $\nu_{\rm AME}[\rm GHz]$  &                         $\Re_{\rm AME}$  &  $-$0.04$\pm$ 0.15  &   0.34$\pm$ 0.25  &  $-$0.33$\pm$ 0.20 \\
                           $W_{\rm AME}$  &        $A_{\rm AME}/\tau_{250}[\rm Jy]$  &   0.66$\pm$ 0.12  &   0.64$\pm$ 0.18  &   0.57$\pm$ 0.15 \\
                           $W_{\rm AME}$  &                         $\Re_{\rm AME}$  &   0.60$\pm$ 0.15  &   0.75$\pm$ 0.27  &   0.67$\pm$ 0.20 \\
        $A_{\rm AME}/\tau_{250}[\rm Jy]$  &                         $\Re_{\rm AME}$  &   0.19$\pm$ 0.15  &   0.43$\pm$ 0.26  &  $-$0.01$\pm$ 0.19 \\
\noalign{\smallskip}
\hline\hline
\end{tabular}
\end{center}
\normalsize
\caption{Spearman rank correlation coefficients (SRCCs) between AME and AME parameters.}
\label{tab:srcc_ame_ame}
\end{table*}

%% file: polametex_table_sources_spearman_AME_TD.txt
\begin{table*}
\begin{center}
\begin{tabular}{llrrr}
\hline\hline
\noalign{\smallskip}
Variable 1 & Variable 2 & SRCC & SRCC & SRCC \\
 &                & selected sample & AME significant & AME semi-significant \\
\noalign{\smallskip}
\hline
\noalign{\smallskip}
                   $A_{\rm AME}[\rm Jy]$  &                    T$_{\rm td}$ or $G_{0}$  &  $-$0.16$\pm$ 0.06  &   0.26$\pm$ 0.10  &  $-$0.33$\pm$ 0.07 \\
                   $A_{\rm AME}[\rm Jy]$  &                        $\beta_{\rm td}$  &   0.49$\pm$ 0.08  &   0.21$\pm$ 0.14  &   0.56$\pm$ 0.09 \\
                   $A_{\rm AME}[\rm Jy]$  &                            $\tau_{250}$  &   0.81$\pm$ 0.04  &   0.53$\pm$ 0.07  &   0.85$\pm$ 0.04 \\
                   $A_{\rm AME}[\rm Jy]$  &               $S_{\rm TD,peak}[\rm Jy]$  &   0.88$\pm$ 0.05  &   0.82$\pm$ 0.07  &   0.91$\pm$ 0.04 \\
                   $A_{\rm AME}[\rm Jy]$  &            $\nu_{\rm TD,peak}[\rm GHz]$  &   0.04$\pm$ 0.11  &   0.34$\pm$ 0.17  &  $-$0.07$\pm$ 0.15 \\
                   $A_{\rm AME}[\rm Jy]$  &                $\Re_{\rm td} x 10^{-4}$  &   0.88$\pm$ 0.05  &   0.85$\pm$ 0.08  &   0.90$\pm$ 0.05 \\
                $\nu_{\rm AME}[\rm GHz]$  &                    T$_{\rm td}$ or $G_{0}$  &   0.40$\pm$ 0.12  &   0.21$\pm$ 0.22  &   0.60$\pm$ 0.15 \\
                $\nu_{\rm AME}[\rm GHz]$  &                        $\beta_{\rm td}$  &  $-$0.35$\pm$ 0.13  &  $-$0.35$\pm$ 0.26  &  $-$0.44$\pm$ 0.16 \\
                $\nu_{\rm AME}[\rm GHz]$  &                            $\tau_{250}$  &  $-$0.25$\pm$ 0.12  &   0.01$\pm$ 0.23  &  $-$0.57$\pm$ 0.14 \\
                $\nu_{\rm AME}[\rm GHz]$  &               $S_{\rm TD,peak}[\rm Jy]$  &  $-$0.14$\pm$ 0.13  &   0.11$\pm$ 0.23  &  $-$0.47$\pm$ 0.14 \\
                $\nu_{\rm AME}[\rm GHz]$  &            $\nu_{\rm TD,peak}[\rm GHz]$  &   0.34$\pm$ 0.13  &   0.19$\pm$ 0.23  &   0.42$\pm$ 0.17 \\
                $\nu_{\rm AME}[\rm GHz]$  &                $\Re_{\rm td} x 10^{-4}$  &  $-$0.11$\pm$ 0.13  &   0.11$\pm$ 0.21  &  $-$0.42$\pm$ 0.14 \\
                           $W_{\rm AME}$  &                    T$_{\rm td}$ or $G_{0}$  &   0.23$\pm$ 0.12  &   0.57$\pm$ 0.18  &   0.07$\pm$ 0.15 \\
                           $W_{\rm AME}$  &                        $\beta_{\rm td}$  &  $-$0.24$\pm$ 0.13  &  $-$0.23$\pm$ 0.25  &  $-$0.21$\pm$ 0.16 \\
                           $W_{\rm AME}$  &                            $\tau_{250}$  &  $-$0.17$\pm$ 0.12  &  $-$0.18$\pm$ 0.23  &  $-$0.00$\pm$ 0.16 \\
                           $W_{\rm AME}$  &               $S_{\rm TD,peak}[\rm Jy]$  &  $-$0.05$\pm$ 0.12  &   0.22$\pm$ 0.22  &   0.08$\pm$ 0.15 \\
                           $W_{\rm AME}$  &            $\nu_{\rm TD,peak}[\rm GHz]$  &   0.18$\pm$ 0.13  &   0.68$\pm$ 0.19  &   0.00$\pm$ 0.17 \\
                           $W_{\rm AME}$  &                $\Re_{\rm td} x 10^{-4}$  &  $-$0.01$\pm$ 0.12  &   0.28$\pm$ 0.22  &   0.12$\pm$ 0.15 \\
        $A_{\rm AME}/\tau_{250}[\rm Jy]$  &                    T$_{\rm td}$ or $G_{0}$  &   0.68$\pm$ 0.08  &   0.87$\pm$ 0.07  &   0.62$\pm$ 0.11 \\
        $A_{\rm AME}/\tau_{250}[\rm Jy]$  &                        $\beta_{\rm td}$  &  $-$0.52$\pm$ 0.09  &  $-$0.20$\pm$ 0.20  &  $-$0.64$\pm$ 0.10 \\
        $A_{\rm AME}/\tau_{250}[\rm Jy]$  &                            $\tau_{250}$  &  $-$0.64$\pm$ 0.08  &  $-$0.64$\pm$ 0.09  &  $-$0.68$\pm$ 0.10 \\
        $A_{\rm AME}/\tau_{250}[\rm Jy]$  &               $S_{\rm TD,peak}[\rm Jy]$  &  $-$0.41$\pm$ 0.08  &  $-$0.13$\pm$ 0.13  &  $-$0.54$\pm$ 0.09 \\
        $A_{\rm AME}/\tau_{250}[\rm Jy]$  &            $\nu_{\rm TD,peak}[\rm GHz]$  &   0.54$\pm$ 0.11  &   0.88$\pm$ 0.12  &   0.40$\pm$ 0.15 \\
        $A_{\rm AME}/\tau_{250}[\rm Jy]$  &                $\Re_{\rm td} x 10^{-4}$  &  $-$0.34$\pm$ 0.09  &  $-$0.01$\pm$ 0.15  &  $-$0.48$\pm$ 0.10 \\
                         $\Re_{\rm AME}$  &                    T$_{\rm td}$ or $G_{0}$  &   0.04$\pm$ 0.16  &   0.45$\pm$ 0.27  &  $-$0.20$\pm$ 0.20 \\
                         $\Re_{\rm AME}$  &                        $\beta_{\rm td}$  &   0.31$\pm$ 0.14  &   0.05$\pm$ 0.26  &   0.36$\pm$ 0.18 \\
                         $\Re_{\rm AME}$  &                            $\tau_{250}$  &   0.54$\pm$ 0.14  &   0.19$\pm$ 0.23  &   0.65$\pm$ 0.17 \\
                         $\Re_{\rm AME}$  &               $S_{\rm TD,peak}[\rm Jy]$  &   0.68$\pm$ 0.14  &   0.60$\pm$ 0.24  &   0.72$\pm$ 0.16 \\
                         $\Re_{\rm AME}$  &            $\nu_{\rm TD,peak}[\rm GHz]$  &   0.20$\pm$ 0.16  &   0.60$\pm$ 0.28  &  $-$0.01$\pm$ 0.20 \\
                         $\Re_{\rm AME}$  &                $\Re_{\rm td} x 10^{-4}$  &   0.70$\pm$ 0.14  &   0.66$\pm$ 0.23  &   0.73$\pm$ 0.17 \\
\noalign{\smallskip}
\hline\hline
\end{tabular}
\end{center}
\normalsize
\caption{Spearman rank correlation coefficients (SRCCs) between AME and Thermal Dust parameters.}
\label{tab:srcc_ame_td}
\end{table*}

%% file: polametex_table_sources_spearman_AME_FF.txt
\begin{table*}
\begin{center}
\begin{tabular}{llrrr}
\hline\hline
\noalign{\smallskip}
Variable 1 & Variable 2 & SRCC & SRCC & SRCC \\
 &                & selected sample & AME significant & free-free semi-significant \\
\noalign{\smallskip}
\hline
\noalign{\smallskip}
                   $A_{\rm AME}[\rm Jy]$  &                       EM [cm$^{-6}/$pc]  &   0.59$\pm$ 0.04  &   0.65$\pm$ 0.11  &   0.55$\pm$ 0.03 \\
                $\nu_{\rm AME}[\rm GHz]$  &                       EM [cm$^{-6}/$pc]  &   0.06$\pm$ 0.11  &   0.27$\pm$ 0.20  &  $-$0.24$\pm$ 0.14 \\
                           $W_{\rm AME}$  &                       EM [cm$^{-6}/$pc]  &   0.12$\pm$ 0.10  &   0.84$\pm$ 0.13  &  $-$0.02$\pm$ 0.14 \\
        $A_{\rm AME}/\tau_{250}[\rm Jy]$  &                       EM [cm$^{-6}/$pc]  &   0.01$\pm$ 0.07  &   0.46$\pm$ 0.14  &  $-$0.21$\pm$ 0.07 \\
                         $\Re_{\rm AME}$  &                       EM [cm$^{-6}/$pc]  &   0.59$\pm$ 0.14  &   0.90$\pm$ 0.24  &   0.42$\pm$ 0.17 \\
\noalign{\smallskip}
\hline\hline
\end{tabular}
\end{center}
\normalsize
\caption{Spearman rank correlation coefficients (SRCCs) between AME and free-free parameters.}
\label{tab:srcc_ame_ff}
\end{table*}

%% file: polametex_table_sources_spearman_TD_FF.txt
\begin{table*}
\begin{center}
\begin{tabular}{llrrr}
\hline\hline
\noalign{\smallskip}
Variable 1 & Variable 2 & SRCC & SRCC & SRCC \\
 &                & selected sample & AME significant & free-free semi-significant \\
\noalign{\smallskip}
\hline
\noalign{\smallskip}
                    T$_{\rm td}$ or $G_{0}$  &                       EM [cm$^{-6}/$pc]  &   0.30$\pm$ 0.06  &   0.49$\pm$ 0.13  &   0.09$\pm$ 0.07 \\
                        $\beta_{\rm td}$  &                       EM [cm$^{-6}/$pc]  &   0.28$\pm$ 0.07  &  $-$0.17$\pm$ 0.16  &   0.41$\pm$ 0.07 \\
                            $\tau_{250}$  &                       EM [cm$^{-6}/$pc]  &   0.43$\pm$ 0.03  &   0.10$\pm$ 0.09  &   0.51$\pm$ 0.03 \\
               $S_{\rm TD,peak}[\rm Jy]$  &                       EM [cm$^{-6}/$pc]  &   0.64$\pm$ 0.03  &   0.50$\pm$ 0.08  &   0.64$\pm$ 0.04 \\
            $\nu_{\rm TD,peak}[\rm GHz]$  &                       EM [cm$^{-6}/$pc]  &   0.45$\pm$ 0.10  &   0.62$\pm$ 0.17  &   0.29$\pm$ 0.15 \\
                $\Re_{\rm td} x 10^{-4}$  &                       EM [cm$^{-6}/$pc]  &   0.66$\pm$ 0.04  &   0.56$\pm$ 0.09  &   0.65$\pm$ 0.04 \\
\noalign{\smallskip}
\hline\hline
\end{tabular}
\end{center}
\normalsize
\caption{Spearman rank correlation coefficients (SRCCs) between thermal dust and free-free parameters.}
\label{tab:srcc_td_ff}
\end{table*}

%% file: polametex_table_sources_spearman_TD_TD.txt
\begin{table*}
\begin{center}
\begin{tabular}{llrrr}
\hline\hline
\noalign{\smallskip}
Variable 1 & Variable 2 & SRCC & SRCC & SRCC \\
 &                & selected sample & AME significant & AME semi-significant \\
\noalign{\smallskip}
\hline
\noalign{\smallskip}
                    T$_{\rm td}$ or $G_{0}$  &                        $\beta_{\rm td}$  &  $-$0.47$\pm$ 0.09  &  $-$0.44$\pm$ 0.18  &  $-$0.47$\pm$ 0.10 \\
                    T$_{\rm td}$ or $G_{0}$  &                            $\tau_{250}$  &  $-$0.52$\pm$ 0.05  &  $-$0.57$\pm$ 0.10  &  $-$0.58$\pm$ 0.07 \\
                    T$_{\rm td}$ or $G_{0}$  &               $S_{\rm TD,peak}[\rm Jy]$  &  $-$0.17$\pm$ 0.07  &   0.11$\pm$ 0.14  &  $-$0.32$\pm$ 0.08 \\
                    T$_{\rm td}$ or $G_{0}$  &            $\nu_{\rm TD,peak}[\rm GHz]$  &   0.92$\pm$ 0.08  &   0.97$\pm$ 0.12  &   0.85$\pm$ 0.13 \\
                    T$_{\rm td}$ or $G_{0}$  &                $\Re_{\rm td} x 10^{-4}$  &  $-$0.07$\pm$ 0.07  &   0.22$\pm$ 0.14  &  $-$0.25$\pm$ 0.09 \\
                        $\beta_{\rm td}$  &                            $\tau_{250}$  &   0.60$\pm$ 0.08  &   0.34$\pm$ 0.18  &   0.69$\pm$ 0.09 \\
                        $\beta_{\rm td}$  &               $S_{\rm TD,peak}[\rm Jy]$  &   0.54$\pm$ 0.08  &   0.06$\pm$ 0.17  &   0.68$\pm$ 0.09 \\
                        $\beta_{\rm td}$  &            $\nu_{\rm TD,peak}[\rm GHz]$  &  $-$0.13$\pm$ 0.12  &  $-$0.31$\pm$ 0.22  &   0.01$\pm$ 0.16 \\
                        $\beta_{\rm td}$  &                $\Re_{\rm td} x 10^{-4}$  &   0.50$\pm$ 0.08  &   0.03$\pm$ 0.17  &   0.65$\pm$ 0.09 \\
                            $\tau_{250}$  &               $S_{\rm TD,peak}[\rm Jy]$  &   0.90$\pm$ 0.04  &   0.69$\pm$ 0.10  &   0.92$\pm$ 0.04 \\
                            $\tau_{250}$  &            $\nu_{\rm TD,peak}[\rm GHz]$  &  $-$0.31$\pm$ 0.11  &  $-$0.51$\pm$ 0.16  &  $-$0.29$\pm$ 0.14 \\
                            $\tau_{250}$  &                $\Re_{\rm td} x 10^{-4}$  &   0.85$\pm$ 0.05  &   0.61$\pm$ 0.13  &   0.88$\pm$ 0.05 \\
               $S_{\rm TD,peak}[\rm Jy]$  &            $\nu_{\rm TD,peak}[\rm GHz]$  &   0.06$\pm$ 0.12  &   0.18$\pm$ 0.19  &  $-$0.01$\pm$ 0.15 \\
               $S_{\rm TD,peak}[\rm Jy]$  &                $\Re_{\rm td} x 10^{-4}$  &   0.99$\pm$ 0.05  &   0.99$\pm$ 0.11  &   0.99$\pm$ 0.05 \\
            $\nu_{\rm TD,peak}[\rm GHz]$  &                $\Re_{\rm td} x 10^{-4}$  &   0.17$\pm$ 0.12  &   0.30$\pm$ 0.19  &   0.06$\pm$ 0.15 \\
\noalign{\smallskip}
\hline\hline
\end{tabular}
\end{center}
\normalsize
\caption{Spearman rank correlation coefficients (SRCCs) between the thermal dust parameters.}
\label{tab:srcc_td_td}
\end{table*}